\documentclass[final]{article}
\usepackage[nonatbib]{neurips_2024}
\usepackage{bm} % fixes boldsymbol
\usepackage{amsmath} % Comment out for acmart
\usepackage{amssymb} % Comment out for acmart
\usepackage{amsthm} % Comment out for acmart
\usepackage[boxed]{algorithm}
\usepackage[noend]{algpseudocode}
\usepackage{braket}

\newcommand{\labs}{\left|}
\newcommand{\rabs}{\right|}
\newcommand{\lnorm}{\left\|}
\newcommand{\rnorm}{\right\|}

\newcommand{\lbrac}{\left[}
\newcommand{\rbrac}{\right]}
\newcommand{\lpar}{\left(}
\newcommand{\rpar}{\right)}

\algnewcommand\algorithmicinput{\textbf{Input:}}
\algnewcommand\Input{\item[\algorithmicinput]}
\algnewcommand\algorithmicoutput{\textbf{Output:}}
\algnewcommand\Output{\item[\algorithmicoutput]}
\algnewcommand\algorithmicalgorithm{\textbf{Algorithm:}}
\algnewcommand\Algorithm{\item[\algorithmicalgorithm]}

\newtheorem{theorem}{Theorem}[section]
\newtheorem{lemma}{Lemma}[section]
\newtheorem{proposition}{Proposition}[section]
\newtheorem{corollary}{Corollary}[section]

\newtheorem{fact}{Fact}[section]
\newtheorem{remark}{Remark}[section]
\newtheorem{definition}{Definition}[section]
\newtheorem{problem}{Problem}[section]

\DeclareMathOperator{\tr}{tr}
\DeclareMathOperator{\trace}{trace}
\DeclareMathOperator{\rank}{rank}

\DeclareMathOperator{\im}{im}
\DeclareMathOperator{\diag}{diag}

\DeclareMathOperator{\poly}{poly}
\DeclareMathOperator{\polylog}{polylog}

\DeclareMathOperator{\gap}{gap}
\DeclareMathOperator{\sgn}{sgn}

\DeclareMathOperator{\round}{round}
\DeclareMathOperator{\grid}{grid}
\DeclareMathOperator{\length}{length}

\newcommand\vece{\boldsymbol{\mathrm{e}}}

\newcommand\vecr{\boldsymbol{\mathrm{r}}}

\newcommand\vecu{\boldsymbol{\mathrm{u}}}
\newcommand\vecv{\boldsymbol{\mathrm{v}}}

\newcommand\vecx{\boldsymbol{\mathrm{x}}}
\newcommand\vecy{\boldsymbol{\mathrm{y}}}
\newcommand\vecz{\boldsymbol{\mathrm{z}}}

\newcommand\matA{\boldsymbol{\mathrm{A}}}
\newcommand\matB{\boldsymbol{\mathrm{B}}}
\newcommand\matC{\boldsymbol{\mathrm{C}}}
\newcommand\matD{\boldsymbol{\mathrm{D}}}
\newcommand\matE{\boldsymbol{\mathrm{E}}}

\newcommand\matG{\boldsymbol{\mathrm{G}}}
\newcommand\matH{\boldsymbol{\mathrm{H}}}
\newcommand\matI{\boldsymbol{\mathrm{I}}}

\newcommand\matK{\boldsymbol{\mathrm{K}}}
\newcommand\matL{\boldsymbol{\mathrm{L}}}
\newcommand\matM{\boldsymbol{\mathrm{M}}}

\newcommand\matP{\boldsymbol{\mathrm{P}}}
\newcommand\matQ{\boldsymbol{\mathrm{Q}}}
\newcommand\matR{\boldsymbol{\mathrm{R}}}
\newcommand\matS{\boldsymbol{\mathrm{S}}}
\newcommand\matT{\boldsymbol{\mathrm{T}}}
\newcommand\matU{\boldsymbol{\mathrm{U}}}
\newcommand\matV{\boldsymbol{\mathrm{V}}}
\newcommand\matW{\boldsymbol{\mathrm{W}}}
\newcommand\matX{\boldsymbol{\mathrm{X}}}
\newcommand\matY{\boldsymbol{\mathrm{Y}}}
\newcommand\matZ{\boldsymbol{\mathrm{Z}}}

\newcommand\matDelta{\boldsymbol{\mathrm{\Delta}}}

\newcommand\matLambda{\boldsymbol{\mathrm{\Lambda}}}

\newcommand\matPi{\boldsymbol{\mathrm{\Pi}}}
\newcommand\matSigma{\boldsymbol{\mathrm{\Sigma}}}
\newcommand\matPhi{\boldsymbol{\mathrm{\Phi}}}
\newcommand\matPsi{\boldsymbol{\mathrm{\Psi}}}

\newcommand\matAtilde{\widetilde{\boldsymbol{\mathrm{A}}}}

\newcommand\matCtilde{\widetilde{\boldsymbol{\mathrm{C}}}}

\newcommand\matGtilde{\widetilde{\boldsymbol{\mathrm{G}}}}
\newcommand\matHtilde{\widetilde{\boldsymbol{\mathrm{H}}}}

\newcommand\matKtilde{\widetilde{\boldsymbol{\mathrm{K}}}}
\newcommand\matLtilde{\widetilde{\boldsymbol{\mathrm{L}}}}
\newcommand\matMtilde{\widetilde{\boldsymbol{\mathrm{M}}}}

\newcommand\matPtilde{\widetilde{\boldsymbol{\mathrm{P}}}}
\newcommand\matQtilde{\widetilde{\boldsymbol{\mathrm{Q}}}}
\newcommand\matRtilde{\widetilde{\boldsymbol{\mathrm{R}}}}
\newcommand\matStilde{\widetilde{\boldsymbol{\mathrm{S}}}}

\newcommand\matUtilde{\widetilde{\boldsymbol{\mathrm{U}}}}
\newcommand\matVtilde{\widetilde{\boldsymbol{\mathrm{V}}}}

\newcommand\matXtilde{\widetilde{\boldsymbol{\mathrm{X}}}}

\newcommand\matZtilde{\widetilde{\boldsymbol{\mathrm{Z}}}}

\newcommand\matPitilde{\widetilde{\boldsymbol{\mathrm{\Pi}}}}

\newcommand\matUbar{\bar{\boldsymbol{\mathrm{U}}}}

\newcommand\matAhat{\widehat{\boldsymbol{A}}}

\newcommand\matChat{\widehat{\boldsymbol{C}}}

\newcommand\matQhat{\widehat{\boldsymbol{Q}}}

\newcommand{\fl}{\boldsymbol{\mathsf{fl}}}
\newcommand{\umach}{\textbf{\textup{u}}}

\usepackage{bm} % fixes boldsymbol
\usepackage{amsmath} % Comment out for acmart
\usepackage{amssymb} % Comment out for acmart
\usepackage{amsthm} % Comment out for acmart
\usepackage[boxed]{algorithm}
\usepackage[noend]{algpseudocode}

\DeclareMathOperator{\countsmaller}{count}

\newcommand{\FLINF}{\mathsf{INF}}
\newcommand{\MM}{\mathsf{MM}}
\newcommand{\MMX}{\mathsf{MMX}}
\newcommand{\LUR}{\mathsf{LUR}}
\newcommand{\INV}{\mathsf{INV}}
\newcommand{\CHOLESKY}{\mathsf{CHOLESKY}}
\newcommand{\EIG}{\mathsf{EIG}}
\newcommand{\DEFLATE}{\mathsf{DEFLATE}}
\newcommand{\OP}{\mathsf{OP}}

\newcommand{\COUNT}{\mathsf{COUNT}}
\newcommand{\QR}{\mathsf{QR}}
\newcommand{\HERM}{\mathsf{HERM}}
\newcommand{\SUB}{\mathsf{SUB}}

\newcommand{\REDUCE}{\mathsf{REDUCE}}
\newcommand{\SAMPLER}{\mathsf{N}}
\newcommand{\REGULARIZE}{\mathsf{REGULARIZE}}
\newcommand{\COND}{\mathsf{COND}}
\newcommand{\GAP}{\mathsf{GAP}}

\newcommand{\PROJECTOR}{\mathsf{PROJECTOR}}
\newcommand{\DENSITY}{\mathsf{DENSITY}}
\newcommand{\SGN}{\mathsf{SGN}}
\newcommand{\SIGMAK}{\mathsf{SIGMAK}}
\newcommand{\PCA}{\mathsf{PCA}}
\newcommand{\PURIFY}{\mathsf{PURIFY}}

\newcommand{\pmult}{\mu_{\MM}}
\newcommand{\pinv}{\mu_{\INV}}

\newcommand{\cinv}{c_\INV}
\newcommand{\err}{\boldsymbol{\mathsf{err}}}
\newcommand{\keig}{\kappa_{\matV}}
\newcommand{\normeinv}{\left\|\matE_1^{\INV}\right\|}

\newcommand{\erronenhalf}[2][\matM]{
    \mathcal{E}_{1}\left(
        \tfrac{n}{#2},#1
    \right)
}
\newcommand{\errinvnhalf}[2][\matM]{
    \mathcal{E}_{\INV}\left(
        \tfrac{n}{#2},#1
    \right)
}

\newcommand{\errormatrix}{\matE}
\newcommand{\geneigmatrix}{\matLambda}
\newcommand{\projectormatrix}{\matPi}
\newcommand{\projectormatrixtilde}{\matPitilde}

\usepackage[utf8]{inputenc}
\usepackage{listings}
\usepackage{enumerate}
\usepackage{graphicx}
\usepackage{xcolor}
\usepackage[pagebackref=true]{hyperref}
\usepackage{setspace}
\hypersetup{
    colorlinks,
    linkcolor={blue},
    citecolor={blue},
    urlcolor={blue}
}

\title{Invariant subspaces and PCA in nearly matrix multiplication time}

\author{%
  Aleksandros Sobczyk \\
  IBM Research and ETH Zurich\\
  \texttt{obc@zurich.ibm.com} \\
  \And
  Marko Mladenovi\'c \\
  ETH Zurich \\
  \texttt{mmladenovic@ethz.ch} \\
  \And
  Mathieu Luisier \\
  ETH Zurich \\
  \texttt{mluisier@iis.ee.ethz.ch}
}

\begin{document}
\maketitle

\begin{abstract}
    Approximating invariant subspaces of generalized eigenvalue problems (GEPs) is a fundamental computational problem at the core of machine learning and scientific computing. It is, for example, the root of Principal Component Analysis (PCA) for dimensionality reduction, data visualization, and noise filtering, and of Density Functional Theory (DFT), arguably the most popular method to calculate the electronic structure of materials. 
    Given Hermitian $\matH,\matS\in\mathbb{C}^{n\times n}$, where $\matS$ is positive-definite,  let $\matPi_k$ be the true spectral projector on the invariant subspace that is associated with the $k$ smallest (or largest) eigenvalues of the GEP $\matH\matC=\matS\matC\matLambda$, for some $k\in[n]$. 
    We show that we can compute a matrix $\projectormatrixtilde_k$ such that $\|\projectormatrix_k-\projectormatrixtilde_k\|_2\leq \epsilon$, in $O\lpar n^{\omega+\eta}\polylog(n,\epsilon^{-1},\kappa(\matS),\gap_k^{-1}) \rpar$ bit operations in the floating point model, for some $\epsilon\in(0,1)$, with probability $1-1/n$. Here, $\eta>0$ is arbitrarily small, $\omega\lesssim 2.372$ is the matrix multiplication exponent, $\kappa(\matS)=\|\matS\|_2\|\matS^{-1}\|_2$, and $\gap_k$ is the gap between eigenvalues $k$ and $k+1$. 
    To achieve such provable ``forward-error'' guarantees, our methods rely on a new $O(n^{\omega+\eta})$ stability analysis for the Cholesky factorization, and a smoothed analysis for computing spectral gaps, which can be of independent interest.
    Ultimately, we obtain new matrix multiplication-type bit complexity upper bounds for PCA problems, including classical PCA and (randomized) low-rank approximation. 
\end{abstract}
\newpage
\tableofcontents

\newpage
\section{Introduction}\label{section:intro}
Generalized eigenvalue problems (GEPs) arise naturally in a plethora of applications in machine learning, scientific computing, and engineering. Given a pair of matrices $\matH$ and $\matS$, often referred to as a \textit{matrix pencil}, the problem of interest has the following form \begin{align}
    \matH\matC=\matS\matC\geneigmatrix,
    \label{eq:gep_main}
\end{align}
where $\matC$ and $\geneigmatrix$ are the unknown eigenvector and eigenvalue matrices, respectively. 
Of particular importance are the so-called ``Hermitian definite'' or simply ``definite'' GEPs/pencils, in which  case $\matH$ is Hermitian and $\matS$ is Hermitian and positive-definite. 
In many important applications, the quantity of interest is an (arbitrarily large) subset of eigenvectors, defining an \textit{invariant subspace}, rather than the entire $\matC$ and $\geneigmatrix$ solutions of the GEP.

In data science and machine learning, invariant subspaces play a central role in many problems, including Spectral Clustering \cite{ng2001spectral,shi2000normalized}, Language Models \cite{hu2022lora}, Image Processing \cite{rufai2014lossy,bakir2004learning}, Recommendation Systems \cite{drineas2002competitive}, Principal Components Analysis (PCA) \cite{diamantaras1996principal,jolliffe2002principal,scholkopf1998nonlinear}, Support Vector Machines \cite{mangasarian2005multisurface}, and many others
\cite{feldman2020turning,cohen2015dimensionality,malhi2004pca,baldi1989neural}. We particularly focus on PCA applications, which can take the form of a GEP as in Eq. \eqref{eq:gep_main} where $\matH$ is the sample covariance and $\matS$ the identity. In more advanced settings, $\matH$ and $\matS$ can be defined over a kernel \cite{scholkopf1998nonlinear,barshan2011supervised}; See Section \ref{section:pca} and Appendix \ref{appendix:pca} for more details. 
Another closely related application comes from Density Functional Theory \cite{kohn1965self} (DFT), which is not a machine learning problem per se, but it is probably the most commonly used method (it was awarded the Nobel prize in Chemistry in 1998) to compute the electronic and structural properties of materials.  
In this case, $\matH$ is the Hamiltonian and $\matS$ the overlap matrix (cf. Appendix \ref{appendix:dft_background}). The spectral projector on the invariant subspace corresponding to the smallest generalized eigenvalues (occupied energies) directly provides the density matrix and the electron density. Obtaining them often presents a challenge from the computational point of view.

\subsection{Problem definition}
The main focus of this work is the computation of \textit{spectral projectors} on invariant subspaces that are associated with a subset of the spectrum of Hermitian definite GEPs. As the Abel-Ruffini theorem excludes exact computation, even in exact arithmetic, we seek for approximate computations, as described in the following Problem \ref{problem:spectral_projector}.
\begin{problem}[Spectral projector]
    \label{problem:spectral_projector}
    Given a Hermitian definite GEP 
    $
        \matH\matC=\matS\matC\geneigmatrix
    $
    of size $n$,
    an integer $1\leq k\leq n-1$, and accuracy $\epsilon\in(0,1)$, compute a matrix $\projectormatrixtilde_k\in\mathbb{C}^{n\times n}$ such that
    \begin{align}
        \lnorm \projectormatrixtilde_k - \projectormatrix_k \rnorm \leq \epsilon,
        \label{eq:invariant_subspace_forward_approximation}
    \end{align}
    where $\projectormatrix_k$ is the true spectral projector on the invariant subspace associated with the $k$ smallest or largest eigenvalues.
\end{problem}
Before proposing algorithms to solve Problem \ref{problem:spectral_projector}, we first make some clarifications and define useful concepts.
\paragraph{Type of approximation:} The approximation of the form of Equation \eqref{eq:invariant_subspace_forward_approximation} is commonly called a ``forward approximation'' or ``forward error'' in numerical analysis. It quantifies the distance between the true solution of the problem and the one returned by an algorithm. It is a stronger and harder to achieve notion of approximation than the related ``backward error.'' For details see Appendix \ref{appendix:forward_and_backward_approximation}.
\paragraph{Model of computation:} While many finite precision models of computation exist in the literature, all algorithms in this work are analyzed in the floating point model of computation, which is also the prominent model implemented in existing computers. Each real number $\alpha$ is rounded to a floating point number $\fl(\alpha) = (1+\theta)\alpha$, where $\theta\in\mathbb{C}$ satisfies $|\theta|\leq \umach \in\mathbb{R}_{>0}$. The machine precision $\umach$ bounds also the errors introduced by arithmetic operations $\{+,-,\times,/\}$, and the expression $\log(1/\umach)$ gives the number of bits required to achieve the desired precision. 
More details can be found in Appendix \ref{appendix:floating_point}.
\paragraph{Bit complexity:} The complexity of numerical algorithms is often measured in terms of the arithmetic operations executed, commonly referred to as \textit{arithmetic complexity}. A more realistic notion is the \textit{bit complexity}, which bounds the number of boolean operations. In the floating point model, it is straightforward to translate the arithmetic to the bit complexity if we have an upper bound on the number of bits. For instance,  arithmetic operations on $b$ bits can be typically carried out in $O(b^2)$ bit operations, or even faster by using more advanced algorithms \cite{schonhage1971fast,furer2007faster,harvey2021integer}. 
\paragraph{Matrix multiplication time:} In two seminal works \cite{demmel2007fastla,demmel2007fastmm}, it was demonstrated that matrix multiplication and other fundamental problems in Numerical Linear Algebra can be solved in the floating point model with nearly $O(n^{\omega+\eta})$ bit-complexity (up to polylogarithmic factors), where $\eta$ is an arbitrarily small positive number and $\omega$ is the matrix multiplication exponent, to-date bounded by $\omega \lesssim 2.372$ \cite{duan2023faster,williams2024new,alman2024more}. Hereafter, we will use the notation  $T_{
    \MM}(n)=n^{\omega+\eta}$.

\subsection{Existing algorithms}
Here we give a brief overview of existing algorithms. We refer to Appendix \ref{appendix:related_work} for more details.
GEPs in general can be solved using classic eigensolvers and related techniques in $\widetilde O(n^3)$ floating operations, e.g., by reducing the matrix (or pencil) to tridiagonal form with similarity transformations and applying the shifted QR algorithm on the tridiagonal matrix, or by using a divide-and-conquer method (see \cite{davies2001analysis,nakatsukasa2013stable,ipsen1997computing,bai1991direct,bai1998using,golub2013matrix,demmel1997applied,bai1997inverse,paige1971computation} and references therein). Significant progresses beyond the $\widetilde O(n^3)$ bit complexity barrier have been made 
\cite{demmel2007fastmm,demmel2007fastla,banks2022pseudospectral,louis2016accelerated,benor2018quasi,demmel2023generalized,musco2018stability,schneider2023fast}. Regarding the computation of eigenvalues, two notable examples are the $\widetilde O(T_{\MM}(n))$ algorithm of \cite{louis2016accelerated} for the largest eigenvalye, and the $\widetilde O(n^2)$ algorithm of \cite{musco2018stability} for the spectral norm.

The first to have addressed the problem of computing invariant subspaces in nearly $O(T_{\MM}(n))$ in floating point is \cite{demmel2007fastla} (see also \cite{ballard2011minimizingLA,ballard2011minimizingSVD}). The authors described an iterative algorithm for the Schur decomposition, and showed that each individual step is numerically stable, and it takes $O(T_{\MM}(n)$ operations. 
An end-to-end bound on the number of iterations to achieve a backward approximate solution was left open. More recently, the seminal work of \cite{banks2022pseudospectral} extended the analysis to obtain an end-to-end $\widetilde O(T_{\MM}(n))$ complexity to approximately diagonalize a matrix, and  \cite{demmel2023generalized,schneider2023fast} provided a rigorous analysis for the generalized eigenproblem case. 
In Corollary 1.7 and Proposition 1.1 of \cite{banks2022pseudospectral}, it was also outlined how to translate the backward diagonalization error to a forward error for the eigenvectors. The reported bound, however, has two main limitations: it relies on simplicity of the spectrum, which is a strict assumption, and it requires as input an over-estimate on the eigenvector condition number of the problem, which is unknown, and \cite{banks2022pseudospectral} does not describe how to compute it (see Appendix \ref{appendix:why_not_diagonalization} for more details). In this work we describe how to overcome these limitations and provide a novel, end-to-end, provably accurate analysis (in the sense of Eq. \eqref{eq:invariant_subspace_forward_approximation}) for arbitrary invariant subspaces of definite GEPs with $\widetilde O(T_{\MM}(n))$ boolean complexity. 

\subsection{Contributions and methods}

Our main contribution, summarized in the following Theorem \ref{theorem:spectral_projector} and Algorithm \ref{alg:spectral_projector}, is an end-to-end analysis to solve Problem \ref{problem:spectral_projector} in nearly $O(T_{\MM}(n))$ time.

\begin{theorem}
    \label{theorem:spectral_projector}
    Let $(\matH,\matS)$ be a Hermitian 
     definite pencil of size $n$ with $\|\matH\|,\|\matS^{-1}\|\leq 1$,  $\lambda_1\leq\lambda_2\leq\ldots\leq \lambda_n$ its eigenvalues,  $\gap_k=\lambda_{k+1}-\lambda_k$ and $\kappa(\matS)=\|\matS\|\|\matS^{-1}\|$. Algorithm \ref{alg:spectral_projector}
    \begin{align*}
        \projectormatrixtilde_k \leftarrow \PROJECTOR(\matH,\matS,k,\epsilon),
    \end{align*} takes as inputs $\matH$, $\matS$, an integer $ k\in [n-1]$, an error parameter $\epsilon\in(0,1)$ and returns a matrix $\projectormatrixtilde_k$ such that
    \begin{align*}
        \Pr\lbrac 
            \lnorm \projectormatrixtilde_k - \projectormatrix_k \rnorm \leq \epsilon
        \rbrac
        \geq 1-1/n,
    \end{align*}
    where $\projectormatrix_k$ is the true spectral projector on the invariant subspace that is associated with the $k$ smallest (or largest) eigenvalues.
    The algorithm executes 
    \begin{align*}
        O\lpar
                T_{\MM}(n)\lpar
                    \log(\tfrac{n}{\gap_k})\log(\tfrac{1}{\gap_k})
                    +
                    \log(n\kappa(\matS))\log(\kappa(\matS))
                    +
                    \log\lpar
                        \log(
                            \tfrac{\kappa(\matS)}{\epsilon\gap_k}
                        )
                    \rpar
                \rpar
            \rpar
    \end{align*}
    floating point operations with 
    $
        O\lpar
                \log(n)
                \lpar
                    \log^4(\tfrac{n}{\gap_k})
                    +
                    \log^4(n\kappa(\matS))
                    +
                    \log^3(\tfrac{1}{\epsilon\gap_k})
                    \log(\tfrac{\kappa(\matS)}{\epsilon\gap_k})
                \rpar
            \rpar
    $
    bits of precision. Internally, the algorithm needs to generate a total of at most $\widetilde O(n)$ standard normal floating point numbers using additional $O(\log(\log(n)))$ bits. 
\end{theorem}

To achieve the results of Theorem \ref{theorem:spectral_projector}, we provide a novel $O(T_{\MM}(n))$-type complexity analysis of several problems in numerical linear algebra that can be of independent interest.

In brief, our methodology is as follows. 
We first observe that if we can determine reasonable ``guesses'' for the spectral gap ($\widetilde\gap_k$) and for the midpoint ($\widetilde\mu_k$) between the $\lambda_k$ and $\lambda_{k+1}$ eigenvalues then we can efficiently compute the spectral projector by approximating the sign function
\begin{align*}
    \sgn(\widetilde\mu_k-\matS^{-1}\matH),
\end{align*}
using the analysis of \cite{banks2022pseudospectral} for the Newton iteration.
The matrix $\frac{1}{2}(\matI+\sgn(\widetilde\mu_k-\matS^{-1}\matH))$ indeed transforms in exact arithmetic  all eigenvalues that are smaller than $\widetilde\mu_k$ to $1$ and the ones larger than $\widetilde\mu_k$ to zero.
As will be proved in Proposition \ref{proposition:alg_purify}, in Section \ref{section:sign_spectral_projector}, this approach is sufficient to provide an accurate spectral projector $\projectormatrixtilde_k$ in floating point. As a consequence, the problem reduces to approximating the aforementioned midpoint and gap. 
As a baseline, in Appendix \ref{appendix:why_not_diagonalization} we prove that this can be done in nearly $O(T_{\MM}(n))$ with iterative inversion \cite{demmel2007fastla} and diagonalization \cite{banks2022pseudospectral} or, similarly, by iteratively calling generalized diagonalization \cite{demmel2023generalized}.
However, this approach presents two drawbacks: It does not take advantage of the inherent symmetry of the problem, and, at the same time, it performs a full diagonalization when we are only interested in the gap between two specific eigenvalues, which is seemingly redundant. 
We formally prove this claim by designing a novel approach that achieves better complexity, typically by a factor of $O(\log(n))$ (cf. Section \ref{section:comparison_with_diagonalization}).
Importantly, no explicit diagonalization is necessary.

To minimize the complexity of our algorithm, it is crucial to leverage symmetry. To that end we use the Cholesky factorization of $\matS$ in the spirit of the Cholesky-QR algorithm \cite{davies2001analysis}. We highlight that, while other factorizations have been solved in $O(T_{\MM}(n))$ in floating point, an end-to-end analysis for Cholesky remains open. In exact arithmetic, for example, the LU of a Hermitian definite matrix directly provides its Cholesky and \cite{demmel2007fastla} showed that the LU factorization of non-symmetric matrices can be obtained in $O(T_{\MM}(n))$. However, when considering arithmetic errors, the relationship between LU and Cholesky does not hold in floating point, as demonstrated by the counter-example of Appendix \ref{appendix:appendix_lur_counterexample}. Other fast Cholesky algorithms have been proposed for special classes of matrices, e.g., for matrices with well-defined separators \cite{george1973nested,lipton1979generalized,gilbert1986analysis} and graph Laplacians \cite{kyng2016sparsified,kyng2016approximate}. However, they do not generalize to arbitrary dense matrices. Our analysis is the first to improve the classic $O(n^3)$ floating point Cholesky algorithms  \cite{kielbasinski1987note,higham2009cholesky} for the general case, with provable error bounds. In the following Theorem \ref{theorem:stable_cholesky_bounds} we summarize our new analysis of the Cholesky factorization Algorithm \ref{alg:stable_cholesky} (see also Appendix \ref{appendix:cholesky_analysis}). We note that the algorithm itself is not new, only its analysis.

\begin{theorem}
    \label{theorem:stable_cholesky_bounds}
    Given a Hermitian positive-definite matrix $\matM$, there exists an algorithm $\matL\leftarrow \CHOLESKY(\matM)$, listed in 
    Algorithm \ref{alg:stable_cholesky}, which requires $O(T_{\MM}(n))$ arithmetic operations. This algorithm is logarithmically stable, in a sense that, there exist global constants $c_1$, $c_2$, $c_3$, such that for all $\epsilon\in(0,1)$, if executed in a floating point machine with precision \begin{align*}
        \umach \leq \umach_{\CHOLESKY} := \epsilon \frac{1}{c_1n^{c_2}\kappa(\matM)^{c_3\log n}},
    \end{align*}
    which translates into
    $
        O\left(
            \log(n)\log(\kappa(\matM)) + \log(\tfrac{1}{\epsilon})
        \right)
    $
    required bits of precision,
    then it does not break down due to arithmetic errors, and the solution returned satisfies
    $
        \|\matL\matL^*-\matM\|\leq \epsilon\|\matM\|.
    $
\end{theorem}
This stand-alone result fulfills the definition of ``logarithmic-stability,'' a notion of numerical stability that is commonly used in the related literature \cite{demmel2007fastla,banks2022pseudospectral}. 
Given this new Cholesky analysis, the following transformation of the GEP to a regular Hermitian eigenvalue problem:
\begin{align*}
    \matH\matC=\matS\matC\geneigmatrix \Rightarrow \matL^*\matH\matL (\matL^{-1}\matC) = (\matL^{-1}\matC)\geneigmatrix,
\end{align*}
can be carried out accurately in $O(T_{\MM}(n))$ in floating point, with provable forward-error bounds for all eigenvalues of the transformed problem. Here, $\matL$ is the Cholesky factor of $\matS^{-1}$ instead of $\matS$. Specifically, in Proposition \ref{proposition:alg_reduce} in Appendix \ref{appendix:reduce}, we prove that the corresponding Algorithm \ref{alg:reduce}, $\matHtilde\leftarrow\REDUCE(\matH,\matS,\epsilon)$, returns a Hermitian matrix $\matHtilde$ such that, 
for any given accuracy $\epsilon\in(0,1)$,
\begin{align*}
    \labs \lambda_i(\matHtilde) - \lambda_i(\matS^{-1}\matH)\rabs \leq \epsilon, \forall i\in[n].
\end{align*}
The symmetry induced by the Cholesky transformation is crucial to design an efficient algorithm for the spectral gap. As described in Section \ref{section:spectral_gap}, and analyzed in Appendices \ref{appendix:regularize} and  \ref{appendix:spectral_gap}, any spectral gap or eigenvalue of a Hermitian definite pencil can be approximated by an iterative algorithm that uses only ``counting-queries'', i.e., queries that ask how many eigenvalues are smaller than a given threshold. This way we completely avoid diagonalization, thus leading to a lower complexity. 

To perform the counting queries efficiently, the transformed matrix $\matHtilde$ must be regularized with small random perturbations, in the spirit of smoothed analysis \cite{spielman2004smoothed}, which has recently drawn attention in the context of matrix algorithms \cite{boutsidis2016optimal,benor2018quasi,banks2022pseudospectral,demmel2023generalized,meyer2024unreasonable} (see Appendix \ref{appendix:regularize} for the analysis). 
These aforementioned works typically require a guarantee on the minimum eigenvalue gap of the perturbed matrix, e.g.,  \cite{meyer2024unreasonable} uses a Minami-type bound \cite{minami1996local}, while in \cite{banks2022pseudospectral,demmel2023generalized} the entire pseudospectrum of the perturbed matrix must be shattered with respect to a grid. 
The latter is even more challenging to achieve than a minimum gap and it requires $\widetilde O(n^2)$ random bits. Our algorithm is significantly less demanding in terms of randomness: 
All we need is the Wegner estimate \cite{wegner1981bounds,aizenman2017matrix} for the density-of-states of random Hermitian operators, and only $\widetilde O(n)$ random bits in total. 

Finally, in Section \ref{section:pca}, we apply our main results to prove the first matrix multiplication-type upper bounds for the bit complexity of PCA algorithms. Specifically, for the standard PCA formulation, we show that we can first compute the spectral projector and then use deflation to obtain a basis for the desired low-dimensional embedding in nearly matrix multiplication time. We then apply similar arguments to the seminal Block-Krylov PCA algorithm of \cite{musco2015randomized}.

\begin{algorithm}[htb]
\centering
\caption{$\PROJECTOR$.}
\label{alg:spectral_projector}
\begin{algorithmic}[1]
\setstretch{0.8}
\Statex \begin{align*}
    \PROJECTOR
\end{align*}
\small
\Input Hermitian definite pencil $\matH\in\mathbb{H}^{n}$, $\matS\in\mathbb{H}^n_{++}$, gap index $k$, accuracy $\epsilon$.
\Require $\|\matH\|\leq 1$, $\|\matS^{-1}\|\leq 1$, $k\in[n-1]$.
\setstretch{1.5}
\Algorithm $\projectormatrixtilde \leftarrow\PROJECTOR(\matH,\matS,k,\epsilon)$.
\State $\widetilde\mu_k,\widetilde\gap_k \leftarrow \GAP(\matH,\matS,k,\tfrac{1}{8}, \tfrac{1}{2n})$.
\State $\widetilde\kappa \leftarrow \COND(\matS,\tfrac{1}{4}, \tfrac{1}{2n})$. \Comment{This is skipped when $\matS=\matI$.}
\State $\projectormatrixtilde\leftarrow \PURIFY(\matH,\matS, \widetilde\mu_k,\widetilde\gap_k,\widetilde\kappa,\epsilon)$.
\State \Return $\projectormatrixtilde$.
\Output Approximate projector $\projectormatrixtilde$ on the invariant subspace associated with the $k$ smallest eigenvalues.
\Ensure $\|\projectormatrixtilde-\projectormatrix\|\leq \epsilon$ with probability at least $1-1/n$.
\end{algorithmic}
\end{algorithm}

\begin{algorithm}[htb]
\caption{$\CHOLESKY(\matM)$.}
\label{alg:stable_cholesky}
\begin{algorithmic}[1]
\setstretch{0.8}
\Statex \begin{align*}
    \CHOLESKY
\end{align*}
\small
\Input Matrix $\matM=
\begin{pmatrix}         
    \matA & \matB^*\\
    \matB & \matC
\end{pmatrix}
\in\mathbb{H}^{n}_{++}
$.
\Require $\matM$ is positive-definite, $\umach\leq\epsilon\frac{1}{c_1n^{c_2}\kappa(\matM)^{c_3\log(n)}}$ for some constants $c_1,c_2,c_3$ and $\epsilon\in(0,1)$.
\setstretch{1.5}
\Algorithm $\matL \leftarrow\CHOLESKY(\matM)$.
\If{$n=1$} 
    \State \Return $\matL=\sqrt{\matM_{1,1}}$.
\Else
    \State $\matL_{11} \leftarrow \CHOLESKY(\matA)$. \Comment{$\matL_{11}\matL_{11}^*=\matA+\errormatrix^{\mathsf{CH}}_{\matA}$.}
    \State $\matB\matA i\leftarrow \MM\left(
            \matB,
            \INV\left(\matA\right)
        \right).$
        \Comment{$
        \matB\matA i
        =
            \matB\left(
                \matA^{-1}+\errormatrix_{1}^{\INV}
            \right)
            +
            \errormatrix_2^{\MM}
        =
            \matB\matA^{-1} + \errormatrix_{\matB\matA i}.
        $}
    \State $\matL_{21} \leftarrow \MM\left(
        \matB\matA i,
        \matL_{11}
    \right).$
    \Comment{
    $
    \matL_{21}
    =
    \left(
        \matB\matA^{-1}+\errormatrix_{\matB\matA i}
    \right)
    \matL_{11}+\errormatrix_3^{\MM}
    =
    \matB\matA^{-1}\matL_{11}+\errormatrix_{\matL_{21}}
    $.}
    \State $\matStilde\leftarrow \matC-\HERM(\MM(\matB\matA i,\matB^*))$. \Comment{$ \matStilde=\matC-
    \left(
        (\matB\matA i)\matB^*+\errormatrix_4^{\MM}
    \right)+
    \errormatrix_5^{\SUB}
    =
    \matS + \errormatrix_{\matS}
    $
    ,}
    \Statex \Comment{where $\errormatrix_{\matS} = \matB\errormatrix_1^{\INV}\matB^* + \errormatrix_2^{\MM}\matB^* + \errormatrix_4^{\MM} + \errormatrix_5^{\SUB}.$}
    \State $\matL_{22} \leftarrow \CHOLESKY(\matStilde)$. \Comment{$ \matL_{22}\matL_{22}^*=\matStilde+\errormatrix_{\matStilde}^{\mathsf{CH}}$.}
    \State \Return $\matL = 
    \begin{pmatrix}
        \matL_{11} & \\
        \matL_{21} & \matL_{22}
    \end{pmatrix}
    $.
\EndIf
\Output Lower triangular Cholesky factor $\matL\in\mathbb{C}^{n\times n}$.
\Ensure $\|\matL\matL^*-\matM\|\leq \epsilon \|\matM\|$, $\matL$ is lower triangular, and $\matL\matL^*$ is Hermitian and positive-definite.
\end{algorithmic}
\end{algorithm}

\subsection{Notation}

Matrices are denoted by bold capital letters and vectors by bold small letters. For real or complex constants we typically use Greek letters, or the Latin letters $c,C$. The vector ${ \vece}_i$ denotes the $i$-th column of the standard basis. $\matA^*$ is the conjugate transpose of $\matA $ and $\matA^\dagger$ denotes the pseudoinverse. The 2-norm is the default for matrices and vectors. $\kappa(\matA)=\|\matA\|\|\matA^\dagger\|$ is the two-norm condition number of $\matA$.
For the error analysis of the various algorithms, we use $\errormatrix_{i}^{\OP}$ to denote the error matrices that are introduced by the floating point errors of the $i$-th operation $\OP$. The letters $\epsilon$ and $\delta$ typically denote (scalar) error quantities and failure probabilities, respectively.
$[n]$ is the set $\{1,2,...,n\}$.  We denote by $\mathbb{H}^n\subset \mathbb{R}^{n}$ the set of Hermitian matrices of size $n\times n$,  $\mathbb{H}^n_{+}$ the set of Hermitian positive semi-definite matrices and  $\mathbb{H}^n_{++}$ the set of Hermitian positive definite matrices. For a matrix $\matA$ and a scalar $z$ we write $z\pm\matA$ as a shorthand for $z\matI\pm\matA$. $\Lambda(\matA)$ and $\Lambda(\matA,\matB)$ denote the spectrum of a matrix $\matA$ and a matrix pencil $(\matA,\matB)$, respectively. The eigenvalues and singular values are always sorted in ascending order by default: $\lambda_1\leq \lambda_2\leq \ldots\leq\lambda_n$. $\Lambda_{\epsilon}(\matA)$ is the $\epsilon$-pseudospectrum of $\matA$ (see Definition \ref{definition:pseudospectrum}). 

\section{Computing spectral projectors with the sign function}
\label{section:sign_spectral_projector}
Given a Hermitian definite pencil $(\matH,\matS)$, our ultimate goal is to compute a forward error approximation of the spectral projector that is associated with the $k$ smallest eigenvalues, as described in Problem \ref{problem:spectral_projector}. 
Algorithm \ref{alg:purify} solves this problem provably and efficiently, but it requires that we already have a suitable approximation of the eigenvalue gap that separates the desired subspace from the rest of the eigenspace. The algorithm is called $\PURIFY$ since it is inspired by ``purification'' techniques in DFT, referring to the removal of the unoccupied orbitals. The computation of the gap and the midpoint is in fact the bottleneck of our main algorithm, however, we still show that they can be computed efficiently in Section \ref{section:spectral_gap}, and, importantly, without diagonalizing any matrices. The properties of Algorithm \ref{alg:purify} are stated in Proposition \ref{proposition:alg_purify}.

\begin{proposition}
    \label{proposition:alg_purify}
    Let $\matH\in\mathbb{H}^n$ with $\|\matH\|\leq 1$,
    $\matS\in\mathbb{H}^{n}_{++}$ with $\|\matS^{-1}\|\leq 1$, $k\in[n-1]$ and $\epsilon\in(0,1).$ Let $\mu_k=\frac{\lambda_k+\lambda_{k+1}}{2}$ and $\gap_k=\lambda_k-\lambda_{k+1}$, where $\lambda_1\leq\ldots\leq\lambda_n$ are the generalized eigenvalues of the Hermitian definite pencil $(\matH,\matS)$ and assume that we want to compute $\projectormatrix_k$ which is the true spectral projector associated with the $k$ smallest eigenvalues. If we have access to 
    \begin{align*}
        \widetilde\mu_k\in\mu_k\pm\tfrac{1}{8}
        \gap_k
        \quad
        \widetilde\gap_k\in(1\pm\tfrac{1}{8})\gap_k,
        \quad
        \widetilde\kappa \in [\kappa(\matS),C\kappa(\matS)],
    \end{align*} for some constant $C>1$, then
    Algorithm \ref{alg:purify} computes  $\projectormatrixtilde_k\leftarrow \PURIFY(\matH,\matS,\widetilde\mu_k,\widetilde\gap_k,\widetilde\kappa,\epsilon)$ such that
    $
        \|\projectormatrixtilde_k-\projectormatrix_k\|\leq \epsilon,
    $
    in 
    $
    O
    \lpar
        T_{\MM}(n) \lpar
            \log(\tfrac{1}{\gap_k})
            +
            \log(\log(\tfrac{\kappa(\matS)}{\epsilon\gap_k}))
        \rpar
    \rpar
    $
    floating point operations using
    $
        O\lpar
            \log(n)\log^3(\tfrac{1}{\gap_k})\log(\tfrac{\kappa(\matS)}{\epsilon\gap_k})
        \rpar
    $
    bits of precision.
    \begin{proof}
        The full proof of Proposition \ref{proposition:alg_purify} can be found in Appendix \ref{appendix:alg_purify}. We briefly summarize it here. The main idea is to use the sign function algorithm from \cite{banks2022pseudospectral}, $\SGN$, to approximate $\sgn(\widetilde\mu_k-\matS^{-1}\matH)$. If we already know that $\widetilde\mu_k$ is a reasonable approximation of $\mu_k$, and that it is located inside the correct eigenvalue gap, then, in exact arithmetic, our problem is equivalent to computing $\sgn(\mu_k-\matS^{-1}\matH)$. 
        The result can be used to filter the desired spectral projector, often referred as ``purification'' in the context of DFT. The main challenge is to ensure that all propagated numerical errors, success probabilities, and input parameters for all algorithms are well-defined and bounded. To obtain the final forward errors we must rely on matrix similarity arguments, the properties of the pseudospectrum, the eigenvalue bounds of Weyl and Kahan from Fact \ref{fact:weyl_kahan}, and Lemma \ref{lemma:sign_function_under_perturbation}, which gives explicit bounds on the sign function under small floating point perturbations.
    \end{proof}
\end{proposition}
The rest of the paper is devoted to the analysis of our new algorithm for the spectral gap and the midpoint based on eigenvalue counting queries, described in Theorem \ref{theorem:alg_spectral_gap}. For comparison purposes, in Appendix \ref{appendix:why_not_diagonalization} we analyze a diagonalization-based algorithm for the same task (which is a new result itself), specifically, using the state-of-the-art $\EIG$ algorithm of \cite{banks2022pseudospectral}. We compare the two algorithms in Section \ref{section:comparison_with_diagonalization}, demonstrating that our counting-based algorithm is indeed faster.

\section{Fast spectral gaps with counting queries}
\label{section:spectral_gap}
Our core algorithm efficiently approximates spectral gaps based on ``eigenvalue counting queries'' only,  thus avoiding an explicit (and expensive) diagonalization. 
To give some intuition on the main idea, consider the following simplified version of the problem.

\begin{problem}[Gap finder]
    \label{problem:gap_finder}
    Let $\lambda_1\leq\ldots\leq\lambda_n$ in $[-1,1]$ be $n$ (unknown) real values  (e.g., they can be the eigenvalues of the original matrix pencil) $\mu_k=\tfrac{\lambda_k+\lambda_{k+1}}{2}$, and $\gap_k=\lambda_{k+1}-\lambda_{k}$, for some $k\in[n-1]$. Given $k$ and some error parameter $\epsilon\in(0,1/2)$ as input, we want to approximate $\mu_k$ and $\gap_k$ up to additive $\epsilon\gap_k$, i.e., we look for $\widetilde\mu_k=\mu_k\pm\epsilon\gap_k$ and $\widetilde\gap_k\in(1\pm\epsilon)\gap_k$. Only queries of the following form can be performed: We fix a parameter $\gamma\in(0,1/2)$, which distorts all $\lambda_i$  to some (unknown) $\lambda_i'\in[\lambda_i-\gamma,\lambda_i+\gamma]$. We can then choose any value $h\in[-1-\gamma,1+\gamma]$ and ask how many values $\lambda_i'$ are smaller than $h$. For each $\gamma$, we can query arbitrarily many different values for $h$, and each $h$-query costs $q(1/\gamma)=O(\polylog(1/\gamma))$.
\end{problem}
The query cost is arbitrary to avoid trivial solutions by setting $\gamma=0$. The following proposition is proved in Appendix \ref{appendix:spectral_gap}:
\begin{proposition}
    \label{proposition:gap_finder}
        Problem \ref{problem:gap_finder} can be solved iteratively  by executing a total of $O(\log(\tfrac{1}{\epsilon\gap_k}))$ iterations and $\Theta(1)$ queries per iteration, where each query costs at most $q(\tfrac{1}{\epsilon\gap_k})$.
\end{proposition}

\subsection{Smoothed analysis of eigenvalue counting}
To use the counting query model of Problem \ref{problem:gap_finder} and Proposition \ref{proposition:gap_finder} to compute the spectral gap of a matrix pencil, we need a ``black-box'' method to count eigenvalues that are smaller than a threshold. We first describe a straightforward, deterministic algorithm $\COUNT(\matXtilde,h,\varepsilon)$ for this task (see Lemma \ref{lemma:eigenvalue_count_less_than_h}), which takes as input a Hermitian matrix $\matXtilde$, a scalar $h$, and a parameter $\varepsilon$, with the requirement that $\sigma_{\min}(h-\matXtilde)>\varepsilon$. It returns the precise number of eigenvalues of $\matXtilde$ that are smaller than $h$. The runtime of the algorithm depends on $\log(1/\varepsilon)$, and must therefore be minimized.
For this we resort to smoothed analysis: We apply a random perturbation to ensure that $\varepsilon$ is at least polynomial in $1/n$, up to some other factors detailed in Appendix \ref{appendix:regularize}.

To build a random ``regularizer,'' in Definition \ref{definition:stable_normal_sampler} we introduce a random oracle that samples numbers from a standard normal distribution and returns their floating point representation using a pre-specified number of bits. 
Based on this simple oracle, we can design a floating point algorithm $\matXtilde\leftarrow \REGULARIZE(\matA,\gamma,\delta)$ which has the following properties described in Proposition \ref{proposition:alg_regularize}:
\begin{proposition}
    \label{proposition:alg_regularize}
    Let $\matA$ with  $\|\matA\|\leq 1$ be a Hermitian matrix,  $\gamma,\delta\in(0,1/4)$ two given parameters, and $\matXtilde\leftarrow \REGULARIZE(\matA,\gamma,\delta)$. Let $\mathsf{g}$ be an arbitrary (but fixed) grid of points in $[-2,2]$ with cardinality $|\mathsf{g}|=T$.
    For every element $h_i\in\mathsf{g}$ consider the matrices $\matM_i=h_i-\matXtilde$ and $\matMtilde=h_i-\matXtilde+\errormatrix_{i}$, where $\errormatrix_{i}$ denote the diagonal floating point error matrices induced by the shift. All the following hold simultaneously with probability $1-2\delta$ if we use 
    $O(\log(\tfrac{Tn}{\gamma\delta}))$
    bits of precision:
    \begin{align*}
        \|\matXtilde\| \leq 4/3,
        \quad
        \labs \lambda_i(\matXtilde)-\lambda_i(\matA) \rabs \leq \tfrac{9}{16}\gamma,
        \quad
        \sigma_{\min}(\matMtilde_i) \geq \tfrac{\gamma\delta}{4nT\sqrt{4\pi\ln(4n/\delta)}}.
    \end{align*}
    \begin{proof}
        The main result that we use in the proof can be traced back to the Wegner estimate for the density-of-states of Hermitian operators under random diagonal disorder \cite{wegner1981bounds}.
        See Appendix \ref{appendix:regularize} and in particular \ref{proposition:alg_regularize_appendix} for more details.
    \end{proof}
\end{proposition}
\subsection{Computing the gap and the midpoint}
We can now describe the algorithm $\GAP$ in Theorem \ref{theorem:alg_spectral_gap}, that computes the $k$-th gap and the midpoint of a Hermitian definite pencil. The same methodology can be extended to approximate any singular value, as described in Proposition \ref{proposition:alg_sigmak} in Appendix \ref{appendix:sigma_k}. 
\begin{theorem}[$\GAP$]
    \label{theorem:alg_spectral_gap}
    Let $\matH\in\mathbb{H}^n$, $\matS\in\mathbb{H}^n_{++}$ and $\|\matH\|,\|\matS^{-1}\|\leq 1$, which define a Hermitian definite pencil $(\matH,\matS)$. Given $k\in[n-1]$, accuracy $\epsilon\in(0,1)$, and failure probability $\delta\in(0,1/2)$, there exists an algorithm
    \begin{align*}
        \widetilde\mu_k,\widetilde\gap_k \leftarrow \GAP(\matH,\matS,k,\epsilon,\delta)
    \end{align*}
    which returns $\widetilde\mu_k=\mu_k\pm\epsilon\gap_k$ and $\widetilde \gap_k=(1\pm\epsilon)\gap_k$, where $\mu_k=\tfrac{\lambda_k+\lambda_{k+1}}{2}$ and $\gap_k=\lambda_{k}-\lambda_{k+1}$. The algorithm requires 
    \begin{align*}
        O\lpar
            T_{\MM}(n)
            \log(\tfrac{1}{\delta\epsilon\gap_k})
            \log(\tfrac{1}{\epsilon\gap_k})
        \rpar
    \end{align*}
    arithmetic operations using $O\lpar\log(n)
    \lpar
        \log^4(\tfrac{n}{\delta\epsilon\gap_k})
        +
        \log(\kappa(\matS))
    \rpar
    \rpar$ bits, where $\lambda_i$ are the eigenvalues of $(\matH,\matS)$. 
    If $\kappa(\matS)$ is unknown, additional $O(T_{\MM}(n)\log(\tfrac{n\kappa(\matS)}{\delta})\log(\kappa(\matS)))$ floating point operations and $O(\log(n)\log^4(\tfrac{n\kappa(\matS)}{\delta}))$ bits are sufficient to compute it with Corollary \ref{corollary:alg_cond}.
    \begin{proof} The full proof builds upon the results that are detailed in Appendices \ref{appendix:regularize} and \ref{appendix:spectral_gap}. A summary is the following.
    We first fix our initial error parameter $\gamma_0=1/8$ and call $\matHtilde=\REDUCE\lpar \matH,\matS,\tfrac{\gamma_0}{4} \rpar$ (Algorithm \ref{alg:reduce}), which internally uses $\CHOLESKY$ to reduce the GEP to a regular Hermitian one. From Proposition \ref{proposition:alg_reduce}, a Hermitian matrix $\matHtilde$ is returned such that
    $|\lambda_i(\matHtilde)-\lambda_i(\matH,\matS)|\leq \tfrac{\gamma_0}{4}$. 
    
    Next, we use the same counting query model as in Proposition \ref{proposition:gap_finder}. 
    We first regularize $\matHtilde$ using $\matXtilde\leftarrow \REGULARIZE(\matHtilde,\tfrac{\gamma_0}{2},\delta_0)$, where $\delta_0=\delta/2$ is the initial failure probability. Conditioning on success of Proposition \ref{proposition:alg_regularize} (with probability $1-\delta_0)$, for all $i\in[n]$, it holds that $|\lambda_i(\matXtilde)-\lambda_i(\matHtilde)|\leq 9\gamma_0/16$. Summing the two eigenvalue error bounds, we conclude that all eigenvalues of $(\matH,\matS)$, which initially lie in $[-1,1]$, are distorted by at most $\gamma_0$ in $\matXtilde$.
    We now have all necessary tools to go back to the counting query model of Proposition \ref{proposition:gap_finder}: In the first step we construct a grid $\mathsf{g}=\{-1, -7/8, -6/8, \ldots,7/8,1,9/8\}$. Clearly, $|\mathsf{g}|=\Theta(1)$. 
    Since we conditioned on the success of Proposition \ref{proposition:alg_regularize}, the regularization ensures that for every $h_j\in\mathsf{g}$ it holds that
    $\sigma_{\min}(h_j-\matXtilde+\errormatrix)\geq \varepsilon_0$ with $\varepsilon_0=\frac{\gamma_0\delta_0}{8|\mathsf{g}|n\sqrt{\pi\ln(4n/\delta_0)}}$. This allows us to efficiently execute $\COUNT(\matXtilde,h_j,\varepsilon_0)$ for every $h_j$. 
    
    At the end of the first iteration, we have computed two intervals $I_k$ and $I_{k+1}$, where $I_k$ contains $\lambda_k$ and $I_{k+1}$ contains $\lambda_{k+1}$, and each interval has size at most $1$, i.e., half the size of $[-1,1]$. We continue by halving at each step both $\gamma$ and $\delta$, constructing the corresponding grids as per the proof of Proposition \ref{proposition:gap_finder}, and counting eigenvalues over the grid. 
    In each iteration after the first one, we keep track of two intervals $I_k$ and $I_{k+1},$ and two corresponding grids $\mathsf{g}_k$ and $\mathsf{g}_{k+1}$ with size $|\mathsf{g}_k|=|\mathsf{g}_{k+1}|=\Theta(1)$. We therefore only need to execute a constant number of $\COUNT$ queries, and in each iteration the size of the intervals $I_k$ and $I_{k+1}$ is halved.
    The algorithm terminates after a total of $m=O\lpar \log(\tfrac{1}{\epsilon\gap_k}) \rpar$ iterations and finally provides the advertised complexity, bit requirements, failure probability, and approximation guarantees.
    \end{proof}

\end{theorem}

\subsection{Sketch proof of Theorem \ref{theorem:spectral_projector}}
The proof of our main Theorem \ref{theorem:spectral_projector} directly follows from Theorem \ref{theorem:alg_spectral_gap} together with Proposition \ref{proposition:alg_purify} as well as the algorithm $\SIGMAK$ (described in Appendix \ref{appendix:sigma_k}) which is used to compute the condition number of $\matS$. The full proof can be found in Appendix \ref{appendix_proofs:theorem:spectral_projector}.

\subsection{Comparison with diagonalization}
\label{section:comparison_with_diagonalization}
We can now compare Theorem \ref{theorem:alg_spectral_gap} with a diagonalization-based approach that is detailed in Proposition \ref{proposition:eig_based_gap}. We fix $\delta=O(1/n)$ so that both algorithms succeed with the same probability.

For $\epsilon,\gap_k,\kappa^{-1}(\matS)\in\Omega(\poly(1/n))$, the total arithmetic complexity of the algorithm of Theorem \ref{theorem:alg_spectral_gap} is $O(T_{\MM}(n)\log^2(n))$ using $O(\log^5(n))$ bits. For the same parameters, Proposition \ref{proposition:eig_based_gap} requires need a total of $O(T_{\MM}(n)\log^3(n))$ arithmetic operations, and $O(\log^5(n))$ bits. Thus, in total, Algorithm \ref{theorem:alg_spectral_gap} is faster by a factor of $O(\log(n))$.
 
In the extreme case where $\epsilon,\gap_k,\kappa(\matS)=\Theta(1)$,  Theorem \ref{theorem:alg_spectral_gap} counts $O(T_{\MM}(n)\log(n))$ arithmetic operations and $O(\log^5(n))$ bits, while Proposition \ref{proposition:eig_based_gap} requires $O(T_{\MM}(n)\log^2(n))$ operations, and $O(\log^5(n))$  bits. Thus, Proposition \ref{proposition:eig_based_gap} is  again slower by a factor of $O(\log(n))$. Interestingly, in this case Theorem \ref{theorem:alg_spectral_gap} is faster than even a single call to $\EIG$, which requires $O(T_{\MM}(n)\log^2(n))$ arithmetic operations.
We conclude that, at least based on the currently existing algorithms, diagonalization is redundant for the computation of spectral gaps and invariant subspaces.

\subsection{Application in DFT}
In Appendix \ref{appendix:dft_background} we demonstrate how our main results can be directly applied to approximate density matrices and electron densities of atomic systems in DFT. Even though is not a machine learning problem per se, we decided to dedicate a section in the Appendix due to its importance: DFT calculations persistently occupy supercomputing clusters and the corresponding  software libraries and literature receive tens of thousands of citations annually at the time of this writing \cite{kresse1996efficient,giannozzi2017advanced,giannozzi2009quantum,soler2002siesta,hutter2014cp2k}. Our work is the first analysis to provide forward-error guarantees in finite precision for these problems in nearly matrix multiplication time.
\section{PCA}
\label{section:pca}

Since its introduction in the early twentieth century \cite{pearson1901liii,hotelling1933analysis}, Principal Component Analysis is one of the most important tools in statistics, data science, and machine learning. It can be used, for example, to visualize data, to reduce dimensionality, or to remove noise from data; cf. \cite{jolliffe2002principal,diamantaras1996principal} for reviews on the vast bibliography. In its simplest formulation, given a (centered) data matrix $\matX\in\mathbb{R}^{m\times n}$, the goal is to learn a $k$-dimensional embedding $\matC_k$, where $k<n$, that maximizes the sample variance, which can be written as an optimization problem
\begin{align}
    \label{eq:pca_optimization}
    \matC_k=\arg \max_{\matC^\top\matC=\matI_{k\times k}}\tr(\matC^\top\matH\matC),
\end{align}
where $\matH=\matX^\top\matX\in\mathbb{R}^{n\times n}$ is the sample covariance. It can be shown that the solution $\matC_k$ corresponds to the principal $k$ singular vectors of $\matH$, i.e. the ones that correspond to the largest $k$ singular values. Evidently, since the sample covariance is always symmetric and positive semi-definite, this can be written as a Hermitian eigenvalue problem
\begin{align*}
    \matH\matC=\matC\matLambda,
\end{align*}
(which is indeed a definite GEP as in Equation \eqref{eq:gep_main} with $\matS=\matI$). By solving for $\matC_k$, we can project the data in $k$ dimensions by computing $\matX\matC_k$, preserving as much of the variance in $k$ dimensions as possible. To compute $\matC_k$ we can directly use our main results. However, the solution of Equation \eqref{eq:pca_optimization} is an actual orthonormal basis for the invariant subspace rather than the spectral projector that Theorem \ref{theorem:spectral_projector} returns. This can be addressed with deflation: Once we have the spectral projector $\projectormatrixtilde_k$, assuming that the approximation is sufficiently tight, we can apply a subsequent deflation step based on a rank-revealing QR factorization to obtain a $k$-dimensional basis. This can be done deterministically in $O(n^3)$ time \cite{gu1996efficient} or in randomized $O(n^\omega)$  \cite{demmel2007fastla,banks2022pseudospectral}.

In Appendix \ref{appendix:vanilla_pca} we prove the following Theorem \ref{theorem:pca} for Algorithm \ref{alg:pca}, which builds upon our main Theorem \ref{theorem:spectral_projector}, the algorithm of Proposition \ref{proposition:alg_sigmak} to approximate $\|\matX-\matX_k\|$, and the $\DEFLATE$ algorithm of \cite{banks2021gaussian}, to solve the standard PCA problem of Eq. \eqref{eq:pca_optimization}. Following the existing literature, the result is stated for real matrices, but it can be trivially adapted to the complex case as well.

\begin{theorem}[PCA]
    \label{theorem:pca}
    Let $\matX\in\mathbb{R}^{m\times n}$ be a centered data matrix, $\matH$ the $n\times n$ symmetric sample covariance matrix, i.e., $\matH=\matX^\top\matX$, $\|\matH\|\leq 1$, $k\in[n]$ a target rank, and $\epsilon\in(0,1)$ an accuracy parameter. Given $\matH$,
    we can compute a matrix $\matCtilde_k$ with $k$ columns such that $\|\matX-\matX\matCtilde_k\matCtilde_k^\top\| \leq (1+\epsilon)\|\matX-\matX\matC_k\matC_k^\top\|$,
    where $\matC_k\in\mathbb{R}^{n\times k}$ contains the top-$k$ (right) singular vectors of $\matX$ in
    \begin{align*}
        O\lpar T_{\MM}(n) \lpar 
            \log(\tfrac{n}{\sigma_{k+1}})\log(\tfrac{1}{\sigma_{k+1}})
            +
            \log(\tfrac{n}{\gap_k})\log(\tfrac{1}{\gap_k})
            +
            \log(\log(\tfrac{n}{\epsilon\sigma_{k+1}\gap_k}))
        \rpar\rpar
    \end{align*}
    arithmetic operations using
    $
        O\lpar
            \log(n)
            \lpar
                \log^4(\tfrac{n}{\epsilon\gap_k})
                +
                \log^4(\tfrac{n}{\sigma_{k+1}})
            \rpar
            +
            \log(\tfrac{1}{\epsilon\sigma_{k+1}})
        \rpar
    $
    bits of precision, with probability at least $1-O(1/n)$.
\end{theorem}

\subsection{Block-Krylov PCA} 
In some applications, the target dimension $k$ might be small, i.e., $k\ll n$. This condition has driven a whole area of research in so-called low-rank approximation algorithms for PCA \cite{frieze1998fast,sarlos2006improved, clarkson2009numerical,halko2011finding,martinsson2011randomized,musco2015randomized,boutsidis2014near,bhojanapalli2014tighter,allen2016lazysvd}. Such approaches are also suitable for kernel PCA, since they rely on matrix-vector products and therefore the kernel matrix does not need to be explicitly formed. 
The techniques from the previous section can be directly applied to obtain new bit complexity upper bounds for existing state-of-the-art algorithms, which are typically analyzed in exact arithmetic. They internally rely on the computation of the principal singular vectors of submatrices, which can be improved with our methods.
Specifically, in Appendix \ref{appendix:low_rank_pca} we summarize a floating point analysis of the Block-Krylov Iteration algorithm (see Algorithm \ref{alg:bkpca}), essentially, providing a matrix multiplication-type upper bound on the bit complexity with only a polylogarithmic dependence on the singular value gap. In a nutshell, we directly obtain the following result:
\begin{theorem}[Bit complexity analysis of Block-Krylov PCA]
    \label{theorem:block_krylov_pca}
        Let $\matX$ be a data matrix $\matX\in\mathbb{R}^{m\times n}$, $\|\matX\|\leq 1$, $k\in[n]$ a target rank, $\epsilon_{\PCA}\in(0,1)$ an accuracy parameter, and $q=\Theta\lpar \frac{\log(n)}{\sqrt{\epsilon_{\PCA}}}\rpar$. Let $T_{\MMX}(k)$ denote the complexity to stably multiply $\matX$ or $\matX^\top$ with a dense matrix with $k$ columns from the right (see Def. \ref{def:stable_mmx}).
        Using the Steps 1-6 that are detailed in Appendix \ref{appendix:block_krylov_pca} as a floating point implementation of Algorithm \ref{alg:bkpca}, we can compute a matrix $\matZtilde_k\in\mathbb{R}^{m\times k}$ that satisfies \begin{align*}
            \lnorm \matZtilde_k\matZtilde_k^\top -\matZ_k\matZ_k^\top \rnorm \leq O(\epsilon_{\PCA}),
        \end{align*}
        with high probability, where $\matZ_k$ is an approximate basis for the top-$k$ principal components of $\matX$, returned by Algorithm \ref{alg:bkpca} in exact arithmetic. The total cost is at most
        \begin{align*}
            O\lpar 
                qT_{\MMX}(k)
                \log(\tfrac{\kappa(\matK)}{\gap_k(\matM)})
                + 
                m(qk)^{\omega-1} 
                \log(\tfrac{1}{\gap_k(\matM)})
                +
                (qk)^\omega\polylog(\tfrac{qk}{\gap_k(\matM)})
            \rpar
        \end{align*}
        floating point operations, using
        $O\lpar 
            \polylog(\tfrac{mq\kappa(\matK)}{\epsilon_{\PCA}\gap_k})
        \rpar$
        bits of precision. $\matK,\matM$ are as in Alg. \ref{alg:bkpca}.
        \begin{proof}
            The full proof can be found in Thm. \ref{theorem:block_krylov_pca_appendix}, Appendix \ref{appendix:block_krylov_pca}. The main idea is to apply the counting query methodology to compute the condition number of the Block-Krylov matrix $\matK$, as well as the $k$-th spectral gap and the midpoint of the reduced matrix $\matM$ in Line 5 of Alg. \ref{alg:bkpca}. Thereafter, we can compute a spectral projector and an approximate basis for the top-$k$ singular vectors of $\matM$ using $\PURIFY$ and $\DEFLATE$, similar to the analysis of classical PCA in the previous section.
        \end{proof}
\end{theorem}

\section{Open problems}
We mention some open problems and interesting future directions. 
\begin{enumerate}[(i)]
    \item \textbf{Bit requirement of $\SGN$:} The major bottleneck for the bit requirements of our main algorithms comes from the $\SGN$ algorithm of \cite{banks2022pseudospectral}. An inverse-free Newton-Schultz iteration \cite{kenney1995matrix}, or the implicit repeated squaring of \cite{demmel2023generalized,schneider2023fast} can potentially give significant improvements.
    \item \textbf{Sparse algorithms:} In applications like DFT it commonly occurs that the matrices have special structure, i.e., they are banded and/or sparse. It remains open whether Problem \ref{problem:spectral_projector} can be provably solved faster than our reported results in finite precision for these special cases (recall that the tridiagonal QR algorithm requires $O(n^3)$ operations to return the eigenvectors). An end-to-end stability analysis of existing fast eigensolvers would be the place to start \cite{gu1995divide,vogel2016superfast}.
    \item \textbf{Distributed PCA:}
    The techniques for Block-Krylov PCA can be potentially applied to distributed or streaming PCA algorithms, which are also based on randomized low-rank approximations. E.g., in the distributed PCA algorithm of   \cite{boutsidis2016optimal}, it is straightforward to replace the SVD computation on the server by a counting query iteration. The full analysis of such an approach is left as future work.
\end{enumerate}
\section{Conclusion}
\label{section:conclusion}
In this work we provided an end-to-end analysis to approximate spectral projectors on $k$-dimensional invariant subspaces of Hermitian definite matrix pencils $(\matH,\matS)$ that require at most $O
\lpar
T_{\MM}(n)\polylog (n,\epsilon^{-1},\kappa(\matS),\gap_k^{-1})
\rpar
$ bit operations in the floating point model of computation. This is the first end-to-end analysis that improves the $\widetilde O(n^3)$ complexity of  classic eigensolvers for both the regular and the generalized case. To achieve this result we introduced a new method to approximate spectral gaps by querying the number of eigenvalues that are smaller than a threshold, and therefore completely avoid an explicit diagonalization of any matrix or pencil. This approach required proving that the Cholesky factorization can be stably computed in $O(T_{\MM}(n))$ floating point operations, a novel result \textit{per se}. Our results have direct implications on PCA problems, providing matrix multiplication type upper bounds for the bit complexity of classical and Block-Krylov PCA.

\section*{Acknowledgements}
We would like to thank Daniel Kressner, Nian Shao, Ryan Schneider, Nicolas Deutschmann, Nikhil Srivastava, Ilse Ipsen, Cameron Musco, David Woodruff, Uria Mor, and Haim Avron for helpful discussions.

\bibliographystyle{plain}

\appendix
\newpage
\section{Preliminaries}
\label{appendix:preliminaries}
In this section the model of computation is defined and basic linear algebra principles are summarized. Further details can be found in standard textbooks such as \cite{higham2002accuracy,golub2013matrix}.

\subsection{Forward and backward approximation}
\label{appendix:forward_and_backward_approximation}
In numerical analysis, a ``forward-error'' often measures the distance between the true solution to a problem and the solution returned by a numerical algorithm. A related notion is the so-called ``backward error'' or ``backward approximation.'' For more details we can refer to the standard textbook of Higham \cite{higham2002accuracy}.  In this case, the solution returned is the exact solution of a ``nearby problem,'' and the backward error quantifies the distance of the original problem to this nearby problem. In PCA, for example, a backward-approximation could be defined by $\matCtilde_k=\arg\min_{\matC^\top \matC=\matI_{k\times k}} \tr(\matC^\top\matHtilde\matC)$. The backward error would be given $\|\matH-\matHtilde\|$ for some norm. An algorithm is \textit{backward stable} if the backward error is always well-defined and bounded. A forward error type of approximation is often harder to achieve. A common rule-of-thumb states that
\begin{align*}
    \text{forward error} \lesssim \text{backward error} \times \text{condition number of the problem}.
\end{align*}
However, this does not generally hold for eigenvalue problems. For example, Proposition 1.1 in \cite{banks2022pseudospectral} can be used to transformed a backward approximate diagonalization error to a forward error for the eigenvectors, but the bound depends on the minimum eigenvalue gap between any eigenvalue pair. As the authors point in Remark 1.2, special treatment is needed in terms of the invariant subspaces in the presence of multiple eigenvalues, which are thoroughly analyzed in this work.

\subsection{Floating point model}
\label{appendix:floating_point}
We assume a standard floating point model of computation and borrow its axioms from \cite[Chapter 2]{higham2002accuracy}. 
There is a fixed number of bits to represent floating point numbers, specifically, one bit is reserved for the sign $s$, $p$ bits are used for the exponent $e$, and $t$ bits are used for the significand $m$. 

A real number $\alpha\in\mathbb{R}$ is rounded to a floating point number 
\begin{align*}
    \fl(\alpha) = s\times 2^{e-t} \times m.
\end{align*}
The sign $s$ is $+$ if the corresponding bit is one, and $-$ if the bit is zero. The exponent $e$ is stored as a binary number in the so-called \textit{biased form}, and its range is $e\in[-M,M]$, where $M=2^{p-1}$. The significand $m$ is an integer that satisfies $2^{t-1}\leq m \leq 2^t-1$, where the lower bound is enforced to ensure that the system is \textit{normalized}, i.e. the first bit of $m$ is always $1$.
We can therefore write $\fl(\alpha)$ in a more intuitive representation
\begin{align*}
    \fl(\alpha) = \pm 2^{e}\times \lpar \tfrac{m_1}{2} + \tfrac{m_2}{2^2} + \ldots + \tfrac{m_t}{2^t} \rpar,
\end{align*}
where the first bit $m_1$ of $m$ is always equal to one for normalized numbers. The range of normalized numbers is therefore $[2^{-M},2^{M}(2-2^{-t})]$. Numbers that are smaller than $2^{-M}$ are called \textit{subnormal} and they will be ignored for simplicity, since we can either add more bits in the exponent, or account for them in the failure probability when the numbers are random (the latter strategy is used for example in Lemma \ref{lemma:diagonal_gaussian_sampler}). Similarly, numbers that are larger than $2^{M}(2-2^{-t})$ are assumed to be numerically equal to infinity, denoted by $\FLINF$. 

From \cite[Theorem 2.2]{higham2002accuracy},
for all real numbers $\alpha$ in the normalized range it holds that \begin{align*}
    \fl(\alpha) = (1+\theta)\alpha,
\end{align*}
where $\theta\in\mathbb{R}$ satisfies $|\theta|\leq 2^{-t}:=\umach$, where $\umach$ is the \textit{machine precision}. Clearly, $t=O(\log(1/\umach))$, in which case we can always obtain a bound for the number of required bits of a numerical algorithm if we have an upper bound for the precision $\umach$. We will write the same for complex numbers which are represented as a pair of normalized floating point numbers.

The floating point implementation of each arithmetic operation $\odot \in\{+,-,\times,/\}$ also satisfies
\begin{align}
    \fl(\alpha\odot\beta) = (1+\theta)(\alpha\odot\beta),\quad |\theta|\leq\umach.
    \label{eq:elementwise_flop_errors}
\end{align}
Divisions and multiplications with $1$ and $2$ do not introduce errors (for the latter we simply increase/decrease the exponent).
We assume that we also have an implementation of $\sqrt{\cdot}$ such that $\fl(\sqrt{\alpha})=(1+\theta)\sqrt{\alpha}$ where $|\theta|\leq \umach$.
From \cite[Lemma 3.1]{higham2002accuracy}, we can bound products of errors as
\begin{align*}
    \prod_{i=1}^n (1+\theta_i)^{\rho_i} = 1+\eta_n,
\end{align*}
where $\rho_i=\pm 1$ and $|\eta_n|\leq \tfrac{n\umach}{1-n\umach}$.

The above can be extended also for complex arithmetic (see \cite[Lemma 3.5]{higham2002accuracy}), where the bound becomes $|\theta|\leq O(\umach)$, but we will ignore the constant prefactor for simplicity. 

Operations on matrices can be analyzed in a similar manner. Let $\otimes$ denote the element-wise multiplication between two matrices and $\oslash$ the element-wise division. The floating point representation of a matrix $\matA$ satisfies
\begin{align*}
    \fl(\matA) = \matA+\matDelta \otimes \matA,\quad |\matDelta_{i,j}|\leq \umach.
\end{align*}
It can be shown that $\|\matDelta\|\leq \umach\sqrt{n}\|\matA\|$. 

For any operation $\odot\in\{+,-,\otimes,\oslash\}$ and matrices $\matA$ and $\matB$ it holds that
\begin{align}
    \label{eq:matrix_flop_errors}
    \fl(\matA\odot \matB) = \matA \odot \matB +\matDelta\otimes(\matA \odot \matB), \quad |\matDelta_{i,j}|\leq \umach, \quad \|\matDelta\otimes(\matA \odot \matB)\|\leq \umach\sqrt{n}\|\matA\odot\matB\|.
\end{align}

\subsection{Spectral decomposition, pseudospectrum, and eigenvalue bounds}
\label{appendix:preliminaries_pseudospectrum}
We first recall the definition of the spectral decomposition of a diagonalizable matrix.
\begin{definition}[Spectrum and spectral decomposition]
    A matrix $\matA$ is diagonalizable if there exist invertible matrix $\matV$ and diagonal matrix $\matLambda$ such that $\matA=\matV\matLambda\matV^{-1}$. This is called the \textit{spectral decomposition} of $\matA$. The set $\Lambda(\matA)=\{\matLambda_{i,i} | i=1,\ldots,n\}$ is the spectrum of $\matA$.
\end{definition}

The spectral theorem states that Hermitian matrices (or, more generally, normal matrices) can be always diagonalized by unitary transformations.
\begin{theorem}[Spectral theorem]
    If $\matA\in\mathbb{C}^{n\times n}$ is Hermitian, then there exists orthogonal matrix $\matQ\in\mathbb{C}^{n\times n}$ such that $\matA=\matQ\matLambda\matQ^*$, where $\matLambda$ is a diagonal matrix with real diagonal elements.
\end{theorem}

In the sections that follow we need to bound the (forward) errors on the eigenvalues of matrices under perturbations that are introduced due to the finite precision arithmetic. Such bounds can be derived by the classic Bauer-Fike theorem the following (cf. \cite{bauer1960norms,eisenstat1998relative} for more details). 

There are many bounds in the literature describing the effect of perturbations on the eigenvalues. We summarize some classic results in the following proposition and refer to Bhatia's monograph for a detailed overview \cite{bhatia2007perturbation}.

\begin{fact}
    \label{fact:weyl_kahan}
    Let $\matA$ and $\matB$ be two $n\times n$ matrices. The following bounds are known between the eigenvalues $\lambda_1(\matA)\leq \lambda_2(\matA)\leq\ldots\leq\lambda_n(\matA)$  of $\matA$ and $\lambda_1(\matB)\leq\ldots \leq \lambda_n(\matB)$ of $\matB$:
    \begin{center}
    \small
    \begin{tabular}{c c  l l}
        $\matA$ & $\matB$ & bound & reference\\\hline
        Hermitian & Hermitian & $ |\lambda_i(\matA)-\lambda_i(\matB)|\leq \|\matA-\matB\|$ &  \cite{weyl1912asymptotische} \\
        & & & \\
        Hermitian & Non-Hermitian &  $|\lambda_i(\matA)-\lambda_i(\matB)|\leq O(\log(n))\|\matA-\matB\|$ & \cite{kahan1975spectra,pokrzywa1981spectra}\\
    \end{tabular}
    \end{center}
\end{fact}

Pseudospectral analysis is useful when the aforementioned bounds are not applicable.
\begin{definition}[Pseudospectrum]
\label{definition:pseudospectrum}
For some $\epsilon>0$, the $\epsilon$-pseudospectrum of a matrix $\matM\in \mathbb{C}^{n\times n}$ is defined as:
\begin{align*}
    \Lambda_\epsilon(\matM) &:= \left\{ \lambda\in \mathbb{C} : \lambda \in \Lambda(\matM + \errormatrix) \text{ for some }\|\errormatrix\| < \epsilon\right\} \\
    &= \left\{\lambda \in \mathbb{C} : \left\|(\lambda\matI - \matM)^{-1}\right\| > 1/\epsilon \right\}
\end{align*}
where $\Lambda(\matM)$ is the spectrum of $\matM$. 
\end{definition}
Recall some useful properties from the seminal book of Trefethen and Embree \cite{trefethen2005spectra}.

\begin{proposition}[Collective results from \cite{trefethen2005spectra}]
    \label{proposition:core_pseudospectrum_properties}
    Let $D(z,r)$ denote the open disk of radius $r$ in the complex plane centered at $z\in\mathbb{C}$, $\matM$ and $\errormatrix$ be two $n\times n$ matrices, and $\epsilon > \|\errormatrix\|$ be a positive real number. The following hold:
    \begin{enumerate}[(i)]
        \item $\Lambda_{\epsilon-\|\errormatrix\|}(\matM) \subseteq \Lambda_{\epsilon}(\matM+\errormatrix) \subseteq \Lambda_{\epsilon+\|\errormatrix\|}(\matM)$ (\cite[Thm 52.4]{trefethen2005spectra}),
        \item Any bounded connected component of $\Lambda_{\epsilon}(\matM)$ has a nonempty intersection with $\Lambda(\matM)$, i.e., it contains at least one eigenvalue (\cite[Thm 4.3]{trefethen2005spectra}),
        \item $\bigcup_i D(\lambda_i,\epsilon)\subseteq \Lambda_{\epsilon}(\matM) \subseteq \bigcup_iD(\lambda_i,\epsilon\keig(\matM))$ (\cite[Thms 4.3 and 52.2]{trefethen2005spectra}).
    \end{enumerate}
\end{proposition}

\subsection{Eigenvector condition number of definite pencils}
\label{appendix:properties_hpd_gep}
Some approximation bounds throughout the paper depend on the eigenvector condition number of the generalized eigenproblem \eqref{eq:gep_main}. For arbitrary GEPs a bound for this quantity might not always exist, but a straightforward bound exists for the Hermitian definite case. To obtain such a bound we define the eigenvector condition number of a diagonalizable matrix as follows:
\begin{align*}
\keig (A):=\inf_{
    \substack{
        \matV\matD\matV^{-1}=\matA\\
        \matD:\text{ diagonal}
    }
}
\|V\|\|V^{-1}\|.
\end{align*}

\begin{proposition}
\label{proposition:kv_from_ks}
Let $\matH\in\mathbb{H}^n$ and $\matS\in\mathbb{H}^n_{++}$, and consider the definite GEP
$
    \matH\matC=\matS\matC\geneigmatrix.
$
Then
\begin{align*}
    \keig (\matS^{-1}\matH)\leq\sqrt{\kappa(\matS)}.
\end{align*}
\begin{proof}
Since $\matS$ is Hermitian positive-definite, it can be written as $\matS=\matL\matL^*$ for some matrix $\matL$. Then we can transform the GEP to a Hermitian eigenproblem, specifically
\begin{align*}
    \matL^{-1}\matH\matL^{-*}\matL^*\matC=\matL^*\matC\geneigmatrix.
\end{align*}
Since $\matL\matL^*=\matS$ we have that $\|\matL\|^2=\|\matL\matL^*\|=\|\matS\|$ and similarly $\|\matL^{-1}\|^2=\|\matS^{-1}\|$, which means that $\kappa(\matL)=\sqrt{\kappa(\matS)}$.
Since $\matL^{-1}\matH\matL^{-*}$ is Hermitian, it can be diagonalized by a unitary matrix, i.e. there exists $\matChat$ such that $\|\matL^*\matChat\|=\|(\matL^*\matChat)^{-1}\|=1$ and $\matL^*\matChat$ diagonalizes $\matL^{-1}\matH\matL^{-*}$. For $\matChat$ we have that \begin{align*}\|\matChat\|=\|\matL^{-*}\matL^{*}\matChat\|\leq \|\matL^{-*}\|\|\matL^*\matChat\|=\|\matL^{-*}\|.\end{align*} Similarly, \begin{align*}\|\matChat^{-1}\|=\|\matChat^{-1}\matL^{-*}\matL^{*}\|\leq \|\matChat^{-1}\matL^{-*}\|\|\matL^*\|=\|(\matL^*\matC)^{-1}\|\|\matL^*\|=\|\matL^*\|.\end{align*}
Undoing the transformation, we can see that $\matChat$ also satisfies:
\begin{align*}
    \matH\matChat=\matS\matChat\geneigmatrix,
\end{align*}
i.e. $\matChat$ diagonalizes $\matS^{-1}\matH$ since $\matChat^{-1}\matS^{-1}\matH\matChat=\geneigmatrix$.
We conclude that $\kappa_{\matV} (\matS^{-1}\matH)\leq \kappa(\matChat)\leq \|\matL\|\|\matL^{-1}\|=\sqrt{\kappa(\matS)}$.
\end{proof}
\end{proposition}

\subsection{Imported floating point algorithms for fast linear algebra}
Before we dive into the details, we recall the concept of ``fast matrix multiplication'', pioneered by Strassen \cite{strassen1969gaussian}, who showed that two square matrices can be multiplied in $O(n^{\omega})$, where $\omega=\log_27\approx 2.807 < 3$, in real arithmetic. Since then, the \textit{matrix multiplication exponent} $\omega$ has been significantly reduced, the record to date being $\omega\leq 2.371552$ \cite{duan2023faster,williams2024new}.
In two seminal works \cite{demmel2007fastmm,demmel2007fastla}, Demmel, Dumitriu, Holtz, and Kleinberg, proved that \textit{any} fast matrix multiplication algorithm can be executed numerically stably in a floating point machine with almost the same arithmetic complexity as in real arithmetic. They also showed that other problems in Numerical Linear Algebra can be reduced to such matrix multiplications, including inversion, LU and QR factorizations, and solving linear systems of equations.  
Our algorithms build on the existing results for stable fast matrix multiplication \cite{demmel2007fastmm}, inversion \cite{demmel2007fastla}, as well as backward-approximate diagonalization and the matrix sign function \cite{banks2022pseudospectral}. We import the corresponding results from the aforementioned works in the following theorems.

\begin{theorem}[$\MM$, stable fast matrix multiplication \cite{demmel2007fastmm}]
\label{theorem:fast_mm}
For every $\eta>0$, there exists a fast matrix multiplication algorithm $\MM$ which takes as input two matrices $\matA,\matB\in\mathbb{C}^{n\times n}$ and a machine precision $\umach>0$ and returns $\matC\leftarrow \MM(\matA,\matB)$ such that
\begin{align*}
    \|\matC-\matA\matB\| \leq \mu_{\MM}(n)\cdot\umach\|\matA\|\|\matB\|,
\end{align*}
on floating point machine with precision $\umach$, where $\mu_{\MM}(n)=n^{c_{\eta}}$, for some constant $c_{\eta}$ independent of $n$. Such an algorithm is called $\mu_{\MM}(n)$-stable, and it requires $T_{\MM}(n)=O(n^{\omega+\eta})$ arithmetic operations, where $\omega$ is the exponent of matrix multiplication in real arithmetic.
\end{theorem}

\begin{theorem}[$\INV$, logarithmically-stable fast inversion \cite{demmel2007fastla}]
\label{theorem:fast_inv}
For every $\eta>0$, there exists a fast inversion algorithm $\INV$ which takes as input an invertible matrix $\matA\in\mathbb{C}^{n\times n}$ and a machine precision $\umach>0$ and returns $\matC\leftarrow \INV(\matA)$ such that
\begin{align*}
    \|\matC-\matA^{-1}\| \leq \mu_{\INV}(n)\cdot\umach\cdot\kappa(\matA)^{\cinv\log(n)}\|\matA^{-1}\|,
\end{align*}
on floating point machine with precision $\umach$, where $\mu_{\INV}(n)=O(n^{c_{\eta}+\log(10)})$, for some constants $\cinv\leq 8$ and $c_{\eta}$ independent of $n$. Such an algorithm is called  $(\mu_{\INV}(n),\cinv)$-stable, and it requires $T_{\INV}(n)=O(T_{\MM}(n))$ arithmetic operations, where $T_{\MM}(n)$ is the same as in Theorem \ref{theorem:fast_mm}.
\end{theorem}

\paragraph{Sign function and deflation.} The next result that we import is a floating point algorithm for the matrix sign function, which, for a diagonalizable matrix $\matA=\matV\matLambda\matV^{-1}$ such that $\Lambda(\matA)$ does not interesect with the imaginary axis, is defined as $\sgn(\matA)=\matV\sgn(\matLambda)\matV^{-1}$, where $\sgn(\matLambda)$ is a diagonal matrix and each diagonal entry contains the sign of the real part of the corresponding eigenvalue. To state the main result, we first recall the definition of the \textit{Circles of Apollonius}.
\begin{definition}[Circles of Apollonius, imported from Section 4.1 of \cite{banks2022pseudospectral}]
    Let $\alpha\in(0,1)$. The Circles of Apollonius $\mathsf{C}_{\alpha}$ are defined as $\mathsf{C}_{\alpha}=\mathsf{C}^+_{\alpha}\cup\mathsf{C}^-_{\alpha}$,
    \begin{align*}
        \mathsf{C}^+_{\alpha}=\{z\in\mathbb{C}: |m(z)|\leq\alpha\},
        \quad
        \mathsf{C}^+_{\alpha}=\{z\in\mathbb{C}: |m(z)|^{-1}\leq\alpha\},
    \end{align*}
    where $m(z)=\frac{1-z}{1+z}$ is the M\"obius transformation taking the right half-plane to the unit disk. The disk $\mathsf{C}^+_{\alpha}$ is centered at $\frac{1+\alpha^2}{1-\alpha^2}$ and has radius $\frac{2\alpha}{1-\alpha^2}$, and $\mathsf{C}^-_{\alpha}$ is its reflection with respect to the imaginary axis.
\end{definition}

\begin{theorem}[$\SGN$, imported Theorem 4.9 from \cite{banks2022pseudospectral}]
\label{theorem:sgn}
There is a deterministic algorithm $\SGN(\matA,\alpha,\eta,\epsilon)$ which takes as input a matrix $\matA\in\mathbb{C}^{n\times n}$, an accuracy parameter $\epsilon\in(0,1/12)$, and parameters $\eta\in(0,1)$,  $0<1-\alpha<1/100$, such that it is guaranteed that $\Lambda_{\eta}(\matA)\subset \mathsf{C}_{\alpha}$. The algorithm returns a matrix $\matStilde$ such that
\begin{align*}
    \|\matStilde-\sgn(\matA)\|\leq\epsilon,
\end{align*}
as long as the machine precision satisfies
\begin{align*}
    \umach\leq\umach_{\SGN}:=\frac{\alpha^{2^{N+1}(\cinv\log(n)+3)}}{\pinv(n)\sqrt{n}N},
\end{align*}
corresponding to at most
\begin{align*}
    O\lpar \log(n)\log^3(\tfrac{1}{1-\alpha})\lpar\log(\tfrac{1}{\epsilon})+\log(\tfrac{1}{\eta})\rpar \rpar
\end{align*}
bits of precision.
Here \begin{align*}
    N=\lceil 
    \log(\tfrac{1}{1-\alpha})+3\log(\log(\tfrac{1}{1-\alpha})) + \log(\log(\tfrac{1}{\epsilon\eta})) +7.79\rceil
\end{align*}
denotes the number of iterations that the algorithm executes. The arithmetic complexity is
\begin{align*}
    O(NT_{\MM}(n))).
\end{align*}
\end{theorem}

In PCA, we will need to be able to compute a basis for the column space of an approximate spectral projector. For this, we recall the following algorithm.

\begin{theorem}[$\DEFLATE$, imported Theorem 5.3 of \cite{banks2022pseudospectral}]
\label{theorem:deflate}
There exists a randomized algorithm $\matCtilde_k\leftarrow \DEFLATE(\projectormatrixtilde_k,k,\beta,\epsilon)$, which takes as input a matrix $\projectormatrixtilde_k$, a rank parameter $k\in[n]$, a parameter $\beta$ such that $\|\projectormatrixtilde_k-\projectormatrix_k\|\leq \beta$, for some projector matrix $\projectormatrix_k$ of rank-$k$, and a desired accuracy $\epsilon$, and returns a (complex) matrix $\matCtilde_k$ such that, there exists a matrix $\matC_k$ with $k$ columns that forms an orthonormal basis for $\im(\projectormatrix_k)$, and
\begin{align*}
    \|\matCtilde_k-\matC_k\|\leq \epsilon.
\end{align*}
The algorithm succeeds with probability at least $1-\tfrac{(20n)^3\sqrt{\beta}}{\epsilon^2}$. The number of arithmetic operations is at most $O(T_{\MM}(n))$, and it requires $O(\log(n)+\log(\tfrac{1}{\epsilon}))$ bits of precision. Internally, the algorithm generates $O(n^2)$ random numbers to form a $n\times n$ complex Ginibre matrix.\footnote{We highlight that our Gaussian sampling oracle in Definition \ref{definition:stable_normal_sampler} is slightly different than the one described in \cite{banks2022pseudospectral}, but it should be straightforward to adapt the analysis using either of the two definitions.}
\end{theorem}

\paragraph{Computing an orthonormal basis and the spectral norm.} 
Lastly, we recall the following two results. The first one is for the QR factorization from \cite{demmel2007fastla}, which we adapt suitably for our analysis, specifically, to compute an orthonormal basis for the column space of a rectangular matrix.

\begin{theorem}[Basis computation, follows from Section 4 in \cite{demmel2007fastla}]
\label{theorem:fast_qr}
Let $\matA\in\mathbb{R}^{m\times n},m\geq n$. There exists an algorithm $\matQtilde,\matRtilde=\QR(\matA)$, which returns a matrix $\matQtilde\in\mathbb{R}^{m\times n}$ and an upper triangular matrix $\matRtilde\in\mathbb{R}^{n\times n}$ in $O(mn^{\omega-1})$ floating point operations using $O(\log(\tfrac{n\kappa(\matA)}{\epsilon_{\QR}}))$ bits, where $\epsilon_{\QR}\in(0,1)$ is a given accuracy. The matrix $\matQtilde$ satisfies
\begin{align*}
    \|\matQ-\matQtilde\matPhi\|\leq \epsilon_{\QR},
\end{align*}
for some orthogonal matrix $\matPhi\in\mathbb{R}^{n\times n}$, where $\matQ$ has orthonormal columns and $\matA=\matQ\matR$ is the true economy-QR of $\matA$.

\begin{proof}
    We first scale $\matA'\leftarrow\matA/M$, where $M$ is the smallest power of two that is larger than $n\|\matA\|_{\max}$. This ensures that $\Omega(1/n)\leq\|\matA'\|\leq 1$. Since we scale by a power of $2$, then there are no floating point errors and the orthonormal basis from the QR factorization remains the same (only the upper triangular factor is scaled). 
    
    We now use the corresponding algorithm of \cite{demmel2007fastla} on $\matA'$. It returns three matrices: an upper triangular matrix $\matRtilde\in\mathbb{R}^{n\times n}$, a matrix $\matW\in\mathbb{R}^{m\times n}$ with $\|\matW\|\in O(n)$, and a matrix $\matY\in\mathbb{R}^{n\times m}$ with $\|\matY\|\in O(n)$. The matrix $\matPsi=\matI-\matW\matY^\top+\matE_{\matPsi}$ exactly satisfies $\matPsi\matPsi^\top=\matI$, and the error matrix satisfies $\|\matE_{\matPsi}\|\in O(\poly(n)\umach)$. 
    Moreover, $\matPsi\begin{pmatrix}
        \matRtilde\\
        0_{m-n\times n}
    \end{pmatrix}
    =\matAhat$, where $\matAhat=\matA'+\matE_{\matA}$ and $\|\matE_{\matA}\|\leq O(\poly(n)\umach)\|\matA'\|\leq O(\poly(n)\umach)$. Note that $\|\matRtilde\|\leq \|\matAhat\|\leq \|\matA'\|+\|\matE_{\matA}\|\leq 1+O(\poly(n)\umach)$. 
    The total cost of the algorithm is $O(mn^{\omega-1})$ floating point operations (ignoring the negligible term $\eta>0$ in the exponent).
    
    Using $\matW$ and $\matY$, we can construct an approximate basis $\matQtilde$ as follows. Let $\matY_{n}\in\mathbb{R}^{n\times n}$ contain the first $n$ columns of $\matY$. We can compute the matrix $\matQtilde=\begin{pmatrix}
        \matI_n\\
        0_{m-n\times n}
    \end{pmatrix}
    -
    \MM(\matW,\matY_n)
    $
    in $O(mn^{\omega-1})$ floating point operations by performing the multiplication in blocks of size $n\times n$. We can write $\matQtilde=\matQhat+\matE_{\matQhat}$ where $\matQhat$ contains the first $n$ columns of $\matPsi$ and $\|\matE_{\matQhat}\|\in O(\poly(n)\umach)$. 

    Note that $\matRtilde$ satisfies $\matQhat\matRtilde=\matA'+\matE_{\matA}$. Since $\matQhat$ has orthonormal columns, the singular values of $\matRtilde$ satisfy $\sigma_i(\matRtilde)=\sigma_i(\matQhat\matRtilde)=\sigma_i(\matA'+\matE_{\matA})\in\sigma_i(\matA')\pm\|\matE_{\matA}\|$, where the last comes from the stability of singular values which is a consequence of Weyl's inequality in Fact \ref{fact:weyl_kahan}. 
    Now, if $\|\matE_{\matA}\|\leq \tfrac{1}{4}\sigma_{\min}(\matA') $, which is achieved by setting $\umach\leq c\frac{1}{\poly(n)\kappa(\matA)}$ for some constant $c$, then $\matRtilde$ is full rank, and it holds that $\|\matRtilde^{-1}\|=1/\sigma_{\min}(\matRtilde)\leq \frac{4}{3\sigma_{\min}(\matA')}
    \leq
    O(n)\kappa(\matA)$. 
    
    Since $\matRtilde$ remains full rank, we have that $\matQhat\matRtilde=\matAhat\Rightarrow \matQhat=\matA'\matRtilde^{-1}+\matE_{\matA}\matRtilde^{-1}$. If we write the true economy-QR of $\matA'$ as $\matA'=\matQ\matR$, then we can see that $\|\matR\matRtilde^{-1}\| = \|\matQ\matR\matRtilde^{-1}\|=\|\matA\matRtilde^{-1}\|=\|\matQhat-\matE_{\matA}\matRtilde^{-1}\|$. This implies that $\|\matR\matRtilde^{-1}\|\in [1- \|\matE_{\matA}\matRtilde^{-1}\|, 1+\|\matE_{\matA}\matRtilde^{-1}\|]$. With similar arguments we obtain 
    \begin{align*}
        \|\matRtilde\matR^{-1}\|\in[1-\|\matE_{\matA}\matR^{-1}\|,1+\|\matE_{\matA}\matR^{-1}\|].
    \end{align*}
    We previously argued that $\|\matR\|,\|\matRtilde\|\leq O(1)$, and $\|\matR^{-1}\|,\|\matRtilde^{-1}\|\leq O(n)\kappa(\matA)$. Thus, if we further enforce $\|\matE_{\matA}\|\leq \epsilon_1c_1\frac{1}{\poly(n)\kappa(\matA)}$ for some constant $c_1$, then
    $\max\{\|\matE_{\matA}\matRtilde^{-1}\|,\|\matE_{\matA}\matR^{-1}\|\}\leq \epsilon_{1}$, for some $\epsilon_1\in(0,1/4)$, which means that all the singular values of $\matR\matRtilde^{-1}$ are inside the interval $[1-2\epsilon_1,1+2\epsilon_1]$. 
    In other words, $\matR\matRtilde^{-1}$ is approximately orthogonal. Indeed, we can write $\matR\matRtilde^{-1}=\matPhi+\matE_{\Phi}$ where $\matPhi\matPhi^\top=\matI$ and all the singular values of $\matE_{\matPhi}$ are inside $[-2\epsilon_1,2\epsilon_1]$. In particular, $\matPhi=\matU\matV^\top$, where $\matU,\matV$ come from the SVD of $\matR\matRtilde^{-1}=\matU\matSigma\matV^\top$.
    
    We can now go back to $\matQtilde$. From the above, $\matQhat$ can be written as
    \begin{align*}
        \matQhat=\matAhat\matRtilde^{-1}+\matE_{\matA}\matRtilde^{-1} 
        =
        \matQ\matPhi+\matQ\matE_{\matPhi}+\matE_{\matA}\matRtilde^{-1},
    \end{align*}
    which ultimately gives
    \begin{align*}
        \matQtilde
        =
        \matQhat + \matE_{\matQhat}
        =
        \matQ\matPhi+\matQ\matE_{\matPhi}+\matE_{\matA}\matRtilde^{-1} + \matE_{\matQhat}.
    \end{align*}
    This means that $\matQtilde$ is just a rotation of the true $\matQ$ plus some additive error terms that we can control with $\umach$. In particular,
    \begin{align*}
        \|\matQtilde-\matQ\matPhi\|
        &\leq
        \|\matQ\matE_{\matPhi}+\matE_{\matA}\matRtilde^{-1} + \matE_{\matQhat}\|
        \\
        &\leq
        \epsilon_1+O(\poly(n)\umach)\|\matRtilde^{-1}\| +O(\poly(n)\umach)
        \\
        &\leq
        O(\epsilon_1),
    \end{align*}
    where in the last we applied the required bound for $\umach\leq \epsilon_1c_1\tfrac{1}{\poly(n)\kappa(\matA)}$.

    Combining everything, we conclude that if the machine precision satisfies
    \begin{align*}
        \umach \leq \epsilon_{\QR}\frac{1}{\poly(n)\kappa(\matA)},
    \end{align*}
    then we can compute a matrix $\matQtilde$ in $O(mn^{\omega-1})$ floating point operations such that there exists orthogonal matrix $\matPhi$ satisfying
    \begin{align*}
        \|\matQtilde-\matQ\matPhi\|\leq \epsilon_{\QR},
    \end{align*}
    using $O(\log(n\kappa(\matA)/\epsilon_{\QR}))$ bits of precision. 
\end{proof}
\end{theorem}

The second result is for the computation of the spectral norm of the matrix from \cite{musco2018stability} using the Lanczos algorithm.
\begin{theorem}[Imported variant of Theorem 18 from the full version of \cite{musco2018stability}]
\label{theorem:alg_norm}
Let $\matA\in\mathbb{R}^{m\times n}$ and $\delta\in(0,1/2)$ a failure probability parameter. We can compute a vector $\vecy$ such that, with probability at least $1-\delta$,
\begin{align*}
    \tfrac{9}{10} \|\matA\| \leq \tfrac{\|\matA\vecy\|}{\|\vecy\|} \leq \|\matA\|,
\end{align*}
in $O\lpar mn \log(n)\log(1/\delta) \rpar$ floating point operations using $O(\log(n))$ bits of precision. Internally the algorithm generates $O(n\log(1/\delta))$ random bits.
\end{theorem}
We can obtain the following corollary.
\begin{corollary}[Spectral norm]
    \label{corollary:spectral_norm}
    Let $\matB\in\mathbb{C}^{m\times n}$. We can compute a value $\widetilde\Sigma\in\Theta(\|\matB\|)$ with probability at least $1-2\delta$ in $O\lpar mn\log(n)\log(1/\delta)\rpar$ floating point operations using $O(\log(n))$ bits of precision.
    \begin{proof}
        Let $\matR$ be the real part of $\matB$ and $\matZ$ be the imaginary part. We can approximate $R\in[\tfrac{9}{10}\|\matR\|, \|\matR\|]$ and $Z\in[\tfrac{9}{10}\|\matZ\|,\|\matZ\|]$ using the algorithm of Theorem \ref{theorem:alg_norm}. We then set $\widetilde \Sigma = R+Z\in[0.9(\|\matR\|+\|\matZ\|),\|\matR\|+\|\matZ\|]$. Using the triangle inequality we have that
        \begin{align*}
            0.9\|\matB\| \leq 0.9(\|\matR\|+\|\matZ\|) \leq \widetilde\Sigma \leq \|\matR\|+\|\matZ\| \leq 2\|\matB\|,
        \end{align*}
        or, in other words, $\widetilde\Sigma \in[0.9\|\matB\|,2\|\matB\|]$.
    \end{proof}
\end{corollary}

\subsection{Symmetrization}
There are certain cases where it is crucial to ensure that the result of a floating point multiplication or inversion remains Hermitian. The following proposition states that we can always ``symmetrize'' a floating point matrix with small additional errors. 

\begin{proposition}[$\HERM$]
\label{proposition:floating_point_symmetrization}
Let $\matC$ be a Hermitian matrix and $\matC'=\matC+\errormatrix$ for some (non-Hermitian) matrix $\errormatrix$, such that $\|\matC-\matC'\|=\|\errormatrix\|$. Consider the matrix $\HERM(\matC')$, where $\HERM(\matA)$ is an algorithm that replaces the strictly lower triangular part of $\matA$ with the strictly upper triangular part of $\matA$.
Then $\HERM(\matC')$ is Hermitian and it holds that:
\begin{align*}
    \|\matC-\HERM(\matC')\|
        &\leq 
   c_{\HERM}\log(n)\cdot \|\errormatrix\|,
\end{align*}
for some constant $c_{\HERM}$.
\begin{proof}
    For any matrix $\matA$, the upper triangular part of $\matA$, denoted as $\Delta_U(\matA)$, satisfies the following inequality:
    \begin{align*}
        \|\Delta_U(\matA)\| \leq c \log(n)\cdot \|\matA\|,
    \end{align*}
    for some constant $c$.
    The same holds for the lower triangular part $\Delta_L(\matA)$. For the diagonal part $\diag(\matA))$
    it holds that $\|\diag(\matA)\|\leq\|\matA\|$ (see \cite{bhatia2000pinching} for proofs). Let $\matC''=\HERM(\matC')$. By the definition of $\HERM(\cdot)$, the matrix $\matC''$ can be written as $\matC''=\Delta_U(\matC')+\Delta_U(\matC')^* - \diag(\matC')=\matC + \left(\Delta_U(\errormatrix)+\Delta_U(\errormatrix)^* - \diag(\errormatrix)\right)$. Therefore
    \begin{align*}
        \| \matC''-\matC \| 
        &= 
        \|\Delta_U(\errormatrix)+\Delta_U(\errormatrix)^* - \diag(\errormatrix)\|
        \leq 
        2c\log (n) \|\errormatrix\| + \|\errormatrix\| 
        \leq 
        c_{\HERM}\log(n)\|\errormatrix\|,
    \end{align*}
    for some constant $c_{\HERM}$.
\end{proof}

\end{proposition}

We can directly use this to derive bounds for matrix multiplication where the result is a Hermitian matrix and for Hermitian matrix inversion.
\begin{corollary}[Symmetrized matrix multiplication]
\label{corollary:symmetrized_matrix_multiplication}
Let $\matA$ and $\matB$ be two matrices and $\matC=\matA\matB$ be a Hermitian matrix. If $\MM$ is a $\pmult$-stable matrix multiplication algorithm then the matrix 
\begin{align*}
    \matC'=\HERM(\MM(\matA,\matB))
\end{align*} 
is Hermitian and it satisfies
\begin{align*}
    \|\matC'-\matC\| \leq  c_{\HERM}\log(n)\cdot \umach\|\matA\|\|\matB\|\cdot \pmult(n).
\end{align*}
\begin{proof}
    Straightforward combination of Theorem \ref{theorem:fast_mm} and Proposition \ref{proposition:floating_point_symmetrization}.
\end{proof}
\end{corollary}

\begin{corollary}[Symmetrized inversion]
\label{corollary:symmetrized_matrix_inversion}
Let $\matA$ be an invertible Hermitian matrix. If $\INV$ is a $(\pinv,\cinv)$-stable inversion algorithm then the matrix 
\begin{align*}
    \matC=\HERM(\INV(\matA))
\end{align*} 
is Hermitian and it satisfies
\begin{align*}
    \|\matC-\matA^{-1}\| \leq  c_{\HERM}\log(n) \cdot \pinv (n)\cdot \umach \cdot \kappa(\matA)^{\cinv \log n}\|\matA^{-1}\|.
\end{align*}
In addition, if $\umach$ satisfies
\begin{align*}
    \umach \leq \epsilon\frac{1}{c_{\HERM}\log(n)\cdot\pinv(n)\cdot\kappa(\matA)^{c_{\INV}\log(n)+1}}
\end{align*}
for some $\epsilon\in(0,1/2)$, then both of the following hold:
\begin{align*}
    \|\matC-\matA^{-1}\|\leq \epsilon \|\matA^{-1}\|,\\
    \frac{1}{2}\kappa(\matA)\leq \kappa(\matC)
    \leq 2\kappa(\matA).
\end{align*}
\begin{proof}
The first part is straightforward combination of Theorem \ref{theorem:fast_inv} and Proposition \ref{proposition:floating_point_symmetrization}. For the second part, we directly bound
\begin{align*}
(1-\epsilon)\|\matA^{-1}\|
\leq
(1-\epsilon\tfrac{1}{\kappa(\matA)}) \|\matA^{-1}\|
&\leq \|\matC\| 
\leq
(1+\epsilon\tfrac{1}{\kappa(\matA)})\|\matA^{-1}\|
\leq (1+\epsilon)\|\matA^{-1}\|
\\
\Rightarrow 
\tfrac{1}{2}\|\matA^{-1}\| &\leq \|\matC\| \leq 2\|\matA^{-1}\|.
\end{align*}
Note that $\|\matC-\matA^{-1}\|\leq \epsilon\frac{1}{\kappa(\matA)}\|\matA^{-1}\|=\epsilon\frac{1}{\|\matA\|}=\epsilon\sigma_{\min}(\matA^{-1})$.
Since $\matC$ and $\matA^{-1}$ are both Hermitian then from Weyl's inequality
\begin{align*}
    (1-\epsilon)\sigma_{\min}(\matA^{-1})
    &\leq 
    \sigma_{\min}(\matC) \leq (1+\epsilon)\sigma_{\min}(\matA^{-1})\\
    \Rightarrow
    \tfrac{1}{2}\sigma_{\min}(\matA^{-1})
    &\leq 
    \sigma_{\min}(\matC) \leq 
    2\sigma_{\min}(\matA^{-1})
\end{align*}
where we conclude that
\begin{align*}
    \tfrac{1}{4}\kappa(\matA) \leq \kappa(\matC) \leq 4\kappa(\matA).
\end{align*}
\end{proof}
\end{corollary}

\begin{remark} 
    Throughout this paper, to simplify our bounds, we assume that \begin{align*}1 \leq \mu_\MM(n), \pinv (n), \cinv\log (n), c_{\HERM}\log(n).\end{align*}
    Moreover, we define a global upper bound $\mu$ such that
    \begin{align*}
        \pmult(n), c_{\HERM}\log(n)\pmult(n),\pinv(n)\leq  \mu(n) \leq O(n^{c_{\eta}'}),
    \end{align*}
    where, from Theorem \ref{theorem:fast_mm}, $c_{\eta}'$ is a constant that does not depend on $n$.
    \label{remark:simplification_mu}
\end{remark}

\subsection{Existing algorithms for invariant subspaces and the (generalized) eigenproblem}
\label{appendix:related_work}

We give a brief overview of some results related to the computation of spectral projectors, eigenvectors, and invariant subspaces for the Hermitian definite generalized eigenproblem. Many more details can be found in standard textbooks and references therein \cite{demmel1997applied,saad2011numerical,golub2013matrix,trefethen2005spectra,parlett1998symmetric}, as well as the recent overview of \cite{srivastava2023complexity} for the complexity of the eigenproblem. For further reading we can highlight some influential works from the enormous bibliography on eigenvalue algorithms and perturbation bounds
\cite{demmel1987computing,
demmel1987three,
bai1991direct,
stewart1972sensitivity,
stewart1973error,
stewart1977perturbation,
stewart1979pertubation,
ipsen1997computing,
nakatsukasa2012condition,
bai1998using,
benner1997new,
ipsen2000absolute,
jia2001analysis,
davies2001analysis,
adachi2017solving,
paige1971computation,
paige1976error}. 

The prominent classic method to solve the Hermitian (generalized or regular) eigenproblem is to use numerically stable Householder transformations to reduce a Hermitian matrix to tridiagonal form, and then apply a tridiagonal shifted QR algorithm to diagonalize the tridiagonal matrix, which gives both the eigenvalues and the eigenvectors. The paramount QR algorithm was originally proposed in \cite{francis1961qr,francis1962qr,kublanovskaya1962some}, its convergence for the symmetric case was analyzed in \cite{wilkinson1968global,dekker1971shifted} in exact arithmetic, and the non-symmetric was recently analyzed in floating point in \cite{banks2021global1,banks2022global2,banks2022global3}. Other classical algorithms for computing invariant subspaces include \cite{malyshev1989computing,malyshev1993parallel,bai1997inverse,bai1998using}.

With the aforementioned procedure, the floating point complexity is about $\widetilde O(n^3)$ floating point operations, potentially up to $\polylog(1/\epsilon,1/\gap_k)$  factors. It is known that for Hermitian definite pencils $(\matH,\matS)$ one can extend this procedure such that, simultanteously, $\matH$ will become (Hermitian) tridiagonal and $\matS$ the identity \cite{crawford1973reduction}. The Hermitian tridiagonal matrix can be further reduced with similarity transformations to a symmetric tridiagonal matrix, in which case one can apply again tridiagonal symmetric QR. Using standard perturbation bounds as the ones used in this work one can potentially obtain provable forward errors invariant subspaces. 

While there exist works that overcome the $\widetilde O(n^3)$ worst-case complexity barrier for some computations related to the eigenproblem, none of them provide end-to-end forward errors for individual invariant subspaces. \cite{pan1999complexity} showed that the eigenvalues of a matrix can be computed in $O(n^\omega)$ arithmetic operations, however, it  becomes $O(n^{\omega+1})$ boolean operations in rational arithmetic (up to some other omitted factors), and therefore slower than standard $\widetilde O(n^3)$ floating point eigensolvers. \cite{benor2018quasi} proposed a quantum-inspired method for approximate diagonalization in a backward-error sense, with $O(n^{\omega+1})$ bit complexity. \cite{banks2022pseudospectral} proved that backward-approximate diagonalization can be solved in $\widetilde O(n^\omega)$ bit complexity in floating point by using smoothed analysis, improving both the $\widetilde O(n^3)$ classic algorithms for the Hermitian case and the $O(\poly(n/\epsilon))$ algorithm of \cite{armentano2018stable} for the non-Hermitian case. This was extended for the generalized eigenproblem case in \cite{demmel2023generalized}. As already mentioned in the introduction, to obtain forward errors for eigenvectors from the backward-approximate solution, the corresponding bounds that are reported in \cite{banks2022pseudospectral,demmel2023generalized} require the existence of a minimum eigenvalue gap in the original matrix, while the analysis for invariant subspaces was left as an open problem. Our work completes this analysis, and takes a step even further, by showing that (at least based on existing results), explicit diagonalization is redundant for approximating individual invariant subspaces. 
\section{Proof of Proposition \ref{proposition:alg_purify}}
\label{appendix:alg_purify}
\begin{algorithm}[htb]
\centering
\caption{$\PURIFY$.}
\label{alg:purify}
\begin{algorithmic}[1]
\setstretch{0.8}
\Statex \begin{align*}
    \PURIFY
\end{align*}
\small
\Input Hermitian definite pencil $\matH\in\mathbb{H}^{n}$, $\matS\in\mathbb{H}^n_{++}$, approximate condition  number $\widetilde\kappa$ of $\matS$, approximate values $\widetilde\mu_k$, $\widetilde\gap_k$, accuracy $\epsilon\in(0,1/12)$.
\Require $\|\matH\|\leq 1$, $\|\matS^{-1}\|\leq 1$, $\widetilde\kappa\in\Theta(\kappa(\matS))$, $\widetilde\mu_k\in\mu_k\pm\tfrac{1}{8}\gap_k$,  and $\widetilde\gap_k\in (1\pm\tfrac{1}{8})\gap_k$, where $\mu_k=\frac{\lambda_k+\lambda_{k+1}}{2}$ and $\gap_k=\lambda_{k}-\lambda_{k+1}$ for some $k\in[n-1]$.
\setstretch{1.5}
\Algorithm $\matPtilde \leftarrow\PURIFY(\matH,\matS,\widetilde\mu_k,\widetilde\gap_k,\widetilde\kappa,\epsilon)$.
\State $\matS_{\INV}\leftarrow\INV(\matS)$. \Comment{$\matS_{
INV}=\matS^{-1}+\errormatrix_1^{\INV}$}
\State $\matHtilde \leftarrow \MM(\matS_{\INV},\matH)$. \Comment{$\matHtilde=\matS_{\INV}\matH+\errormatrix_2^{\MM}$}
\State $\matMtilde \leftarrow \widetilde\mu_k-\matHtilde+\errormatrix_3^{(-)}$.
\State $\matCtilde\leftarrow 
    \SGN\lpar
        \matMtilde,\tfrac{\epsilon}{16\cdot199},\tfrac{\widetilde\gap_k}{32}
    \rpar$.
    \Comment{$\matCtilde = \sgn(\matMtilde)+\errormatrix_4^{\SGN}$}
\State $\projectormatrixtilde\leftarrow \frac{1}{2}\lpar 1+\matCtilde\rpar + \errormatrix_5^{(+)}$
\State \Return $\projectormatrixtilde$.
\Output Approximate spectral projector $\projectormatrixtilde$.
\Ensure $\|\projectormatrixtilde-\projectormatrix\|\leq \epsilon$.
\end{algorithmic}
\end{algorithm}

For the analysis of Algorithm \ref{alg:purify} we first recall that, in exact arithmetic, small perturbations do not have a severe influence in the sign function of a matrix. 

\begin{lemma}
    \label{lemma:sign_function_under_perturbation}
    Let $\matH\in\mathbb{H}^n$, $\matS\in\mathbb{H}^{n}_{++}$, where $\mu$ is a scalar, and $\matM=\mu-\matS^{-1}\matH$ and $\matMtilde=\mu-\matS^{-1}\matH+\errormatrix$.
    If 
    $
        \|\errormatrix\|\leq \epsilon \tfrac{|\lambda_{\min}(\matM)|^2\pi}{128\keig(\matM)}
    $ for some $\epsilon\in(0,1)$, where $|\lambda_{\min}(\matM)|$ is the smallest eigenvalue of $\matM$ in absolute value,
    then $\|\sgn(\matM)-\sgn(\matMtilde)\|\leq \epsilon$.
    \begin{proof}
        The proof uses standard techniques from holomorphic functional calculus and the properties of the pseudospectrum, but it is stated for completeness. A similar proof, for example, can be found in \cite{bai1998using} for general matrices $\matA$ (i.e. in that proof $\matA$ can have complex eigenvalues).
        
        Recall that for a matrix $\matA$ that has no eigenvalues on the imaginary axis, it holds that 
        \begin{align*}
            \sgn(\matA) = \matP_+(\matA) - \matP_-(\matA) ,
        \end{align*}
        where $\matP_-(\matA)$ is the spectral projector on the subspace spanned by the eigenvectors corresponding to eigenvalues with negative real part, and $\matP_+(\matA)$ is the corresponding spectral projector to the positive halfplane. 
        Then
        \begin{align*}
            \|\sgn(\matM)-\sgn(\matMtilde)\|
            &=
            \|\matP_+(\matM)
            +
            \matP_-(\matM) 
            - 
            \matP_+(\matM+\errormatrix)
            -
            \matP_-(\matM+\errormatrix)\|
            \\
            &\leq
            \|
                \matP_+(\matM)
                - 
                \matP_+(\matM+\errormatrix)
            \|
            +
            \|
                \matP_-(\matM)
                -
                \matP_-(\matM+\errormatrix)
            \|.
        \end{align*}

        $\matS^{-1}\matH$ has real eigenvalues because it is similar to a Hermitian matrix, i.e. $\Lambda(\matS^{-1}\matH)=\Lambda(\matL^*\matH\matL)$ where $\matL$ is the lower triangular Cholesky factor of $\matS^{-1}$. 
        Thus $\matM$ has also real eigenvalues. From Proposition \ref{proposition:core_pseudospectrum_properties} (i) and (iii)
        \begin{align*}
            \Lambda_{\|\errormatrix\|}(\matMtilde) \subseteq \Lambda_{2\|\errormatrix\|}(\matM) \subseteq \bigcup_i D\bigg(
                \lambda_i(\matM), 2\|\errormatrix\|\keig(\matM)
            \bigg).
        \end{align*}
        If $\|\errormatrix\|\leq \epsilon_1\frac{|\lambda_{\min}(\matM)|}{4\keig(\matM)}$ for some $\epsilon_1\in(0,1)$ then $|\lambda_{\min}(\matMtilde)|\geq |\lambda_{\min}(\matM)|/2$ and $|\lambda_{\max}(\matMtilde)|\leq |\lambda_{\max}(\matM)|+|\lambda_{\min}(\matM)|/2$.
        
        Let $R$ be a rectangle whose bottom-left corner is located at $-i|\lambda_{\min}(\matM)|$ and its top-right corner at $2+|\lambda_{\min}(\matM)|+i|\lambda_{\min}(\matM)|$. Any point on the boundary of $R$ has a distance of at least $|\lambda_{\min}(\matM)|$ from $\Lambda(\matM)$ and at least at least $|\lambda_{\min}(\matM)|/2$ from $\Lambda(\matMtilde)$. From holomorphic functional calculus we have that
        \begin{align*}
            \matP_+(\matA) = \oint_{\partial R}(z-\matM)^{-1}dz,
        \end{align*}
        where in this case $\partial R$ denotes a positively oriented rectifiable curve over the boundary of the rectangle.
        Using the resolvent identity $(z-\matM)^{-1}-(z-\matM')^{-1}=(z-\matM)^{-1}(\matM-\matM')(z-\matM')^{-1}$, for $\matP_+$ we have
        \begingroup
        \allowdisplaybreaks
        \begin{align*}
            \|
                \matP_+(\matM)
                - 
                \matP_+(\matM+\errormatrix)
            \|
            &=
            \left\|
                \frac{1}{2\pi i}
                \oint_{\partial R}(z-\matM)^{-1}dz
                - 
                \frac{1}{2\pi i}
                \oint_{\partial R}(z-\matM+\errormatrix)^{-1}dz
            \right\|
            \\
            &=
            \left\|
                \frac{1}{2\pi i}
                \oint_{\partial R}(z-\matM)^{-1}
                - 
                (z-\matM+\errormatrix)^{-1}dz
            \right\|
            \\
            &=
            \left\|
                \frac{1}{2\pi i}
                \oint_{\partial R}
                (z-\matM)^{-1}
                (z-\matM-(z-\matM+\errormatrix))
                (z-\matM+\errormatrix)^{-1}dz
            \right\|
            \\
            &=
            \left\|
                \frac{1}{2\pi i}
                \oint_{\partial R}
                (z-\matM)^{-1}
                \errormatrix
                (z-\matM+\errormatrix)^{-1}dz
            \right\|
            \\
            &\leq
            \frac{1}{2\pi}
            \oint_{\partial R}
                \left\|
                (z-\matM)^{-1}
                \errormatrix
                (z-\matM+\errormatrix)^{-1}
                \right\|
            dz
            \\
            &\leq
            \frac{1}{2\pi}
            \length(R)
            \cdot
            \|\errormatrix\|
            \cdot
            \sup_{z\in\partial R}\|(z-\matM)^{-1}\|
            \sup_{z\in\partial R}\|(z-\matM+\errormatrix)^{-1}\| 
            \\
            &\leq
            \frac{1}{2\pi}
            (6|\lambda_{\min}(\matM)|+4)
            \cdot 
            \epsilon_1\frac{|\lambda_{\min}(\matM)|}{4\keig(\matM)}
            \cdot
            \frac{1}{|\lambda_{\min}(\matM)|}
            \cdot 
            \frac{2}{|\lambda_{\min}(\matM)|}
            \\
            &\leq
            \epsilon_1\frac{4}{\pi\keig(\matM)|\lambda_{\min}(\matM)|}
            \\
        \end{align*}
        \endgroup
        Since $|\lambda_{\min}(\matM)|\leq2$, then if we set $\epsilon_1\leq \epsilon \frac{|\lambda_{\min}(\matM)|\pi}{16} $ (which satisfies $\epsilon_1< 1$), where $\epsilon$ is the desired final accuracy, this gives
        \begin{align*}
            \|
                \matP_+(\matM)
                - 
                \matP_+(\matM+\errormatrix)
            \|
            \leq \epsilon/2.
        \end{align*}
        The same can be obtained for $\matP_-$. Putting everything together, if 
        $\|\errormatrix\|\leq \epsilon \frac{|\lambda_{\min}(\matM)|^2\pi}{128\keig(\matM)}$ for some $\epsilon\in(0,1)$ ensures that $\lnorm \sgn(\matM+\errormatrix)-\sgn(\matM) \rnorm \leq \epsilon$. From Proposition \ref{proposition:kv_from_ks} $\keig(\matM)\leq \sqrt{\kappa(\matS)}$, which gives the final bound.
    \end{proof}
\end{lemma}

We can now prove Proposition \ref{proposition:alg_purify}, which we restate for readability.
\begin{proposition}[Restatement of Proposition \ref{proposition:alg_purify}]
    \label{proposition:alg_purify_appendix}
    Let $\matH\in\mathbb{H}^n$ with $\|\matH\|\leq 1$,
    $\matS\in\mathbb{H}^{n}_{++}$ with $\|\matS^{-1}\|\leq 1$, $k\in[n-1]$ and $\epsilon\in(0,1).$ Let $\mu_k=\frac{\lambda_k+\lambda_{k+1}}{2}$ and $\gap_k=\lambda_k-\lambda_{k+1}$, where $\lambda_1\leq\ldots\leq\lambda_n$ are the generalized eigenvalues of the Hermitian definite pencil $(\matH,\matS)$ and assume that we want to compute $\projectormatrix_k$ which is the true spectral projector associated with the $k$ smallest eigenvalues. If we have access to 
    \begin{align*}
        \widetilde\mu_k\in\mu_k\pm\tfrac{1}{8}
        \gap_k
        \quad
        \widetilde\gap_k\in(1\pm\tfrac{1}{8})\gap_k,
        \quad
        \widetilde\kappa \in [\kappa(\matS),C\kappa(\matS)],
    \end{align*} for some constant $C>1$, then
    Algorithm \ref{alg:purify} computes  $\projectormatrixtilde_k\leftarrow \PURIFY(\matH,\matS,\widetilde\mu_k,\widetilde\gap_k,\widetilde\kappa,\epsilon)$ such that
    $
        \|\projectormatrixtilde_k-\projectormatrix_k\|\leq \epsilon,
    $
    in 
    $
    O
    \lpar
        T_{\MM}(n) \lpar
            \log(\tfrac{1}{\gap_k})
            +
            \log(\log(\tfrac{\kappa(\matS)}{\epsilon\gap_k}))
        \rpar
    \rpar
    $
    floating point operations using
    $
        O\lpar
            \log(n)\log^3(\tfrac{1}{\gap_k})\log(\tfrac{\kappa(\matS)}{\epsilon\gap_k})
        \rpar
    $
    bits of precision.
    \begin{proof}

        Using the notation of Algorithm \ref{alg:purify}, we have that $\matHtilde=\matS^{-1}\matH+\errormatrix_1^{\INV}\matH+\errormatrix_2^{\MM}$.
        Let us initially set 
        \begin{align}
            \umach 
            =
            \epsilon_0\frac{1}{\mu(n)\widetilde\kappa^{\cinv\log(n)}}
            \leq 
            \epsilon_0\frac{1}{\mu(n)\kappa(\matS)^{\cinv\log(n)}},
            \label{equation:alg_purify_proof_umach_bound_1}
        \end{align}
        for some $\epsilon_0\in(0,1/8)$ (to be determined later).
        From Theorems \ref{theorem:fast_inv} and \ref{theorem:fast_mm}, and from the assumption that $\|\matH\|,\|\matS^{-1}\|\leq 1$ we have that \begin{align*}
            \|\errormatrix_1^{\INV}\| &\leq \umach \pinv(n) \kappa(\matS)^{\cinv\log(n)}\|\matS^{-1}\|
            \leq \epsilon_0,
            \\
            \|\errormatrix_2^{\MM}\| &\leq \umach \pmult(n)\|\matH\| \|\matS_{\INV}\| \leq \umach \pmult(n)\|\matH\| \lpar \|\matS^{-1}\| + \|\errormatrix_1^{\INV}\|\rpar \leq \epsilon_0(1+\epsilon_0) \leq 2\epsilon_0.
        \end{align*}
        
        We next have
        $
            \matMtilde\leftarrow \widetilde\mu_k-\matHtilde+\errormatrix_3^{(-)}
            =
            \widetilde\mu_k-\matS^{-1}\matH+ \errormatrix_1^{\INV}\matH+\errormatrix_2^{\MM}+\errormatrix_3^{(-)} = \widetilde\mu_k-\matS^{-1}\matH+\matB
        $, where $\errormatrix_3^{(-)}$ is a diagonal error matrix with \begin{align*}
            \|\errormatrix_3^{(-)}\|\leq \umach \|\widetilde\mu_k|\|\matHtilde\| \leq \umach \|\matS^{-1}\matH+\errormatrix_1^{\INV}\matH+\errormatrix_2^{\MM}\| \leq \umach (1+\epsilon_0+2\epsilon_0) \leq 4\umach \ll 4\epsilon_0,
        \end{align*}
        and $\matB=\errormatrix_1^{\INV}\matH+\errormatrix_2^{\MM}+\errormatrix_3^{(-)}$.
        We can bound the norm of $\matB$ as:
        \begin{align*}
            \|\matB\|\leq 7\epsilon_0.
        \end{align*}
        
        Now we need to apply Lemma \ref{lemma:sign_function_under_perturbation} to argue that $\lnorm
                \sgn(\mu_k-\matS^{-1}\matH)-
                \sgn(\widetilde\mu_k-\matS^{-1}\matH+\matB)
            \rnorm
            \leq
            \epsilon_1$, 
        for some $\epsilon_1\in(0,1)$. To satisfy the requirements of the lemma, we need to ensure that $\|\matB\| \leq \epsilon_1\frac{\pi|\lambda_{\min}^2(\widetilde\mu_k-\matS^{-1}\matH)|}{128\sqrt{\kappa(\matS)}}$.  This can be achieved by setting 
        \begin{align*}
            \epsilon_0 = \epsilon_1\frac{\pi|\lambda_{\min}^2(\widetilde\mu_k-\matS^{-1}\matH)|}{7\cdot 128\sqrt{\kappa(\matS)}}.
        \end{align*}
        However, we do not know $|\lambda_{\min}(\widetilde\mu_k-\matS^{-1}\matH)|$ and $\kappa(\matS)$. To circumvent this, we can use the assumption that $\widetilde\mu_k$ is well-placed close to the center of $\gap_k$, which implies that 
        \begin{align*}
            |\lambda_{\min}(\widetilde\mu_k-\matS^{-1}\matH)| \geq \tfrac{3}{8}\gap_k\geq \tfrac{1}{3}\widetilde\gap_k,
        \end{align*}
        and also use the assumption that $\kappa(\matS)\leq \widetilde\kappa$.
        This means that we can set
        \begin{align}
            \epsilon_0 = \epsilon_1\frac{\pi}{7\cdot 128\sqrt{\widetilde\kappa}}\cdot \frac{1}{9}\widetilde\gap_k^2,
            \label{equation:alg_purify_proof_umach_bound_2}
        \end{align}
        which implies the desired bound for $\matB$:
        \begin{align*}
            \|\matB\| 
            \leq 
            7\cdot
            \epsilon_1\frac{\pi}{7\cdot 128\sqrt{\widetilde\kappa}}\cdot \frac{1}{9}\widetilde\gap_k^2
            \leq 
            \epsilon_1\frac{\pi|\lambda_{\min}^2(\widetilde\mu_k-\matS^{-1}\matH)|}{128\sqrt{\kappa(\matS)}}
            .
        \end{align*}
        We can now apply Lemma \ref{lemma:sign_function_under_perturbation} to argue that 
        \begin{align*}
             \lnorm
                \sgn(\widetilde\mu_k-\matS^{-1}\matH)-
                \sgn(\widetilde\mu_k-\matS^{-1}\matH+\matB)
            \rnorm
            \leq
            \epsilon_1.
        \end{align*}
        By assumption we know that $\widetilde\mu_k=\mu_k\pm\tfrac{1}{8}\gap_k$ which means that $\sgn(\widetilde\mu_k-\matS^{-1}\matH)=\sgn(\mu_k-\matS^{-1}\matH)$, and therefore 
        \begin{align*}
             \lnorm
                \sgn(\mu_k-\matS^{-1}\matH)-
                \sgn(\widetilde\mu_k-\matS^{-1}\matH+\matB)
            \rnorm
            \leq
            \epsilon_1.
        \end{align*}
        
        To not interrupt the flow, we will skip for now the analysis of $\SGN$ and we will leave it for the end of the proof. 
        From Theorem \ref{theorem:sgn} we know that the result returned by $\SGN$ satisfies 
        $\|\matCtilde-\sgn(\matMtilde)\| \leq \epsilon_{\SGN}$, where we have full control over $\epsilon_{\SGN}$ since it is passed as an argument to $\SGN$, and therefore it can be set to $\epsilon_{\SGN}=\epsilon_1$. Denoting $\matC=\sgn(\mu_k-\matS^{-1}\matH)$ and $\errormatrix_{\matC}=\matCtilde-\matC$, we have
        \begin{align*}
            \lnorm
                \matCtilde - \sgn(\mu_k-\matS^{-1}\matH)
            \rnorm
            =
            \lnorm
                \errormatrix_{\matC}
            \rnorm
            \leq \epsilon_1 + \epsilon_{\SGN} =2\epsilon_1.
        \end{align*}
        
        We now proceed to the line 5 of Algorithm \ref{alg:purify}, which computes $\projectormatrixtilde\leftarrow \frac{1}{2}\lpar 1+\matCtilde \rpar +\errormatrix_5^{(+)}$, where $\errormatrix_5^{(+)}$ is a diagonal error matrix with norm bounded by 
        \begin{align*}
            \|\errormatrix_5^{(+)}\| \leq \umach \|\matCtilde\| \leq \umach(1+\epsilon_1+\epsilon_{\SGN})\leq 3\umach \ll 3\epsilon_1.
        \end{align*}
        Then we can write $\projectormatrixtilde=\frac{1}{2}\lpar1+\matC+\errormatrix_{\matC}\rpar +\errormatrix_5^{(+)}=\projectormatrix+\frac{1}{2}\errormatrix_{\matC}+\errormatrix_5^{(+)}=\projectormatrix+\errormatrix_{\projectormatrix},$
        where $\projectormatrix=\frac{1}{2}(1+\matC)$ and $\errormatrix_{\projectormatrix}=\frac{1}{2}\errormatrix_{\matC}+\errormatrix_5^{(+)}$. Combining with the above:
        \begin{align*}
            \|\errormatrix_{\projectormatrix}\| \leq 4\epsilon_1,
        \end{align*}
        in which case we can set $\epsilon_1 =\frac{\epsilon}{4}$, where $\epsilon$ is the desired accuracy, to guarantee forward error of at most $\epsilon$ for the spectral projector $\projectormatrixtilde$.

        Gathering all the requirements for the machine precision (except for $\SGN$, which is detailed below) from Equations \eqref{equation:alg_purify_proof_umach_bound_1} and \eqref{equation:alg_purify_proof_umach_bound_2}, and the assumption on $\widetilde\kappa $ and $\widetilde\gap_k$ it suffices to set
        \begin{align*}
            \umach &\leq  
             \frac{\epsilon}{4}\frac{\pi}{7\cdot 128\sqrt{\widetilde\kappa}}\cdot \frac{1}{9}\widetilde\gap_k^2\frac{1}{\mu(n)\widetilde\kappa^{\cinv\log(n)}},
        \end{align*}
        which translates to
        \begin{align}
            \label{eq:proposition_alg_purify_bit_bound_without_sgn}
            O\lpar
                \log(1/\umach)
            \rpar
            =
            O\lpar
                \log\lpar
                \tfrac{
                    \sqrt{\widetilde\kappa}\mu(n)\widetilde\kappa^{\cinv\log(n)}
                }{
                    \epsilon\widetilde\gap_k^2
                }
                \rpar
            \rpar
            =
            O\lpar
                \log(n)\log(\kappa(\matS)) + \log\lpar
                    \tfrac{n}{\epsilon\gap_k}
                \rpar
            \rpar
        \end{align}
        bits of precision.
        
        We finally proceed with the analysis of $\SGN$. To use Theorem \ref{theorem:sgn} for the convergence of $\SGN$, we need to find the appropriate parameters $\alpha_{\SGN},\eta_{\SGN}$ to call it. These parameters must be such that $\Lambda_{\eta_{\SGN}}(\matMtilde) \subseteq \mathsf{C}_{\alpha_{\SGN}}$, and $\tfrac{99}{100}<\alpha_{\SGN}<1$.
        From the properties of the pseudospectrum, specifically, from Proposition \ref{proposition:core_pseudospectrum_properties} $(i)$, we know that for any $\eta>0$
        \begin{align*}
            \Lambda_{\eta}(\matMtilde)
            =
            \Lambda_{\eta}(\widetilde\mu_k-\matS^{-1}\matH+\matB) 
            \subseteq 
            \Lambda_{\eta+\|\matB\|}(\widetilde\mu_k-\matS^{-1}\matH)
            \subseteq 
            \Lambda_{\eta+B_{\max}}(\widetilde\mu_k-\matS^{-1}\matH),
        \end{align*}
        where $B_{\max}$ is the upper bound for $\|\matB\|$ that we obtained above:
        \begin{align*}
            \|\matB\| 
            \leq 
            \epsilon_1\frac{\pi \widetilde\gap_k^2}{9\cdot 128\sqrt{\widetilde\kappa}}
            := B_{\max}.
        \end{align*}
        
        For $\eta_{\SGN}=B_{\max}$ and from Proposition \ref{proposition:core_pseudospectrum_properties} $(iii)$ we obtain
        \begin{align*}
            \Lambda_{B_{\max}}(\matMtilde) 
            &\subseteq 
            \Lambda_{2B_{\max}}(\widetilde\mu_k-\matS^{-1}\matH)
            \\
            &\subseteq
            \bigcup_iD\lpar
                \widetilde\mu_k-\lambda_i(\matS^{-1}\matH),
                2B_{\max}\keig(\widetilde\mu_k-\matS^{-1}\matH)
            \rpar
            \\
            &\subseteq
            \bigcup_iD\lpar
                \widetilde\mu_k-\lambda_i(\matS^{-1}\matH),
                2\epsilon_1\tfrac{\pi \widetilde\gap_k^2}{9\cdot 128\sqrt{\widetilde\kappa}}\keig(\matS^{-1}\matH)
            \rpar.
        \end{align*}
        where we used the fact that $\keig(\matS^{-1}\matH)\leq \sqrt{\kappa(\matS)}$. We can get a rough upper bound
        \begin{align*}
            2\epsilon_1\tfrac{\pi \widetilde\gap_k^2}{9\cdot 128\sqrt{\widetilde\kappa}}\keig(\matS^{-1}\matH)
            &\leq
            \epsilon_1\tfrac{2\pi \cdot 9^2}{9\cdot 128\cdot 8^2\sqrt{\kappa(\matS)}}\sqrt{\kappa(\matS)} \gap_k^2
            \leq
            \epsilon_1\tfrac{\gap_k^2}{128},
        \end{align*}
        which gives
        \begin{align*}
            \Lambda_{B_{\max}}(\matMtilde) 
            &\subseteq
            \bigcup_iD\lpar
                \widetilde\mu_k-\lambda_i(\matS^{-1}\matH),
                \epsilon_1\tfrac{\gap_k^2}{128}
            \rpar.
        \end{align*}
        
        It suffices to find the appropriate $\alpha_{\SGN}$ such that the Apollonian circles $\mathsf{C}_{\alpha_{\SGN}}$ will contain the disks of the pseudospectrum. Since the smallest singular value of $\widetilde\mu_k-\matS^{-1}\matH$ is lower bounded by $\frac{3}{8}\gap_k$, the leftmost positive point where any pseudospectral disk intersects with the real axis is 
        $\frac{3}{8}\gap_k-\epsilon_1\frac{\gap_k^2}{128} 
        \geq 
        \gap_k
        (\tfrac{3}{8}-\tfrac{1}{128})
        = 
        \gap_k
        \tfrac{47}{128}
        >
        \tfrac{\widetilde\gap_k}{4}
        $. Similarly for the negative halfplane, the rightmost negative point where a disk intersects with the real axis is $-\frac{\widetilde\gap_k}{4}$. Denoting by $\zeta=\frac{\widetilde\gap_k}{8}$, it suffices to set $\alpha_{\SGN}=\frac{1-\zeta}{1+\zeta}=\frac{8-\widetilde\gap_k}{8+\widetilde\gap_k}$. Clearly, $\alpha_{\SGN}<1$ for all $\infty>\gap_k>0$. For the lower bound, to ensure $\alpha_{\SGN}>99/100$ we need that $\widetilde\gap_k<\tfrac{8}{199}$. To achieve this we can simply scale the input matrices by a constant, since, by assumption, $\|\matS^{-1}\matH\|\leq 1$, and therefore $\gap_k\leq 2$. To conclude, we have all the required parameters to call $\SGN$.
        
        We can now go back to the complexity analysis.
        We are running
        \begin{align*}
            \matCtilde \leftarrow \SGN\lpar \matMtilde, \alpha_{\SGN}, \eta_{\SGN}, \epsilon_{\SGN}\rpar,
        \end{align*}
        where 
        \begin{align*}
            \alpha_{\SGN} &= \frac{8-\widetilde\gap_k}{8+\widetilde\gap_k},
            \qquad
            \eta_{\SGN} = B_{\max} = 
            \epsilon_1\frac{\pi \widetilde\gap_k^2}{9\cdot 128\sqrt{\widetilde\kappa}}
            ,
            \qquad
            \epsilon_{\SGN} = \epsilon_1 = \frac{\epsilon}{4}.
        \end{align*}
        We now invoke Theorem \ref{theorem:sgn}. The number of iterations is bounded by
        \begin{align*}
            N &= O\lpar
                \log(\tfrac{1}{1-\alpha_{\SGN}})+\log(\log(\tfrac{1}{\eta_{\SGN}\epsilon_{\SGN}}))
            \rpar
            \\
            &= O\lpar
                \log(\tfrac{1}{\widetilde\gap_k})
                +\log(\log(\tfrac{\widetilde\kappa}{\epsilon\widetilde\gap_k}))
            \rpar
            \\
            &= O\lpar
                \log(\tfrac{1}{\gap_k})
                +\log(\log(\tfrac{\kappa(\matS)}{\epsilon\gap_k}))
            \rpar.
        \end{align*}
        The number of required bits is
        \begin{align*}
            O\lpar
                \log(n)\log^3(\tfrac{1}{1-\alpha_{\SGN}}) \log(\tfrac{1}{\eta_{\SGN}\epsilon_{\SGN}})
            \rpar
            &=
            O\lpar
                \log(n)\log^3(\tfrac{1}{\widetilde\gap_k})
                \log \lpar 
                    \tfrac{\widetilde\kappa}{\epsilon\widetilde\gap_k}
                \rpar
            \rpar
            \\
            &=
            O\lpar
                \log(n)\log^3(\tfrac{1}{\gap_k})
                \log \lpar 
                    \tfrac{\kappa(\matS)}{\epsilon\gap_k}
                \rpar
            \rpar,
        \end{align*}
        which dominates the previous bit bound of Equation \ref{eq:proposition_alg_purify_bit_bound_without_sgn}.
        
        It remains to bound the arithmetic complexity. There is a constant number of inversions, matrix multiplications, and scalar-matrix operations, which in total take at most $O(T_{\MM}(n))$ floating point operations. Thus, the dominant factor is the call to $\SGN$, which amounts to
        \begin{align*}
            O\lpar T_{\MM}(n)
                \lpar
                    \log(\tfrac{1}{\gap_k})+\log(\log(\tfrac{\kappa(\matS)}{\epsilon\gap_k}))
                \rpar
            \rpar
        \end{align*}
        arithetic operations using the aforementioned number of bits.
    \end{proof}
\end{proposition}

\subsection{Spectral gaps with diagonalization}
\label{appendix:why_not_diagonalization}
The straightforward approach to compute the desired spectral gaps is to iteratively compute $\matS^{-1}\matH$ and diagonalize it, until the eigenvalues are well approximated. If we compute $\matA\leftarrow \MM(\INV(\matS),\matH)$, then we can write $\matA=\matS^{-1}\matH+\errormatrix$, for some error matrix $\errormatrix$. The next step is to approximate $\gap_k(\matS^{-1}\matH)$ and $\mu_k(\matS^{-1}\matH)$. For this one could use the recent state-of-the-art backward-approximate diagonalization algorithm of \cite{banks2022pseudospectral}. 

\begin{theorem}[$\EIG$, imported Theorem 1.6 from \cite{banks2022pseudospectral}]
\label{theorem:eig}
There exists a randomized algorithm $\EIG(\matA,\epsilon_{\EIG})$ which takes any matrix $\matA\in\mathbb{C}^{n\times n}$ with $\|\matA\|\leq 1$ and a desired accuracy parameter $\epsilon_{\EIG}>0$ as inputs and returns a diagonal $\matD$ and an invertible matrix $\matV$ such that
\begin{align*}
    \|\matA-\matV\matD\matV^{-1}\|\leq \epsilon_{\EIG} \text{\quad and\quad} \kappa(\matV)\leq 32n^{2.5}/\epsilon_{\EIG},
\end{align*}
in 
\begin{align*}
O\left(T_{\MM}(n)\log^2(\tfrac{n}{\epsilon_{\EIG}})\right)
\end{align*}
arithmetic operations on a floating point machine with
\begin{align*}
    O(\log^4(n/\epsilon_{\EIG})\log (n))
\end{align*}
bits of precision, with probability at least $1-14/n$.
\end{theorem}

Applying $\EIG$ on $\matA$ we can obtain a backward-approximate diagonalization. But we are not finished yet, since we are interested in each individual eigenvalue. To translate the backward error to a forward error for the eigenvalues, and, ultimately, the spectral gap, one can try to use Corollary 1.7 and Proposition 1.1 of \cite{banks2022pseudospectral}. However, this approach has two main limitations. First, it relies on simplicity of the spectrum, i.e., it assumes that the minimum gap between any pair of eigenvalues is larger than zero. This assumption is quite restrictive, since the desired gap might be well-defined even at the presence of other multiple eigenvalues. For example, in DFT applications it is not uncommon to have eigenenergies with algebraic multiplicity larger than one, and at the same time have a large band-gap that separates the occupied from the unoccupied orbitals. The second limitation is that the aforementioned Corollary 1.7 requires as an input parameter an actual over-estimate for the minimum eigenvalue gap. Even if such a gap exists, it is not described how to estimate it.

For diagonalizable matrices, we can leverage the following Corollary \ref{corollary:diagonalizable_matrix_backward_error_to_forward_eigenvalue_error}, which is an immediate consequence of Kahan's inequality (Fact \ref{fact:weyl_kahan}), and it overcomes the aforementioned limitations.
\begin{corollary}
    \label{corollary:diagonalizable_matrix_backward_error_to_forward_eigenvalue_error}
    If $\matX$ is diagonalizable and it has real eigenvalues then for any $\matZ$ the following bound holds for the eigenvalues of $\matX$:
    \begin{align*}
        \labs \lambda_i(\matX)-\lambda_i(\matZ) \rabs \leq O(\log(n)) \keig(\matX)\|\matX-\matZ\|.
    \end{align*}
    \begin{proof}
        Write $\matX=\matW\matLambda\matW^{-1}$ where $\matW$ diagonalizes $\matX$ and is chosen such that $\kappa(\matW)=\keig(\matX)$. This is always possible since if $\matW$ is any matrix that diagonalizes $\matX$ then $\matLambda$ is similar to $\matX$ and since $\matX$ has real eigenvalues then $\Lambda$ has to be real, and therefore symmetric. Then we can write
        \begin{align*}
            \|\matLambda-\matW^{-1}\matZ\matW\|  \leq
            \|\matW\|\|\matW^{-1}\| \|\matX-\matZ \|
            = \kappa(\matW)  \|\matX-\matZ \|.
        \end{align*}
        Since $\matLambda$ is symmetric, and since $\matW^{-1}\matZ\matW$ is similar to $\matZ$ then from Kahan's inequality (Fact \ref{fact:weyl_kahan}) \begin{align*}
            \labs \lambda_i(\matX)-\lambda_i(\matZ) \rabs 
            =
            \labs \lambda_i(\matLambda) - \lambda_i(\matW^{-1}\matZ\matW) \rabs 
            &\leq 
            O(\log(n))\|\matLambda-\matW^{-1}\matZ\matW\| \\
            &\leq
            O(\log(n)) \kappa(\matW)\|\matX-\matZ\|\\
            &= 
            O(\log(n))\keig(\matX)\|\matX-\matZ\|.
        \end{align*}
    \end{proof}
\end{corollary}
Now we need to use this to get a bound for the computed generalized eigenvalues after $\INV$ and $\EIG$. 
\begin{proposition}[$\EIG$-based gap]
    \label{proposition:eig_based_gap}
    Given a definite pencil $(\matH,\matS)$ with $\|\matH\|,\|\matS^{-1}\|\leq 1$, we can compute $\widetilde\gap_k\in(1\pm\epsilon)\gap_k$ and $\widetilde\mu_k\in\mu_k\pm\epsilon\gap_k$ by iteratively calling $\INV$, $\MM$, and $\EIG$, using
    \begin{align*}
        O\lpar
            T_{\MM}(n)\log(\tfrac{1}{\epsilon\gap_k})\log^2(\tfrac{n\kappa(\matS)}{\epsilon\gap_k})
        \rpar
    \end{align*}
    floating point operations using
    \begin{align*}
        O\lpar
            \log^4(\tfrac{n\kappa(\matS)}{\epsilon\gap_k})\log(n)
        \rpar
    \end{align*}
    bits of precision with probability at least $1-O(\tfrac{\log(1/\epsilon\gap_k)}{n})$.
    \begin{proof}
        Let $\matD,\matV\leftarrow \EIG(\matA,\epsilon_{\EIG})$ be the solution returned by $\EIG$ when applied to $\matA=\matS^{-1}\matH+\errormatrix$. Note that
        \begin{align*}
            \|\matS^{-1}\matH-\matV\matD\matV^{-1}\| 
            = 
            \|\matS^{-1}\matH + \errormatrix - \errormatrix -\matV\matD\matV^{-1}\|
            \leq
            \|\matS^{-1}\matH + \errormatrix  -\matV\matD\matV^{-1}\| + \|\errormatrix\|
            \leq
            \epsilon_{\EIG} + \|\errormatrix\|.
        \end{align*}
        From Corollary \ref{corollary:diagonalizable_matrix_backward_error_to_forward_eigenvalue_error}, since $\matS^{-1}\matH$ is diagonalizable with real eigenvalues then we conclude that
        \begin{align*}
            \labs \lambda_i(\matS^{-1}\matH)-\matD_{i,i} \rabs
            =
            \labs \lambda_i(\matS^{-1}\matH)-\lambda_i(\matV\matD\matV^{-1}) \rabs
            &\leq
            O(\log(n))\keig(\matS^{-1}\matH)\|\matS^{-1}\matH-\matV\matD\matV^{-1}\|\\
            &\leq
            O(\log(n))\keig(\matS^{-1}\matH)\lpar \epsilon_{\EIG} + \|\errormatrix\| \rpar.
        \end{align*}
        We can tune the machine precision such that $\|\errormatrix\|=\epsilon_{\EIG}=\epsilon'\frac{1}{c\log(n)\sqrt{\kappa(\matS)}}$ for some chosen $\epsilon',$ and some global constant $c$, then finally
        \begin{align*}
                \labs \lambda_i(\matS^{-1}\matH)-\lambda_i(\matV\matD\matV^{-1}) \rabs
                \leq
                \epsilon'.
        \end{align*}
        
        We can now consider an iterative scheme, where we call $\EIG$ on $\matS^{-1}\matH$, halving $\epsilon'$ at each step. 
        We need to keep halving $\epsilon'$ until it reaches $\epsilon'=\Theta(\epsilon\gap_k)$, in which case we have a total of $O(\log(\tfrac{1}{\epsilon\gap_k}))$ calls to $\EIG$.
        In the worst case every call costs
        \begin{align*}
            O\lpar
                T_{\MM}(n)\log^2(\tfrac{n}{\epsilon_{\EIG}})
            \rpar
            =
            O\lpar
                T_{\MM}(n)\log^2(\tfrac{n\kappa(\matS)}{\epsilon\gap_k})
            \rpar
        \end{align*}
        arithmetic operations using $O\lpar \log^4(\tfrac{n\kappa(\matS)}{\epsilon\gap_k})\log(n) \rpar$ bits (note that the bits required by $\MM$ and $\INV$ to achieve accuracy $\epsilon'$ are dominated by those required for $\EIG$, and we therefore ignore them). Since we do not know $\gap_k$, we can set the termination criterion to be $\epsilon'\approx \Theta(\epsilon\widetilde\gap_k)$, where $\widetilde\gap_k$ is the approximate gap that we obtain from $\EIG$. 
        
        Each iteration succeeds with high probability $1-1/n$, in which case a union bound gives $1-O(\log(1/(\epsilon\gap_k))/n$, which can potentially be improved to $1-O(1/n)$ without impacting the complexity, but we do not expand further.

        A subtle detail in the analysis above is that we need an estimate for $\kappa(\matS)$ in order to be able to use $\INV$ and to set the machine precision. Since $\kappa(\matS)$ is generally unknown, we need to compute it.
        We will later show in Appendix \ref{appendix:spectral_gap} how to compute the condition number quickly.

        When $\matS=\matI$, we can avoid the inversion and the computation of $\kappa(\matS)$. The arithmetic complexity and the bit requirement are the same by setting $\kappa(\matS)=1$.
    \end{proof} 
\end{proposition}

\begin{remark}
    We highlight that a similar result can be obtained by using the more recent pencil diagonalization algorithm of \cite{demmel2023generalized}. The authors the latter mention that their algorithm should be ``more numerically stable'' than $\EIG$ of \cite{banks2022pseudospectral}, since it uses an inverse-free iteration internally, and they do provide strong theoretical and experimental evidence for this statement. However, as there is no formal, end-to-end proof for the bit complexity at the time of this writing, we choose to compare against $\EIG$ of \cite{banks2022pseudospectral}. Note that, any improvements on the bit requirements for the matrix sign function, directly provide the same improvements for our main Theorems \ref{theorem:spectral_projector} and \ref{theorem:alg_spectral_gap}.
\end{remark}
\section{Analysis of Cholesky}
\label{appendix:cholesky_analysis}

This section is devoted to the analysis of Algorithm \ref{alg:stable_cholesky} and the proof of Theorem \ref{theorem:stable_cholesky_bounds}. 
Let $\matM=
    \begin{pmatrix}
        \matA & \matB^*\\
        \matB & \matC
    \end{pmatrix}$ be a Hermitian matrix. It can be factorized in the form
\begin{align}
    \matM = 
    \begin{pmatrix}
        \matI & \\
        \matB\matA^{-1} & \matI
    \end{pmatrix}
    \begin{pmatrix}
        \matA & \\
         & \matS
    \end{pmatrix}
    \begin{pmatrix}
        \matI & \matA^{-1}\matB^*\\
              & \matI
    \end{pmatrix}
    = \matW\matY\matW^{*},
    \label{eq:block_cholesky_factorization}
\end{align}
where $\matS=\matC-\matB\matA^{-1}\matB^*$ is the Schur complement and
\begin{align*}
    \matW =     \begin{pmatrix}
        \matI & \\
        \matB\matA^{-1} & \matI
    \end{pmatrix},
    \qquad 
    \matY = \begin{pmatrix}
        \matA & \\
         & \matS
    \end{pmatrix}.
\end{align*}

\begin{proposition}
\label{proposition:norm_bounds_posdef}
Let $\matM\in\mathbb{H}^{n}_{++}$ and consider the partitioning    $\matM=
    \begin{pmatrix}
        \matA & \matB^*\\
        \matB & \matC
    \end{pmatrix}$.
The following hold 
\begin{enumerate}[(i)]
\item $\matA,\matC$ and the Schur complement $\matS=\matC-\matB\matA^{-1}\matB^*$ are all positive definite;
\item For all $\matX \in \{\matA,\matC,\matS\}$ we have that
$\|\matX\|\leq \|\matM\|$ and $\|\matX^{-1}\| \leq \|\matM^{-1}\|$;
\item $\|\matB\|\leq \|\matM\|/2.$
\end{enumerate}
\begin{proof}
It is easy to see that $\matA$ is positive definite: since $\matM\succ 0$, then it must also hold that $\matA\succ 0$ since the quadratic form $\vecx^*\matA\vecx$ can be written as $\vecy^*\matM\vecy$ for some vector $\vecy$. For the Schur complement we recall the factorized form of Equation \eqref{eq:block_cholesky_factorization}.
Consider the quadratic form $\vecx^*(\matC-\matB\matA^{-1}\matB^*)\vecx$. Let 
$\vecy = \begin{pmatrix}
    -\matA^{-1}\matB^*\vecx \\
    \vecx
\end{pmatrix}$. Then $\vecy^*\matM\vecy=\vecx^*(\matC-\matB\matA^{-1}\matB^*)\vecx$, which means that every quadratic form for the Schur complement can be written as a quadratic form for the matrix $\matM$, and therefore they are both positive definite.

For the spectral norm bounds, since $\matA$ and $\matC$ are both positive definite then their norms are equal to the largest absolute eigenvalues. 
From the variational characterizaiton of eigenvalues and the discussion above it is easy to see that $\|\matA\|\leq \|\matM\|$. For the Schur complement, let $\vecz$ be the eigenvector such that $\|\matC-\matB\matA^{-1}\matB^*\|= \vecz^* (\matC-\matB\matA^{-1}\matB^*)\vecz$. Then
\begin{align*}
    \vecz^* (\matC-\matB\matA^{-1}\matB^*)\vecz = \vecz^*\matC\vecz - \vecz^*\matB\matA^{-1}\matB^*\vecz \leq \vecz^*\matC\vecz \leq \|\matM\|,
\end{align*}
where we used the fact that $\matB\matA^{-1}\matB^*$ is positive semi-definite (since $\matA^{-1}$ is positive definite) and therefore for all $\vecx$
\begin{align*}
\vecx^*\matB\matA^{-1}\matB^*\vecx\geq 0,
\end{align*} 
where equality with zero holds only when $\vecz\in\ker(\matB^*).$ 

We finally prove the bound for $\|\matB\|$. Let $\vecu$ be the top left singular vector of $\matB$ and $\vecv$ be the top right singular vector. Specifically, if $\matB=\matU\matSigma\matV^*$ is the economy SVD of $\matB$, then $\vecu$ is the first column of $\matU$ and $\vecv^*$ is the first row of $\matV^*$. Then $\vecu^*\matB\vecv=\sigma_{\max}(\matB)\vecv^*\vecv=(\matB)\vecu^*\vecu=\|\matB\|$. Consider the vector $\vecz=\begin{pmatrix}
    \vecv\\
    \vecu
\end{pmatrix}$
. Then
\begin{align*}
    \|\matM\|\geq \vecz^*\matM\vecz =
    \begin{pmatrix}
        \vecv^* & \vecu^*
    \end{pmatrix}
    \begin{pmatrix}
        \matA & \matB^*\\
        \matB & \matC
    \end{pmatrix}
    \begin{pmatrix}
        \vecv \\ \vecu
    \end{pmatrix}
    =
    \vecv^*\matA\vecv + 2\vecu^*\matB\vecv + \vecu^*\matC\vecu 
    \geq 
    2\vecu^*\matB\vecv
    =
    2\|\matB\|,
\end{align*}
where we used the fact that $\matA,\matC$ are positive-definite. We can then obtain bounds for $\|\matC^{-1}\|,\|\matA^{-1}\|,$ and $\|\matS^{-1}\|$, by observing that $\sigma_{\min}(\matX)=\min_{\|\vecz\|=1}\vecz^*\matX\vecz$ and lower bound it by $\sigma_{\min}(\matM)$ using similar argumnents.
\end{proof}
\end{proposition}

Since $\matA$ and $\matS$ are Hermitian and positive-definite, we can recursively  compute their Cholesky factors. This gives rise to the recursive Algorithm \ref{alg:stable_cholesky}. The complexity is as follows.
\begin{proposition}
\label{proposition:stable_cholesky_complexity}
Algorithm \ref{alg:stable_cholesky} requires $O(T_{\MM}(n))$ arithmetic operations.
\begin{proof}
Let $T(n)$ be the number of operations executed for a matrix of size $n\times n$. Steps 4 and 8 require time $T(n/2)$. Step 5 requires time $T_{\MM}(n/2)+T_{\INV}(n/2)$, and step 6 requires $T_{\MM}(n/2)$. The Schur complement in Step 7 requires $T_{\MM}(n/2)+T_{\HERM}(n/2)+(n/2)^2 = T_{\MM}(n/2)+2(n/2)^2$. This becomes
\begin{align*}
    T(n) = 2T(\tfrac{n}{2}) + 3T_{\MM}(\tfrac{n}{2}) + T_{\INV}(\tfrac{n}{2}) + 2(\tfrac{n}{2})^2 \leq 2T(\tfrac{n}{2}) + O(T_{\MM}(\tfrac{n}{2}))\leq O(T_{\MM}(n)).
\end{align*}
\end{proof}
\end{proposition}

\subsection{Error analysis}

The goal is to show that $\CHOLESKY(\matM)$ will return a backward-approximate Cholesky factor $\matL$ such that the error
\begin{align*}
    \|\matL\matL^*-\matA\| \leq f(\umach,\|\matM\|,\|\matM^{-1}\|,n),
\end{align*}
where $f$ is some function, and use this bound to argue about the number of bits that are required to achieve it.  Recall that
    \begin{align*}
        \matM =
        \begin{pmatrix}
            \matA 
            &
            \matB^*
            \\
            \matB
            &
            \matC
        \end{pmatrix},
        \quad
        \matL=
        \begin{pmatrix}
            \matL_{11} 
            &
            \\
            \matL_{21}
            &
            \matL_{22}
        \end{pmatrix},
        \quad
        \matL\matL^* =
        \begin{pmatrix}
            \matL_{11}\matL_{11}^* 
            &
            \matL_{11}\matL_{21}^*
            \\
            \matL_{21}\matL_{11}^*
            &
            \matL_{21}\matL_{21}^* + \matL_{22}\matL_{22}^*
        \end{pmatrix}.
    \end{align*}

It is easy to see that 
\begin{align}
    \|\matL\matL^*-\matM\| &= 
    \left\|
        \begin{pmatrix}
            \matL_{11}\matL_{11}^* 
            &
            \matL_{11}\matL_{21}^*
            \\
            \matL_{21}\matL_{11}^*
            &
            \matL_{21}\matL_{21}^* + \matL_{22}\matL_{22}^*
        \end{pmatrix}
        -
        \begin{pmatrix}
            \matA 
            &
            \matB^*
            \\
            \matB
            &
            \matC
        \end{pmatrix}
    \right\|
    \nonumber
    \\
    &\leq
    \left\| 
        \matL_{11}\matL_{11}^*
        -
        \matA
    \right\|
    +
    \left\| 
        \matL_{21}\matL_{11}^*
        -
        \matB
    \right\|
    +
    \left\| 
        \matL_{21}\matL_{21}^*
        +
        \matL_{22}\matL_{22}^*
        -
        \matC
    \right\|,
\label{eq:main_error_stable_cholesky}
\end{align}
which means that it suffices to bound the individual terms in the sum. 
The first term is the error of the first recursive call. Following the notation of \cite{demmel2007fastla}, let us denote $\err(n)$ the norm-wise error of the algorithm for size $n$. 
We can then write 
\begin{align}
    \label{eq:bound_error_l11}
    \|\matL_{11}\matL_{11}^* - \matA\| \leq \|\errormatrix_0^{\mathsf{CH}}\| \leq \err(n/2).
\end{align}
To simplify the various inequalities in the proofs we use the global bound $\mu$ from Remark \ref{remark:simplification_mu}, where all the terms $\pmult(n)$ and $\pinv(n)$ are bounded by $\mu(n)$. Finally, we also define
\begin{align}
    \mathcal{E}_1(\matM,n) &:= \mu(n)\cdot\kappa(\matM),\nonumber\\
    \mathcal{E}_{\INV}(\matM,n) &:= \mu(n)\cdot \kappa(\matM)^{\cinv \log (n)}.
    \label{eq:einv_and_e1}
\end{align}

We can now get a first expression for the bounds of the terms in Equation \eqref{eq:main_error_stable_cholesky}.
\begin{lemma}
    \label{lemma:main_error_bounds_stable_cholesky}
    In Algorithm \ref{alg:stable_cholesky}, the blocks $\matL_{11},\matL_{21},\matL_{22}$ of the returned matrix $\matL$ satisfy:
    \begin{align*}
    \|\matL_{11}\matL_{11}^*-\matA\| &\leq \err(n/2),
    \\
    \left\| 
        \matL_{21}\matL_{11}^* - \matB
    \right\|
    &\leq
        \left(
            \kappa(\matM)
            +
            \|\errormatrix_{\matB\matA i}\|
        \right)
        \err(n/2)
        +
        \|\errormatrix_{\matB\matA i}\|
        \|\matM\|
        +
        \|\errormatrix_{3}^{\MM}\|
        \|\matL_{11}^*\|,
    \\
    \left\| 
        \matL_{21}\matL_{21}^*
        +
        \matL_{22}\matL_{22}^*
        -
        \matC
    \right\|
    &\leq
    2\kappa(\matM)^2 
        \cdot
        \err(n/2)
        +
        2
        \kappa(\matM)
        \|
            \errormatrix_{\matL_{21}}
        \|
        \|\matL_{11}\|
        +
        \|
        \errormatrix_{\matL_{21}}
        \|^2
        +
        \|
        \errormatrix_{\matS} 
        \|.
    \end{align*}
\begin{proof}
The error for $\matL_{11}\matL_{11}^*$ was already described. For the second inequality, we first expand $\matL_{21}\matL_{11}^*$ as follows:
\begin{align*}
    \matL_{21}\matL_{11}^*
    &=
    (\matB\matA^{-1}+\errormatrix_{\matB\matA i})\matL_{11}\matL_{11}^* + \errormatrix_3^{\MM}\matL_{11}^*
    \\
    &=
    (\matB\matA^{-1}+\errormatrix_{\matB\matA i})
    (\matA+\errormatrix_{\matA}^{\mathsf{CH}})
    + \errormatrix_3^{\MM}\matL_{11}^*
    \\
    &=
    \matB
    +
    \matB\matA^{-1}\errormatrix_{\matA}^{\mathsf{CH}}
    +
    \errormatrix_{\matB\matA i}\matA
    +
    \errormatrix_{\matB\matA i}\errormatrix_{\matA}^{\mathsf{CH}}
    + 
    \errormatrix_3^{\MM}\matL_{11}^*.
\end{align*}
Then
\begin{align*}
    \left\| 
        \matL_{21}\matL_{11}^* - \matB
    \right\|
    &=
    \left\|
    \matB\matA^{-1}\errormatrix_{\matA}^{\mathsf{CH}}
    +
    \errormatrix_{\matB\matA i}\matA
    +
    \errormatrix_{\matB\matA i}\errormatrix_{\matA}^{\mathsf{CH}}
    + 
    \errormatrix_3^{\MM}\matL_{11}^*
    \right\|
    \\
    &\leq
        \|
        \matB
        \|
        \|
        \matA^{-1}
        \|
        \|\errormatrix_{\matA}^{\mathsf{CH}}\|
        +
        \|
        \errormatrix_{\matB\matA i}
        \|
        \|
        \matA
        \|
        +
        \|
        \errormatrix_{\matB\matA i}
        \|
        \|
        \errormatrix_{\matA}^{\mathsf{CH}}
        \|
        + 
        \|
        \errormatrix_3^{\MM}
        \|
        \|
        \matL_{11}^*
        \|
    \\
    &\leq
        \kappa(\matM)
        \err(n/2)
        +
        \|
        \errormatrix_{\matB\matA i}
        \|
        \|
        \matM
        \|
        +
        \|
        \errormatrix_{\matB\matA i}
        \|
        \err(n/2)
        + 
        \|
        \errormatrix_3^{\MM}
        \|
        \|
        \matL_{11}^*
        \|.
\end{align*}    
With a similar procedure we can derive the third bound. We first write
$\matL_{21}=\matB\matA^{-1}\matL_{11}+\errormatrix_{\matL_{21}}$ where $\errormatrix_{\matL_{21}}$ contains several error matrices from previous computations.
Then
\begin{align*}
    \matL_{21}\matL_{21}^*
    &=
    \left(
        \matB\matA^{-1}\matL_{11}+\errormatrix_{\matL_{21}}
    \right)
    \left(
        \matB\matA^{-1}\matL_{11}+\errormatrix_{\matL_{21}}
    \right)^*
    \\
    &=
    \left(
        \matB\matA^{-1}\matL_{11}+\errormatrix_{\matL_{21}}
    \right)
    \left(
        \matB\matA^{-1}\matL_{11}+\errormatrix_{\matL_{21}}
    \right)^*
    \\
    &=
        \matB\matA^{-1}\matL_{11}
        \matL_{11}^*\matA^{-1}\matB^*
        +
        \errormatrix_{\matL_{21}}
        \matB\matA^{-1}\matL_{11}
        +
        \matL_{11}^*
        \matA^{-1}
        \matB^*
        \errormatrix_{\matL_{21}}^*
        +
        \errormatrix_{\matL_{21}}
        \errormatrix_{\matL_{21}}^*
    \\
    &=
        \matB\matA^{-1}(\matA + \errormatrix_{\matA}^{\mathsf{CH}})\matA^{-1}\matB^*
        +
        \errormatrix_{\matL_{21}}
        \matB\matA^{-1}\matL_{11}
        +
        \matL_{11}^*
        \matA^{-1}
        \matB^*
        \errormatrix_{\matL_{21}}^*
        +
        \errormatrix_{\matL_{21}}
        \errormatrix_{\matL_{21}}^*
    \\
    &=
        \matB\matA^{-1}\matB^*
        +
        \matB\matA^{-1}\errormatrix_{\matA}^{\mathsf{CH}}
        \matA^{-1}\matB^*
        +
        \errormatrix_{\matL_{21}}
        \matB\matA^{-1}\matL_{11}
        +
        \matL_{11}^*
        \matA^{-1}
        \matB^*
        \errormatrix_{\matL_{21}}^*
        +
        \errormatrix_{\matL_{21}}
        \errormatrix_{\matL_{21}}^*.
\end{align*}
Similarly
\begin{align*}
    \matL_{22}\matL_{22}^*
    &=
    \matStilde +\errormatrix_{\matStilde}^{\mathsf{CH}}
    \\
    &=
    \matS + \errormatrix_{\matS} + \errormatrix_{\matStilde}^{\mathsf{CH}}
    \\
    &=
    \matC-\matB\matA^{-1}\matB^* + \errormatrix_{\matS} + \errormatrix_{\matStilde}^{\mathsf{CH}}.
\end{align*}
Then
\begin{align*}
    \left\|
        \matL_{21}\matL_{21}^*
        \right.
        &+
        \left.
        \matL_{22}\matL_{22}^*
        -
        \matC
    \right\|\\
    &=
    \left\|
        \matB\matA^{-1}\errormatrix_{\matA}^{\mathsf{CH}}
        \matA^{-1}\matB^*
        +
        \errormatrix_{\matL_{21}}
        \matB\matA^{-1}\matL_{11}
        +
        \matL_{11}^*
        \matA^{-1}
        \matB^*
        \errormatrix_{\matL_{21}}^*
        +
        \errormatrix_{\matL_{21}}
        \errormatrix_{\matL_{21}}^*
        +
        \errormatrix_{\matS} 
        + 
        \errormatrix_{\matStilde}^{\mathsf{CH}}
    \right\|
    \\
    &\leq
        \|
            \matB\matA^{-1}
        \|^2
        \|
            \errormatrix_{\matA}^{\mathsf{CH}}
        \|
        +
        2
        \|
            \errormatrix_{\matL_{21}}
        \|
        \|
            \matB\matA^{-1}
        \|
        \|\matL_{11}\|
        +
        \|
        \errormatrix_{\matL_{21}}
        \|^2
        +
        \|
        \errormatrix_{\matS} 
        \|
        + 
        \|
        \errormatrix_{\matStilde}^{\mathsf{CH}}
        \|
    \\
    &\leq
        \kappa(\matM)^2 
        \cdot
        \err(n/2)
        +
        2
        \kappa(\matM) 
        \|
            \errormatrix_{\matL_{21}}
        \|
        \|\matL_{11}\|
        +
        \|
        \errormatrix_{\matL_{21}}
        \|^2
        +
        \|
        \errormatrix_{\matS} 
        \|
        + 
        \err(n/2)
    \\
    &\leq
        2\kappa(\matM)^2 
        \cdot
        \err(n/2)
        +
        2
        \kappa(\matM)
        \|
            \errormatrix_{\matL_{21}}
        \|
        \|\matL_{11}\|
        +
        \|
        \errormatrix_{\matL_{21}}
        \|^2
        +
        \|
        \errormatrix_{\matS} 
        \|.
\end{align*}    
\end{proof}

\end{lemma}

With this, the task reduces to finding appropriate bounds for the norms of the matrices $\errormatrix_{3}^{\MM},\errormatrix_{\matB\matA i},\errormatrix_{\matS}$ and $\errormatrix_{\matL_{21}}$.
We first derive bounds for the norms of the error matrices $\errormatrix_{(\cdot)}$.

\begin{lemma}
    \label{lemma:cholesky_error_matrix_bounds}
    In Algorithm \ref{alg:stable_cholesky}, the following bounds hold for the error matrices
    \begin{enumerate}[(i)]
        \item \quad$\normeinv \leq \umach\cdot\errinvnhalf{2}\|\matM^{-1}\|$,
        \item \quad$\left\|\errormatrix_2^{\MM}\right\| \leq 
        \umach\cdot
        \erronenhalf{2} \left(
                    1+ \umach\cdot\errinvnhalf{2}
                \right),$
        \item \quad$
        \left\|\errormatrix_{\matB\matA i}\right\| 
            \leq 
            \umach\cdot\errinvnhalf{2}
            \cdot 
            \kappa(\matM)
            +
            \umach
            \cdot
            \erronenhalf{2}
            (1+\umach\cdot\errinvnhalf{2})$,
        \item \quad$\left\|\errormatrix_3^{\MM}\right\| \leq
                \umach\cdot
                \erronenhalf{2}
                \left(
                    1
                    +
                    \umach\cdot\errinvnhalf{2}
                \right)^2
                \|\matL_{11}\|,$
        \item \quad$\left\|\errormatrix_4^{\MM}\right\| \leq 
            \umach\cdot\erronenhalf{2}
                \left(
                    1
                    +
                    \umach\cdot\errinvnhalf{2}
                \right)^2
                \|\matM\|,$
        \item \quad$
        \left\|\errormatrix_5^{\SUB} \right\|
            \leq
            \umach
            \cdot
            \sqrt{n/2} 
            \cdot \|\matM\|
            \cdot
            \left(
                1
                +
                \left(
                    \kappa(\matM)
                    +
                    \umach\cdot\erronenhalf{2}
                \right)
                \left(
                    1
                    +
                    \umach\cdot\errinvnhalf{2}
                \right)^2
            \right),$
        \item \quad$\left\|
            \errormatrix_{\matL_{21}}
        \right\|
            \leq
            \umach\cdot\errinvnhalf{2}
            \cdot
            \left[
                \kappa(\matM)
                +
                \left(
                    2 + \umach\cdot \errinvnhalf{2}
                \right)^2
            \right]
            \cdot
            \|
                \matL_{11}
            \|, $
        \item \quad$
        \left\|
            \errormatrix_{\matS}
        \right\|
            \leq
            \umach\cdot\|\matM\|\cdot\errinvnhalf{2}\cdot
            \left(
                2\kappa(\matM)
                +
                \left(
                    1+ \umach\cdot\errinvnhalf{2}
                \right)^2
                \left(
                    3\kappa(\matM)
                    +
                    \umach\cdot\erronenhalf{2}
                \right)
            \right).$ 
    \end{enumerate}
\begin{proof}
Each term is bounded as follows.
\begin{enumerate}[(i)]
    \item For $\|\errormatrix_1^{\INV}\|$ we have: \begin{align*}
            \|\errormatrix_{1}^{\INV}\| &\leq \pinv (n/2)\cdot \umach \cdot \kappa(\matA)^{\cinv \log \frac{n}{2}}\|\matA^{-1}\| \qquad\text{(...from Thm. \ref{theorem:fast_inv}})\\
            &\leq
            \mu (n/2)\cdot \umach \cdot \kappa(\matA)^{\cinv \log \frac{n}{2}}\|\matM^{-1}\|\\
            &=
            \umach\cdot\errinvnhalf{2}\|\matM^{-1}\|.
        \end{align*}
    \item Similarly, for $\|\errormatrix_2^{\MM}\|$: \begin{align*}
        \|\errormatrix_2^{\MM}\| &\leq \pmult(n/2) \cdot \umach  \|\matB\|\|\matA^{-1}+\errormatrix_1^{\INV}\|
                \qquad\text{(...from Thm. \ref{theorem:fast_mm}})
                \\
                &\leq \pmult(n/2) \cdot \umach  \|\matB\|\|\matA^{-1}\|+ \pmult(n/2) \cdot \umach  \|\matB\|\|\errormatrix_1^{\INV}\|.
                \\
                &\leq \pmult(n/2) \cdot \umach \cdot \|\matM\|\|\matM^{-1}\|
                    + 
                    \pmult(n/2) \cdot \umach \cdot \|\matM\| \cdot
                        \umach\cdot\errinvnhalf{2}\|\matM^{-1}\|
                    \\
                \\
                &\leq \umach\cdot\erronenhalf{2}+ \umach^2 \cdot \erronenhalf{2}
                \cdot \errinvnhalf{2}
                \\
                &= \umach\cdot\erronenhalf{2}\left( 1+ \umach \cdot \errinvnhalf{2}\right)
                .
        \end{align*}
    \item Using these two inequalities we can bound the norm of $\errormatrix_{\matB\matA i}$:
        \begin{align*}
            \|
            \errormatrix_{\matB\matA i}
            \|
            &= 
            \|
                \matB\errormatrix_1^{\INV} + \errormatrix_2^{\MM}
            \|
            \\
            &\leq
            \|
                \matB
            \|
            \|
            \errormatrix_1^{\INV} 
            \|
            +
            \|
            \errormatrix_2^{\MM}
            \|
            \\
            &\leq
            \|
                \matM
            \|
            \cdot \umach\cdot\errinvnhalf{2}
            \|\matM^{-1}\|
            +
            \umach
            \cdot
            \erronenhalf{2}
            (1+\umach\cdot\errinvnhalf{2})
            \\
            &\leq
            \kappa(\matM) 
            \cdot \umach\cdot\errinvnhalf{2}
            +
            \umach
            \cdot
            \erronenhalf{2}
            (1+\umach\cdot\errinvnhalf{2}).
        \end{align*}
    \item Next is the error term of the matrix multiplication between the result of $\MM(\matB,\INV(A))$ and $\matL_{11}$. Expanding Theorem \ref{theorem:fast_mm} gives
        \begin{align*}
            \|\errormatrix_3^{\MM}\| &\leq
            \pmult(n/2) \cdot \umach \cdot\|\matL_{11}\|
            \left\|
                \matB\matA i
            \right\|.
        \end{align*}
        It suffices to bound $
            \left\|
                \matB\matA i
            \right\|=
            \left\|
                \matB\left(
                \matA^{-1}+\errormatrix_{1}^{\INV}
                \right)
                +
                \errormatrix_2^{\MM}
            \right\|
            \leq
            \|\matB\matA^{-1}\|
                +
                \|\matB\errormatrix_{1}^{\INV}\|
                +
                \|\errormatrix_2^{\MM}\|
            $. 
        Using Proposition \ref{proposition:norm_bounds_posdef}, the triangle inequality, and the previous bounds, the two unknown norms in the sum are bounded as follows
        \begin{align*}
            \|\matB\matA^{-1}\| &\leq \|\matB\|\|\matA^{-1}\| \leq \|\matM\|\|\matM^{-1}\| = \kappa(\matM),\\
            \|\matB\errormatrix_{1}^{\INV}\| &\leq  \|\matB\|\|\errormatrix_{1}^{\INV}\| \\
                &\leq 
                \|\matM\|\cdot \umach \cdot \errinvnhalf{2} \|\matM^{-1}\| 
                \\
                &=
                \umach \cdot \errinvnhalf{2}\kappa(\matM).
        \end{align*}
        This gives
        \begin{align*}
            \|\matB\matA i\|
            &\leq
                \kappa(\matM)
                +
                \umach \cdot \errinvnhalf{2}\kappa(\matM)
                +
                \umach\cdot\erronenhalf{2}\left( 1+ \umach \cdot \errinvnhalf{2}\right)
                \\
            &=\kappa(\matM)
                \left(
                    1
                    +
                   \umach \cdot \errinvnhalf{2}
                \right)
                \left(
                1
                +
                \umach\cdot \erronenhalf{2}
                \right)
            \\
            &\leq\kappa(\matM)
                \left(
                    1
                    +
                   \umach \cdot \errinvnhalf{2}
                \right)^2,
        \end{align*}
        where in the last inequality we simplified $\left(1+\umach \cdot\erronenhalf{2}\right) \leq \left(1+\umach \cdot\errinvnhalf{2}\right)$.
        Then
        \begin{align*}
        \|\errormatrix_3^{\MM}\|
            &\leq
                \pmult(n/2) \cdot \umach \cdot \|\matL_{11}\|
                \cdot
                \kappa(\matM) 
                \left(
                    1
                    +
                   \umach \cdot \errinvnhalf{2}
                \right)
                \left(
                1
                +
                \umach\cdot \erronenhalf{2}
                \right)
            \\
            &\leq
            \umach\cdot
            \erronenhalf{2}
            \cdot
            \left(
                1
                +
                \umach\cdot\errinvnhalf{2}
            \right)^2
            \|\matL_{11}\|,
        \end{align*}
        concluding the bound for $\errormatrix_3^{\MM}$.
    \item The norm of $\errormatrix_{4}^{\MM}$ is very similar to that of $\errormatrix_{3}^{\MM}$ since they both involve a multiplication with $\matB\matA i$:
        \begin{align*}
            \|\errormatrix_{4}^{\MM}\|
            &\leq
            c_{\HERM}\log(n/2)\cdot\mu_{\MM}(n/2) \cdot\umach \cdot \|\matB\matA i\|\|\matB\| \qquad\text{(...from Corollary \ref{corollary:symmetrized_matrix_multiplication})}
            \\
            &\leq
            c_{\HERM}\log(n/2)\cdot \mu_{\MM}(n/2) \cdot\umach 
            \cdot
            \kappa(\matM)
                \left(
                    1
                    +
                   \umach \cdot \errinvnhalf{2}
                \right)^2
                \|\matM\|
            \\
            &\leq
            \umach \cdot\erronenhalf{2}
                \left(
                    1
                    +
                    \umach\cdot\errinvnhalf{2}
                \right)^2
                \|\matM\|.
        \end{align*}
    \item         Next is $\errormatrix_5^{\SUB}$. From Equation \eqref{eq:matrix_flop_errors}:
        \begin{align*}
            \|\errormatrix_{5}^{\SUB}\|
            &\leq
            \umach\sqrt{n/2}
            \left\| 
                \matC
                -
                \left(
                    (\matB\matA i)\matB^*+\errormatrix_4^{\MM}
                \right)
            \right\|
            \\
            &\leq
            \umach\sqrt{n/2}
            \left(
                \|
                    \matC
                \|
                +
                \|
                    (\matB\matA i)\matB^*
                \|
                +
                \|
                    \errormatrix_4^{\MM}
                \|
                \right\|
            \\
            &\leq
            \umach\sqrt{n/2}
            \left(
                \|
                    \matM
                \|
                +
                \|
                    (\matB\matA i)\matB^*
                \|
                +
                \|
                    \errormatrix_4^{\MM}
                \|
            \right)
            \\
            &\leq
            \umach\sqrt{n/2}
            \left(
                \|
                    \matM
                \|
                +
                \kappa(\matM) 
                \left(
                    1
                    +
                    \umach\cdot\errinvnhalf{2}
                \right)^2
                \|\matM\|
                +
                \umach\cdot\erronenhalf{2}
                \left(
                    1
                    +
                    \umach\cdot\errinvnhalf{2}
                \right)^2
                \|\matM\| 
            \right)
            \\
            &=
            \umach\sqrt{n/2}
            \cdot \|\matM\|
            \cdot
            \left(
                1
                +
                \kappa(\matM) 
                \left(
                    1
                    +
                    \umach\cdot\errinvnhalf{2}
                \right)^2
                +
                \umach\cdot\erronenhalf{2}
                \left(
                    1
                    +
                    \umach\cdot\errinvnhalf{2}
                \right)^2
            \right)
            \\
            &=
            \umach\sqrt{n/2}
            \cdot \|\matM\|
            \cdot
            \left(
                1
                +
                \left(
                    \kappa(\matM) 
                    +
                    \umach\cdot\erronenhalf{2}
                \right)
                \left(
                    1
                    +
                    \umach\cdot\errinvnhalf{2}
                \right)^2
            \right).
        \end{align*}
    \item  The next term is $\errormatrix_{\matL_{21}}$, which we can bound as follows:
        \begingroup
        \allowdisplaybreaks
        \begin{align*}
            \|
            \errormatrix_{\matL_{21}}
            \|
            &=
            \|
                \errormatrix_{\matB\matA i}
                \matL_{11}
                +
                \errormatrix_3^{\MM}
            \|
            \\
            &\leq
            \|
                \errormatrix_{\matB\matA i}
            \|
            \cdot
            \|
                \matL_{11}
            \|
            +
            \|
                \errormatrix_3^{\MM}
            \|
            \\
            &\leq
                \left[
                    \kappa(\matM) 
                    \cdot \umach\cdot\errinvnhalf{2}
                    +
                    \umach
                    \cdot
                    \erronenhalf{2}
                    (1+\umach\cdot\errinvnhalf{2})
                \right]
                \cdot
                \|
                    \matL_{11}
                \|
                +
                \ldots
                \\
                &\qquad\ldots
                +
                \umach\cdot\erronenhalf{2}
                    \left(
                        1
                        +
                        \umach\cdot\errinvnhalf{2}
                    \right)^2
                    \|\matL_{11}\|
            \\
            &=
            \left[
                \kappa(\matM) 
                \cdot \umach\cdot\errinvnhalf{2}
                +
                \umach
                \cdot
                \erronenhalf{2}
                \cdot
                \left(
                    1+\umach\cdot\errinvnhalf{2}
                \right)
                \cdot
                \left(
                    2 + \umach\cdot \errinvnhalf{2}
                \right)
            \right]
            \cdot
            \|
                \matL_{11}
            \|
            \\
            &\leq
            \umach\cdot\errinvnhalf{2}
            \cdot
            \left[
                \kappa(\matM) 
                +
                \left(
                    2 + \umach\cdot \errinvnhalf{2}
                \right)^2
            \right]
            \cdot
            \|
                \matL_{11}
            \|.
        \end{align*}
        \endgroup
    \item        

        The final and most involved error term is the norm of $\errormatrix_{\matS}$. Recall that from Line 7 of Algorithm \ref{alg:stable_cholesky} the matrix can be written as
        \begin{align*}
            \errormatrix_{\matS} = \matB\errormatrix_1^{\INV}\matB^* + \errormatrix_2^{\MM}\matB^* + \errormatrix_4^{\MM} + \errormatrix_5^{\SUB}.
        \end{align*}
        Once more, the norm of each term is bounded separately.
        \begin{align*}
            \|\matB\errormatrix_1^{\INV}\matB^*\|
            &\leq
            \|\matB\|\|\errormatrix_1^{\INV}\|\|\matB^*\|
            \\
            &\leq
            \|\matM\|^2\cdot \umach\cdot\errinvnhalf{2}\|\matM^{-1}\|
            \\
            &=
            \kappa(\matM) \cdot \|\matM\|\cdot \umach\cdot\errinvnhalf{2}.
        \end{align*}
        Similarly,
        \begin{align*}
            \|\errormatrix_2^{\MM}\matB^*\|
            &\leq
            \umach\cdot
        \erronenhalf{2} \left(
                    1+ \umach\cdot\errinvnhalf{2}
                \right)
                \cdot
                \|\matM\|.
        \end{align*}

        The final bound for $\errormatrix_{\matS}$ is given by the sum of the four bounds that were derived.
        \begingroup
        \allowdisplaybreaks
        \begin{align*}
            \|\errormatrix_{\matS}\| 
            &=
            \|\matB\errormatrix_1^{\INV}\matB^* + \errormatrix_2^{\MM}\matB^* + \errormatrix_4^{\MM} + \errormatrix_5^{\SUB}\|
            \\
            &\leq
            \|\matB\errormatrix_1^{\INV}\matB^*\| + \|\errormatrix_2^{\MM}\matB^*\| + \|\errormatrix_4^{\MM}\| + \|\errormatrix_5^{\SUB}\|
            \\
            &\leq
            \umach\cdot\errinvnhalf{2} \cdot \kappa(\matM)\cdot \|\matM\| \ldots
                \\
                &\qquad\ldots+ 
                \umach\cdot
                \erronenhalf{2} \left(
                    1+ \umach\cdot\errinvnhalf{2}
                \right)
                \cdot
                \|\matM\|
                \\
                &\qquad\ldots+ 
                \umach\cdot\erronenhalf{2}
                \left(
                    1
                    +
                    \umach\cdot\errinvnhalf{2}
                \right)^2
                \|\matM\|
                \\
                &\qquad\ldots+
                \umach\sqrt{n/2}
                \cdot \|\matM\|
                \cdot
                \left(
                    1
                    +
                    \left(
                        \kappa(\matM)
                        +
                        \umach\cdot\erronenhalf{2}
                    \right)
                    \left(
                        1
                        +
                        \umach\cdot\errinvnhalf{2}
                    \right)^2
                \right)
            \\
            &=
            \umach\cdot\|\matM\|\cdot
            \bigg[
                \errinvnhalf{2} \cdot \kappa(\matM) 
                \ldots
                \\
                &\qquad\ldots+
                \erronenhalf{2} \left(
                    1+ \umach\cdot\errinvnhalf{2}
                \right)
                \\
                &\qquad\ldots+
                \erronenhalf{2}
                \left(
                    1
                    +
                    \umach\cdot\errinvnhalf{2}
                \right)^2
                \\
                &\qquad\ldots+
                \sqrt{n/2}
                \left(
                    1
                    +
                    \left(
                        \kappa(\matM)
                        +
                        \umach\cdot\erronenhalf{2}
                    \right)
                    \left(
                        1
                        +
                        \umach\cdot\errinvnhalf{2}
                    \right)^2
                \right)
            \bigg]
            \\
            &=
            \umach\cdot\|\matM\|\cdot\errinvnhalf{2}\cdot
            \bigg[
                \kappa(\matM) 
                +
                \left(
                    1+ \umach\cdot\errinvnhalf{2}
                \right)\ldots
                \\
                &\qquad\ldots+
                \left(
                    1
                    +
                    \umach\cdot\errinvnhalf{2}
                \right)^2
                \\
                &\qquad\ldots+
                \left(
                    1
                    +
                    \left(
                        \kappa(\matM)
                        +
                        \umach\cdot\erronenhalf{2}
                    \right)
                    \left(
                        1
                        +
                        \umach\cdot\errinvnhalf{2}
                    \right)^2
                \right)
            \bigg]
            \\
            &=
            \umach\cdot\|\matM\|\cdot\errinvnhalf{2}\cdot
            \bigg[
                \kappa(\matM) 
                +
                \left(
                    1+ \umach\cdot\errinvnhalf{2}
                \right)
                \ldots
                \\
                &\qquad\ldots+
                \left(
                    1
                    +
                    \umach\cdot\errinvnhalf{2}
                \right)^2
                \\
                &\qquad\ldots+
                \left(
                    1
                    +
                    \left(
                        \kappa(\matM)
                        +
                        \umach\cdot\erronenhalf{2}
                    \right)
                    \left(
                        1
                        +
                        \umach\cdot\errinvnhalf{2}
                    \right)^2
                \right)
            \bigg]
            \\
            &\leq
            \umach\cdot\|\matM\|\cdot\errinvnhalf{2}\cdot
            \left[
                \kappa(\matM) 
                +
                1
                +
                \left(
                    1+ \umach\cdot\errinvnhalf{2}
                \right)^2
                \left(
                    1
                    +
                    1
                    +
                    \left(
                        \kappa(\matM)
                        +
                        \umach\cdot\erronenhalf{2}
                    \right)
                \right)
            \right]
            \\
            &\leq
            \umach\cdot\|\matM\|\cdot\errinvnhalf{2}\cdot
            \left[
                2\kappa(\matM) 
                +
                \left(
                    1+ \umach\cdot\errinvnhalf{2}
                \right)^2
                \left(
                    3\kappa(\matM)
                        +
                    \umach\cdot\erronenhalf{2}
                \right)
            \right].
        \end{align*}
        \endgroup
\end{enumerate}
\end{proof}
\end{lemma}

\subsection{Maintaining positive-definiteness, norms, and condition numbers throughout the recursion}
Given the bounds of Lemma \ref{lemma:cholesky_error_matrix_bounds}, we can now calculate the appropriate machine precision $\umach$ and the corresponding number of bits such that the algorithm will not break down due to loss of positive-definiteness of the submatrices. 
Hereafter, we will denote by $\matM$ the matrix that is passed as an argument in any of the recursive calls of Algorithm \ref{alg:stable_cholesky}, and $\matM_0$ will denote the original matrix that needs to be factorized.

\begin{lemma}
    \label{lemma:stable_cholesky_avoid_breakdown}
    Let $\matM_0\in\mathbb{H}^{n}_{++}$ be a matrix that is factorized using Algorithm \ref{alg:stable_cholesky} and $\kappa$ be its condition number. Then there exist constants $c_1, c_2, c_3, c_4\geq 1$, such that if $n>c_1$ and
    \begin{align*}
    \umach \leq \umach_{++}
        :=
        \frac{1}{c_2\cdot n^{c_3} \cdot\kappa^{c_4\log n}},
    \end{align*}
    then every matrix that is constructed in Line 7 during the recursion, that is, every Schur complement $\matStilde$ and every upper-left block $\matA$, will be Hermitian and positive-definite. Moreover, for each such matrix $\matX$ it holds that $\|\matX\|\leq 2\|\matM_0\|$ and $\kappa(\matX)\leq 2\kappa(\matM_0)$. This value of $\umach$ translates to
    \begin{align*}
        \log(1/\umach) = O(\log(n) \log(\kappa))
    \end{align*}
    required bits of precision.
\begin{proof}
At each recursive step, two matrices need to remain positive-definite: $\matA$ and $\matStilde$. If $\matM$ is positive-definite then so is $\matA$. It remains to ensure the same for the Schur complement. 
To prove this we first fix $\umach$ be bounded by the value
\begin{align}
    \umach \leq \umach_{++} := \frac{1}{n^{2\cinv}\cdot \errinvnhalf[\matM_0]{2}\cdot72\kappa(\matM_0)^2}.
    \label{eq:lemma_avoid_breakdown_umach_upper_bound}
\end{align}

Assume that we are in the first level of recursion, i.e. $\matM=\matM_0$. Per Lemma \ref{lemma:cholesky_error_matrix_bounds}, we can write $\matStilde=\matS+\errormatrix_{\matS}$, where
\begin{align*}
    \|\errormatrix_{\matS}\|\leq \umach\cdot\|\matM_0\|\cdot\errinvnhalf[\matM_0]{2}\cdot
            \left(
                2\kappa(\matM_0)
                +
                \left(
                    1+ \umach\cdot\errinvnhalf[\matM_0]{2}
                \right)^2
                \left(
                    3\kappa(\matM_0)
                        +
                    \umach\cdot\erronenhalf[\matM_0]{2}
                \right)
            \right).
\end{align*}

For the assumed value of $\umach$, both $\umach\cdot\erronenhalf[\matM_0]{2}\leq 1$ and $\umach\cdot\errinvnhalf[\matM_0]{2}\leq 1$, and we also know that $1\leq \kappa(\matM_0)$, in which case the bound simplifies to
\begin{align}
    \label{eq:stable_cholesky_norm_ES_bound}
    \|\errormatrix_{\matS}\|
    &\leq 
    \umach\cdot\|\matM_0\|\cdot\errinvnhalf[\matM_0]{2}\cdot
            \left(
                2\kappa(\matM_0)  
                +
                \left(
                    1+ 1
                \right)^2
                \left(
                    3\kappa(\matM_0) 
                    +
                    1
                \right)
            \right)
    \nonumber
    \\
    &\leq
    \umach\cdot\|\matM_0\|\cdot\errinvnhalf[\matM_0]{2}\cdot
            \left(
                2\kappa(\matM_0)
                +
                4
                \left(
                    3\kappa(\matM_0)
                        +
                    \kappa(\matM_0)
                \right)
            \right)
    \nonumber
    \\
    &\leq
    \umach\cdot\|\matM_0\|\cdot\errinvnhalf[\matM_0]{2}\cdot
            18\kappa(\matM_0)
    \nonumber
    \\
    &\leq
    \frac{1}{n^{2\cinv}\cdot \errinvnhalf[\matM_0]{2}\cdot72\kappa(\matM_0)^2}
    \cdot
    \|\matM_0\|\cdot\errinvnhalf[\matM_0]{2}\cdot
            18\kappa(\matM_0)
    \nonumber
    \\
    &\leq
    \frac{1}{n^{2\cinv}}
    \cdot
    \lambda_{\min}(\matM_0)
    .
\end{align}

Using this, we derive the following useful inequalities
\begin{align*}
    \lambda_{\max}(\matStilde) &\leq \|\matS\|+\|\errormatrix_{\matS}\| \leq (1+1/n^{2\cinv})\|\matM_0\|,
    \\
    \lambda_{\min}(\matStilde) &\geq \lambda_{\min}(\matS)-\|\errormatrix_{\matS} \|
        \geq \lambda_{\min}(\matS) - \frac{1}{n^{2\cinv}}\lambda_{\min}(\matM_0)
        \geq (1-1/n^{2\cinv})\lambda_{\min}(\matM_0),
    \\
    \kappa(\matStilde)&\leq \frac{n^{2\cinv}+1}{n^{2\cinv}-1}\cdot \kappa(\matM_0),
    \\
    \errinvnhalf[\matStilde]{4}
    &=
    \mu(n/4)\kappa(\matStilde)^{\cinv\log(n/4)}
    \\
    &\leq 
    \mu(n/4)\cdot \left(
        \frac{n^{2\cinv}+1}{n^{2\cinv}-1}
    \right)^{\cinv\log(n/4)}
    \cdot
    \kappa(\matM_0)^{\cinv\log(n/4)}
    \\
    &\leq 
    n^{\cinv}
    \cdot
    \mu(n/4)\cdot
    \kappa(\matM_0)^{\cinv\log(n/4)}
    \\
    &\leq
    n^{\cinv}\errinvnhalf[\matM_0]{4}.
\end{align*}
In the above, we used the fact that for $\cinv\geq 1$ we have that $\frac{n^{2\cinv}+1}{n^{2\cinv}-1}\leq 2$, for all $n\geq 3$, and therefore $2^{\cinv\log(n/4)}=(n/4)^{\cinv}\leq n^{\cinv}$.
Since the smallest eigenvalue of $\matStilde$ is larger than a positive value, we can conclude that it is Hermitian and positive-definite, and its condition number and spectral norm are appropriately bounded.

In a similar manner, we can now get a bound for the Schur complement of the Schur complement. 
In this recursive step we call Algorithm \ref{alg:stable_cholesky} with $\matM=\matStilde$. Let
\begin{align*}
\matStilde =
\begin{pmatrix}
    \matA_1 & \matB_1^* \\
    \matB_1 & \matC_1
\end{pmatrix}.
\end{align*}
Let $\matS_1$ be the true Schur complement of $\matStilde$, i.e. $\matS_1=\matC_1-\matB_1\matA_1^{-1}\matB_1^*$, and let $\matStilde_1$ be the approximate Schur complement of $\matStilde$ that is constructed in line 7 of Algorithm \ref{alg:stable_cholesky} when executed on $\matStilde$. Let $\errormatrix_{\matS_1}$ be the corresponding error matrix, i.e. $\matStilde_1 = \matS_1 + \errormatrix_{\matS_1}$. Using Lemma \ref{lemma:cholesky_error_matrix_bounds} and simplifying the terms $(1+1/n^{\cinv})\leq 2$ and $\frac{n^{2\cinv}+1}{n^{2\cinv}-1}\leq 2$ we have that

\begin{align*}
    \|&\errormatrix_{\matS_1}\|
    \leq 
        \umach
        \overbrace{\|\matStilde\|}^{\leq 2\|\matM_0\|}
        \underbrace{
            \errinvnhalf[\matStilde]{4}
        }_{
            \leq n^{\cinv}\errinvnhalf[\matM_0]{4}
        }
        \left[
            2
            \overbrace{
                \kappa(\matStilde)  
            }^{
               \leq 2\kappa(\matM_0)
            }
            +
            \left(
                1+ \umach\cdot
                \underbrace{
                    \errinvnhalf[\matStilde]{4}
                }_{
                    \leq n^{\cinv}\errinvnhalf[\matM_0]{4}
                }
            \right)^2
            \left(
                3
                \overbrace{
                    \kappa(\matStilde)  
                }^{
                   \leq 2\kappa(\matM_0)
                }
                +
                \umach
                \overbrace{
                    \erronenhalf[\matStilde]{4}
                }^{
                    \leq 2\erronenhalf[\matM_0]{2}
                }
            \right)
        \right]
    \\
    &\leq 
        \umach\cdot
        2\|\matM_0\|
        \cdot
        n^{\cinv}\errinvnhalf[\matM_0]{4}
        \left(
            4\kappa(\matM_0)
            +
            \left(
                1+ \umach
                \cdot
                n^{\cinv}
                \errinvnhalf[\matM_0]{4}
            \right)^2
            \left(
                6\kappa(\matM_0)
                +
                \umach\cdot
                2\erronenhalf[\matM_0]{4}
            \right)
        \right)
    \\
    &\leq 
        \umach\cdot
        2\|\matM_0\|
        \cdot
        n^{\cinv}\errinvnhalf[\matM_0]{4}
        \left(
            4\kappa(\matM_0)
            +
            \left(
                1+1
            \right)^2
            \left(
                6\kappa(\matM_0)
                +
                2
            \right)
        \right)
    \\
    &\leq 
        \umach\cdot
        \|\matM_0\|
        \cdot
        n^{\cinv}\errinvnhalf[\matM_0]{4}
        \cdot
        72\kappa(\matM_0)
    \\
    &\leq 
        \frac{1}{n^{2\cinv}\cdot \errinvnhalf[\matM_0]{2}\cdot72\kappa(\matM_0)^2}
        \cdot
        \|\matM_0\|
        \cdot
        n^{\cinv}\errinvnhalf[\matM_0]{4}
        \cdot
        72\kappa(\matM_0)
    \\
    &= 
        \frac{1}{n^{\cinv}}
        \cdot
        \lambda_{\min}(\matM_0).
\end{align*}

The corresponding bounds for $\matStilde_1$ can be derived:
\begin{align*}
    \lambda_{\max}(\matStilde_1) 
    &\leq \|\matStilde\|+\|\errormatrix_{\matS_1}\|
        \\
        &\leq
        \|\matS\|+\|\errormatrix_{\matS}\| + \|\errormatrix_{\matS_1}\|
        \\
        &\leq \|\matM_0\| + \frac{1}{n^{2\cinv}}\|\matM_0\|+ \frac{1}{n^{\cinv}}\|\matM_0\|
        \\
        &\leq
        (1+2/n^{\cinv})\|\matM_0\|,
    \\
    \lambda_{\min}(\matStilde_1) 
        &\geq (1-2/n^{\cinv})\lambda_{\min}(\matM_0),
    \\
    \kappa(\matStilde_1)
        &\leq \frac{n^{\cinv}+2}{n^{\cinv}-2}\cdot \kappa(\matM_0),
    \\
    \errinvnhalf[\matStilde_1]{8}
        &=
        \mu(n/8)\kappa(\matStilde)^{\cinv\log(n/8)}
        \\
        &\leq 
        \mu(n/8)\cdot \left(
            \frac{n^{\cinv}+2}{n^{\cinv}-2}
        \right)^{\cinv\log(n/8)}
        \cdot
        \kappa(\matM_0)^{\cinv\log(n/8)}
        \\
        &\leq 
        n^{\cinv}
        \cdot
        \mu(n/8)\cdot
        \kappa(\matM_0)^{\cinv\log(n/8)}
        \\
        &=
        n^{\cinv}\errinvnhalf[\matM_0]{8}.
\end{align*}
We can therefore identify that $\matStilde_1$ is also Hermitian positive-definite and its extremal eigenvalues are similarly bounded with those of $\matStilde$.

If we keep applying the same analysis for all $\log(n)$ iterations, as long as $n$ is greater than some constant such that $(n^{\cinv}+\log n) / (n^{\cinv}-\log n)\leq 2$, then for each $i=1,...,\log n$ it holds that
\begin{align*}
    \|\errormatrix_{\matS_i}\|\leq \frac{1}{n^{\cinv}}\cdot\lambda_{\min}(\matM_0) \leq \frac{1}{n^{\cinv}}\|\matM_0\|,
\end{align*}
which implies that
\begin{align*}
    \lambda_{\max}(\matStilde_i) 
    &\leq
        (1+i/n^{\cinv})\|\matM_0\|,
    \\
    \lambda_{\min}(\matStilde_i) 
        &\geq (1-i/n^{\cinv})\lambda_{\min}(\matM_0),
    \\
    \kappa(\matStilde_i)
        &\leq \frac{n^{\cinv}+i}{n^{\cinv}-i}\cdot \kappa(\matM_0),
    \\
    \errinvnhalf[\matStilde_i]{2^i}
        &\leq
        n^{\cinv}\errinvnhalf[\matM_0]{2^{i}}.
\end{align*}
The same bounds hold for the matrices $\matA_i$, since they always originate from the top-left corner of a matrix $\matStilde_j$, where $j< i$, i.e., $\matA_i$ either has no errors or it inherits the errors from a matrix $\matStilde_j$.
We can therefore conclude that, for the value of $\umach$ in Inequality \eqref{eq:lemma_avoid_breakdown_umach_upper_bound}, every matrix $\matA_i$ and $\matStilde_i$ that is constructed during the recursion of Algorithm \ref{alg:stable_cholesky} will be Hermitian and positive-definite and its condition number and spectral norm will be at most $2\kappa(\matM_0)$ and $2\|\matM_0\|$ respectively.
The required number of bits is
\begin{align*}
    \log(1/\umach_{++})
    =
    \log
    \left(
        n^{2\cinv}\cdot \errinvnhalf[\matM_0]{2}\cdot72\kappa(\matM_0)^2
    \right)
    =
    O\left(
        \log(n)\log(\kappa(\matM_0))
    \right).
\end{align*}

\end{proof}
\end{lemma}

\subsection{Final backward-approximation bounds and proof of Theorem \ref{theorem:stable_cholesky_bounds}}

Having safeguarded the possibility of a breakdown due to loss of positive-definiteness, we can now revisit the bounds of Lemma \ref{lemma:main_error_bounds_stable_cholesky} and finalize the proof of Theorem \ref{theorem:stable_cholesky_bounds}.

\begin{theorem}[Restatement of Theorem \ref{theorem:stable_cholesky_bounds}]
    \label{theorem:stable_cholesky_bounds_restatement}
    Given a Hermitian positive-definite matrix $\matM$, there exists an algorithm $\matL\leftarrow \CHOLESKY(\matM)$, listed in 
    Algorithm \ref{alg:stable_cholesky}, which requires $O(T_{\MM}(n))$ arithmetic operations. This algorithm is logarithmically stable, in a sense that, there exist global constants $c_1$, $c_2$, $c_3$, such that for all $\epsilon\in(0,1)$, if executed in a floating point machine with precision \begin{align*}
        \umach \leq \umach_{\CHOLESKY} := \epsilon \frac{1}{c_1n^{c_2}\kappa(\matM)^{c_3\log n}},
    \end{align*}
    which translates into
    $
        O\left(
            \log(n)\log(\kappa(\matM)) + \log(\tfrac{1}{\epsilon})
        \right)
    $
    required bits of precision,
    then it does not break down due to arithmetic errors, and the solution returned satisfies
    $
        \|\matL\matL^*-\matM\|\leq \epsilon\|\matM\|.
    $

\begin{proof}
The arithmetic complexity of the algorithm was already bounded in Proposition \ref{proposition:stable_cholesky_complexity}.

For the error proof we assume the bound for $\umach\leq\umach_{++}$ from Inequality \eqref{eq:lemma_avoid_breakdown_umach_upper_bound}.
Combining the inequalities for a single recursion step from Lemma \ref{lemma:cholesky_error_matrix_bounds} with the results of the previous section and the bound for $\umach$, we have the following inequalities (recall once more that $\matM$ is the input of the algorithm for the specific recursive call, while $\matM_0$ is the matrix in the original call):
\begingroup
\allowdisplaybreaks
\begin{align*}
    \left\|\errormatrix_3^{\MM}\right\| &\leq
            \umach\cdot
            \erronenhalf{2}
            \left(
                1
                +
                \umach\cdot\errinvnhalf{2}
            \right)^2
            \|\matL_{11}\|
    \\
        &\leq \umach \cdot 8\cdot  \erronenhalf[\matM_0]{2}\|\matL_{11}\|,
    \\
    \left\|\errormatrix_{\matB\matA i}\right\| 
        &\leq 
        \umach\cdot\errinvnhalf{2}
        \cdot 
        \kappa(\matM)
        +
        \umach
        \cdot
        \erronenhalf{2}
        (1+\umach\cdot\errinvnhalf{2})
    \\
        &\leq
        \umach\cdot 
        \left(
            2n^{\cinv}\errinvnhalf[\matM_0]{2}
            \cdot 
            \kappa(\matM_0)
            +
            4\erronenhalf[\matM_0]{2}
        \right)
    \\
        &\leq
        \umach\cdot
        6n^{\cinv}\errinvnhalf[\matM_0]{2}
        \cdot 
        \kappa(\matM_0),
    \\
    \left\|
        \errormatrix_{\matL_{21}}
    \right\|
        &\leq
        \umach\cdot\errinvnhalf{2}
        \cdot
        \left[
            \kappa(\matM)
            +
            \left(
                2 + \umach\cdot \errinvnhalf{2}
            \right)^2
        \right]
        \cdot
        \|
            \matL_{11}
        \|
    \\
        &\leq
        \umach\cdot n^{\cinv}\errinvnhalf[\matM_0]{2}
        \cdot
        (
            2\kappa(\matM_0)
            +
            9
        )
        \cdot
        \|
            \matL_{11}
        \|
    \\
        &\leq
        \umach\cdot n^{\cinv}\errinvnhalf[\matM_0]{2}
        \cdot 11\kappa(\matM_0)
        \cdot
        \|
            \matL_{11}
        \|,
    \\
    \left\|
        \errormatrix_{\matS}
    \right\|
        &\leq
        \umach\cdot\|\matM\|\cdot\errinvnhalf{2}\cdot
        \left(
            2\kappa(\matM)  
            +
            \left(
                1+ \umach\cdot\errinvnhalf{2}
            \right)^2
            \left(
                3\kappa(\matM)
                    +
                \umach\cdot\erronenhalf{2}
            \right)
        \right)
    \\
        &\leq
        \umach\cdot 2\|\matM_0\|\cdot n^{\cinv}\errinvnhalf[\matM_0]{2}\cdot
        \left(
            4\kappa(\matM_0)  
            +
            \left(
                1+ 1
            \right)^2
            \left(
                6\kappa(\matM_0)
                    +
                1
            \right)
        \right)
    \\
        &\leq
        \umach\cdot n^{\cinv}\errinvnhalf[\matM_0]{2}\cdot
            64\kappa(\matM_0)\cdot \|\matM_0\|.
\end{align*}
\endgroup
Using those inequalities, the first term that we need to bound from Lemma \ref{lemma:main_error_bounds_stable_cholesky} is the following
\begingroup
\allowdisplaybreaks
\begin{align*}
    \left\| 
        \matL_{21}\matL_{11}^* - \matB
    \right\|
    &\leq
        \left(
            \kappa(\matM)
            +
            \|\errormatrix_{\matB\matA i}\|
        \right)
        \err(n/2)
        +
        \|\errormatrix_{\matB\matA i}\|
        \|\matM\|
        +
        \|\errormatrix_{3}^{\MM}\|
        \|\matL_{11}^*\|
    \\
    &\leq
        \left(
            2\kappa(\matM_0) 
            +
            \umach\cdot
            6n^{\cinv}\errinvnhalf[\matM_0]{2}
            \cdot 
            \kappa(\matM_0)
        \right)
        \err(n/2)
        \\
        &\qquad\ldots
        +
        \umach\cdot
            6n^{\cinv}\errinvnhalf[\matM_0]{2}
            \cdot 
            \kappa(\matM_0)
        \cdot 2\|\matM_0\|
        \\
        &\qquad\ldots
        +
        \umach \cdot 8\cdot  \erronenhalf[\matM_0]{2} \cdot
        \underbrace{
            \|\matL_{11}\|
            \|\matL_{11}^*\|
        }_{
            \leq \|\matA\|+\|\errormatrix_{\matA}^{\mathsf{CH}}\| \leq 2\|\matM_0\|+\err(n/2)
        }
    \\
    &\leq
        \left(
            2\kappa(\matM_0) 
            +
            \umach\cdot
            6n^{\cinv}\errinvnhalf[\matM_0]{2}
            \cdot 
            \kappa(\matM_0)
        \right)
        \err(n/2)
        \\
        &\qquad\ldots
        +
        \umach\cdot
            12n^{\cinv}\errinvnhalf[\matM_0]{2}
            \cdot 
            \kappa(\matM_0)
        \|\matM_0\|
        \\
        &\qquad\ldots
        +
        \umach \cdot 16\cdot  \erronenhalf[\matM_0]{2} \cdot \|\matM_0\|
        \\
        &\qquad\ldots
        +
        \umach \cdot 8\cdot  \erronenhalf[\matM_0]{2} \cdot \err(n/2)
    \\
    &\leq
        \left(
            2\kappa(\matM_0) 
            +
            \umach\cdot
            6n^{\cinv}\errinvnhalf[\matM_0]{2}
            \cdot 
            \kappa(\matM_0)
            +
            \umach \cdot 8\cdot  \erronenhalf[\matM_0]{2}
        \right)
        \err(n/2)
        \\
        &\qquad\ldots
        +
        \umach\cdot
        \left(
            12n^{\cinv}\errinvnhalf[\matM_0]{2}
            \cdot 
            \kappa(\matM_0)
            +
            \umach \cdot 16\cdot  \erronenhalf[\matM_0]{2}
        \right)
        \cdot \|\matM_0\|.
    \\
    &\leq
        22\cdot \kappa(\matM_0) 
        \cdot
        \err(n/2)
        +
        \umach\cdot
            28n^{\cinv}\errinvnhalf[\matM_0]{2}
            \cdot 
            \kappa(\matM_0)
        \cdot \|\matM_0\|,
\end{align*}
\endgroup
where in the last inequality we used the assumed bound for $\umach$, namely Inequality \eqref{eq:lemma_avoid_breakdown_umach_upper_bound}.

Similarly
\begingroup
\allowdisplaybreaks
\begin{align*}
    \left\|
        \matL_{21}\matL_{21}^*
        \right.
        &+
        \left.
        \matL_{22}\matL_{22}^*
        -
        \matC
    \right\|
    \leq
        2\kappa^2(\matM)
        \cdot
        \err(n/2)
        +
        2
        \kappa(\matM)
        \|
            \errormatrix_{\matL_{21}}
        \|
        \|\matL_{11}\|
        +
        \|
        \errormatrix_{\matL_{21}}
        \|^2
        +
        \|
        \errormatrix_{\matS} 
        \|
    \\
    &\leq
        8\kappa^2(\matM_0)
        \cdot
        \err(n/2)
        \\
        &\qquad\ldots
        +
        4
        \kappa(\matM_0)
        \cdot
        \umach\cdot n^{\cinv}\errinvnhalf[\matM_0]{2}
            \cdot 11\kappa(\matM_0)
            \cdot
            \|
                \matL_{11}
            \|
        \|\matL_{11}\|
        \\
        &\qquad\ldots
        +
        \left(
        \umach\cdot n^{\cinv}\errinvnhalf[\matM_0]{2}
            \cdot 11\kappa(\matM_0)
            \cdot
            \|
                \matL_{11}
            \|
        \right)
        ^2
        \\
        &\qquad\ldots
        +
        \umach\cdot n^{\cinv}\errinvnhalf[\matM_0]{2}\cdot
                64\kappa(\matM_0)\cdot \|\matM_0\|
    \\
    &\leq
        8\kappa^2(\matM_0)
        \cdot
        \err(n/2)
        \\
        &\qquad\ldots
        +
        44
        \kappa(\matM_0)^2
        \cdot
        \umach\cdot n^{\cinv}\errinvnhalf[\matM_0]{2}
            \cdot
        \left(
            2\|\matM_0\| + \err(n/2)
        \right)
        \\
        &\qquad\ldots
        +
        \left(
        \umach\cdot n^{\cinv}\errinvnhalf[\matM_0]{2}
            \cdot 11\kappa(\matM_0)
        \right)
        ^2
        \cdot
        \left(
            2\|\matM_0\| + \err(n/2)
        \right)
        \\
        &\qquad\ldots
        +
        \umach\cdot n^{\cinv}\errinvnhalf[\matM_0]{2}\cdot
                64\kappa(\matM_0)\cdot \|\matM_0\|
    \\
    &\leq
        \bigg[
            8\kappa^2(\matM_0)
            \\
            &\qquad\ldots
            +
            44
            \kappa(\matM_0)^2
            \cdot
            \umach\cdot n^{\cinv}\errinvnhalf[\matM_0]{2}
            \\
            &\qquad\ldots
            +
            121\left(
            \umach\cdot n^{\cinv}\errinvnhalf[\matM_0]{2}
                \cdot \kappa(\matM_0)
            \right)
            ^2
        \bigg]
        \cdot \err(n/2)
        \\
        &\ldots
        +
        \bigg[       
            88
            \kappa(\matM_0)^2
            \cdot
            \umach\cdot n^{\cinv}\errinvnhalf[\matM_0]{2}
            \\
            &\qquad\ldots
            +
            242
            \left(
            \umach\cdot n^{\cinv}\errinvnhalf[\matM_0]{2}
                \cdot\kappa(\matM_0)
            \right)
            ^2
            \\
            &\qquad\ldots
            +
            \umach\cdot n^{\cinv}\errinvnhalf[\matM_0]{2}\cdot
                64\kappa(\matM_0)
        \bigg]
        \cdot
        \|\matM_0\|
    \\
    &=
        \bigg[
            8
            +
            \left(
                44
                +
                121
                \umach\cdot n^{\cinv}\errinvnhalf[\matM_0]{2}
            \right)
            \cdot
            \umach\cdot n^{\cinv}\errinvnhalf[\matM_0]{2}
        \bigg]
        \cdot
        \kappa(\matM_0)^2
        \cdot \err(n/2)
        \\
        &\ldots
        +
         \umach\cdot \|\matM_0\|\cdot 
         \kappa(\matM_0)\cdot n^{\cinv}\errinvnhalf[\matM_0]{2} 
        \cdot
        \bigg[       
            88
            \kappa(\matM_0)
            +64+\ldots
            \\
            &\qquad\ldots
            +
            242
            \kappa(\matM_0)
                \umach\cdot n^{\cinv}\errinvnhalf[\matM_0]{2}
        \bigg]
    \\
    &\leq
        173
        \cdot
        \kappa(\matM_0)^2
        \cdot \err(n/2)
        +
        \umach
        \cdot
        394 
        \kappa(\matM_0)^2n^{\cinv}\errinvnhalf[\matM_0]{2} 
        \cdot
        \|\matM_0\|
        ,
\end{align*}
\endgroup
where in the last inequality we used again \eqref{eq:lemma_avoid_breakdown_umach_upper_bound}.

We can now derive a recursive formula for the main error bounds, Inequality \eqref{eq:main_error_stable_cholesky}.
\begin{align}
    \err(n) 
    &= \|\matL\matL^*-\matM\| \nonumber
    \\
    &\leq 
        \left\| 
            \matL_{11}\matL_{11}^*
            -
            \matA
        \right\|
        +
        \left\| 
            \matL_{21}\matL_{11}^*
            -
            \matB
        \right\|
        +
        \left\| 
            \matL_{21}\matL_{21}^*
            +
            \matL_{22}\matL_{22}^*
            -
            \matC
        \right\|
        \nonumber
    \\
    &\leq 
        \err(n/2)
        +\ldots
        \nonumber
        \\
        &\qquad\ldots+
        22\cdot \kappa(\matM_0) 
        \cdot
        \err(n/2)
        +
        \umach\cdot
            28n^{\cinv}\errinvnhalf[\matM_0]{2}
            \cdot 
            \kappa(\matM_0)
        \cdot \|\matM_0\|
        +
        \ldots
        \nonumber
        \\
        &\qquad\ldots+
        173
        \cdot
        \kappa(\matM_0)^2
        \cdot \err(n/2)
        +
        \umach
        \cdot
        394 
        \kappa(\matM_0)^2n^{\cinv}\errinvnhalf[\matM_0]{2} 
        \cdot
        \|\matM_0\|
    \nonumber
    \\
    &\leq 
        196
        \kappa(\matM_0)^2
        \cdot \err(n/2)
        +
        \umach
        \cdot
        422
        \kappa(\matM_0)^2
        n^{\cinv}
        \errinvnhalf[\matM_0]{2} 
        \cdot
        \|\matM_0\|.
\end{align}
The base case $n=1$ is trivial since $\matM$ is just a positive real number\footnote{It has to be real otherwise a complex 1-by-1 matrix is not Hermitian.} and we can compute $\matL=\fl(\sqrt{\matM})=(1+\delta)\sqrt{\matM}$ with $|\delta|\leq\umach$, which means that $\matL\matL^*=\matL^2=(1+\delta)^2\matM>0$ and
\begin{align*}
    |\matM-\matL\matL^*| = |\matM(2\delta+\delta^2)| \leq 3\umach\|\matM\| \leq 6\umach \|\matM_0\|.
\end{align*}
If we denote $\alpha=196\kappa(\matM_0)^2$ and $\beta = \umach\cdot422\kappa(\matM_0)^2n^{\cinv}\errinvnhalf[\matM_0]{2} $, then the solution of the recursion can be written as
\begin{align*}
    \err(n) 
        &\leq \alpha^{\log n} \err(1) + \beta\|\matM_0\|\sum_{i=0}^{\log n-1}\alpha^i\\
        &\leq \alpha^{\log n} 6\umach\|\matM_0\| + \log (n) \alpha^{\log n}\beta\|\matM_0\|
        \\
        &=
        \umach n^{c_1}\kappa(\matM_0)^{2\log n}\|\matM_0\| + \log (n)n^{c_1}\kappa(\matM_0)^{2\log n} \umach\cdot422\kappa(\matM_0)^2n^{\cinv}\errinvnhalf[\matM_0]{2}\cdot \|\matM_0\|
        \\
        &=
        \umach \cdot c_4 n^{c_2}\kappa(\matM_0)^{c_3\log n}\cdot \|\matM_0\|.
\end{align*}

We can now absorb the remaining terms inside $\umach$, by tuning the constants of Inequality \eqref{eq:lemma_avoid_breakdown_umach_upper_bound}, to argue that we can achieve any desired (multiplicative) backward-accuracy $\epsilon\in(0,1)$ and if we set 
\begin{align*}
    \umach\leq \epsilon \frac{1}{c_1'n^{c_2'}\kappa(\matM_0)^{c_3'\log n}}
\end{align*} for some constants $c_1',c_2',c_3'$. This translates to the advertised 
\begin{align*}
    \log(1/\umach)
        &=
        O\left(
            \log (n^{c_2'}\kappa(\matM_0)^{c_3'\log n}/\epsilon)
        \right) 
        = 
        O\left(
            \log(n)\log(\kappa(\matM_0)) + \log(\tfrac{1}{\epsilon})
        \right)
\end{align*}
bits of precision.
\end{proof}

\end{theorem}

\subsection{Reducing the GEP to a regular Hermitian eigenproblem}

\label{appendix:reduce}

We can now use $\CHOLESKY$ to reduce the definite GEP to a regular Hermitian eigenproblem
For the rest of the paper, we assume that $\|\matH\|,\|\matS^{-1}\|,\|\matS^{-1}\matH\|\leq 1$. This is not a limitation since we can approximate the norms of  $\matH$ and $\matS^{-1}$ in floating point using the algorithm $\SIGMAK$, which is described later, and then scale accordingly. Formally, let $\eta \gtrsim \|\matH\|$ and $\sigma\gtrsim \|\matS^{-1}\|$. Then we can rewrite the generalized eigenproblem
\begin{align*}
\matH\matC=\matS\matC\geneigmatrix
\quad
\Leftrightarrow
\quad
(\tfrac{1}{\eta}\matH)\matC = (\sigma\matS) \matC (\geneigmatrix\tfrac{1}{\eta\sigma})
\quad
\Leftrightarrow
\quad
\matH'\matC = \matS'\matC\geneigmatrix',     
\end{align*}
i.e. it is the same generalized eigenproblem only with scaled eigenvalues. We can thus safely make the unit-norms assumption. In Proposition \ref{proposition:alg_reduce} we prove the properties of the reduction, which, in brief, states that we can compute  in $O(T_{\MM}(n))$ floating point operations using $O(\log(n)\log(\kappa(\matS))+\log(1/\epsilon))$ bits the matrix
$
    \matHtilde \leftarrow \matLtilde^*\matH\matLtilde,
$,
where $\matLtilde$ is returned by $\CHOLESKY(\INV(\matS))$. The eigenvalues of $\matHtilde$ provably approximate the pencil eigenvalues:
\begin{align}
    \label{equation:reduce}
    \labs
        \lambda_i(\matHtilde) - \lambda_i(\matH,\matS)
    \rabs
    \leq
    \epsilon,
\end{align}
where $\epsilon\in(0,1)$ is a given accuracy.
Algorithm \ref{alg:reduce} details the reduction.

\begin{algorithm}[htb]
\caption{$\REDUCE$.}
\label{alg:reduce}
\begin{algorithmic}[1]
\setstretch{0.8}
\Statex \begin{align*}
    \REDUCE
\end{align*}
\small
\Input Matrix $\matH
\in\mathbb{H}^{n}
$,
matrix $\matS\in\mathbb{H}^{n}_{++}$, accuracy parameter $\epsilon\in(0,1)$.
\Require $\matH$ is Hermitian and $\|\matH\|\leq 1$, $\matS$ is Hermitian and positive definite and $\|\matS^{-1}\|\leq 1$.
\setstretch{1.5}
\Algorithm $\matHtilde \leftarrow\REDUCE(\matH,\matS)$.
\State $\matS_{\INV} \leftarrow \HERM(\INV(\matS))$.
    \Comment{$\matS_{\INV} = \matS^{-1} + \errormatrix^{\INV}_{1}$}.
\State $\matLtilde\leftarrow \CHOLESKY(\matS_{\INV})$.
    \Comment{$\matLtilde\matLtilde^* = \matS_{\INV} + \errormatrix^{\mathsf{CH}}_{2}$}.
\State $\matHtilde \leftarrow \HERM(\MM(\MM(\matLtilde^*,\matH),\matLtilde)).$
    \Comment{$\matHtilde = (\matLtilde^*\matH+\errormatrix^{\MM}_3)\matLtilde + \errormatrix_4^{\MM} = \matLtilde^*\matH\matLtilde + \errormatrix_5$}.
\State \Return $\matHtilde$.
\Output Hermitian matrix $\matHtilde$.
\Ensure $|\lambda_i(\matHtilde)-\lambda_i(\matS^{-1}\matH)|\leq \epsilon$ in $O(T_{\MM}(n))$ floating point operations using $O(\log(n)\log(\kappa(\matS))+\log(1/\epsilon))$ bits.
\end{algorithmic}
\end{algorithm}

\begin{proposition}
    \label{proposition:alg_reduce}
    Given $\matH\in\mathbb{H}^n$,
    $\matS\in\mathbb{H}^{n}_{++}$, and $\epsilon\in(0,1),$ Algorithm \ref{alg:reduce} executes $O(T_{\MM}(n))$ floating point operations and returns a matrix $\matHtilde\leftarrow\REDUCE(\matH,\matS,\epsilon)$, such that, if the machine precision satisfies
    \begin{align*}
        \umach \leq \umach_{\REDUCE}:= \epsilon 
                \frac{1}{
                    \rho_{1}n^{\rho_{2}}(4\kappa(\matS))^{\rho_{3}\log(n)}
                },
    \end{align*}
    for some constants $\rho_1,\rho_2,\rho_3$, and $\epsilon\in(0,1)$, which translates to 
    \begin{align*}
        O\lpar \log(1/\umach_{\REDUCE}) \rpar
        =
        O\lpar
            \log(n)\log(\kappa(\matS)) + \log(1/\epsilon)
        \rpar
    \end{align*}
    required bits of precision,
    then for all $i\in[n]$
    \begin{align*}
        |\lambda_i(\matHtilde)-\lambda_i(\matH,\matS)| \leq \epsilon \|\matS^{-1}\|\|\matH\|,
    \end{align*}
    where $\lambda_i(\matH,\matS)$ are the eigenvalues of the Hermitian definite pencil $(\matH,\matS)$. 
    \begin{proof}
        Let $\matL$ be the (exact) lower triangular Cholesky factor of $\matS^{-1}$, i.e. $\matL\matL^*=\matS^{-1}$ (which is unique up to column phases).
        Expanding the equation in line 3 of Algorithm \ref{alg:reduce} we have
        \begin{align*}
            \Lambda(\matHtilde) &= \Lambda\lpar     
                \matLtilde^*\matH\matLtilde+\errormatrix_5
            \rpar
            \\
            &= \Lambda\lpar 
                \matLtilde\matLtilde^*\matH + \matLtilde\errormatrix_5\matLtilde^{-1}
            \rpar
            \\
            &= \Lambda\lpar
            (\matS^{-1}+\errormatrix^{\INV}_1+\errormatrix_2^{\mathsf{CH}})\matH+\matLtilde\errormatrix_5\matLtilde^{-1}
            \rpar
            \\
            &= \Lambda\lpar
                \matS^{-1}\matH 
                + 
                \errormatrix^{\INV}_1\matH 
                +
                \errormatrix_2^{\mathsf{CH}}\matH
                + 
                \matLtilde\errormatrix_5\matLtilde^{-1}
                \rpar
                \\
            &=\Lambda\lpar
                \matL^*\matH\matL 
                + 
                \matL^{-1}\errormatrix^{\INV}_1\matH\matL 
                +
                \matL^{-1}\errormatrix^{\mathsf{CH}}_2\matH\matL 
                + 
                \matL^{-1}\matLtilde\errormatrix_5\matLtilde^{-1}\matL
            \rpar.
        \end{align*}
        At the same time, $\Lambda(\matS^{-1}\matH)=\Lambda(\matL^*\matH\matL)$.
        Combining this with Kahan's bound from Fact \ref{fact:weyl_kahan} and denoting \begin{align*}    
        \matB=\matL^{-1}\errormatrix^{\INV}_1\matH\matL 
        +
        \matL^{-1}\errormatrix^{\mathsf{CH}}_2\matH\matL 
        + 
        \matL^{-1}\matLtilde\errormatrix_5\matLtilde^{-1}\matL,
        \end{align*}
        we have that
        \begin{align}
            \labs
                \lambda_i(\matH,\matS)
                -
                \lambda_i(\matHtilde)
            \rabs
            =
            \labs 
                \lambda_i(\matL^*\matH\matL)-\lambda_i(\matL^*\matH\matL+ \matB)
            \rabs
            \leq
            C\|\matB\|\log(n),
            \label{eq:bound_eigenvalues_transh}
        \end{align}
        for some constant $C$.
        It remains to bound for $\|\matB\|$.
        \begin{enumerate}[(i)]
            \item $\errormatrix_1^{\INV}$ originates from the inversion of $\matS_{\INV}$. Let $\matS_{\INV}=\HERM(\INV(\matS))=\matS^{-1}+\errormatrix^{\INV}_1$. From Corollary \ref{corollary:symmetrized_matrix_inversion}, we know that as long as the machine precision satisfies
            \begin{align*}
                \umach_{\REDUCE} \leq \delta \frac{1}{
                    c_{\HERM}\log(n)\pinv(n)\kappa(\matS)^{\cinv\log(n)+1}
                },
            \end{align*}
            for some $\delta\in(0,1/2)$,
            then all of the following hold: $\|\errormatrix^{\INV}_1\|\leq \delta\|\matS^{-1}\|$, $\tfrac{1}{2}\|\matS^{-1}\|\leq\|\matS_{\INV}\|\leq 2\|\matS^{-1}\|$, $\tfrac{1}{4}\kappa(\matS)\leq \kappa(\matS_{\INV})\leq 4\kappa(\matS)$.
        \item As a second step, the Cholesky factor $\matLtilde=\CHOLESKY(\matS_{\INV})$ is computed, such that $\matLtilde\matLtilde^*=\matS_{\INV}+\errormatrix^{\mathsf{CH}}_2$. From Theorem \ref{theorem:stable_cholesky_bounds} we know that as long as 
            \begin{align*}
                \umach_{\REDUCE} \leq \delta \frac{1}{
                    c_{1}n^{c_2}\kappa(\matS_{\INV})^{c_3\log(n)}
                },
            \end{align*}
            for some constants $c_1,c_2,c_3$, then $\|\errormatrix^{\mathsf{CH}}_2\|\leq \delta\|\matS_{\INV}\|\leq2\delta\|\matS^{-1}\|$. From the bound of $\kappa(\matS_{\INV})$ in $(i)$, the following is sufficient:
            \begin{align*}
                \umach_{\REDUCE} \leq \delta \frac{1}{
                    c_{1}n^{c_2}\lpar 4\kappa(\matS) \rpar^{c_3\log(n)}
                },
            \end{align*}
            \item Finally, we form the matrix $\matHtilde=\HERM\left(
                \MM(\MM(\matLtilde,\matH),\matLtilde^*)
            \right) = \matLtilde^*\matH\matLtilde+\errormatrix^{\MM}_3\matLtilde + \errormatrix^{\MM}_4 = \errormatrix_3$. Using Theorem \ref{theorem:fast_mm} and Corollary \ref{corollary:symmetrized_matrix_multiplication}, as long as 
            \begin{align*}
                \umach_{\REDUCE} \leq \delta \frac{1}{
                    \mu(n)
                }
                =
                \delta \frac{1}{
                    c_{\HERM}\log(n)\mu_{\MM}(n)
                },
            \end{align*}
            \begin{align*}
            \|\errormatrix_5\|=\|\matHtilde-\matLtilde^*\matH\matLtilde\| 
            &= 
            \left\|
                \matLtilde^*\matH\matLtilde + \errormatrix^{\MM}_3\matLtilde + \errormatrix^{\MM}_4 - \matLtilde^*\matH\matLtilde
            \right\|
            \\
            &=
            \left\|
                \errormatrix^{\MM}_3\matLtilde + \errormatrix^{\MM}_4
            \right\|
            \\
            &\leq
            \overbrace{\|\errormatrix^{\MM}_3\|}
            ^
            {\leq \umach\cdot\pmult(n)\|\matLtilde^*\|\|\matH\|}
            \|\matLtilde\| 
            + 
            \overbrace{\|\errormatrix^{\MM}_4\|}
            ^
            {\leq c_{\HERM}\log(n) \cdot \umach\pmult(n) \|\matLtilde^*+\errormatrix^{\MM}_3\|\cdot\|\matLtilde\|}
            \\
            &\leq
            \umach\cdot\pmult(n)\|\matLtilde\|^2\|\matH\|
            +
            c_{\HERM}\log(n) \cdot \umach\pmult(n) \|\matLtilde^*\matH+\errormatrix^{\MM}_3\|\cdot\|\matLtilde\|
            \\
            &\leq
            \umach\cdot\mu(n)\|\matLtilde\|^2\|\matH\|
            +
            \umach\mu(n) \|\matLtilde^*\matH\|\|\matLtilde\|
            +
            \umach\mu(n) \|\errormatrix^{\MM}_3\|\cdot\|\matLtilde\|
            \\
            &\leq
            2\umach\cdot\mu(n)\|\matLtilde\|^2\|\matH\|
            +
            \umach\mu(n) \cdot \umach\cdot\pmult(n)\|\matLtilde^*\|\|\matH\| \cdot\|\matLtilde\|
            \\
            &=
            \umach\cdot\mu(n)\|\matLtilde\|^2\|\matH\|
            \left(
                2
                +
                \umach\cdot\pmult(n)
            \right)
            \\
            &=
            3\umach\cdot\mu(n)
            \overbrace{\|\matLtilde\|^2}^{\leq \|\matS_{\INV}\|+\|\errormatrix_2^{\mathsf{CH}}\|}
            \|\matH\|
            \\
            &\leq
            3\umach\cdot\mu(n)
            \left(
                \|\matS^{-1}\|
                +
                \|\errormatrix^{\INV}_1\|
                +
                \|\errormatrix_2^{\mathsf{CH}}\|
            \right)
            \|\matH\|
            \\
            &\leq
            3\umach\cdot\mu(n)
            \left(
                2\|\matS^{-1}\|
                +
                2\delta\|\matS^{-1}\|
            \right)
            \|\matH\|
            \\
            &\leq
            12\umach
            \mu(n)
            \|\matS^{-1}\|
            \|\matH\|
            \\
            &\leq
            12\delta
            \|\matS^{-1}\|
            \|\matH\|.
        \end{align*}
        \item For the final desired error we need to bound $\|\matLtilde\|\|\matLtilde^{-1}\|$. Taking the square:
        \begin{align*}
            (\|\matLtilde\|\|\matLtilde^{-1}\|)^2
            = 
            \|\matLtilde\|^2\|\matLtilde^{-1}\|^2
            =
            \|\matLtilde\matLtilde^*\|\|(\matLtilde\matLtilde^*)^{-1}\|
            &=
            \|\matS^{\INV}+\errormatrix_2^{\mathsf{CH}}\|\|(\matS^{\INV}+\errormatrix_2^{\mathsf{CH}})^{-1}\|
            \\
            &=
            \kappa(\matS^{\INV}+\errormatrix_2^{\mathsf{CH}}).
        \end{align*}
        For the spectral norm we have that $\|\matS_{\INV}+\errormatrix_2^{\mathsf{CH}}\|\leq \|\matS_{\INV}\|+\|\errormatrix_2^{\mathsf{CH}}\| \leq 2\|\matS^{-1}\|+2\delta\|\matS^{-1}\| \leq 3\|\matS^{-1}\|$. For the spectral norm of the inverse we need to bound the smallest singular value of $\matS^{\INV}+\errormatrix_2^{\mathsf{CH}}$ from below:
        \begin{align*}
            \sigma_{\min}(\matS^{\INV}+\errormatrix_2^{\mathsf{CH}}) 
            \geq 
            \sigma_{\min}(\matS^{\INV}) - \|\errormatrix_2^{\mathsf{CH}}\|
            \geq 
            \tfrac{1}{2}\sigma_{\min}(\matS^{-1}) - 2\delta\|\matS^{-1}\|
            &=
            \tfrac{1}{2\|\matS\|} - 2\delta\|\matS^{-1}\|
            \\
            &=
            \tfrac{1-4\delta \kappa(\matS)}{2\|\matS\|}.
        \end{align*}
        Combining the last two bounds, that  we have
        \begin{align*}
            \kappa(\matS^{\INV}+\errormatrix_2^{\mathsf{CH}}) \leq \frac{3\|\matS^{-1}\|}{\frac{1-4\delta \kappa(\matS)}{2\|\matS\|}} = \frac{6\kappa(\matS)}{1-4\delta\kappa(\matS)}.
        \end{align*}
        \end{enumerate}

        We now have bounds for all the required quantities to bound $\|\matB\|:$ 
        \begin{align*}
            \|\matB\|
            &= 
            \lnorm
            \matL^{-1}\errormatrix^{\INV}_1\matH\matL 
            +     
            \matL^{-1}\errormatrix^{\mathsf{CH}}_2\matH\matL 
            + 
            \matL^{-1}\matLtilde\errormatrix_5\matLtilde^{-1}\matL
            \rnorm
            \\
            &\leq
            \sqrt{\kappa(\matS)}
            \lnorm
                \errormatrix_1^{\INV}\matH
                +
                \errormatrix^{\mathsf{CH}}_2\matH
                +
                \matLtilde\errormatrix_5\matLtilde^{-1}
            \rnorm\\ 
            &\leq
            \sqrt{\kappa(\matS)}
            \lpar
                \|\errormatrix_1^{\INV}\|\|\matH\| 
                + 
                \|\errormatrix^{\mathsf{CH}}_2\matH\|
                +
                \|\errormatrix_5\|\|\matLtilde\|\|\matLtilde^{-1}\| 
            \rpar
            \\
            &\leq
            \sqrt{\kappa(\matS)}
            \lpar 
                \delta\|\matS^{-1}\|\|\matH\| 
                + 
                2\delta\|\matS^{-1}\|\|\matH\|
                +
                12\delta\|\matS^{-1}\|\|\matH\|
                \sqrt{\frac{6\kappa(\matS)}{1-4\delta\kappa(\matS)}}
            \rpar
            \\
            &\leq
            \delta \sqrt{\kappa(\matS)}\|\matS^{-1}\|\|\matH\|
            \lpar 
                3
                +
                12
                \sqrt{\frac{6\kappa(\matS)}{1-4\delta\kappa(\matS)}}
            \rpar.
        \end{align*}

        Setting $\delta = \epsilon \frac{1}{64C\log(n)\kappa(\matS)}$, where $C$ is the constant from Inequality \eqref{eq:bound_eigenvalues_transh} and $\epsilon\in(0,1)$, this finally gives
        \begingroup
        \allowdisplaybreaks
        \begin{align*}
            \|\matB\| 
            &
            \leq 
            \epsilon \frac{1}{64C\log(n)\kappa(\matS)} 
            \cdot \sqrt{\kappa(\matS)}\|\matS^{-1}\|\|\matH\|
            \lpar
                3 + 12\sqrt{
                    \frac{6\kappa(\matS)}{1-4\epsilon \frac{1}{64C\log(n)\kappa(\matS)}\kappa(\matS)}
                }
            \rpar
            \\
            &=
            \epsilon \frac{1}{64C\log(n)\sqrt{\kappa(\matS)}} \|\matS^{-1}\|\|\matH\|
            \lpar
                3 + 12\sqrt{
                    \frac{6\kappa(\matS)}{1-\epsilon \frac{1}{16}}
                } 
            \rpar\\
            &\leq
            \epsilon \frac{1}{64C\log(n)\sqrt{\kappa(\matS)}} \|\matS^{-1}\|\|\matH\|
            \lpar
            3
            +
            12
            \sqrt{
                \frac{96}{15}\kappa(\matS)
            }
            \rpar
            \\
            &\leq
            \epsilon \frac{1}{64C\log(n)\sqrt{\kappa(\matS)}} \|\matS^{-1}\|\|\matH\| (3+36\sqrt{\kappa(\matS)})\\
            &\leq
            \epsilon \tfrac{1}{C\log(n)}\|\matS^{-1}\|\|\matH\|,
        \end{align*}
        \endgroup
        which we can use in Inequality \eqref{eq:bound_eigenvalues_transh} to obtain
        \begin{align*}
            |\lambda_i(\matH,\matS)-\lambda_i(\matHtilde)|\leq \epsilon\|\matS^{-1}\|\|\matH\|.
        \end{align*}
        We can now gather all the requirements for $\umach_{\REDUCE}$. From the above, $\umach_{\REDUCE}$ needs to satisfy
        
        \begin{align*}
            \umach_{\REDUCE} 
            &\leq
                \epsilon \frac{1}{64C\log(n)\kappa(\matS)} \min
                \left\{
                    \frac{1}{
                        c_{\HERM}\log(n)\pinv(n)\kappa(\matS)^{\cinv\log(n)+1}
                    },\right.
                    \\&\qquad\qquad\qquad\qquad
                    \left.
                    \frac{1}{
                        c_{1}n^{c_2}(4\kappa(\matS))^{c_3\log(n)}
                    },
                    \frac{1}{
                        c_{\HERM}\log(n)\mu_{\MM}(n)
                    }
                \right\}
                \\
            &\leq
                \epsilon 
                \frac{1}{
                    \rho_{1}n^{\rho_{2}}(4\kappa(\matS))^{\rho_{3}\log(n)}
                },
        \end{align*}
        for some suitably chosen constants $\rho_{1},\rho_{2},\rho_{3}$. This translates to
        \begin{align*}
            O\lpar \log(1/\umach_{\REDUCE}) \rpar
            =
            O\lpar
                \log(n)\log(\kappa(\matS)) + \log(1/\epsilon)
            \rpar
        \end{align*}
        required bits of precision. 
        Having established the eigenvalue bounds, note that $\lambda_i(\matH,\matS)=\lambda_i(\matL^*\matH\matL,\matI)$. Since $\matHtilde$ and $\matL^*\matH\matL$ are both Hermitian, the eigenvalue bound holds also for the singular values.
        For scaled matrices $\|\matS^{-1}\|,\|\matH\|\leq 1$ we obtain a maximum $\epsilon$ additive error for the eigenvalues.
    \end{proof}
\end{proposition}

\subsection{Counterexample for floating point LU}
\label{appendix:appendix_lur_counterexample}
In this section provide a counter-example which illustrates why the backward-stable LU factorization algorithm of \cite{demmel2007fastla} is not sufficient to obtain a Cholesky factor for a Hermitian postive-definite matrix.
We apply one-by-one the steps of the $\LUR$ algorithm of \cite{demmel2007fastla} on the matrix:
\begin{align*}
    \matA = \begin{pmatrix} 3 & 1 \\ 1 & 3 \end{pmatrix}.
\end{align*}
\begin{enumerate}[(a)]
    \item In the first step the matrices $\matL_L$ and $\matU_L$ are computed by calling $\LUR(\matA_{1:2,1})$, to  compute the LU factorization of the first column of $\matA$. It is easy to see that this returns $\matL_L=\begin{pmatrix}
        1\\
        \tfrac{1}{3}(1+\theta_1)
    \end{pmatrix}$
    where $|\theta_1|\leq \umach$, and $\matU_L=3$ (the $\matU_L$ factor is a 1-by-1 matrix in this case).
    
    \item The second step updates the upper right corner of $\matA$, by multiplying from the left with the inverse of the top-left block of $\matL_L$. Assuming that multiplication and inversion with the number $1$ does not incur errors, then for this specific example $\matA_{1,2}$ is not modified in this step, i.e., it remains equal to $1$, without errors.

    \item The third step updates the Schur complement, which becomes $\matA_{2,2}\leftarrow \fl(\matA_{2,2}-\tfrac{1}{3}(1+\theta_1)) = (3-\tfrac{1}{3}(1+\theta_1))(1+\theta_2)$, where $|\theta_2|\leq \umach$ is the error term incurred by subtraction.

    \item The next step computes the $\matL_R,\matU_R\leftarrow \LUR(\matA_{2,2})$ of the (updated) Schur complement. In this specific case it trivially returns $\matL_R=1$ and $\matU_R=\matA_{2,2}=(3-\tfrac{1}{3}(1+\theta_1))(1+\theta_2)$.

    \item The final steps combine together the left and right LU factors to finally return
    $
        \matL=\begin{pmatrix}
            1 & 0\\
            \tfrac{1}{3}(1+\theta_1) & 1
        \end{pmatrix},
    $
    \item and
    $    \matU=\begin{pmatrix}
            3 & 1 \\
            0 & (3-\tfrac{1}{3}(1+\theta_1))(1+\theta_2)
        \end{pmatrix}.
    $
\end{enumerate}
Then, in exact arithmetic, $\matL\matU=\begin{pmatrix}
    3 & 1 \\
    1+\theta_1 & 3(1+\theta_2) + \theta_2(1+\theta_1)\tfrac{1}{3}
\end{pmatrix},$
which is not symmetric.

\section{Regularization with diagonal disorder}
\label{appendix:regularize}
We now arrive to the more interesting part of the analysis.
In Section \ref{section:sign_spectral_projector} we argued that if we know the gap and the midpoint then the sign function can yield the desired spectral projector. To compute the gap and midpoint efficiently,
we will take advantage of the symmetry induced by the Cholesky-based reduction, and use the Wegner estimate \cite{wegner1981bounds} for Hermitian diagonal perturbations to regularize the problem in the spirit of smoothed analysis. We use the following variant of the Wegner estimate.
\begin{proposition}
    \label{proposition:wegner_diagonal_gaussian_matrix}
    Let $\matG$ be a random diagonal matrix with independent diagonal elements sampled from $\mathcal{N}(0,\sigma^2)$, for $\sigma=\frac{1}{2\sqrt{2\ln(4n/\delta)}}$ and some $\delta\in(0,1/2)$. Then for any interval $I\subset \mathbb{R}$ 
    \begin{align*}
        \Pr\lbrac
            |\Lambda(\matH+\matG)\cap I|\geq 1
        \rbrac
        &\leq \sqrt{4\pi\ln(4n/\delta)}n|I|,
    \end{align*}  
    where $\matH$ is a fixed Hermitian matrix and $I\subset \mathbb{R}$ is a fixed interval.
    \begin{proof}
        This directly comes from the well-known result of  Wegner \cite{wegner1981bounds}, which states that  that for any interval $I\subset \mathbb{R}$
        \begin{align*}
            \Pr\lbrac
                |\Lambda(\matH+\matG)|\cap I| \geq 1
            \rbrac
            &\leq
            \pi\|\rho\|_\infty n|I|
            \leq
            \sqrt{\frac{\pi}{2\sigma^2}} n |I|,
        \end{align*}
    where $\|\rho\|_\infty=\frac{1}{\sqrt{2\pi\sigma^2}}$ is the supremum the probability density.
    \end{proof}
\end{proposition}
Similar results can be obtained for other classes of random matrices $\matG$. Such an example is the Gaussian Unitary Ensemble (GUE) \cite{aizenman2017matrix}, in which case $\matG$ is dense, but it is invariant under rotations which might be useful for other applications.

\subsection{Sampling Gaussians and perturbing in floating point}

The next step is to describe how to use such random diagonal perturbations in floating point. Briefly, we first assume that we have a fixed number of bits for the floating point exponent and for the mantissa. If we sample a standard normal Gaussian (in infinite precision), it can happen that it is too large to fit in the given number of bits. This is accounted for in the failure probability, which is very small due to the decay of the normal distribution. We describe our sampling method in the following definition, which can be thought of as sampling only the most significant bits of the floating point representation of a Gaussian.
\begin{definition}
    A floating point standard normal sampler $\SAMPLER(p,t)$ takes as input the number of exponent bits $p$, and the number of mantissa bits $t$, and returns a floating point number $\widetilde g\leftarrow \fl(g)$ where $g$ is sampled from $\mathcal{N}(0,1)$. Following the definitions of the floating point model in Appendix \ref{appendix:floating_point}, if $|g|\in[2^{-M}, 2^{M}(2-2^{-t})]$, where $M=2^{p-1}$, then $|\widetilde g-g|\leq 2^{-t}|g|=\umach |g|$. If $|g|> 2^{M}(2-2^{-t})$ the sampler returns $\widetilde g=\pm\FLINF$, i.e. the floating point representation of infinity, and if $|g|< 2^{-M}$ it returns zero.
    \label{definition:stable_normal_sampler}
\end{definition}

\begin{lemma}[Diagonal Gaussian sampler]
    \label{lemma:diagonal_gaussian_sampler}
    Let $\gamma\in(0,1/2)$ and $\sigma>0$, and assume that we want to obtain a floating representation of the matrix $\gamma\matG$, where $\matG$ is a diagonal matrix with independent diagonal entries from $\mathcal{N}(0,\sigma^2)$.
    Let $\SAMPLER(p,t)$ be a floating point standard normal sampler as in Definition \ref{definition:stable_normal_sampler}, and $\umach=2^{-t}$. Let $\widetilde\sigma$ be such that $|\widetilde\sigma-\sigma|\leq \eta_c\sigma $, where $|\eta_c|\leq\tfrac{c\umach}{1-c\umach}$ for some (integer) constant $c>1$, and
    \begin{align*}
        \matGtilde=\diag(\widetilde g_1, \widetilde g_2, \ldots, \widetilde g_n),
    \end{align*} 
    where 
    $\widetilde g_i=\SAMPLER(p,t)$. If 
    $p\geq C\log(\log(\tfrac{n}{\delta}))$ for some global constant $C$, then
    the diagonal matrix $\matVtilde =\fl(\gamma\cdot\widetilde\sigma\cdot \matGtilde)=\gamma\matG+\errormatrix$, has all the following properties with probability at least $1-\delta$:
    \begin{enumerate}[(i)]
        \item $\matVtilde_{i,i}\notin \{0,\FLINF\}$,
        \item  $\errormatrix$ is diagonal and $\|\errormatrix\|\leq \gamma\eta_{c+2}\sigma\sqrt{2\ln(4n/\delta)}$,
        \item $\|\matVtilde \|
        \leq 
        \gamma
        \lpar 1+\eta_{c+2}\rpar 
            \sigma\sqrt{2\ln(4n/\delta)}.
        $
    \end{enumerate}
    
    \begin{proof}
        Let us first analyze the conditions that are necessary such that all the sampled numbers lie within the floating point bounds from Appendix \ref{appendix:floating_point}, specifically, $ |g_i|\in[2^{-M},2^{M}(2-2^{-t})]$, $i\in[n]$, where $M=2^{p-1}$. For simplicity we consider $|g_i|\in [2^{-M},2^{M}]$.

        For the lower bound, we have that
        \begin{align*}
            \Pr[|g_i|\leq 2^{-M}] \leq 2\frac{1}{\sqrt{2\pi}}2^{-M},
        \end{align*}
        where we naively upper bounded the standard normal probability density in the interval $[0,2^{-M}]$ by the constant function $\frac{1}{\sqrt{2\pi}}$. For $M=\log_2\lpar\tfrac{2n}{\delta}\sqrt{\tfrac{2}{\pi}}\rpar$ this implies that there are no subnormal $g_i$ for all $i\in[n]$ simultaneously. This value of $M$ translates to $p=O\lpar\log(\log(\tfrac{n}{\delta}))\rpar$ bits.
        
        For the upper bound, a standard normal tail bound gives
        \begin{align}
            \Pr[|g_i|\geq 2^{M}] \leq 2\exp(-2^{2M}/2).
            \label{eq:gaussian_tail}
        \end{align}
        For the aforementioned value of $M$, the probability that each one of the $g_i$ is larger than $2^{M}$ is exponentially small in $\delta/n$, therefore we can conclude that the sampler does not return any $\FLINF$ values with exponentially high probability.
        
        This is already enough to argue that the sampler returns floating point numbers that are not subnormal and not $\FLINF$. However, we will need a tighter bound for the magnitude of the $g_i$ to bound the norm of the diagonal random matrix.
        Analyzing  Eq \eqref{eq:gaussian_tail} for 
        $M=\tfrac{1}{2}\log_2(2\ln(\tfrac{4n}{\delta}))$,
        and taking a union bound over all $i\in[n]$ we have that $|g_i|\leq \sqrt{2\ln(\tfrac{4n}{\delta})}\ll 2^{M}$ holds for all $i$ simultaneously with probability at least $1-\delta/2$.
 
        Conditioning on the event that all the $g_i$ are in the correct range, from Definition \ref{definition:stable_normal_sampler}, $|\widetilde g_i-g_i|\leq \umach |g_i|$ holds for all $i$. Then
        \begin{align*}
            \widetilde v_i = \fl( \widetilde\sigma \cdot \widetilde g_i) = (1+\theta)\widetilde\sigma\widetilde g_i = \sigma g_i(1+\eta_{c+1}),
        \end{align*}
        where $c>1$ is a constant and $|\eta_{c+1}|\leq \tfrac{(c+1)\umach}{1-(c+1)\umach}$. This means that every diagonal element of $\fl(\widetilde\sigma\matGtilde)$ is $(1+\eta_{c+1})$-far from a random variable that is sampled from $\mathcal{N}(0,\sigma^2)$. Multiplying each diagonal element of $\fl(\widetilde\sigma\matGtilde)$ with $\gamma$ to form $\matVtilde $ (similar to the Step 4 of Algorithm \ref{alg:regularize}) simply increases the (relative) error to $(1+\eta_{c+2})$. Then we can write
        \begin{align*}
            \matVtilde  = \gamma\matV + \errormatrix,
        \end{align*}
        where $\matV$ is diagonal with independent diagonal entries from $\mathcal{N}(0,\sigma^2)$ and $\errormatrix$ is a diagonal matrix with diagonal elements $\errormatrix_{i,i}$ bounded in magnitude by
        \begin{align*}
            |\errormatrix_{i,i}| \leq \frac{(c+2)\umach}{1-(c+2)\umach}\gamma|\matV_{i,i}|.
        \end{align*}
        
        It also holds that $\|\matV\|\leq \sigma\sqrt{2\ln(4n/\delta)}$ since the diagonal elements $\matV_{i,i}$ since they are just the $g_i$'s scaled by $\sigma$, and we already bounded $|g_i|\leq \sqrt{2\ln(4n/\delta)}$, which implies that $\|\errormatrix\|\leq \gamma \tfrac{(c+2)\umach}{1-(c+2)\umach}\sigma\sqrt{2\ln(4n/\delta)}$. Finally
        \begin{align*}
            \|\matVtilde\|\leq \gamma\|\matV\|+\|\errormatrix\| \leq \gamma \lpar
               1+\tfrac{(c+2)\umach}{1-(c+2)\umach}\rpar 
            \sigma\sqrt{2\ln(4n/\delta)}.
        \end{align*}
        We conditioned only on two random events, where each holds with probability at least $1-\delta/2$, giving the final success probability of at least $1-\delta$.
    \end{proof}
\end{lemma}

\begin{algorithm}[htb]
\caption{$\REGULARIZE$.}
\label{alg:regularize}
\begin{algorithmic}[1]
\setstretch{0.8}
\Statex \begin{align*}
    \REGULARIZE
\end{align*}
\small
\Input Hermitian matrix $\matA\in\mathbb{H}^{n}$, perturbation scale factor $\gamma\in(0,1/4)$, failure probability parameter $\delta\in(0,1/4)$.
\Require $\|\matA\|\leq 1$.
\setstretch{1.5}
\Algorithm $\matXtilde \leftarrow \REGULARIZE(\matA,\gamma,\delta)$.
\State $\widetilde\sigma = \fl\lpar  \tfrac{1}{2\sqrt{2\ln(4n/\delta)}}\rpar$.   
\State $p\leftarrow \lceil C\log(\log(\tfrac{n}{\delta})) \rceil$. 
\State $\matGtilde = \diag(\widetilde g_1,\ldots,\widetilde g_n)$, $\widetilde g_i\leftarrow \SAMPLER(p,\log(1/\umach))$
\State $\matVtilde = \fl(\gamma\cdot\widetilde\sigma\cdot\matGtilde)=\gamma\sigma \matG + \errormatrix$. \Comment{Where $\matG=\diag(g_1,\ldots,g_n),g_i\leftarrow\mathcal{N}(0,1)$.}
\State $\matXtilde = \matA+\matVtilde +\errormatrix^{(+)}$. 
\Output Hermitian perturbed matrix $\matXtilde$.
\Ensure See Proposition \ref{proposition:alg_regularize_appendix}.
\end{algorithmic}
\end{algorithm}

We can now use the Wegner estimate to get a minimum singular value bound for diagonal shifts over a grid, defined as follows.
\begin{definition}[Grid]
    A $1$-d grid in the real line is a set of $s$ points defined as
    \begin{align*}
        \grid(l,r,h) = \{l+jh|j\in\mathbb{Z}_{\geq 0}, l+jh\leq r\}.
    \end{align*}
    The cardinality of a grid $\mathsf{g}=\grid(l,r,h)$ is denoted as $|\mathsf{g}|$.
\end{definition}

The following Proposition \ref{proposition:alg_regularize} summarizes the properties of Algorithm \ref{alg:regularize}, $\REGULARIZE$, that we will use to perform efficient eigenvalue counting queries to compute the spectral gap.

\begin{proposition}
    \label{proposition:alg_regularize_appendix}
    Let $\matA$ with  $\|\matA\|\leq 1$ be a Hermitian matrix,  $\gamma,\delta\in(0,1/4)$ two given parameters, and $\matXtilde\leftarrow \REGULARIZE(\matA,\gamma,\delta)$. Let $\mathsf{g}$ be an arbitrary (but fixed) grid of points in $[-2,2]$ with size $|\mathsf{g}|=T$.
    For every element $h_i\in\mathsf{g}$ consider the matrices $\matM_i=h_i-\matXtilde$ and $\matMtilde=h_i-\matXtilde+\errormatrix_{i}$, where $\errormatrix_{i}$ denote the diagonal floating point error matrices induced by the shift. If the machine precision $\umach$ satisfies
    \begin{align*}
        \umach \leq \tfrac{\gamma\delta}{32(c+2)nT\sqrt{\pi\ln(4n/\delta)}},
    \end{align*}
    which translates to
    $O(\log(\tfrac{Tn}{\gamma\delta}))$
    bits, then all the following hold simultaneously with probability $1-2\delta:$
    \begin{align*}
        \|\matXtilde\| \leq 4/3,
        \quad
        \labs \lambda_i(\matXtilde)-\lambda_i(\matA) \rabs \leq \tfrac{9}{16}\gamma,
        \quad
        \sigma_{\min}(\matMtilde_i) \geq \tfrac{\gamma\delta}{4nT\sqrt{4\pi\ln(4n/\delta)}}.
    \end{align*}
        \begin{proof}
        Let $\matX=\matA+\gamma\sigma\matG$, where $\sigma\matG$ is a diagonal matrix with independent elements from $\mathcal{N}(0,\sigma^2)$ where $\sigma=\frac{1}{2\sqrt{2\ln(4n/\delta)}}$.
        Let $\matB=\matX/\gamma=\matA/\gamma+\sigma\matG$. 
        Denote by $E_{\matG}$ the event $\|\sigma\matG\|\leq \frac{1}{2}$. From a tail bound as in Lemma \ref{lemma:diagonal_gaussian_sampler}, $\Pr[E_{\matG}]\geq 1-\delta$. 
        Conditioning on $E_{\matG}$, we obtain $\|\matB\|=\|\matA/\gamma+\sigma\matG\|\leq 1/\gamma+1/2 :=\beta$. This directly implies that $\Lambda(\matB)\subseteq (-1/\gamma-1/2,1/\gamma+1/2)$. 
        Let $h_i'=h_i/\gamma$ for all $i=1,\ldots,T$.
        
        From the Wegner estimate (Proposition \ref{proposition:wegner_diagonal_gaussian_matrix}), for every fixed interval $I$,
        \begin{align*}
            \Pr\lbrac |\Lambda(h'-\matB)\cap I| \geq 1 \rbrac
            \leq
            \sqrt{4\pi\ln(4n/\delta)}n|I|.
        \end{align*}
        Setting $I_i$ to be an interval centered at $h_i'$ with size $|I_i|=\frac{\delta}{Tn\sqrt{4\pi\ln(4n/\delta)}}$, 
        we have that the probability that $I_i$ contains any eigenvalues is at most $\delta/T$. 
        Since there are no eigenvalues in $I_i$, this means that the smallest singular value of $h_i'-\matB$ is at least half the size of $I_i$, i.e. $\sigma_{\min}(h_i'-\matB)\geq |I_i|/2=\frac{\delta}{2Tn\sqrt{4\pi\ln(4n/\delta)}}$. 
        Hence, $\sigma_{\min}(h_i-\matX)=\gamma\sigma_{\min}(h_i'-\matB)\geq \frac{\gamma\delta}{2Tn\sqrt{4\pi\ln(4n/\delta)}}$ holds as well.

        From Lemma \ref{lemma:diagonal_gaussian_sampler}, if the diagonal Gaussian sampler is called with $p=\Theta(\log(\log(\tfrac{n}{\delta}))$ and $t=\log(1/\umach)$ then 
        \begin{align*}
            \|\errormatrix\| &\leq \gamma \eta_{c+2}\frac{1}{2},\\
            \|\matVtilde\| &\leq \gamma(1+\eta_{c+2})\frac{1}{2},
        \end{align*}
        where $|\eta_{c+2}|\leq\frac{(c+2)\umach}{1-(c+2)\umach}$. The elements of the diagonal error matrix $\errormatrix^{(+)}$ in Algorithm \ref{alg:regularize} are bounded by
        \begin{align*}
            |\errormatrix^{(+)}_{i,i}| \leq \umach|\matA_{i,i}+\matVtilde_{i,i}| \leq \umach (\|\matA\|+\|\matVtilde\|)\leq \umach\lpar 1 + \tfrac{1}{2}\gamma(1+\eta_{c+2})\rpar.
        \end{align*}
        Then 
        \begin{align*}
            \|\matXtilde-\matX\| 
            &=
            \|\errormatrix^{(+)}+\errormatrix\|\\
            &\leq \umach \|\matA+\matVtilde\| + \tfrac{1}{2}\gamma \eta_{c+2}
            \\
            &\leq \umach \lpar 1 + \tfrac{1}{2}\gamma(1+\eta_{c+2})\rpar + \tfrac{1}{2}\gamma \eta_{c+2}.
        \end{align*}
        If $\umach \leq \tfrac{\epsilon}{2(c+2)}$ for some $\epsilon\in(0,1)$ then $|\eta_{c+2}|\leq \epsilon$ and since $\gamma<1/2$ we also have that
        $\|\matXtilde-\matX\|\leq \epsilon$. Recall that $\|\matX\|=\|\matA+\gamma\sigma\matG\|\leq 1+\gamma/2\leq 5/4$
        and thus 
        \begin{align*}
        \|\errormatrix_i\|
        \leq 
        \umach(|h_i|+\|\matXtilde\|)
        \leq
        \umach(|h_i|+\|\matX\|+\epsilon)
        \leq 
        \umach(2+(1+\gamma/2)+\epsilon)\leq 5\umach \leq 5\frac{\epsilon}{2(c+2)}\leq \epsilon,
        \end{align*}
        assuming that $c\geq 1$. 
        We now collect all the error bounds and apply Weyl's inequality to argue that
        \begin{align*}
            |\lambda_j(h_i-\matX)-\lambda_j(h_i-\matXtilde+\errormatrix_i)|
            \leq
            \|\matX-\matXtilde\|+\|\errormatrix_i\|
            \leq 
            2\epsilon.
        \end{align*}
        The same holds for the singular values $\sigma_j$ since the matrices are Hermitian.
        Thus if \begin{align*}
            \epsilon 
            \leq 
            \frac{\gamma\delta}{8Tn\sqrt{4\pi\ln(4n/\delta)}} 
            \leq
            \sigma_{\min}(h_i - \matX ) / 4
        \end{align*}
        then  
        \begin{align*}
            \labs
            \sigma_{\min}(h_i-\matXtilde+\errormatrix_i)
            -
            \sigma_{\min}(h_i-\matX)
            \rabs
            \leq 
            \sigma_{\min}(h_i-\matX)/2,
        \end{align*}
        which implies
        \begin{align*}
            \sigma_{\min}(h_i-\matXtilde+\errormatrix_i)
            \geq 
            \sigma_{\min}(h_i-\matX)/2
            =
            \frac{\gamma\delta}{4Tn\sqrt{4\pi\ln(4n/\delta)}}.
        \end{align*}
        
        The aforementioned bound on $\epsilon$ also gives two (loose) bounds that are useful for the analysis of algorithms later:
        \begin{align*}
            \|\matXtilde\| \leq \|\matX\|+\epsilon \leq 5/4 + 1/32 \leq 4/3,
        \end{align*} 
        and also from Weyl's inequality
        \begin{align*}
            \labs \lambda_i(\matA)-\lambda_i(\matXtilde) \rabs \leq \|\matA-\matX\|+\|\matX-\matXtilde\| \leq \tfrac{\gamma}{2} + \tfrac{\gamma}{16} = \tfrac{9}{16}\gamma.
        \end{align*} 
        
        The requirement for the machine precision is
        \begin{align*}
            \umach \leq \frac{\epsilon}{2(c+2)} \leq \frac{\gamma\delta}{32(c+2)Tn\sqrt{\pi\ln(4n/\delta)}},
        \end{align*}
        which translates to 
        \begin{align*}
            O\lpar
                \log ( 
                    \tfrac{Tn}{\gamma\delta}
                )
            \rpar
        \end{align*}
        bits of precision (recall that we also required $p=\Theta(\log(\log(n/\delta)))$ for the exponent of the Gaussian sampler, but this does not affect the total number of bits).

        For the complexity, we have $n$ calls to $\SAMPLER(p,t)$ and $O(n)$ scalar additions and multiplications thereafter, which accumulate to $O(n)$ total floating point operations using $O\lpar\log(\tfrac{Tn}{\gamma\delta})\rpar$ bits.

        For the success probability, there are two types random events that we took into account: $E_{\matG}$, which fails with probability at most $\delta$, and the Wegner estimate for $T$ points $h_i$ in the grid $\mathsf{g}$, where each one fails with probability at most $\delta/T$. A union bound over all random events gives a success probability of at least $1-\delta-T\delta/T=1-2\delta$.
    \end{proof}
\end{proposition}

\section{Fast spectral gaps and eigenvalues with counting queries}
\label{appendix:spectral_gap}
The core of our methods is an algorithm to efficiently approximate spectral gaps using only ``eigenvalue counting queries,'' avoiding an explicit (and expensive) diagonalization. 

\subsection{Counting eigenvalues}

The main subroutine of our algorithm is used to count the eigenvalues that are smaller than a given threshold efficiently.

\begin{lemma}
    \label{lemma:eigenvalue_count_less_than_h}
    Let $\matXtilde$ be Hermitian, $\|\matXtilde\|\leq 2$, and $h\in[-2,2]$ is a fixed point that can be exactly represented in floating point, i.e. $\fl(h)=h$. There exists an algorithm $\COUNT(\matXtilde,h,\varepsilon)$ which takes as input $h,\matXtilde,$ and  $\varepsilon\in(0,4/199)$ with the guarantee that $\sigma_{\min}(h-\matXtilde)\geq \varepsilon$, and it returns an integer $n_{(-)} $, which is the precise number of eigenvalues of $\matXtilde$ that are smaller than $h$. The algorithm requires 
    \begin{align*}
        O\lpar
            T_{\MM}(n)
            \lpar
                \log(\tfrac{1}{\varepsilon}) + \log(\log(\tfrac{n}{\varepsilon}))
            \rpar
        \rpar
    \end{align*}
    arithmetic operations using
    \begin{align*}
        O\lpar
            \log(n)\log^3(\tfrac{1}{\varepsilon})\log(\tfrac{n}{\varepsilon})
        \rpar
    \end{align*}
    bits of precision.
    \begin{proof}
        Let $\matM=h-\matXtilde$ and $\matMtilde=h-\matXtilde+\errormatrix$ be its floating point counterpart. As the shift only distorts the diagonal elements, $\|\errormatrix\|=\max_i|\errormatrix_{i,i}| \leq \umach\max_i|h-\matXtilde_{i,i}|\leq 4\umach$. By assumption $\sigma_{\min}(\matM)\geq \varepsilon$, hence if $\umach\leq \varepsilon/8$ then Weyl's inequality implies $\sigma_{\min}(\matM+\errormatrix) \geq \sigma_{\min}(\matM)-\varepsilon/2\geq \varepsilon/2$.

        Assume that we want to approximate $n_{(-)} $ by the expression 
        \begin{align*}
            \widetilde n_{(-)} 
            =
            \COUNT(\matXtilde,h,\varepsilon)
            =
            \round\lpar
                \fl(\tr(\SGN(\matMtilde,\alpha_{\SGN},\eta_{\SGN},\epsilon_{\SGN})))
            \rpar,
        \end{align*} for some $\alpha_{\SGN},\eta_{\SGN},\epsilon_{\SGN}$ (to be determined). For simplicity we will use $\SGN(\matMtilde)$ to denote $\SGN(\matMtilde,\alpha_{\SGN},\eta_{\SGN},\epsilon_{\SGN})$. 
        Assuming for now that $\SGN$ converged to the requested error $\epsilon_{\SGN}$, we know that $\|\sgn(\matMtilde)-\SGN(\matMtilde)\|\leq \epsilon_{\SGN}$.
        To proceed recall that for any matrix $\matA$
        \begin{align*}
            |\fl(\tr(\matA))-\tr(\matA)| \leq \frac{\log(n)\umach}{1-\log(n)\umach} \|\matA\|,
        \end{align*}
        which can be achieved with a binary tree-type addition.

        Now consider the error
        \begin{align}
            |n_{(-)} -\widetilde n_{(-)} | 
            &= 
            \labs
                \tr(\sgn(\matM)) -\tr(\sgn(\matMtilde)) + \tr(\sgn(\matMtilde)) - \tr(\SGN(\matMtilde)) + \tr(\SGN(\matMtilde)) - \fl(\tr(\SGN(\matMtilde)))
            \rabs
            \nonumber
            \\
            &\leq
            \labs
                \tr(\sgn(\matM)) -\tr(\sgn(\matMtilde))
            \rabs
            +
            \labs
                \tr(\sgn(\matMtilde)) - \tr(\SGN(\matMtilde))
            \rabs
            + 
            \labs
                \tr(\SGN(\matMtilde)) - \fl(\tr(\SGN(\matMtilde)))
            \rabs
            \nonumber
            \\
            &\leq
            \labs
                \tr(\sgn(\matM)) -\tr(\sgn(\matMtilde))
            \rabs
            +
            n\epsilon_{\SGN}
            +
            \tfrac{\log(n)\umach}{1-\log(n)\umach} (1+\epsilon_{\SGN})
            \nonumber
            \\
            &\leq
            n
            \lnorm
                \sgn(\matM) -\sgn(\matMtilde)
            \rnorm
            +
            n\epsilon_{\SGN}
            +
            \tfrac{\log(n)\umach}{1-\log(n)\umach} (1+\epsilon_{\SGN}).
            \label{equation:bound_eigenvalue_counts}
        \end{align}
        It remains to bound $\|\sgn(\matM)-\sgn(\matMtilde)\|$. For this we can directly use Lemma \ref{lemma:sign_function_under_perturbation} where we set $\matS=\matI$, $\matH=\matXtilde$, and $\mu=h$. The lemma states that if $
            \|\errormatrix\|
            \leq 
            \epsilon_{\mathsf{shift}}\frac{
                |\lambda_{\min}(h-\matXtilde)|^2\pi
            }{128}
        $ then
        $\|\sgn(h-\matXtilde)-\sgn(h-\matXtilde+\errormatrix)\| \leq \epsilon_{\mathsf{shift}}$ for some $\epsilon_{\mathsf{shift}}\in(0,1)$ that we can choose. By assumption $|\lambda_{\min}(h-\matXtilde)|\geq \varepsilon$, which means that 
        \begin{align*}
            4\umach\leq \epsilon_{\mathsf{shift}}\frac{\varepsilon^2\pi}{128n}
        \end{align*} is sufficient to guarantee that
        \begin{align*}
            n\lnorm
                \sgn(\matM) -\sgn(\matMtilde)
            \rnorm
            \leq \epsilon_{\mathsf{shift}.}
        \end{align*}

        For $\SGN$,  we set $\eta_{\SGN}=\tfrac{\varepsilon}{4}$, $\alpha_{\SGN}=\tfrac{4-\varepsilon}{4+\varepsilon}$ and $\epsilon_{\SGN}=\tfrac{1}{8n}$. Note that $\alpha_{\SGN}$ satisfies the requirement $99/100<\alpha_{\SGN}<1$ as long as $\varepsilon<4/199$). Then based on Theorem \ref{theorem:sgn} $\SGN$ requires
        \begin{align*}
            N = O\lpar
                \log(\tfrac{1}{1-\alpha_{\SGN}})+ \log(\log(\tfrac{1}{\epsilon_{\SGN}\eta_{\SGN}}))
            \rpar
            =
            O\lpar
                \log(\tfrac{1}{\varepsilon}) + \log(\log(\tfrac{n}{\varepsilon}))
            \rpar            
        \end{align*}
        iterations and
        \begin{align*}
         O\lpar \log(n)\log^3(\tfrac{1}{1-\alpha_{\SGN}})\lpar\log(\tfrac{1}{\epsilon_{\SGN}})+\log(\tfrac{1}{\eta_{\SGN}})\rpar \rpar
         =
            O\lpar
                \log(n)\log^3(\tfrac{1}{\varepsilon})\log(\tfrac{n}{\varepsilon})
            \rpar            
        \end{align*}
        bits of precision.

        If we set $\epsilon_{\mathsf{shift}}=1/8$, then the bound for $\umach$ becomes
        \begin{align*}
            \umach \leq \frac{\epsilon^2\pi}{32\cdot 128n},
        \end{align*}
        in which case 
        \begin{align*}
            \frac{\log(n)\umach}{1-\log(n)\umach}(1+\epsilon_{\SGN}) \ll \frac{1}{8}. 
        \end{align*}
        Then
        \eqref{equation:bound_eigenvalue_counts} becomes
        \begin{align*}
            |n_{(-)} -\widetilde n_{(-)} | 
            &<
            \tfrac{1}{8}
            +
            \tfrac{1}{8}
            +
            \tfrac{1}{8}
            =3/8 < 1/2,
        \end{align*}   
        which means that if we round to the closest integer then $\round(\widetilde n_{(-)})=n_{(-)}$. Note that
        $\log(1/\umach)=O\lpar  \log(\frac{n}{\epsilon}) \rpar $ which is dominated by the bit requirements of $\SGN$.

    \end{proof}
\end{lemma}

\subsection{Approximate midpoint and gap}

Having an efficient way to count eigenvalues smaller than a threshold, we show how to approximate spectral gaps. We can now prove Proposition \ref{proposition:gap_finder}, which we restate below for readability.

\begin{proposition}
    \label{proposition:gap_finder_appendix}
        Problem \ref{problem:gap_finder} can be solved in time $O\lpar \log(\tfrac{1}{\epsilon\gap_k})q(\tfrac{1}{\epsilon\gap_k})\rpar $.
    \begin{proof}
        We start with $\gamma_0=1/4$, and make a grid $\mathsf{g}=\grid(-1,1+\gamma,\gamma)$ which consists of the points $\{-1, -3/4, -2/4, -1/4, 0, 1/4, 2/4, 3/4, 1, 5/4\}$ (a useful note for later is that all the numbers in the grid can be exactly represented in floating point). Let $\countsmaller(h_j)$ denote the number of $\gamma_0$-distorted values $\lambda_i'$ that are smaller than $h_j$, for some $h_j\in\mathsf{g},j=1,\ldots,10$. Query all $h_j\in\mathsf{g}$ for $\countsmaller(h_j)$. 
        Set 
        $j_k=\arg\min_j \{h_j | \countsmaller(h_j)\geq k\}$ 
        and 
        $j_{k+1}=\arg\min_j \{h_j | \countsmaller(h_j)\geq k+1\}$. 
        Now $\lambda'_k\in[h_{j_k-1}, h_{j_k}]$ and $\lambda'_{k+1}\in[h_{j_{k+1}-1}, h_{j_{k+1}}]$. 
        The points in the grid are equispaced so each interval has length $\gamma$. 
        Since by definition $\lambda'_k\in[\lambda_k-\gamma_0,\lambda_k+\gamma_0]$, then $\lambda_k\in[h_{j_{k}-1}-\gamma_0, h_{j_{k}}+\gamma_0]=[h_{j_k-2}, h_{j_k+1}]$. 
        Similarly for $\lambda_{k+1}$. 
        We have now restricted both $\lambda_k$ and $\lambda_{k+1}$ inside some intervals of length $3\gamma_0$. 
        Let us denote those intervals as  $I_k=[h_{j_k-2}, h_{j_k+1}]$ and $I_{k+1}=[h_{j_{k+1}-2}, h_{j_{k+1}+1}]$. 
        In the next step we halve $\gamma_0$ to $\gamma_1=1/8$. 
        We now create two grids, one inside $I_k$ and one inside $I_{k+1}$: $\mathsf{g}_k=\grid(h_{j_k-2}, h_{j_k+1}+\gamma_1, \gamma_1)$ and similar for $\mathsf{g}_{k+1}$. 
        Each of the new grids has exactly $8$ points and each point can be exactly represented in floating point. 
        As in the previous iteration, we query all the $8$ points $h_j$ of $\mathsf{g}_{k}$ for $\countsmaller(h_j)$, and we pick a new $j_k=\arg\min\{h_j | \countsmaller(h_j)\geq k\}$, and we do the same for $j_{k+1}$. There are a total of 16 queries. 
        As before, we have now restricted $\lambda_k$ and $\lambda_{k+1}$ inside two new intervals $I_k$ and $I_{k+1}$ where $|I_k|=|I_{k+1}|=3\gamma_1$, which is half the size of the corresponding intervals of the previous iterations. If we set $\widetilde\lambda_k$ equal to the midpoint of $I_k$, then $|\widetilde\lambda_k-\lambda_k|\leq 3\gamma/2$. The same for $\lambda_{k+1}$. We keep repeating the same procedure by halving $\gamma_i$ in each step $i$.

        It remains to find a proper termination condition. Let us denote $\widetilde\mu_k=\frac{\widetilde\lambda_k+\widetilde\lambda_k}{2}$ and $\widetilde\gap_k=\widetilde\lambda_k-\widetilde\lambda_{k+1}$, where $\widetilde\lambda_k$ and $\widetilde\lambda_{k+1}$ are as above. Then, after $m$ steps,
        \begin{align*}
            \widetilde\mu_k&=\mu_k\pm 3\gamma_m/2,\\
            \widetilde\gap_k&=\gap_k\pm 3\gamma_m.
        \end{align*}
        Thus, $\gamma_m\approx \epsilon\gap_k/3$ is a sufficient terminating criterion. We don't know $\gap_k$, but we can use $\widetilde\gap_k$ instead, i.e., $\gamma\leq \epsilon\frac{\widetilde\gap_k-3\gamma_m}{3}$ is also a sufficient, which gives a quantifiable termination criterion $\gamma_m \leq \epsilon\widetilde\gap_k/6$. But since $\widetilde\gap_k\geq \gap_k-3\gamma_m$, then the terminating criterion will be reached in the worst case when
        \begin{align*}
            \gamma_m \leq \epsilon\frac{\gap_k-3\gamma_m}{6} \Rightarrow \gamma_m\leq\frac{\epsilon\gap_k}{9},
        \end{align*}
        meaning that the algorithm will halt and return an $\epsilon$-accurate $\widetilde\mu_k, \widetilde\gap_k$ in at most $O\lpar \log(\tfrac{9}{\gamma_m})\rpar = O\lpar \log(\tfrac{1}{\epsilon\gap_k})\rpar$ iterations.

        Each $h_j$-query at iteration $i$ costs $O(\polylog(1/\gamma_i))$ by assumption, and there are exactly 16 $h_j$-queries in each iteration, which gives a total query cost of $O(\polylog(1/\gamma_i))$ per iteration, which is maximized for the smallest $\gamma_m=\Theta(\epsilon\gap_k)$.
    \end{proof}
\end{proposition}

We can now describe the algorithm $\GAP$ that computes the $k$-th gap and the midpoint of a Hermitian definite pencil.
\begin{theorem}[$\GAP$, Restatement of Theorem \ref{theorem:alg_spectral_gap}]
    \label{theorem:alg_spectral_gap_appendix}
    Let $\matH\in\mathbb{H}^n$, $\matS\in\mathbb{H}^n_{++}$ and $\|\matH\|,\|\matS^{-1}\|\leq 1$, which define a Hermitian definite pencil $(\matH,\matS)$. Given $k\in[n-1]$, accuracy $\epsilon\in(0,1)$, and failure probability $\delta\in(0,1/2)$, there exists an algorithm
    \begin{align*}
        \widetilde\mu_k,\widetilde\gap_k \leftarrow \GAP(\matH,\matS,k,\epsilon,\delta)
    \end{align*}
    which returns $\widetilde\mu_k=\mu_k\pm\epsilon\gap_k$ and $\widetilde \gap_k=(1\pm\epsilon)\gap_k$, where $\mu_k=\tfrac{\lambda_k+\lambda_{k+1}}{2}$ and $\gap_k=\lambda_{k}-\lambda_{k+1}$. The algorithm requires 
    \begin{align*}
        O\lpar
            T_{\MM}(n)
            \log(\tfrac{1}{\delta\epsilon\gap_k})
            \log(\tfrac{1}{\epsilon\gap_k})
        \rpar
    \end{align*}
    arithmetic operations using $O\lpar\log(n)
    \lpar
        \log^4(\tfrac{n}{\delta\epsilon\gap_k})
        +
        \log(\kappa(\matS))
    \rpar
    \rpar$ bits, where $\lambda_i$ are the eigenvalues of $(\matH,\matS)$. 
    If $\kappa(\matS)$ is unknown, additional $O(T_{\MM}(n)\log(\tfrac{n\kappa(\matS)}{\delta})\log(\kappa(\matS)))$ floating point operations and $O(\log(n)\log^4(\tfrac{n\kappa(\matS)}{\delta}))$ bits are suficient to compute it with Corollary \ref{corollary:alg_cond}.
    \begin{proof}
        We first set $\gamma_0=1/8$ and call $\matHtilde=\REDUCE\lpar \matH,\matS,\tfrac{\gamma_0}{4} \rpar$. From Proposition \ref{proposition:alg_reduce}, $\REDUCE$ requires $O(T_{\MM}(n))$ floating point operations using $O(\log(n)\log(\kappa(\matS))+\log(1/\gamma_0))$ bits, and it returns a matrix $\matHtilde$ that satisfies
        $|\lambda_i(\matHtilde)-\lambda_i(\matH,\matS)|\leq \tfrac{\gamma_0}{4}$. 
        
        We then call $\matXtilde\leftarrow \REGULARIZE(\matHtilde,\tfrac{\gamma_0}{2},\delta_0)$, where $\delta_0=\delta/2$ is the initial failure probability. From Proposition \ref{proposition:alg_regularize}, for all $i$ it holds that $|\lambda_i(\matXtilde)-\lambda_i(\matHtilde)|\leq 9\gamma_0/16$.
        Therefore, all the eigenvalues, which initially lie in $[-1,1]$, are distorted by at most $\gamma_0$.
        
        Next, recall the counting query model of Proposition \ref{proposition:gap_finder}: in the first step we construct a grid $\mathsf{g}=\grid(-1,1+\gamma_0,\gamma_0)$, where $|\mathsf{g}|=10$. 
        From Proposition \ref{proposition:alg_regularize}, the regularization ensures that for every $h_j\in\mathsf{g}$ it holds that
        $\sigma_{\min}(h_j-\matXtilde+\errormatrix)\geq \varepsilon_0$ for $\varepsilon_0=\frac{\gamma_0\delta_0}{8|\mathsf{g}|n\sqrt{\pi\ln(4n/\delta_0)}}$ (this is included in the success probability of $1-\delta_0$ of Proposition \ref{proposition:alg_regularize}). Then
        we execute $\COUNT(\matXtilde,h_j,\varepsilon_0)$ for every $h_j$. We continue by halving at each step both $\gamma$ and $\delta$, constructing the corresponding grids as per Proposition \ref{proposition:gap_finder}, and counting eigenvalues over the grid. In each iteration after the first one, we keep track of two grids $\mathsf{g}$ with size $|\mathsf{g}|=\Theta(1)$ each, and therefore we can ignore $|\mathsf{g}|$ in the complexity.
        
        After a total of $m=O\lpar \log(\tfrac{1}{\epsilon\gap_k}) \rpar$ iterations, we have that $\gamma_m=\Theta(\epsilon\gap_k)$ and $\delta_m=\Theta(\delta_0\epsilon\gap_k)$.
        Invoking Lemma \ref{lemma:eigenvalue_count_less_than_h}, and plugging in the value of $\varepsilon_m$, each call to $\COUNT$ during the entire algorithm costs at most
        \begin{align*}
            O\lpar
                T_{\MM}(n)
                \lpar 
                    \log(\tfrac{1}{\varepsilon_m})
                    +
                    \log(\log(\tfrac{n}{\varepsilon_m}))
                \rpar
            \rpar
            =
            O\lpar
                T_{\MM}(n)\log(\tfrac{n}{\gamma_m\delta_m})
            \rpar
        \end{align*}
        arithmetic operations using $
        O\lpar 
            \log(n)\log^3(\tfrac{1}{\varepsilon_m})
            \log(\tfrac{n}{\varepsilon_m})
        \rpar
        =
        O\lpar \log(n)\log^4(\tfrac{n}{\gamma_m\delta_m})\rpar$ bits. The last iteration where  $\gamma_m=\Theta(\epsilon\gap_k)$ and $\delta_m=\Theta(\delta_0\epsilon\gap_k)$ is the most expensive, which gives a total of
        \begin{align*}
            O\lpar
                T_{\MM}(n)\log(\tfrac{n}{\delta_0\epsilon\gap_k})
                \cdot
                m
            \rpar
            =
            O\lpar
                T_{\MM}(n)\log(\tfrac{n}{\delta_0\epsilon\gap_k})
                \log(\tfrac{1}{\epsilon\gap_k})
            \rpar
        \end{align*}
        arithmetic operations
        and
        \begin{align*}
        O\lpar
            \log(n)
            \log^4(\tfrac{n}{\delta_0\epsilon\gap_k})
            \rpar
        \end{align*}
        bits. The failure probability is $\delta_0$ in the first iteration and halved at each subsequent iteration, in which case a union bound converges to $2\delta_0=\delta$.

        To be able to call $\REDUCE$ in every step, we need to approximate $\kappa(\matS)$ in order to set the machine precision appropriately. For this we can use Corollary \ref{corollary:alg_cond}, which is detailed in the next section, and it returns $\widetilde\kappa\in\Theta(\kappa(\matS))$ with probability $1-\delta$. It requires $O(T_{\MM}(n)\log(\tfrac{n\kappa(\matS)}{\delta})\log(\kappa(\matS)))$ floating point operations using $O(\log(n)\log^4(\tfrac{n\kappa(\matS)}{\delta}))$ bits.
    \end{proof}
\end{theorem}

\subsection{Singular values, singular gaps, and condition number}
\label{appendix:sigma_k}
The gap finder can be extended to compute eigenvalues and singular values and singular gaps or arbitrary matrices instead of eigenvalue gaps of Hermitian matrices. The following algorithm in Proposition \ref{proposition:alg_sigmak} computes the $k$-th singular value of a matrix, for arbitrary $k\in[n]$, using counting queries.

\begin{proposition}[$\SIGMAK$]
    \label{proposition:alg_sigmak}
        Given a matrix $\matA\in\mathbb{C}^{m\times n},n\geq m$ with $\|\matA\|\leq 1$, an integer $k\in[n]$ such that $\rank(\matA)\geq k$, an accuracy $\epsilon\in(0,1)$, and failure probability $\delta\in(0,1)$, there exists an algorithm
        \begin{align*}
            \widetilde\sigma_{k} \leftarrow \SIGMAK(\matA,k,\epsilon,\delta),
        \end{align*}
        which returns a value $\widetilde\sigma_{k}\in (1\pm\epsilon)\sigma_k(\matA)$ with probability at least $1-\delta$. The algorithm executes
        \begin{align*}
            O\lpar
            T_{\MM}(n)
            \lpar
                \tfrac{m}{n}
                +
                \log(\tfrac{n}{\delta\epsilon\sigma_k})
            \rpar
            \log(\tfrac{1}{\epsilon\sigma_k})
        \rpar
        \end{align*}
        floating point operations and requires $O\lpar
        \log(n)
        \log^4(\tfrac{n}{\delta\epsilon\sigma_k})
        \rpar$ bits.\\
        {\bf Notes:} If $m> n$ then we can apply the algorithm on $\matA^\top$. If $\|\matA\|>1$, we can use Theorem \ref{theorem:alg_norm} to approximate $\|\matA\|$ and scale accordingly. The $\rank(\matA)\geq k$ assumption can potentially be omitted by a more sophisticated counting query strategy.
    \begin{proof}
    We will solve the problem with counting queries, similar to Theorem \ref{theorem:alg_spectral_gap}, halving $\gamma$ and $\delta$ in every iteration. We start with $\gamma_0=1/8$ and $\delta_0=\delta/2$.
    First, we divide $\matA$ by two, to ensure that $\|\matA\|\leq 1/2$. Then we construct $\matAtilde=\HERM(\MM(\matA,\matA^*))=\matA\matA^*/4+\errormatrix^{\MM}$. To perform this rectangular multiplication in practice, we partition $\matA$ in $m/n$ blocks of size $n\times n$ each, perform the individual multiplications and sum the results. We can ensure that $\|\errormatrix^{\MM}\| \leq \gamma_0/4$ if we use $O(\log(\tfrac{n}{\gamma_0}))$ bits. Then from Weyl's inequality $|\lambda_k(\matAtilde)-\lambda_k(\matA\matA^*/4)| \leq \gamma_0/4$.
    We then call  $\matXtilde\leftarrow \REGULARIZE(\matAtilde,\gamma_0,\delta_0)$. Conditioning on success of Proposition \ref{proposition:alg_regularize}, for all $i$ it holds that $|\lambda_i(\matXtilde)-\lambda_i(\matAtilde)|\leq 9\gamma_0/16$, and therefore $|\lambda_i(\matXtilde)-\lambda_i(\matA\matA^*/4)|\leq \gamma_0$.

    Let $\lambda_i'=\lambda_i(\matXtilde)$ and $\lambda_i=\lambda_i(\matA\matA^*/4)$.
    Since all the $\lambda_i$ are non-negative, then all the $\lambda_i'\in[-\gamma_0,1+\gamma_0]$. We thus construct grid $\mathsf{g}=\grid(0,1+\gamma_0,\gamma_0)=\{0,1/8,2/8,3/8,\ldots,1,9/8\}$. We now need to perform counting queries similar to Theorem \ref{theorem:alg_spectral_gap}. Since we conditioned on success of Proposition \ref{proposition:alg_regularize}, $\REGULARIZE$ ensures that for every $h_j\in\mathsf{g}$ it holds that 
    $\sigma_{\min}(h_j-\matXtilde+\errormatrix_{h_j})\geq \varepsilon_0$ for $\varepsilon_0=\frac{\gamma_0\delta_0}{8|\mathsf{g}|n\sqrt{\pi\ln(4n/\delta_0)}}$. 
    Then we execute $C_j\leftarrow \COUNT(\matXtilde,h_j,\varepsilon_0)$ for every $h_j\in\mathsf{g}$ (in total $\Theta(1)$ calls to $\COUNT$). We set $j_k$ to be the smallest $j$ such that $C_j\geq k$. 
    Now $\lambda'_{1}\in[h_{j_k}, h_{j_k+1}]$.  
    Since $\lambda'_{k}\in[\lambda_k-\gamma_0,\lambda_k+\gamma_0]$, then $\lambda_k\in[h_{j_{1}-1}-\gamma_0, h_{j_{k}}+\gamma_0]=[h_{j_k-2}, h_{j_k+1}]:=I_k$, where $|I_k|=3\gamma_0$ (we can always ignore the negative portion of $I_k$, if there is any).
    
    In the next step we halve $\gamma_0$ to $\gamma_1=1/8$, and we create another grid inside $I_k$, $\mathsf{g}_1=\grid(h_{j_k-2}, h_{j_k+1}+\gamma_1, \gamma_1)$. The new grid has at most $8$ points. 
    We query all the points $h_j$ of $\mathsf{g}_{1}$ for $C_j\leftarrow \COUNT(\matXtilde,h_j,\epsilon_1)$, and we pick a new $j_1=\arg\min\{h_j | C_j\geq k\}$. 
    Now $\lambda_k$ lies inside a new interval $I_k$ with $|I_k|=3\gamma_1$, which is half the size of the interval of the previous iteration. If we set $\widetilde\lambda_k$ equal to the midpoint of $I_k$ (which is always positive), then $|\widetilde\lambda_k-\lambda_k|\leq 3\gamma/2$. We keep repeating the same procedure by halving $\gamma_i$ and $\delta_i$ in each step $i$.
    
    After $m$ steps we have that
    \begin{align*}
        \widetilde\lambda_k&=\lambda_k\pm3\gamma_m/2.
    \end{align*}
    A sufficient termination criterion is $\gamma_m\approx \epsilon\lambda_k$. Instead of $\lambda_k$ (which is unknown) we use $\widetilde \lambda_k$ instead. Then $\gamma\leq \epsilon\frac{\widetilde\lambda_k-3\gamma_m}{3}$ is also a sufficient, which gives a  termination criterion $\gamma_m \leq \epsilon\widetilde\lambda_k/6$ that we can actually calculate. In turn, this termination criterion will be reached in the worst case when
    \begin{align*}
        \gamma_m \leq \epsilon\frac{\lambda_k-3\gamma_m/2}{6} \Rightarrow \gamma_m\leq\frac{\epsilon\lambda_k}{9},
    \end{align*}
    meaning that the algorithm will halt and return an $\epsilon$-accurate $\lambda_k$ in at most $m=O\lpar \log(\tfrac{9}{\gamma_m})\rpar = O\lpar \log(\tfrac{1}{\epsilon\lambda_k})\rpar$ iterations. 

    Note that in each iteration we make a total of $\Theta(1)$ calls to $\COUNT$. From Lemma \ref{lemma:eigenvalue_count_less_than_h}, if we plug in the corresponding $\varepsilon_i$, each call to $\COUNT$ during the entire algorithm costs at most
    \begin{align*}
        O\lpar
            T_{\MM}(n)\log(\tfrac{n}{\gamma_i\delta_i})
        \rpar
    \end{align*}
    arithmetic operations using $O\lpar \log(n)
    \log^4(\tfrac{n}{\gamma_i\delta_i})
    \rpar$ bits. The cost is maximized in the last iteration where $\gamma_m=\Theta(\epsilon\lambda_k)$ and $\delta_m=\Theta(\delta_0\epsilon\lambda_k)$, which gives
    \begin{align*}
        O\lpar
            T_{\MM}(n)\log(\tfrac{n}{\delta_0\epsilon\lambda_k})
            \cdot
            m
        \rpar
        =
        O\lpar
            T_{\MM}(n)\log(\tfrac{n}{\delta_0\epsilon\lambda_k})
            \log(\tfrac{1}{\epsilon\lambda_k})
        \rpar
    \end{align*}
    arithmetic operations
    and
    \begin{align*}
    O\lpar
        \log(n)
        \log^4(\tfrac{n}{\delta_0\epsilon\lambda_k})
        \rpar
    \end{align*}
    bits. As in Theorem \ref{theorem:alg_spectral_gap}, since the failure probability is halved in each iteration, a union bound converges to $2\delta_0=\delta$.
    
    Now we have that $|\lambda_k-\widetilde\lambda_k| \leq \epsilon \lambda_k$. 
    Recall that $\sigma_k(\matA)=\sqrt{\lambda_k(\matA\matA^*)}=2\sqrt{\lambda_k}$, since we computed $\matA\matA^*$ and divided by two. 
    If we set $\widetilde\sigma_k=2\sqrt{\widetilde\lambda_k}$ then
    \begin{align*}
        (1-\epsilon)\frac{\sigma_k^2}{4}
        \leq 
        \frac{\widetilde\sigma_k^2}{4}
        \leq (1+\epsilon)\frac{\sigma_k^2}{4}
        \Rightarrow
        (1-\epsilon)\sigma_k \leq \widetilde\sigma_k \leq (1+\epsilon)\sigma_k,
    \end{align*}
    where the last holds since $\sqrt{1+\epsilon}\leq 1+\epsilon$ and $1-\epsilon\leq \sqrt{1-\epsilon}$ for $\epsilon\in(0,1)$. Note that we lazily assumed that we computed $\widetilde\sigma_k=2\sqrt{\widetilde\lambda_k}$ exactly, which does not hold, the square root introduces a machine precision error. The calibration is left as an exercise. If we replace $\lambda_k$ with $\sigma_k^2/4$ and $\delta_0=\delta/2$ in the complexity bounds we get the desired
    \begin{align*}
        O\lpar
            T_{\MM}(n)\log(\tfrac{n}{\delta\epsilon\sigma_k})
            \log(\tfrac{1}{\epsilon\sigma_k})
        \rpar
    \end{align*}
    arithmetic operations
    and
    $
    O\lpar
        \log(n)
        \log^4(\tfrac{n}{\delta\epsilon\sigma_k})
        \rpar
    $
    bits.
    \end{proof}
\end{proposition}

We now have a tool for the condition number.
\begin{corollary}[$\COND$]
    \label{corollary:alg_cond}
    Let $\matA\in\mathbb{H}^n_{++}$. Given $\delta \in(0,1/2)$, we can compute $\widetilde\kappa$ such that $\kappa(\matA)\leq \widetilde\kappa\leq 32\kappa(\matA)$, with an algorithm $\widetilde\kappa\leftarrow \COND(\matA,\delta)$ in 
    \begin{align*}
        O\lpar
                T_{\MM}(n)
                    \log(\tfrac{n\kappa(\matA)}{\delta})
                    \log(\kappa(\matA)))
            \rpar
    \end{align*}
    using 
    $O\lpar
            \log(n)
            \log^4(\tfrac{n\kappa(\matA)}{\delta})
        \rpar$
    bits of precision with probability at least $1-\delta$.
    \begin{proof}
        We first compute $\widetilde\Sigma\in[0.9\|\matA\|,2\|\matA\|]$ using Corollary \ref{corollary:spectral_norm} with parameter $\delta/3$ in $O(n^2\log(n)\log(1/\delta))$ floating point operations using $O(\log(n))$ bits of precision. The algorithm succeeds with probability at least $1-\tfrac{2\delta}{3}$. 
        We then scale $\matA'\leftarrow \matA/M$ where $M$ is the smallest power of two that is larger than $4\widetilde\Sigma$. This implies $\frac{1}{16}\leq \|\matA'\|\leq \frac{1}{3.6}$, and also $\sigma_{\min}(\matA')
        \in
        \lbrac \tfrac{\sigma_{\min}(\matA)}{16\|\matA\|}, \tfrac{\sigma_{\min}(\matA)}{3.6\|\matA\|} \rbrac = 
        \lbrac 
        \tfrac{1}{16\kappa(\matA)}, \tfrac{1}{3.6\kappa(\matA)} \rbrac
        $.
        Thus, it suffices to approximate $\sigma_{\min}(\matA')$ and scale it to obtain the desired approximation for $\kappa(\matA)$.
        
        We now  call $\widetilde\sigma_{\min}'\leftarrow \SIGMAK(\matA',1,1/2,\delta/3)$, which succeeds probability $1-\delta/3$ and returns $\widetilde\sigma_{\min}'\in(1\pm \tfrac{1}{2})\sigma_{\min}(\matA')$. It requires
        \begin{align*}
            O\lpar
                T_{\MM}(n)
                \lpar 
                    \log(\tfrac{n}{\delta\sigma_{\min}(\matA')})
                    \log(\tfrac{1}{\sigma_{\min}(\matA')})
                \rpar
            \rpar
            =
            O\lpar
                T_{\MM}(n)
                    \log(\tfrac{n\kappa(\matA)}{\delta})
                    \log(\kappa(\matA))
            \rpar
        \end{align*}
        arithmetic operations
        and
        \begin{align*}
        O\lpar
            \log(n)
            \log^4(\tfrac{n}{\delta\sigma_{\min}(\matA')})
        \rpar
        =
        O\lpar
            \log(n)
            \log^4(\tfrac{n\kappa(\matA)}{\delta})
        \rpar
        \end{align*}
        bits. 
        Then we can set $\widetilde\kappa= \frac{1}{\widetilde\sigma_{\min}'}\in \lbrac
            \kappa(\matA), 32\kappa(\matA)
        \rbrac$.
        The success probability is $1-\tfrac{2\delta}{3}-\tfrac{\delta}{3}=1-\delta$.
    \end{proof}
\end{corollary}

\subsection{Proof of Theorem \ref{theorem:spectral_projector}}
\label{appendix_proofs:theorem:spectral_projector}
We now have all the prerequisites to prove that Algorithm \ref{alg:spectral_projector} provides the guarantees of our main Theorem \ref{theorem:spectral_projector}, which we restate for readability.

\begin{theorem}[Restatement of Theorem \ref{theorem:spectral_projector}]
    \label{theorem:spectral_projector_restatement}
    Let $(\matH,\matS)$ be a Hermitian 
     definite pencil of size $n$, with $\|\matH\|,\|\matS^{-1}\|\leq 1$, and  $\lambda_1\leq\lambda_2\leq\ldots\leq \lambda_n$ its eigenvalues. There exists an algorithm
    \begin{align*}
        \projectormatrixtilde_k \leftarrow \PROJECTOR(\matH,\matS,k,\epsilon),
    \end{align*} which takes as input $\matH$, $\matS$, an integer $1\leq k\leq n-1$, an error parameter $\epsilon\in(0,1)$ and returns a matrix $\projectormatrixtilde_k$ such that
    \begin{align*}
        \Pr\lbrac 
            \lnorm \projectormatrixtilde_k - \projectormatrix_k \rnorm \leq \epsilon
        \rbrac
        \geq 1-1/n,
    \end{align*}
    where $\projectormatrix_k$ is the true spectral projector on the invariant subspace that is associated with the $k$ smallest eigenvalues.
    The algorithm executes 
    \begin{align*}
        O\lpar
                T_{\MM}(n)\lpar
                    \log(\tfrac{n}{\gap_k})\log(\tfrac{1}{\gap_k})
                    +
                    \log(n\kappa(\matS))\log(\kappa(\matS))
                    +
                    \log\lpar
                        \log(
                            \tfrac{\kappa(\matS)}{\epsilon\gap_k}
                        )
                    \rpar
                \rpar
            \rpar
    \end{align*}
    floating point operations, using \begin{align*}
        O\lpar
                \log(n)
                \lpar
                    \log^4(\tfrac{n}{\gap_k})
                    +
                    \log^4(n\kappa(\matS))
                    +
                    \log^3(\tfrac{1}{\epsilon\gap_k})
                    \log(\tfrac{\kappa(\matS)}{\epsilon\gap_k})
                \rpar
            \rpar
    \end{align*}
    bits of precision, where $\kappa(\matS)=\|\matS\|\|\matS^{-1}\|$ and $\gap_k=\lambda_{k+1}-\lambda_k$. Internally, the algorithm needs to generate a total of at most $\widetilde O(n)$ standard normal floating point numbers using additional $O(\log(\log(n)))$ bits. In the regular case, when $\matS=\matI$, the $O(\log(n\kappa(\matS))\log(\kappa(\matS)))$ term in the arithmetic complexity and the $O(\log^4(n\kappa(\matS)))$ term in the number of bits are removed.
\begin{proof}
We first compute $\widetilde\mu_k$, $\widetilde\gap_k$,$\widetilde\kappa$ which satisfy the requirements of Proposition \ref{proposition:alg_purify} using $\COND$ and $\GAP$. Afterwards we can use them to call $\PURIFY$ to obtain the spectral projector $\projectormatrixtilde_k$ such that $\|\projectormatrixtilde_k-\projectormatrix_k\|\leq \epsilon$. 
We first call $\COND(\matS,\tfrac{1}{4},\tfrac{1}{2n})$, which, from Corollary \ref{corollary:alg_cond}, requires 
        \begin{align*}
            O\lpar
                T_{\MM}(n)\log(n\kappa(\matS))\log(\kappa(\matS)))
            \rpar
        \end{align*}
        arithmetic operations and
        \begin{align*}
            O\lpar
                \log(n)\log^4(n\kappa(\matS))
            \rpar
        \end{align*}
        bits.
Then, given the result of $\COND,$ we call $\GAP(\matH,\matS,k,\tfrac{1}{8},\tfrac{1}{2n})$ 
 which requires \begin{align*}
            O\lpar
                T_{\MM}(n)\log(\tfrac{n}{\gap_k})\log(\tfrac{1}{\gap_k}))
            \rpar
        \end{align*}
        arithmetic operations and
        \begin{align*}
            O\lpar
                \log(n)\log^4(\tfrac{n}{\gap_k})
                +\log(\kappa(\matS))
            \rpar
        \end{align*}
        bits.  They both succeed at the same time with $1-1/n$ probability. $\PURIFY(\matH,\matS,\widetilde\mu_k,\widetilde\gap_k,\widetilde\kappa,\epsilon)$ requires
        \begin{align*}
            O\lpar
                T_{\MM}(n)\lpar
                    \log(\tfrac{1}{\gap_k})+\log(\log(\tfrac{\kappa(\matS)}{\epsilon\gap_k}))
                \rpar
            \rpar
        \end{align*} floating point operations and
        \begin{align*}
            O\lpar
                \log(n)
                \log^3(\tfrac{1}{\gap_k})\log(\tfrac{\kappa(\matS)}{\epsilon\gap_k})
            \rpar
        \end{align*}
        bits.
        The total arithmetic complexity is therefore
        \begin{align*}
            O\lpar
                T_{\MM}(n)\lpar
                    \log(\tfrac{n}{\gap_k})\log(\tfrac{1}{\gap_k})
                    +
                    \log(n\kappa(\matS))\log(\kappa(\matS))
                    +
                    \log\lpar
                        \log(
                            \tfrac{\kappa(\matS)}{\epsilon\gap_k}
                        )
                    \rpar
                \rpar
            \rpar
        \end{align*}
        and the bit requirement is
        \begin{align*}
            O\lpar
                \log(n)
                \lpar
                    \log^4(\tfrac{n}{\gap_k})
                    +
                    \log^4(n\kappa(\matS))
                    +
                    \log^3(\tfrac{1}{\gap_k})
                    \log(\tfrac{\kappa(\matS)}{\epsilon\gap_k})
                \rpar
            \rpar.
        \end{align*}
        Notice that the $\epsilon$ term appears only inside $\log\log$ in the arithmetic complexity and in only one of the $\log$ factors of the number of bits.
\end{proof}
\end{theorem}

\section{DFT background}
\label{appendix:dft_background}
Density Functional Theory (DFT) \cite{kohn1965self}, which was awarded the Nobel prize in Chemistry in 1998, is considered as one of the most common methods to perform electronic structure calculations thanks to its \textit{ab initio} character. That means that no input besides the initial atomic structure is required to determine the relaxed geometry, energy levels, and the electron structure of a given material, in their bulk or nanostructured form. The main idea behind DFT is to describe the system and its properties by the electron density only. DFT calculations are widely used in industry and in academia to predict, for example, the properties of novel materials
\cite{maurer2019} or to optimize the performance of batteries \cite{zhao2019}, solar cells \cite{yan2017}, or nanoelectronic devices \cite{klinkert2020}. Many scientific libraries implementing DFT algorithms have been developed, which persistently occupy supercomputing clusters and receive up to tens of thousands of citations annually at the time of this writing \cite{kresse1996efficient,giannozzi2017advanced,soler2002siesta,hutter2014cp2k}. Despite the success of DFT calculations, important theoretical aspects of their underlying algorithms remain unclear.

Assume a system that is composed of a set of $n_a\in\mathbb{N}$ atoms positioned inside a fixed three-dimensional domain. The Kohn-Sham equation in real-space describes the electronic wave function $\psi(\vecr)$ of the system 
\begin{align}
    \mathcal{H}(\vecr) \psi(\vecr) = E \psi(\vecr),
\label{eq:kohn_sham}
\end{align}
where $\vecr=(x,y,z)$ is any position in the domain. Here $\mathcal{H}(\vecr)$ is the so-called single-particle Hamiltonian operator of the system, which can be written in the following form:
\begin{align*}
\mathcal{H}(\vecr) = -\frac{1}{2}\nabla^2 + E_{ion}(\vecr) + \int \frac{n(\vecr')}{\| \vecr - \vecr' \|}d\vecr' + E_{xc}(\vecr),
\end{align*}
where the first term is the kinetic term, the second term accounts for electron-ion interactions, the third term represents the electron-electron interactions, which are solved at the Hartree level through Poisson's equation, and the fourth term is the so-called exchange-correlation term. 
The electron density $n(\vecr)$ is the main observable of interest in DFT. It is related to the solutions $E_i$ and $\psi_i(\vecr)$ of Eq. \eqref{eq:kohn_sham} through 
\begin{align}
    n(\vecr) = \sum_{i} f(E_i;E_F) |\psi_{i}(\vecr)|^2,
    \label{eq:electron_density_main_equation}
\end{align}
where $E_F$ is the Fermi level of the system, $E_i$ is the energy level (eigenvalue) corresponding to the wave function $\psi_i(\vecr)$, and $f(E_i;E_F)$ is the occupation term. At zero temperature $(T=0K)$, $f(E_i;E_F)$ is equal to one for occupied states ($E_i<E_F$) and  zero otherwise.\footnote{The definition of $f(E;E_F)$ can vary to allow for partial occupations, i.e. at nonzero temperature $T>0$ it can be replaced by the Fermi-Dirac distribution $1-\frac{1}{1+\exp((E_i-E_F)/k_BT)}$, where $k_B$ is Boltzmann's constant. This case is not considered in this work.} 
The $E_F$, $E_i$, and $\psi_i$  are the unknown quantities for which there exists no analytical solution, except for some special cases. Approximate numerical solutions can be sought after expanding the wave functions into a suitable basis. In this case, the 
$\psi_i$'s are approximated by a set of $n=\Theta(n_a)$ basis functions $\chi_j(\vecr),j\in[n]$, typically localized around atomic positions.\footnote{A common alternative is to approximate the electronic wave function using Plane Waves. In this work we focus on localized basis sets.} 
Common basis sets that lead to efficient algorithms to solve the eigenvalue problem of Eq. \eqref{eq:kohn_sham} include Linear Combination of Atomic Orbitals (LCAO) \cite{clark1999}, Gaussian-Type Orbitals (GTO) \cite{basch1969}, or Maximally Localized Wannier Functions (MLWF) \cite{marzari2012}. In this context, the wave functions take the form of linear combinations of the $n$ localized orbitals. In particular, for all $i\in[n]$
\begin{align}
    \psi_i(\vecr) = \sum_{i=1}^{n}c_{ij}\chi_j(\vecr),
    \label{eq:expansion_of_wave_functions}
\end{align}
where  $c_{ij}$ are the (unknown) complex coefficients. This expansion can also be written in matrix notation as
\begin{align}
    \label{eq:basis_expansion}
    \vecy(\vecr) = \begin{pmatrix}
        \psi_1(\vecr)\\
        \psi_2(\vecr)\\
        \vdots\\
        \psi_{n}(\vecr)
    \end{pmatrix}
    = 
    \begin{pmatrix}
        c_{11} & c_{12} & \ldots & c_{1n} \\
        c_{21} & c_{22} & \ldots & c_{2n} \\
        \vdots & \vdots & \ddots & \vdots \\
        c_{n1} & c_{n2} & \ldots & c_{nn} \\
    \end{pmatrix}
    \begin{pmatrix}
        \chi_1(\vecr)\\
        \chi_2(\vecr)\\
        \vdots\\
        \chi_{n}(\vecr)
    \end{pmatrix}
    =
    \matC\vecx(\vecr),
\end{align}
where $\vecy(\vecr)\in\mathbb{C}^{n}$ and $\vecx(\vecr)\in\mathbb{C}^{n}$  are (complex) vectors and $\matC$ is a $n\times n$ matrix.
By expanding the wave functions into LCAO, GTO, or MLWF basis, the Hamiltonian operator becomes a matrix
\begin{align*}
    \matH_{i,j} = \langle \chi_i | \mathcal{H} | \chi_j \rangle = \int \chi_i^*(\vecr) \mathcal{H}(\vecr) \chi_j(\vecr)dr.
\end{align*}
Similarly, an overlap matrix $\matS$ must be introduced
\begin{align*}
    \matS_{i,j} = \langle \chi_i | \chi_j \rangle = \int \chi_i^*(\vecr) \chi_j(\vecr)dr,
\end{align*}
which is equal to the identity matrix $\matI$ if the basis elements are orthogonal (e.g. $\matS=\matI$ for MLWF). The entries of $\matH$ and $\matS$ represent the interaction between two basis elements that can be located on the same or on different atoms. Furthermore, due to the complex conjugate property of the inner product, both $\matH$ and $\matS$ are Hermitian matrices. Moreover, $\matS$ is  positive-definite (\cite[Chapter 3]{mayer2003simple}). Putting everything together, the expansion coefficients $c_{ij}$ are obtained by solving the following generalized eigenvalue problem as already formulated in Eq. \eqref{eq:gep_main}:
\begin{align*}
    \matH\matC=\matS\matC\geneigmatrix.
\end{align*}

The so-called density matrix $\matP$ can be derived from the  $\matC$ and $\geneigmatrix$ solutions
\begin{align}
    \matP=\matC f(\geneigmatrix; E_F)\matC^*.
    \label{eq:density_matrix}
\end{align}
The density matrix possesses interesting properties. The matrix $\matP\matS $ is a projector matrix and therefore it is idempotent, i.e. $\matP \matS =(\matP \matS )^2$. Moreover, it can be used to derive the electron density at any position $r$
\begin{align}
    n(\vecr) = \sum_{i,j\in n} \matP_{ij}\chi_i(\vecr) \chi_j^*(\vecr) = \vecx^*(\vecr) \matP  \vecx(\vecr),
    \label{eq:electron_density}
\end{align}
where $\vecx(\vecr)\in\mathbb{C}^{n}$ is a vector such that $\vecx_i(\vecr)=\chi_i(\vecr)$, i.e. $\vecx(\vecr)$ contains the values of all the basis functions at position $r$. Given these basic definitions, we next give an overview of existing algorithms.

\subsection{Linear scaling methods}
Because of the strong localization of the $\chi_i(\vecr)$'s, the $(i,j)$ interactions rapidly decay with the distance between the atoms and can be discarded beyond a pre-defined cut-off radius. As a consequence, $\matH$ and $\matS$ are often banded matrices with sometimes a sparse band. It should also be noted that at equilibrium, the number of negatively charged electrons present in the system (which corresponds to the integral of $n(\vecr)$ over space) must exactly compensate the number of positively charged protons $Z_p$. This allows for the computation of the Fermi level: from Eq. \eqref{eq:electron_density} and the definition of $\matS$, it follows that $\trace(\matP\matS)=Z_p$, which means that one can perform a binary search on the Fermi level $E_F$ until $\trace(\matP\matS)=Z_p$ is satisfied.
These observation has led to the development of so-called linear scaling methods, which aim to exploit the sparsity of the matrices to obtain faster solutions \cite{yang1991direct,galli1992large,kohn1993density,kohn1996density,galli1996linear,goedecker1999linear,goedecker2003linear}. Typically, the density matrix and the Fermi level are iteratively computed,
until all required constraints are satisfied, i.e., the density matrix should be idempotent and $\trace(\matP\matS)=Z_p$. This can be done, for example, by approximating the matrix sign function using a Newton iteration \cite{vandevondele2012linear}.
To optimize performance, heuristics such as truncating matrix elements with negligible magnitude are applied, at the cost of decreasing the solution accuracy. Empirical evidence shows that such methods tend to scale nearly linearly to the system size, however, they do not exhibit provable theoretical guarantees. Similar methods have been studied based on finite-differences, which aim to approximate the Fermi distribution with polynomial expansions \cite{zhou2006self,zhou2006parallel}. Finally, the work of \cite{gu1995divide,vogel2016superfast} offer the possibility to achieve lower complexities than $O(n^3)$ for both the eigenvalues and eigenvectors of structured matrices, by leveraging the fast multipole method. However, an end-to-end stability analysis of those algorithms remains open.
Indeed, we are not aware of any end-to-end analysis with provable approximation guarantees that has lower than $O(n^3)$  worst-case complexity for any of the aforementioned algorithms.

\subsection{Density matrix}
\label{appendix:density_matrix}

We can directly apply Theorem \ref{theorem:spectral_projector} to compute density matrices in DFT.

The following Theorem summarizes the analysis of Algorithm \ref{alg:density} for the density matrix.

\begin{algorithm}[htb]
\centering
\caption{$\DENSITY$.}
\label{alg:density}
\begin{algorithmic}[1]
\setstretch{0.8}
\Statex \begin{align*}
    \DENSITY
\end{align*}
\small
\Input Hermitian definite pencil $\matH\in\mathbb{H}^{n}$, $\matS\in\mathbb{H}^n_{++}$, number of occupied orbitals $k$.
\Require $\|\matH\|\leq 1$, $\|\matS^{-1}\|\leq 1$, $k\in[n-1]$.
\setstretch{1.5}
\Algorithm $\matPtilde \leftarrow\DENSITY(\matH,\matS,k,\epsilon)$.
\State $\projectormatrixtilde\leftarrow \PROJECTOR(\matH,\matS,k,\tfrac{\epsilon}{32})$.
\State $\matS_{\INV} \leftarrow \INV(\matS)$. \Comment{$\matS_{\INV}=\matS^{-1}+\errormatrix_1^{\INV}$.}
\State $\matPtilde \leftarrow \HERM\lpar \MM(\projectormatrixtilde,\MM(\matS_{\INV},\projectormatrixtilde^*))\rpar$.
\Comment{$\matPtilde= \projectormatrixtilde\lpar \matS_{\INV}\projectormatrixtilde^*+\errormatrix_2^{\MM}\rpar + \errormatrix_3^{\MM}$.}
\State \Return $\matPtilde$.
\Output Approximate density matrix $\matPtilde$.
\Ensure $\|\matPtilde-\matP\|\leq \epsilon$ with probability at least $1-O(1/n)$.
\end{algorithmic}
\end{algorithm}

\begin{theorem}[$\DENSITY$]
    \label{theorem:alg_density}
    Given a Hermitian definite pencil $(\matH,\matS)$ with $\|\matH\|,\|\matS\|\leq 1$, an integer $k\in[n-1]$ denoting the number of occupied states in the system, and $\epsilon\in(0,1)$, Algorithm \ref{alg:density} returns a matrix $\matPtilde\leftarrow \DENSITY(\matH,\matS,k,\epsilon)$ such that
    \begin{align*}
        \|\matPtilde-\matP\| \leq \epsilon,
    \end{align*}
    where $\matP$ is the true density matrix of the system and succeeds with probability at least $1-O\left(
                1/n
        \right)$.
    The floating point arithmetic complexity is 
    \begin{align*}
        O\lpar
            T_{\MM}(n)\lpar
                \log(\tfrac{n}{\gap_k})\log(\tfrac{1}{\gap_k})
                +
                \log(n\kappa(\matS))\log(\kappa(\matS))
            \rpar
        \rpar,
    \end{align*}
    and it requires
    \begin{align*}
        O\lpar
                \log(n)
                \lpar
                    \log^4(\tfrac{n}{\gap_k})
                    +
                    \log^4(n\kappa(\matS))
                    +
                    \log^3(\tfrac{1}{\epsilon\gap_k})
                    \log(\tfrac{\kappa(\matS)}{\epsilon\gap_k})
                \rpar
            \rpar
    \end{align*}
    bits of precision.
    \begin{proof}
        In the first step of the algorithm we compute $\projectormatrixtilde$ such that
        \begin{align*}
            \|\projectormatrix-\projectormatrixtilde\|\leq \tfrac{\epsilon}{32}:=\epsilon_{\projectormatrix},
        \end{align*}
        where $\projectormatrix$ is the spectral projector on the invariant subspace of the definite pencil associated with the $k$ smallest eigenvalues. 
    
        Then we invert $\matS$, and the final step of the algorithm computes $\matPtilde$. Unrolling all the computations, $\matPtilde$ can be written as $\projectormatrix\matS^{-1}\projectormatrix^* + \errormatrix_{\matP}$ where $\errormatrix_{\matP}$ is a Hermitian error matrix (yet to be detailed). It is not hard to verify that $\projectormatrix\matS^{-1}\projectormatrix^*$ is equal to $\matP$, i.e. the true density matrix.

        It thus remains to bound $\|\errormatrix_{\matP}\|.$ We can directly force $\|\errormatrix_1^{\INV}\| \leq \epsilon_{\projectormatrix} \|\matS^{-1}\|\leq \epsilon_{\projectormatrix}$ by setting $\umach\leq  \epsilon_{\projectormatrix} \tfrac{1}{\pinv(n)\kappa(\matS)^{\cinv\log(n)}}$. Then from line 5 of Algorithm \ref{alg:density}
        \begin{align*}
            \matPtilde 
            &= \projectormatrixtilde\matS_{\INV}\projectormatrixtilde^*+\projectormatrixtilde\errormatrix_2^{\MM}+\errormatrix_3^{\MM}
            \\
            &= \lpar
                \projectormatrix+\errormatrix_{\projectormatrix}
            \rpar 
            \lpar
                \matS^{-1}+\errormatrix_1^{\INV}
            \rpar
            \lpar
                \projectormatrix+\errormatrix_{\projectormatrix}
            \rpar^*
            +
            \lpar
                \projectormatrix+\errormatrix_{\projectormatrix}
            \rpar
            \errormatrix_2^{\MM}
            +
            \errormatrix_3^{\MM}
            \\
            &=
            \projectormatrix\matS^{-1}\projectormatrix^*
            +
            \projectormatrix\errormatrix_1^{\INV}\projectormatrix^*
            +
            \projectormatrix
            \lpar
                \matS^{-1}+\errormatrix_1^{\INV}
            \rpar
            \errormatrix_{\projectormatrix}
            ^*
            +
            \errormatrix_{\projectormatrix}
            \lpar
                \matS^{-1}+\errormatrix_1^{\INV}
            \rpar
            \projectormatrix
            ^*
            +
            \ldots
            \\
            &\qquad\qquad\qquad\ldots+
            \errormatrix_{\projectormatrix}
            \lpar
                \matS^{-1}+\errormatrix_1^{\INV}
            \rpar
            \errormatrix_{\projectormatrix}
            ^*
            +
            \lpar
                \projectormatrix+\errormatrix_{\projectormatrix}
            \rpar
            \errormatrix_2^{\MM}
            +
            \errormatrix_3^{\MM}.
        \end{align*}

        We bound each term as follows.
        \begin{enumerate}[(i)]
            \item $\|\projectormatrix\errormatrix_1^{\INV}\projectormatrix^*\|\leq \|\errormatrix_1^{\INV}\|\leq \epsilon_{\projectormatrix} $,
            \item $\lnorm
                \projectormatrix\lpar \matS^{-1}+\errormatrix_1^{\INV} \rpar \errormatrix_{\projectormatrix}^*
            \rnorm
            \leq
            (1+\epsilon_{\projectormatrix} )\cdot \epsilon_{\projectormatrix} \leq 2\epsilon_{\projectormatrix},
            $
            \item $\lnorm
                \errormatrix_{\projectormatrix}\lpar \matS^{-1}+\errormatrix_1^{\INV} \rpar \errormatrix_{\projectormatrix}^*
            \rnorm
            \leq
            (1+\epsilon_{\projectormatrix} )\cdot \epsilon_{\projectormatrix}^2 \leq 2\epsilon_{\projectormatrix}^2,
            $
            \item $\|\errormatrix_2^{\MM}\| 
            \leq 
            \umach\mu(n)\|\matS_{\INV}\|\|\projectormatrixtilde^*\| 
            \leq 
            \umach\mu(n)\lpar \|\matS^{-1}\|+\|\errormatrix_1^{\INV}\|\rpar (1+\|\errormatrix_{\projectormatrix}\|)
            \ll
            \epsilon_{\projectormatrix} (1+\epsilon_{\projectormatrix} )(1+\epsilon_{\projectormatrix})
            \leq 4\epsilon_{\projectormatrix} 
            $,
            \item $\|\errormatrix_3^{\MM}\| 
            \leq \umach\mu(n)\|\projectormatrixtilde\|\|\matS_{\INV}\projectormatrixtilde^*+\errormatrix_2^{\MM}\|
            \ll
            \epsilon_{\projectormatrix} (1+\epsilon_{\projectormatrix})\lbrac (1+\epsilon_{\projectormatrix})(1+\epsilon_{\projectormatrix})+4\epsilon_{\projectormatrix} )
            \rbrac
            \leq
            \epsilon_{\projectormatrix} \cdot 2 \cdot (4+4\epsilon_{\projectormatrix}) \leq 16\epsilon_{\projectormatrix}.
            $
        \end{enumerate}
        Putting everything together we have that
        \begin{align*}
            \lnorm \matP-\matPtilde \rnorm
            &\leq
            \epsilon_{\projectormatrix}
            +
            2\epsilon_{\projectormatrix}
            +
            2\epsilon_{\projectormatrix}^2
            +
            4\epsilon_{\projectormatrix}
            +
            16\epsilon_{\projectormatrix}
            \leq 
            25\epsilon_{\projectormatrix}
            = \epsilon\frac{25}{32}
            \leq \epsilon.
        \end{align*}
        Note that the aforementioned value of $\umach$ is already dominated by the requirement of $\PROJECTOR$. The same holds for the arithmetic complexity. Therefore, the arithmetic complexity, the success probability, and the number of required bits, similar to those of Theorem \ref{theorem:spectral_projector} up to a constant factor.
    \end{proof}
\end{theorem}

\subsection{Electron density queries}
\label{appendix:electron_density}
Having a good approximation of the density matrix, we can use it to query for the electron density at any point of interest. A direct application of Theorem \ref{theorem:alg_density} gives the following.
\begin{corollary}
    \label{corollary:electron_density_appendix}
    Given a position $\vecr$ in the atomic domain, let $\vecx$ be a vector $\vecx=\begin{pmatrix}
        \chi_1(\vecr) & \chi_2(\vecr) & \ldots & \chi_n(\vecr)
    \end{pmatrix},$
    where $\chi_i$ are the basis functions that were used to construct the Hamiltonian and overlap matrices $\matH,\matS$. Let $\matPtilde\leftarrow\DENSITY(\matH,\matS,\epsilon,k)$ where $k$ is the number of occupied states in the atomic system, and let $\matP$ be the true density matrix. Let $n(\vecr)=\vecx^*\matP\vecx$ be the true electron density at $\vecr$ and $\widetilde n(\vecr)=\vecx^*\matPtilde\vecx$. Then as long as $\DENSITY$ succeeds,
    \begin{align*}
        |n(\vecr)-\widetilde n(\vecr)| \leq 8\epsilon\|\vecx\|^2.
    \end{align*}
    \begin{proof}
        First we recall that $\epsilon\in(0,1/12)$ and note that $\|\matPtilde\|\leq \|\matP\|+\epsilon \leq 1+\epsilon.$ The quadratic form $ \vecx^*\matPtilde\vecx$ is computed in two steps: $\vecy\leftarrow\fl(\matPtilde\vecx)$ and $\widetilde n(\vecr)=\fl(\vecx^*\vecy)$. From \cite{higham2002accuracy}, Eq. (3.12) and Lemma 6.6, we know that the matrix-vector product satisfies $\vecy= \matPtilde\vecx +\vece$ where $\|\vece\|=\|\matPtilde\vecx-\vecy\|\leq \sqrt{n}\eta_n\|\matPtilde\|\|\vecx\|$, and  $\eta_n:=\frac{n\umach}{1-n\umach}$. For the chosen $\umach$ we have $\sqrt{n}\eta_n\leq2\epsilon$. From the same equations, the subsequent dot product satisfies $\widetilde n(\vecr)=\fl(\vecx^*\vecy)=\vecx^*\vecy+\epsilon $ where $
        |\epsilon|
        \leq \eta_n\|\vecx\|\|\vecy\|\leq \eta_n\|\vecx\|\left(
            \|\matPtilde\vecx\|+\|\vece\|
        \right)
        \leq
        \eta_n\|\matPtilde\|\|\vecx\|^2\left(
            1+\sqrt{n}\eta_n 
        \right)
        \leq
        4\epsilon\|\vecx\|^2
        $. Then $| n(\vecr)-\widetilde n(\vecr)| = |\vecx^*\matP\vecx - (\vecx^*\vecy+\epsilon)|
        =
        |\vecx^*\matP\vecx-\vecx^*\matPtilde\vecx-\vecx^*\vece-\epsilon|
        \leq
        |\vecx^*\matP\vecx-\vecx^*\matPtilde\vecx|
        +
        |\vecx^*\vece|
        +
        |\epsilon|
        \leq
        2\epsilon\|\vecx\|^2
        +
        2\epsilon \|\vecx\|^2
        +
        4\epsilon\|\vecx\|^2
        =
        8\epsilon\|\vecx\|^2.
        $
    \end{proof}
\end{corollary}
This can be generalized for many different points $\vecr$. If  $N$ is the number of those points and $N=\Theta(n)$ (which is typically the case in applications), then one can stack the vectors $\vecx(\vecr_1), \vecx(\vecr_2),\ldots,\vecx(\vecr_N)$ as the rows of a matrix $\matX$ and use fast matrix multiplication to compute all $n(\vecr_i)$ simultaneously, instead of querying each $n(\vecr_i)$ one-by-one.

If $N$ is asymptotically larger than the system size, i.e., $N=O(\poly(n))$, the query complexity can be reduced by applying the Johnson-Lindenstrauss (JL) lemma \cite{johnson1984extensions}, at the cost of a polynomial dependence on the accuracy. In brief, since the matrix $\matP\matS$ is an orthogonal projector, and therefore idempotent, it can be deduced that \begin{align*}
    \vecx^*\matP\vecx=\vecx^*\matP\matS\matP\vecx = \vecx^*\matP\matL^{-*}\matL^{-1}\matP\vecx = \|\vecx^*\matP\matL^{-*}\|^2,
\end{align*}
where $\matL$ is the lower triangular Cholesky factor of $\matS^{-1}$. Then $\|\vecx^*\matP\matL^{-*}\|^2$ can be approximated by $\|\vecx^*\matP\matL^{-*}\matT\|^2$, where $\matT$ is a $n\times r$ sparse random sign matrix scaled by $1/\sqrt{r}$ \cite{achlioptas2001database}, with $r=O(\log(N)/\epsilon^2)$ columns. The representation of $\matT$ requires $O(n\times r)=O(n\log(N)/\epsilon^2)$ random bits. Alternative random matrix distributions satisfying the JL property, with varying sparsity, runtime, and dimension requirements exist in the literature \cite{ailon2006approximate,kane2014sparser,nelson2013osnap}. 
The value of $r$ is worst-case optimal \cite{larsen2017optimality}, however, tighter bounds have been studied in the literature  \cite{bartal2011dimensionality,sobczyk2022approximate,baston2022stochastic,epperly2024xtrace}, thus potentially leading to more practical algorithms. The full analysis is omitted.

\section{Deflation and bit complexity of PCA}
\label{appendix:pca}

Since its introduction in the early twentieth century \cite{pearson1901liii,hotelling1933analysis}, Principal Component Analysis is one of the most important tools in statistics, data science, and machine learning. It can be used, for example, to visualize data, to reduce dimensionality, or to remove noise in measurements; cf. \cite{jolliffe2002principal,diamantaras1996principal} for reviews on the vast bibliography. In its simplest formulation, given a (centered) data matrix $\matX\in\mathbb{R}^{m\times n}$, the goal is to find a $k$-dimensional embedding $\matC_k$, where $k<n$, that maximizes the sample variance, which can be written as an optimization problem
\begin{align}
    \matC_k=\arg \max_{\matC^\top\matC=\matI_{k\times k}}\tr(\matC^\top\matH\matC),
\end{align}
where $\matH=\matX^\top\matX\in\mathbb{R}^{n\times n}$ is the sample covariance. It can be shown that the solution $\matC_k$ corresponds to the principal $k$ singular vectors of $\matH$, i.e. the ones that correspond to the largest $k$ singular values. Evidently, since the sample covariance is always symmetric and positive semi-definite, this can be written as a Hermitian eigenvalue problem
\begin{align*}
    \matH\matC=\matC\matLambda,
\end{align*}
(which is indeed a definite GEP as in Equation \eqref{eq:gep_main} with $\matS=\matI$). This way we can project the data in $k$ dimensions by computing $\matX\matC_k$, preserving as much of the variance in $k$ dimensions as possible. 
Classically, PCA can be solved by approximating the SVD of $\matH$ via diagonalization.

\subsection{Vanilla PCA}
\label{appendix:vanilla_pca}
With this short introduction, we can now describe how to enhance the analysis of PCA algorithms by using our results for spectral projectors to obtain forward-error guarantees for all the SVD-related computations. 
Using our Theorem \ref{theorem:spectral_projector}, the $\SIGMAK$ algorithm of Proposition \ref{proposition:alg_sigmak}, and the $\DEFLATE$ algorithm of \cite{banks2021gaussian}, as described in Algorithm \ref{alg:pca}, we can state the following result for the classic (``vanilla'') PCA, which provides forward error guarantees in matrix multiplication time regardless how large $k$ is. This is can serve as the backbone for more advanced algorithms that are analyzed in the next section, which typically first use random projections to reduce the matrix size to approximately $O(k\times k)$ and then perform the SVD computation on the smaller matrix.
\begin{remark}
    For simplicity, we have assumed that the covariance matrix contains no errors. This might not be true if $\matH$ is computed numerically from the data matrix $\matX$. It is not hard to extend the algorithm to take this into account as well, and nothing changes in the analysis (except some negligible constant factors). To handle errors in the input matrix one can adapt the analysis from the next Section \ref{appendix:low_rank_pca}.
\end{remark}

\begin{algorithm}[htb]
\centering
\caption{$\PCA$.}
\label{alg:pca}
\begin{algorithmic}[1]
\setstretch{0.8}
\Statex \begin{align*}
    \PCA
\end{align*}
\small
\Input Sample Covariance Matrix $\matH\in\mathbb{R}^{n\times n}$ of a centered data matrix $\matX\in\mathbb{R}^{m\times n}$, target rank $k$, accuracy $\epsilon\in(0,1)$.
\Require $\|\matH\|\leq 1$, $k\in[n-1]$.
\setstretch{1.5}
\Algorithm $\matCtilde_k \leftarrow\PCA(\matH,k,\epsilon)$.
\State $\widetilde\sigma_k\leftarrow \SIGMAK(\matH,k,\tfrac{1}{2},\tfrac{1}{3n})$
\State $\epsilon'\leftarrow \epsilon\widetilde\sigma_k/4$.
\State $\projectormatrixtilde_k\leftarrow \PROJECTOR(\matH,\matI,k,(\tfrac{\epsilon'^2}{20^3n^4})^2)$.
\State $\matCtilde_k\leftarrow \DEFLATE(\projectormatrixtilde_k,k,(\tfrac{\epsilon'^2}{20^3n^4})^2,\epsilon')$.
\State \Return $\matCtilde_k$.
\Output Approximate principal component matrix $\matCtilde_k$.
\Ensure $\|\matX-\matX\matCtilde_k\matCtilde_k^\top\|\leq (1+\epsilon)\|\matX-\matX\matC_k\matC_k^T\|$ with probability at least $1-O(1/n)$.
\end{algorithmic}
\end{algorithm}

\begin{theorem}[PCA]
    Let $\matH$ be an $n\times n$ symmetric sample covariance matrix of a centered data matrix $\matX\in\mathbb{R}^{m\times n}$, i.e. $\matH=\matX^\top\matX$, $\|\matH\|\leq 1$, $k\in[n]$ is a target rank, and $\epsilon\in(0,1)$ an accuracy parameter. 
    Then we can compute a matrix $\matCtilde_k$ with $k$ columns such that $\|\matX-\matX\matCtilde_k\matCtilde_k^\top\| \leq (1+\epsilon)\|\matX-\matX\matC_k\matC_k^\top\|$,
    where $\matC_k\in\mathbb{R}^{n\times k}$ contains the top-$k$ (right) singular vectors of $\matX$ in
    \begin{align*}
        O\lpar T_{\MM}(n) \lpar 
            \log(\tfrac{n}{\sigma_{k+1}})\log(\tfrac{1}{\sigma_{k+1}})
            +
            \log(\tfrac{n}{\gap_k})\log(\tfrac{1}{\gap_k})
            +
            \log(\log(\tfrac{n}{\epsilon\sigma_{k+1}\gap_k}))
        \rpar\rpar
    \end{align*}
    arithmetic operations using
    $
        O\lpar
            \log(n)
            \lpar
                \log^4(\tfrac{n}{\epsilon\gap_k})
                +
                \log^4(\tfrac{n}{\sigma_{k+1}})
            \rpar
            +
            \log(\tfrac{1}{\epsilon\sigma_{k+1}})
        \rpar
    $
    bits of precision, with probability at least $1-O(1/n)$.
    \begin{proof}
        We first compute 
        $\widetilde\sigma_{k+1}\leftarrow \SIGMAK(\matH,k+1,\tfrac{1}{2},\tfrac{1}{n})$ such that, from Proposition \ref{proposition:alg_sigmak}, $\widetilde\sigma_{k+1}\in(1\pm\tfrac{1}{2})\sigma_{k+1}(\matH)$ with probability $1-1/n$. It requires 
        \begin{align*}
            O\lpar
            T_{\MM}(n)\log(\tfrac{n}{\sigma_{k+1}})
            \log(\tfrac{1}{\sigma_{k+1}})
        \rpar
        \end{align*}
        floating point operations and $O\lpar
        \log(n)
        \log^4(\tfrac{n}{\sigma_{k+1}})
        \rpar$ bits.
        Then we set $\epsilon'\leftarrow \epsilon\widetilde\sigma_{k+1}/4$. In the next step, we compute        
        $\matPitilde_k\leftarrow \PROJECTOR(\matH,\matI,k,(\tfrac{\epsilon'^2}{20^3n^4})^2),$
        which, from Theorem \ref{theorem:spectral_projector}, returns a spectral projector $\projectormatrixtilde_k$ that satisfies $\|\matPitilde_k-\matPi_k\|\leq (\tfrac{\epsilon'^2}{20^3n^4})^2$ with probability $1-1/n$. 
        It requires \begin{align*}
            &O\lpar T_{\MM}(n) \lpar 
                    \log(\tfrac{n}{\gap_k})\log(\tfrac{1}{\gap_k})+\log(\log(\tfrac{n}{\epsilon'\gap_k}))
                \rpar
            \rpar\\
            &\qquad =
            O\lpar T_{\MM}(n) \lpar 
                    \log(\tfrac{n}{\gap_k})\log(\tfrac{1}{\gap_k})+\log(\log(\tfrac{n}{\epsilon\sigma_{k+1}\gap_k}))
                \rpar
            \rpar
        \end{align*}
        arithmetic operations and
        $
            O\lpar
                \log(n)
                    \log^4(\tfrac{n}{\epsilon'\gap_k}) 
            \rpar
            =
            O\lpar
                \log(n)
                    \log^4(\tfrac{n}{\epsilon\sigma_{k+1}\gap_k}) 
            \rpar
        $
        bits. Then we use $\DEFLATE$ to compute the matrix $\matC'_k\leftarrow \DEFLATE(\matPitilde_k,k,(\tfrac{\epsilon'^2}{20^3n^4})^2,\epsilon')$. From Theorem \ref{theorem:deflate}, $\DEFLATE$ succeeds with probability 
        \begin{align*}
            1-\tfrac{(20n)^3\sqrt{(\epsilon'^2/(20^3n^4))^2}}{\epsilon'^2}
            =
            1-1/n,
        \end{align*} and it requires $O(T_{\MM}(n))$ floating point operations using $O(\log(n/\epsilon'))=O(\log(n/(\sigma_{k+1}\epsilon)))$ bits, and it internally generates $O(n^2)$ random complex normal variables. On success it returns $\matC_k'\in\mathbb{C}^{n\times k}$ such that $\|\matC_k-\matC'_k\|\leq \epsilon'$, where $\matC_k$ is a matrix whose columns form an orthonormal basis for $\matPi_k$. We can then keep only the real part of $\matC'_k$, and set $\matCtilde_k\leftarrow Re(\matC'_k)$. The spectral norm can only decrease by removing the imaginary part, therefore $\|\matCtilde_k-\matC_k\|\leq \epsilon'$ holds as well. Now we can write $\matCtilde_k=\matC_k+\matE$ where $\|\matE\|\leq \epsilon'$, which means that
        \begin{align*}
            \lnorm \matX-\matX\matCtilde_k\matCtilde_k^\top\rnorm
            &=
            \lnorm \matX-\matX(\matC_k+\matE)(\matC_k+\matE)^\top \rnorm
            \\
            &\leq
            \lnorm \matX-\matX\matC_k\matC_k^\top\rnorm
            +
            \lnorm \matX\matE\matC_k^\top \rnorm
            +
            \lnorm \matX\matC_k\matE^\top  \rnorm
            +
            \lnorm \matX\matE\matE^\top \rnorm
            \\
            &\leq
            \lnorm \matX-\matX\matC_k\matC_k^\top\rnorm
            +
            2\epsilon'\|\matX\|
            +
            \epsilon'^2\|\matX\|
            \\
            &\leq
            \lnorm \matX-\matC_k\matC_k^\top\matX\rnorm
            +
            3\frac{\epsilon\widetilde\sigma_{k+1}}{4}
            \\
            &\leq
            (1+\tfrac{9}{8}\epsilon)\lnorm \matX-\matX\matC_k\matC_k^\top \rnorm,
        \end{align*}
        where in the last we used the fact that 
        \begin{align*}
            \widetilde\sigma_{k+1}(\matH)\leq 3\sigma_{k+1}(\matH)/2=3\sigma_{k+1}^2(\matX)/2\leq 3\sigma_{k+1}(\matX)/2=3\|\matX-\matX\matC_k\matC_k^\top\|/2.
        \end{align*}
        The result follows by rescaling $\epsilon$ by a constant and by summing together the individual algorithm complexities and bits.
    \end{proof}
\end{theorem}

\subsection{Low-rank PCA}
\label{appendix:low_rank_pca}
In many applications of PCA the parameter $k$ is chosen to be small, i.e. $k\ll d$. This has led to an extensive research area on the so-called \textit{low-rank approximation} algorithms for PCA. 
One of the earliest works which introduced this type of randomized low-rank approximations is \cite{frieze1998fast,frieze2004fast}. Some other landmark works in the field include the  analysis of randomized PCA and low-rank approximation \cite{martinsson2011randomized,halko2011finding,sarlos2006improved,clarkson2009numerical,clarkson2017low}, and the pioneering Block-Krylov PCA of \cite{musco2015randomized}, which is essentially optimal in the matrix-vector query model \cite{simchowitz2018tight}. 
The approximation accuracy of low-rank approximation-based PCA methods is often measured with respect to the spectral or the Frobenius norm error, i.e. the matrix $\matCtilde_k$ that is returned should satisfy $\matC_k^\top\matC_k=\matI_{k\times k}$ and:
\begin{align}
    \lnorm \matX - \matX\matCtilde_k\matCtilde_k^\top\rnorm_{\{2,F\}}
    \leq
    (1+\epsilon)
    \lnorm \|\matX-\matX\matC_k\matC_k^\top
    \rnorm_{\{2,F\}}.
    \label{eq:low_rank_approx_main_bound}
\end{align}

Most of the aforementioned low-rank approximation type algorithms assume exact arithmetic. A subtle detail is that they often rely on the computation of an (exact) SVD of some smaller submatrix, which is not realistic due to the Abel-Ruffini theorem. This can be justified for practical reasons since SVD can be approximated in polynomial time to arbitrary accuracy using classical solvers \cite{golub2013matrix}. Quoting \cite{halko2011finding}:
\begin{quote}\textit{``Techniques for computing the SVD are iterative by necessity, but they converge so fast that we can treat them as finite for practical purposes.''}
\end{quote}
However, if one wants to rigorously prove forward error approximations and end-to-end complexity upper bounds, there is necessarily a dependence on the singular value gap that separates the principal invariant subspace from the rest of the spectrum (consider a small perturbation of the $2$-by-$2$ identity matrix and $k=1$ as a straightforward example). 

Assuming an exact SVD algorithm, the seminal analysis of the Block-Krylov PCA of \cite{musco2015randomized} can in fact provide per-singular vector guarantees, which are much stronger than the classical norm-wise bounds. 
Then arithmetic complexity of Block-Krylov PCA depends on $\poly(\tfrac{k}{\sqrt{\epsilon}})$.
It is favorable for coarse accuracy $\epsilon$ (i.e. when $\epsilon=\Theta(1)$) and small rank $k$. It is not suitable, however, for larger $k$. E.g., it can be the case in applications where we need to keep $k=n/20$ of the original dimensions. The same holds when higher accuracy is required, e.g. when $\epsilon=1/\poly(n)$. Then the complexity of these methods is already higher than standard eigensolvers.

In finite precision, the landscape is even less clear. At the time of this submission, the only  work 
related to low-rank approximation PCA that we are aware of with end-to-end bit complexity bounds is \cite{musco2018stability} for the approximation of matrix functions applied on vectors. They prove that the Lanczos method can be stably applied to compute a vector $\vecu$ such that the quantity $\|\matA\vecu\|$ approximates $\|\matA\|$. This is a backward-approximate solution for the top-1 singular vector. Concurrently with this work, and independently, \cite{kacham2024faster} analyzed the bit complexity of Block-Krylov PCA and achieved similar bounds as ours, albeit with different techniques. However, \cite{kacham2024faster} did not describe how to compute the condition number which is required in order to adjust the machine precision. Our analysis covers the computation of all the involved parameters.

For general invariant subspaces $(k>1)$, consider one of the simplest randomized low-rank approximation algorithms, often referred to as ``subspace iteration'' \cite{halko2011finding} or ``simultaneous iteration'' \cite{musco2015randomized}. The algorithm first samples a matrix $\matG\in\mathbb{R}^{n\times l}$ with i.i.d standard normal elements, where $l=\Theta(k)$, and computes $\matY\leftarrow (\matA\matA^T)^q\matA\matG$, where typically $q\approx \log(\min\{m,n\})$. It then returns the $\matQ$ factor from the economy QR factorization of $\matY$. It can be shown (see e.g. Corollary 10.10 of \cite{halko2011finding}) that
\begin{align*}
    \|(\matI-\matQ\matQ^\top)\matA\| \lesssim \|\matA-\matA_k\|.
\end{align*}
The returned matrix has slightly more than $k$ columns, often referred to as ``oversampling,'' but, importantly, there is no explicit SVD involved in the computation. Even in this case, floating point arithmetic already spoils the approximation guarantees: we cannot compute $\matQ$ without rounding errors. 
Stability analysis of QR factorization is typically carried out in the backward-error sense \cite{higham2002accuracy,demmel2007fastla}, which makes the analysis of low-rank approximation PCA algorithms even more complicated.

\subsection{Bit complexity analysis of Block-Krylov PCA}
\label{appendix:block_krylov_pca}
We now analyze the seminal Block-Krylov iteration algorithm of \cite{musco2015randomized}, which is listed in Algorithm \ref{alg:bkpca} for convenience (in exact arithmetic). For the floating point analysis the following methodology is used. The main result is stated in Theorem \ref{theorem:block_krylov_pca}. We shall denote by $T_{\MM}(\matX,q)$ the cost of multiplying $\matX$ with a dense matrix with $q$ columns with a numerically stable multiplication algorithm, like the one of Theorem \ref{theorem:fast_mm}. Formally, for the rest of this section, we assume the following subroutine. 
\begin{definition}[$\MMX$]
\label{def:stable_mmx}
Let $\matX\in\mathbb{R}^{m\times n}$ be the input matrix of Algorithm \ref{alg:bkpca}. We assume a subroutine $\MMX(\matB)$, which takes as input a matrix $\matB$ with $k$ columns. It returns a matrix $\matC\in\mathbb{R}^{m\times r}$ such that
\begin{align*}
    \lnorm \matC - \matX\matB \rnorm \leq \umach \poly(m,k) \|\matX\|\|\matB\|,
\end{align*}    
in a total of $T_{\MMX}(k)$ floating point operations, using at most $O(\log(1/\umach))$ bits of precision. We also assume $\MMX^\top(\matB)$ which approximates the product $\matX^\top \matB$ with the same cost and approximation bounds.
\end{definition}
The standard inner-product based algorithm as well as a block variant of the $\MM$ algorithm of Theorem \ref{theorem:fast_mm} satisfy this definition.

\begin{algorithm}[htb]
\centering
\caption{Block-Krylov Iteration (Alg. 2 of \cite{musco2015randomized})}
\label{alg:bkpca}
\begin{algorithmic}[1]
\setstretch{0.8}
\Statex \begin{align*}
    \text{Block-Krylov Iteration (exact arithmetic)}
\end{align*}
\small
\Input Data matrix $\matX\in\mathbb{R}^{m\times n}$, target rank $k$, accuracy $\epsilon\in(0,1)$.
\Require Exact arithmetic.
\setstretch{1.5}
\Algorithm $\matZ_k \leftarrow$ Block-Krylov Iteration ($\matX,k,\epsilon)$.
\State $q\leftarrow \Theta(\tfrac{\log(n)}{\sqrt{\epsilon}}).$
\State $\matG\leftarrow \mathcal{N}(0,1)^{n\times k}.$
\State $\matK \leftarrow \lbrac \matX\matG,(\matX\matX^\top)\matX\matG,\ldots,(\matX\matX^\top)^q\matX\matG \rbrac$.
\State $\matQ,\matR\leftarrow$ Economy-QR$(\matK)$. \Comment{$\matQ\in\mathbb{R}^{m\times qk}.$}
\State $\matM\leftarrow \matQ^\top\matX\matX^\top\matQ\in\mathbb{R}^{qk\times qk}$.
\State $\matUbar_k\leftarrow$ top-$k$ singular vectors of $\matM$.
\State \Return $\matZ_k\leftarrow \matQ\matUbar_k$.
\Output Approximate principal component matrix $\matZ$.
\Ensure $\|\matX-\matZ_k\matZ_k^\top\matX\|\leq (1+\epsilon)\|\matX-\matX_k\|$ with high probability (in exact arithmetic).
\end{algorithmic}
\end{algorithm}

\paragraph{Step 1: Constructing the Block-Krylov matrix.}
We first observe that we can scale the matrix $\matG$ in Algorithm \ref{alg:bkpca} to have norm at most one, since this does not affect the Krylov basis.
We assume that the (scaled) Gaussian matrix $\matG$ in Algorithm \ref{alg:bkpca} is given exactly. This is not realistic, but it greatly simplifies the analysis and it can be addressed by using the floating point Gaussian sampler of Definition \ref{definition:stable_normal_sampler} to obtain a small error with high probability. 

Given $\matG$ and $\matX$, we can construct the Krylov matrix $\matK$ using $O(q)$ calls to $\MMX(\matG)$. The first call $\matX_1\leftarrow \MMX(\matG)$ returns a matrix $\matX_1=\matX\matG+\matE_1$ where $\|\matE_1\|\leq \umach \cdot O(\poly(mk))\|\matX\|\|\matG\|\leq \umach\cdot  O(\poly(mk))$, since we assumed constant norms. If we keep performing multiplications recursively to build the Krylov matrix $\|\matKtilde\|$, after $O(q)$ multiplications it holds that $\|\matKtilde-\matK\|\leq \umach \cdot O(\poly(mkq))$. 
The total cost is $O(qT_{\MMX}(k))$ floating point operations. If $\matX$ is sparse and we use the standard inner product-based algorithm, $T_{\MMX}(k)=O(k{\tt nnz}(\matX))$. If $\matX$ is dense and we use Theorem \ref{theorem:fast_mm} then $O(qmk^{\omega-2})$ floating point operations are sufficient.

\paragraph{Step 2: Condition number of the Block-Krylov matrix.}
So far we have approximated $\matK$ by $\matKtilde=\matK+\matE_{\matK}$. 

In the following steps we will need an approximation for the condition number of $\matK$. We can get such an approximation using a variant of the $\COND$ algorithm (Corollary \ref{corollary:alg_cond}), which internally uses the $\SIGMAK$ algorithm of Proposition \ref{proposition:alg_sigmak} based on counting queries to approximate the condition number. In that case, we start with some arbitrary error $\|\matE_{\matK}\|$, and we keep keep halving it until we have a sufficiently good approximation. 
At each iteration $t$, the algorithm executes 
\begin{align*}
    O\lpar qT_{\MMX}(k))\rpar
\end{align*} floating point operations to construct the Krylov matrix using $O(\polylog(m,q,\tfrac{1}{\epsilon_t})$ bits. At most $O(\log(\kappa(\matK)))$ iterations are required. The most expensive iteration is the last one, when $\epsilon_t\approx 1/\kappa(\matK)$, which gives a total cost of at most
\begin{align*}
    O\lpar q\log(\kappa(\matK))T_{\MMX}(k) \rpar    
\end{align*}
floating point operations using at most $O(\polylog(m,q,\kappa(\matK)))$ bits.
It returns a value $\widetilde\kappa\in\Theta(\kappa(\matK))$, and it succeeds with high probability.

\paragraph{Step 3: Computing a basis.}
The computation of the basis in Step 4 of Algorithm \ref{alg:bkpca} is arguably the hardest part in the analysis of the Block-Krylov iteration.
With similar arguments as in the proof of Theorem \ref{theorem:fast_qr}, if we write the economy-QR factorizations $\matKtilde=\matQ_{\matKtilde}\matR_{\matKtilde}$ and $\matK=\matQ\matR$, there exists an orthogonal matrix $\matPhi_1$ such that $\matQ-\matQ_{\matKtilde}\matPhi_1=\matE_{1}$ and $\|\matE_{1}\|\leq \epsilon'$, for some $\epsilon'$. 
This requires that $\|\matKtilde-\matK\|\leq \epsilon'\frac{1}{4}\frac{1}{\poly(mnqk)\kappa(\matK)}$, which implies a requirement
\begin{align*}
  \umach\leq \epsilon'\frac{c}{\poly(mq)\kappa(\matK)},
\end{align*}
for some constant $c$. Instead of $\kappa(\matK)$ we can use $\widetilde\kappa$ from the previous Step 2.
We can now use Theorem \ref{theorem:fast_qr} (which computes a basis via QR) on $\matKtilde$ to approximate $\matQ_{\matKtilde}$ by the matrix
\begin{align*}
\matQtilde
=
\matQ_{\matKtilde}\matPhi_2+\matE_{\QR}
=
(\matQ\matPhi_1+\matE_{1})\matPhi_2+\matE_{\QR}
=
\matQ\matPhi_1\matPhi_2 + \matE_{1}\matPhi_2+\matE_{\QR}
.
\end{align*}
Let $\matE_{\matQ}=\matE_1\matPhi_2+\matE_{\QR}$.
We have that $\|\matE_{\matQ}\|=\|\matE_{1}\matPhi_2+\matE_{\QR}\|\leq \epsilon'+\matE_{\QR}.$ From Theorem \ref{theorem:fast_qr} we can ensure that $\|\matE_{\QR}\|\leq \epsilon'$ if we use $O(\log(\tfrac{mq\kappa(\matKtilde)}{\epsilon'}))$ bits.
Thus, $\matQtilde$ is an approximate orthogonal basis for the range of the true Block-Krylov matrix $\matK$. 

To summarize, as long as 
$\umach\leq \epsilon'\frac{c}{\poly(mq)\kappa(\matK)}$, 
which can be achieved by replacing $\kappa(\matK)$ with $\widetilde\kappa$ above, then we can compute $\matQtilde=\matQ\matPhi+\matE_{\matQ}$, where $\matQ$ is the true orthonormal basis from the QR factorization of $\matK$, $\matPhi$ is a $qk\times qk$ orthogonal matrix, and $\|\matE_{\matQ}\|\leq \epsilon'$. 
The total cost is $O(m(qk)^{\omega-1})$ floating point operations for Theorem \ref{theorem:fast_qr}, using $O\lpar \log(\tfrac{mq\kappa(\matK)}{\epsilon'}) \rpar$ bits.

\paragraph{Step 4: Computing the reduced matrix.}
In Step 5, Algorithm \ref{alg:bkpca} forms the matrix $\matM$ to compute its top-$k$ singular vectors. To analyze the computation of $\matM$, the first observation is that $\matQ$ in lines 4-7 can be replaced by any basis for the column space of $\matK$. In particular, we replace it by $\matQtilde$. We then perform the multiplication in two steps: $\matMtilde_1^\top=\MMX^\top(\matQtilde)$, which returns $\matMtilde_1=\matQtilde^\top\matX+\matE_1^{\MM}$, where $\|\matE_1^{\MM}\|
\leq 
\umach\cdot O(\poly(qknm))\|\matQtilde\|\|\matX\|
\in
\umach \cdot O(\poly(qm))
$.
Then we compute $\matMtilde \leftarrow \MM(\matMtilde_1,\matMtilde_1^\top)=\matMtilde_1\matMtilde_1^\top+\matE_2^{\MM}$
where $\|\matE_2^{\MM}\| \leq \umach\cdot O(\poly(qkn)) \|\matM_1\|^2\in\umach\cdot  O(\poly(qkn))$, where the last is implied if $\umach$ is sufficiently smaller than $1/\poly(qm)$. 
Putting everything together we can write 
\begin{align*}
    \matMtilde
    =\matPhi_2^\top\matPhi_1^\top\matQ^\top\matX\matX^\top\matQ\matPhi_1\matPhi_2+\matE_{\matM},
\end{align*} where $\matE_{\matM}$ contains the errors in $\matQtilde$ and also the ones from the multiplication in floating point, and $\|\matE_{\matM}\|\leq O(\|\matE_{\matQ}\|)\in O(\epsilon')$. $\epsilon'$ is as in Step 3. So far we have assumed $O(\log(\tfrac{mq\kappa(\matKtilde)}{\epsilon'}))$ bits.

\paragraph{Step 5: Spectral gap and midpoint.}
Now we have written $\matMtilde=\matPhi^\top\matQ^\top\matX\matX^\top\matQ\matPhi+\matE_{\matM}$, where $\matPhi=\matPhi_2\matPhi_1$ is orthogonal and $\|\matE_{\matM}\|\leq O(\epsilon')$ for some $\epsilon'\in(0,1)$. We can use the counting queries in the spirit of Theorem \ref{theorem:alg_spectral_gap} to compute the spectral gap and the midpoint of $\matM$.

In particular, we start with some desired bound $\epsilon'_0\in(0,1)$ for $\|\matE_{\matM}\|$ by setting the number of bits to 
$O(\log(\tfrac{qm\kappa(\matKtilde)}{\epsilon'_0}))$. 
At each iteration we repeat Steps 1, 3, and 4 (we do not need to compute again the condition number of $\matK$ in Step 2). 
After at most $m=O(\log(1/\gap_k(\matM)))$ iterations, the error $\epsilon'_m$ satisfies $\epsilon'_m=\Theta(\gap_k(\matM))$, and we obtain two quantities $\widetilde\gap_k\in (1\pm\tfrac{1}{8})\gap_k(\matM)$ and $\widetilde\mu_k\in \mu_k(\matM)+\tfrac{1}{8}\gap_k(\matM)$ with high probability. 
In each iteration we need to construct $\matMtilde$ using Steps 1, 3, and 4 with the specified number of bits. 

Step 1 costs 
$O(qT_{\MMX}(k))$ operations. 
Step 3 requires $O(m(qk)^{\omega-1})$ operations.
Step 4 executes $O(T_{\MMX}(qk) + m(qk)^{\omega-1})$ operations.
The result is the $qk\times qk$ matrix $\matMtilde$ on which we call $\COUNT$ on each iteration. From Lemma \ref{lemma:eigenvalue_count_less_than_h}, assuming that we have regularized $\matMtilde$ appropriately using $\REGULARIZE$, $\COUNT$ costs at most $O((qk)^\omega\polylog(qk/\epsilon'_m))=O((qk)^\omega\polylog(qk/\gap_k(\matM)))$.
The maximum number of bits is in the last iteration, which is equal to $O\lpar
    \polylog\lpar
        \tfrac{mq\kappa(\matK)}{\gap_k(\matM)}
    \rpar
\rpar$.

\paragraph{Step 6: Principal singular vectors.}

Given a suitable approximation for the gap and the midpoint, we next use Lemma \ref{lemma:sign_function_under_perturbation} to prove forward error bounds between the true $k$-spectral projector of $\matM=\matPhi^\top\matQ^\top\matX\matX^\top\matQ\matPhi$ and the one of $\matMtilde=\matM+\matE_{\matM}$. From Lemma \ref{lemma:sign_function_under_perturbation}, if we set $\mu=\widetilde\mu_k$, we have that $\|\sgn(\matM-\widetilde\mu_k)-\sgn(\matMtilde-\widetilde\mu_k)\|\leq\epsilon_{\SGN}$ if $\|\matE_{\matM}\|\leq \epsilon_{\SGN}\tfrac{|\lambda_{\min}(\widetilde\mu_k-\matM)|^2\pi}{128}=\epsilon_{\SGN}\Theta(\gap_k(\matM)^2)$. The same holds for the spectral projectors, i.e. $\|\matPi_k(\matM)-\matPi_k(\matMtilde)\|\leq \epsilon_{\SGN}$. The bound for $\|\matE_{\matM}\|$ is achieved by setting $\umach\leq \frac{\epsilon_{\SGN}\widetilde\gap_k^2}{\poly(mq)\widetilde\kappa}$, which translates to $O\lpar \log(\tfrac{mq\kappa(\matK)}{\epsilon_{\SGN}\widetilde\gap_k})  \rpar$ bits.

It remains to approximate $\matPi_k(\matMtilde)$, denoted as $\matPi_k$ for simplicity, and then use deflation, similar to the vanilla PCA Algorithm \ref{alg:pca}.
We first set $\epsilon_{\PURIFY}=(\frac{\epsilon_{\PCA}^2}{20^3(qk)^4})^2$, and then use Algorithm \ref{alg:purify}, 
\begin{align*}
    \projectormatrixtilde_k\leftarrow \PURIFY(\matMtilde,\matI,\widetilde\mu_k,\widetilde\gap_k,\epsilon_{\PURIFY}),
\end{align*}
followed by  $\matUtilde_k\leftarrow \DEFLATE(\projectormatrixtilde_k,k,\epsilon_{\PURIFY},\epsilon_{\PCA})$. From Proposition \ref{proposition:alg_purify}, $\PURIFY$ costs $O((qk)^\omega\log(qk/\gap_k))$ floating point operations using  \begin{align*}
    O(\polylog(\tfrac{qk}{\epsilon_{\PURIFY}\gap_k}))
    =
    O(\polylog(\tfrac{qk}{\epsilon_{\PCA}\gap_k}))
\end{align*} bits. 
From Theorem \ref{theorem:deflate}, $\DEFLATE$ costs $O((qk)^\omega)$ operations using $O(\polylog(qk/\epsilon_{\PCA}))$ bits, and succeeds with high probability. The returned matrix $\matUtilde_k$ satisfies $\|\matUtilde_k-\matU_k\|\leq \epsilon_{\PCA}$, where $\matU_k$ is a matrix whose columns form an orthonormal basis for the span of the top-$k$ singular vectors of $\matMtilde$.
This also implies that $\|\matUtilde_k\matUtilde_k^\top-\matU_k\matU_k^\top\|\leq 3\epsilon_{\PCA}$.
Then 
\begin{align*}
    \|\projectormatrix_k(\matM)-\matUtilde_k\matUtilde_k^\top\| 
    &\leq 
    \|\projectormatrix_k(\matM)-\projectormatrix_k(\matMtilde)\|
    +
    \|\projectormatrix_k(\matMtilde) -\matUtilde_k\matUtilde_k^\top\|
    \leq 
    \epsilon_{\SGN} + 3\epsilon_{\PCA}.
\end{align*}

We finally return $\matZtilde_k=\MM(\matQtilde,\matUtilde_k)$. It holds that  where $\|\matZtilde_k-\matQtilde\matUtilde_k\|\leq \umach\cdot O(\poly(mq))$. We can then write 
\begin{align*}
    \matZtilde_k\matZtilde_k^\top &= \matQtilde\matUtilde_k\matUtilde_k^\top\matQtilde^\top + \matE_{\matZ}
    \\
    &=
    \matQtilde(\projectormatrix_k(\matM)+\matE_{\projectormatrix})\matQtilde^\top + \matE_{\matZ}
    \\
    &=
    \matQtilde\projectormatrix_k(\matM)\matQtilde^\top+\matQtilde\matE_{\projectormatrix}\matQtilde^\top + \matE_{\matZ}
    \\
    &=
    (\matQ\matPhi+\matE_{\matQ})\projectormatrix_k(\matM)(\matQ\matPhi+\matE_{\matQ})^\top+\matQtilde\matE_{\projectormatrix}\matQtilde^\top + \matE_{\matZ}
    \\
    &=
    \matQ\matPhi\projectormatrix_k(\matM)\matPhi^\top \matQ^\top+\matE'+\matQtilde\matE_{\projectormatrix}\matQtilde^\top + \matE_{\matZ},
\end{align*}
where $\matQ\matPhi\projectormatrix_k(\matM)\matPhi^\top\matQ^\top=\matZ_k\matZ_k^\top$ is equivalent to the true, exact arithmetic $\matZ_k$ of Algorithm \ref{alg:bkpca}. The error matrices satisfy $\|\matE_{\matZ}\|\leq \umach\cdot O(\poly(mq))$, $\|\matQtilde\matE_{\projectormatrix}\matQtilde^\top\| \leq O(\epsilon_{\SGN}+\epsilon_{\PCA})$, and $\|\matE'\|\leq O(\|\matE_{\matQ}\|)\in O(\|\matE_{\matM}\|) \in O(\epsilon_{\SGN}\gap_k(\matM)^2)$.

\paragraph{Putting everything together.}
We can now state the main result by summarizing the Steps 1-6.
\begin{theorem}[Restatement of Theorem \ref{theorem:block_krylov_pca}]
    \label{theorem:block_krylov_pca_appendix}
        Let $\matX$ be a data matrix $\matX\in\mathbb{R}^{m\times n}$, $\|\matX\|\leq 1$, $k\in[n]$ a target rank, $\epsilon_{\PCA}\in(0,1)$ an accuracy parameter, and $q=\Theta\lpar \frac{\log(n)}{\sqrt{\epsilon_{\PCA}}}\rpar$. Let $T_{\MMX}(k)$ denote the complexity to stably multiply $\matX$ or $\matX^\top$ with a dense matrix with $k$ columns from the right (see Def. \ref{def:stable_mmx}).
        Using the Steps 1-6 that are detailed in Appendix \ref{appendix:block_krylov_pca} as a floating point implementation of Algorithm \ref{alg:bkpca}, we can compute a matrix $\matZtilde_k\in\mathbb{R}^{m\times k}$ that satisfies \begin{align*}
            \lnorm \matZtilde_k\matZtilde_k^\top -\matZ_k\matZ_k^\top \rnorm \leq O(\epsilon_{\PCA}),
        \end{align*}
        with high probability, where $\matZ_k$ is an approximate basis for the top-$k$ principal components of $\matX$, returned by Algorithm \ref{alg:bkpca} in exact arithmetic. The total cost is at most
        \begin{align*}
            O\lpar 
                qT_{\MMX}(k)
                \log(\tfrac{\kappa(\matK)}{\gap_k(\matM)})
                + 
                m(qk)^{\omega-1} 
                \log(\tfrac{1}{\gap_k(\matM)})
                +
                (qk)^\omega\polylog(\tfrac{qk}{\gap_k(\matM)})
            \rpar
        \end{align*}
        floating point operations, using
        $O\lpar 
            \polylog(\tfrac{mq\kappa(\matK)}{\epsilon_{\PCA}\gap_k})
        \rpar$
        bits of precision. $\matK,\matM$ are the same as in Alg. \ref{alg:bkpca}. The only parameters that are initially required are $k$ and $\epsilon_{\PCA}$.
    \begin{proof}
        We first compute $\widetilde\kappa\in\Theta(\kappa(\matK))$ in Steps 1 and 2. This costs 
        \begin{align*}
            O\lpar q\log(\kappa(\matK))T_{\MMX}(k) \rpar 
        \end{align*} floating point operations using $O(\polylog(qm\kappa(\matK)))$ bits.

        Then, in Step 5 we iteratively use Steps 1, 3, and 4 to compute the midpoint and the gap, in a total of $O(\log(1/\gap_k))$ iterations. The total cost is
        \begin{align*}
            O\lpar 
                \lpar 
                    qT_{\MMX}(k) + m(qk)^{\omega-1} 
                \rpar
                \log(\tfrac{1}{\gap_k(\matM)})
                +
                (qk)^\omega\polylog(\tfrac{qk}{\gap_k(\matM)})
            \rpar
        \end{align*}
        floating point operations using at most
        \begin{align*}
            O\lpar
                \polylog(\tfrac{mq\kappa(\matK)}{\gap_k(\matM)})
            \rpar
        \end{align*}
        bits of precision. The only parameter that we require to know beforehand, except for the matrix sizes, is $\widetilde\kappa$ from the previous step.

        In the last Step 6, assuming that $\epsilon_{\SGN}=\epsilon_{\PCA}$, we need to run Steps 1, 3, and 4, using $O\lpar 
            \polylog(\tfrac{mq\kappa(\matK)}{\epsilon_{\PCA}\gap_k})
        \rpar$
        bits. The number of arithmetic operations does not exceed the one from the previous step. Thereafter the costs of $\DEFLATE,$ $\PURIFY$, and $\MM$ are negligible compared to the costs of Steps 1, 3, and 4. The final matrix $\matZtilde_k$ satisfies
        \begin{align*}
            \lnorm \matZtilde_k\matZtilde_k^\top -\matZ_k\matZ_k^\top \rnorm 
            \leq 
            O(\epsilon_{\PCA}),
        \end{align*}
        where $\matZ_k$ is the true, exact arithmetic projector in Algorithm \ref{alg:bkpca}.

        Summarizing everything, we can compute $\matZtilde_k$ as advertised above in a total of at most
        \begin{align*}
            O\lpar 
                qT_{\MMX}(k)
                \log(\tfrac{\kappa(\matK)}{\gap_k(\matM)})
                + 
                m(qk)^{\omega-1} 
                \log(\tfrac{1}{\gap_k(\matM)})
                +
                (qk)^\omega\polylog(\tfrac{qk}{\gap_k(\matM)})
            \rpar
        \end{align*}
        floating point operations, using
        $O\lpar 
            \polylog(\tfrac{mq\kappa(\matK)}{\epsilon_{\PCA}\gap_k})
        \rpar$
        bits of precision.
    \end{proof}
\end{theorem}


\begin{thebibliography}{100}

\bibitem{achlioptas2001database}
Dimitris Achlioptas.
\newblock Database-friendly random projections.
\newblock In {\em Proc. Twentieth ACM SIGMOD-SIGACT-SIGART Symposium on Principles of Database Systems}, pages 274--281, 2001.

\bibitem{adachi2017solving}
Satoru Adachi, Satoru Iwata, Yuji Nakatsukasa, and Akiko Takeda.
\newblock Solving the trust-region subproblem by a generalized eigenvalue problem.
\newblock {\em SIAM Journal on Optimization}, 27(1):269--291, 2017.

\bibitem{ailon2006approximate}
Nir Ailon and Bernard Chazelle.
\newblock Approximate nearest neighbors and the fast {Johnson-Lindenstrauss} transform.
\newblock In {\em Proc. Thirty-Eighth Annual ACM Symposium on Theory of Computing}, pages 557--563, 2006.

\bibitem{aizenman2017matrix}
Michael Aizenman, Ron Peled, Jeffrey Schenker, Mira Shamis, and Sasha Sodin.
\newblock Matrix regularizing effects of {G}aussian perturbations.
\newblock {\em Communications in Contemporary Mathematics}, 19(03):1750028, 2017.

\bibitem{allen2016lazysvd}
Zeyuan Allen-Zhu and Yuanzhi Li.
\newblock {LazySVD: Even faster SVD decomposition yet without agonizing pain}.
\newblock {\em Advances in Neural Information Processing Systems}, 29, 2016.

\bibitem{alman2024more}
Josh Alman, Ran Duan, Virginia~Vassilevska Williams, Yinzhan Xu, Zixuan Xu, and Renfei Zhou.
\newblock More asymmetry yields faster matrix multiplication.
\newblock {\em arXiv preprint arXiv:2404.16349}, 2024.

\bibitem{armentano2018stable}
Diego Armentano, Carlos Beltr{\'a}n, Peter B{\"u}rgisser, Felipe Cucker, and Michael Shub.
\newblock A stable, polynomial-time algorithm for the eigenpair problem.
\newblock {\em Journal of the European Mathematical Society}, 20(6):1375--1437, 2018.

\bibitem{bai1991direct}
Zhaojun Bai and James Demmel.
\newblock {\em On a direct algorithm for computing invariant subspaces with specified eigenvalues}.
\newblock University of Tennessee, Computer Science Department, 1991.

\bibitem{bai1998using}
Zhaojun Bai and James Demmel.
\newblock Using the matrix sign function to compute invariant subspaces.
\newblock {\em SIAM Journal on Matrix Analysis and Applications}, 19(1):205--225, 1998.

\bibitem{bai1997inverse}
Zhaojun Bai, James Demmel, and Ming Gu.
\newblock An inverse free parallel spectral divide and conquer algorithm for nonsymmetric eigenproblems.
\newblock {\em Numerische Mathematik}, 76(3):279--308, 1997.

\bibitem{bakir2004learning}
G{\"o}khan~H Bak{\i}r, Jason Weston, and Bernhard Sch{\"o}lkopf.
\newblock Learning to find pre-images.
\newblock {\em Advances in Neural Information Processing Systems}, 16:449--456, 2004.

\bibitem{baldi1989neural}
Pierre Baldi and Kurt Hornik.
\newblock Neural networks and principal component analysis: Learning from examples without local minima.
\newblock {\em Neural networks}, 2(1):53--58, 1989.

\bibitem{ballard2011minimizingSVD}
Grey Ballard, James Demmel, and Ioana Dumitriu.
\newblock {Minimizing Communication for Eigenproblems and the Singular Value Decomposition}.
\newblock {\em Technical Report UCB/EECS-2011-14}, February 2011.

\bibitem{ballard2011minimizingLA}
Grey Ballard, James Demmel, Olga Holtz, and Oded Schwartz.
\newblock {Minimizing Communication in Numerical Linear Algebra}.
\newblock {\em SIAM Journal on Matrix Analysis and Applications}, 32(3):866--901, 2011.

\bibitem{banks2022pseudospectral}
Jess Banks, Jorge Garza-Vargas, Archit Kulkarni, and Nikhil Srivastava.
\newblock Pseudospectral {S}hattering, the {S}ign {F}unction, and {D}iagonalization in {N}early {M}atrix {M}ultiplication {T}ime.
\newblock {\em Foundations of Computational Mathematics}, pages 1--89, 2022.

\bibitem{banks2022global2}
Jess Banks, Jorge Garza-Vargas, and Nikhil Srivastava.
\newblock Global {C}onvergence of {H}essenberg {S}hifted {QR} {II}: {N}umerical {S}tability.
\newblock {\em arXiv preprint arXiv:2205.06810}, 2022.

\bibitem{banks2022global3}
Jess Banks, Jorge Garza-Vargas, and Nikhil Srivastava.
\newblock Global {C}onvergence of {H}essenberg {S}hifted {QR} {III}: {A}pproximate {R}itz {V}alues via {S}hifted {I}nverse {I}teration.
\newblock {\em arXiv preprint arXiv:2205.06804}, 2022.

\bibitem{banks2021global1}
Jess Banks, Jorge Garza-Vargas, and Nikhil Srivastava.
\newblock Global {C}onvergence of {H}essenberg {S}hifted {QR} {I}: {E}xact {A}rithmetic.
\newblock {\em Foundations of Computational Mathematics}, pages 1--34, 2024.

\bibitem{banks2021gaussian}
Jess Banks, Archit Kulkarni, Satyaki Mukherjee, and Nikhil Srivastava.
\newblock Gaussian {R}egularization of the {P}seudospectrum and {D}avies’ {C}onjecture.
\newblock {\em Communications on Pure and Applied Mathematics}, 74(10):2114--2131, 2021.

\bibitem{barshan2011supervised}
Elnaz Barshan, Ali Ghodsi, Zohreh Azimifar, and Mansoor~Zolghadri Jahromi.
\newblock {Supervised principal component analysis: Visualization, classification and regression on subspaces and submanifolds}.
\newblock {\em Pattern Recognition}, 44(7):1357--1371, 2011.

\bibitem{bartal2011dimensionality}
Yair Bartal, Ben Recht, and Leonard~J Schulman.
\newblock Dimensionality reduction: beyond the {Johnson-Lindenstrauss bound}.
\newblock In {\em Proc. 2011 Annual ACM-SIAM Symposium on Discrete Algorithms}, pages 868--887. SIAM, 2011.

\bibitem{basch1969}
Harold Basch, C.~J. Hornback, and J.~W. Moskowitz.
\newblock {Gaussian-Orbital Basis Sets for the First-Row Transition-Metal Atoms}.
\newblock {\em The Journal of Chemical Physics}, 51(4):1311--1318, 1969.

\bibitem{baston2022stochastic}
Robert~A Baston and Yuji Nakatsukasa.
\newblock Stochastic diagonal estimation: probabilistic bounds and an improved algorithm.
\newblock {\em arXiv preprint arXiv:2201.10684}, 2022.

\bibitem{bauer1960norms}
Friedrich~L Bauer and Charles~T Fike.
\newblock Norms and exclusion theorems.
\newblock {\em Numerische Mathematik}, 2:137--141, 1960.

\bibitem{benor2018quasi}
Michael Ben-Or and Lior Eldar.
\newblock {A Quasi-Random Approach to Matrix Spectral Analysis}.
\newblock In {\em Proc. 9th Innovations in Theoretical Computer Science Conference}, pages 6:1--6:22. Schloss Dagstuhl--Leibniz-Zentrum fuer Informatik, 2018.

\bibitem{benner1997new}
Peter Benner, Volker Mehrmann, and Hongguo Xu.
\newblock A new method for computing the stable invariant subspace of a real {H}amiltonian matrix.
\newblock {\em Journal of computational and applied mathematics}, 86(1):17--43, 1997.

\bibitem{bhatia2000pinching}
Rajendra Bhatia.
\newblock Pinching, trimming, truncating, and averaging of matrices.
\newblock {\em The American Mathematical Monthly}, 107(7):602--608, 2000.

\bibitem{bhatia2007perturbation}
Rajendra Bhatia.
\newblock {\em Perturbation bounds for matrix eigenvalues}.
\newblock SIAM, 2007.

\bibitem{bhojanapalli2014tighter}
Srinadh Bhojanapalli, Prateek Jain, and Sujay Sanghavi.
\newblock Tighter low-rank approximation via sampling the leveraged element.
\newblock In {\em Proc. Twenty-Sixth Annual ACM-SIAM Symposium on Discrete Algorithms}, pages 902--920. SIAM, 2014.

\bibitem{boutsidis2014near}
Christos Boutsidis, Petros Drineas, and Malik Magdon-Ismail.
\newblock Near-optimal column-based matrix reconstruction.
\newblock {\em SIAM Journal on Computing}, 43(2):687--717, 2014.

\bibitem{boutsidis2016optimal}
Christos Boutsidis, David~P Woodruff, and Peilin Zhong.
\newblock Optimal principal component analysis in distributed and streaming models.
\newblock In {\em Proc. Forty-Eighth Annual ACM Symposium on Theory of Computing}, pages 236--249, 2016.

\bibitem{clark1999}
Tim Clark and Rainer Koch.
\newblock {\em Linear Combination of Atomic Orbitals}, pages 5--22.
\newblock Springer Berlin Heidelberg, 1999.

\bibitem{clarkson2009numerical}
Kenneth~L Clarkson and David~P Woodruff.
\newblock Numerical linear algebra in the streaming model.
\newblock In {\em Proc. Forty-First Annual ACM Symposium on Theory of Computing}, pages 205--214, 2009.

\bibitem{clarkson2017low}
Kenneth~L Clarkson and David~P Woodruff.
\newblock Low-rank approximation and regression in input sparsity time.
\newblock {\em Journal of the ACM}, 63(6):1--45, 2017.

\bibitem{cohen2015dimensionality}
Michael~B Cohen, Sam Elder, Cameron Musco, Christopher Musco, and Madalina Persu.
\newblock Dimensionality reduction for k-means clustering and low rank approximation.
\newblock In {\em Proc. Forty-Seventh Annual ACM Symposium on Theory of Computing}, pages 163--172, 2015.

\bibitem{crawford1973reduction}
Charles~R. Crawford.
\newblock Reduction of a band-symmetric generalized eigenvalue problem.
\newblock {\em Communications of the ACM}, 16(1):41--44, 1973.

\bibitem{davies2001analysis}
Philip~I Davies, Nicholas~J Higham, and Fran{\c{c}}oise Tisseur.
\newblock Analysis of the cholesky method with iterative refinement for solving the symmetric definite generalized eigenproblem.
\newblock {\em SIAM Journal on Matrix Analysis and Applications}, 23(2):472--493, 2001.

\bibitem{dekker1971shifted}
Theodorus~J Dekker and Joseph~F Traub.
\newblock The shifted {QR} algorithm for {Hermitian} matrices.
\newblock {\em Linear Algebra and its Applications}, 4(3):137--154, 1971.

\bibitem{demmel2007fastla}
James Demmel, Ioana Dumitriu, and Olga Holtz.
\newblock Fast linear algebra is stable.
\newblock {\em Numerische Mathematik}, 108(1):59--91, 2007.

\bibitem{demmel2007fastmm}
James Demmel, Ioana Dumitriu, Olga Holtz, and Robert Kleinberg.
\newblock Fast matrix multiplication is stable.
\newblock {\em Numerische Mathematik}, 106(2):199--224, 2007.

\bibitem{demmel2023generalized}
James Demmel, Ioana Dumitriu, and Ryan Schneider.
\newblock Generalized {P}seudospectral {S}hattering and {I}nverse-{F}ree {M}atrix {P}encil {D}iagonalization.
\newblock {\em arXiv preprint arXiv:2306.03700}, 2023.

\bibitem{demmel1997applied}
James~W Demmel.
\newblock {\em Applied numerical linear algebra}.
\newblock SIAM, 1997.

\bibitem{demmel1987three}
James~Weldon Demmel.
\newblock Three methods for refining estimates of invariant subspaces.
\newblock {\em Computing}, 38(1):43--57, 1987.

\bibitem{demmel1987computing}
James~Weldon Demmel and Bo~K{\aa}gstr{\"o}m.
\newblock Computing stable eigendecompositions of matrix pencils.
\newblock {\em Linear Algebra and its Applications}, 88:139--186, 1987.

\bibitem{diamantaras1996principal}
Konstantinos~I Diamantaras and Sun~Yuan Kung.
\newblock {\em Principal component neural networks: theory and applications}.
\newblock John Wiley \& Sons, Inc., 1996.

\bibitem{drineas2002competitive}
Petros Drineas, Iordanis Kerenidis, and Prabhakar Raghavan.
\newblock Competitive recommendation systems.
\newblock In {\em Proc. Thiry-Fourth Annual ACM Symposium on Theory of Computing}, pages 82--90, 2002.

\bibitem{duan2023faster}
Ran Duan, Hongxun Wu, and Renfei Zhou.
\newblock Faster matrix multiplication via asymmetric hashing.
\newblock In {\em 2023 IEEE 64th Annual Symposium on Foundations of Computer Science}, pages 2129--2138. IEEE, 2023.

\bibitem{eisenstat1998relative}
Stanley~C Eisenstat and Ilse~CF Ipsen.
\newblock Relative perturbation results for eigenvalues and eigenvectors of diagonalisable matrices.
\newblock {\em BIT Numerical Mathematics}, 38(3):502--509, 1998.

\bibitem{epperly2024xtrace}
Ethan~N Epperly, Joel~A Tropp, and Robert~J Webber.
\newblock {XTrace: Making the Most of Every Sample in Stochastic Trace Estimation}.
\newblock {\em SIAM Journal on Matrix Analysis and Applications}, 45(1):1--23, 2024.

\bibitem{feldman2020turning}
Dan Feldman, Melanie Schmidt, and Christian Sohler.
\newblock Turning big data into tiny data: Constant-size coresets for k-means, {PCA}, and projective clustering.
\newblock {\em SIAM Journal on Computing}, 49(3):601--657, 2020.

\bibitem{francis1961qr}
John~GF Francis.
\newblock The {QR} transformation a unitary analogue to the {LR} transformation—{P}art 1.
\newblock {\em The Computer Journal}, 4(3):265--271, 1961.

\bibitem{francis1962qr}
John~GF Francis.
\newblock The {QR} transformation—{P}art 2.
\newblock {\em The Computer Journal}, 4(4):332--345, 1962.

\bibitem{frieze1998fast}
Alan Frieze, Ravi Kannan, and Santosh Vempala.
\newblock Fast monte-carlo algorithms for finding low-rank approximations.
\newblock In {\em Proc. 39th Annual Symposium on Foundations of Computer Science}, page 370, 1998.

\bibitem{frieze2004fast}
Alan Frieze, Ravi Kannan, and Santosh Vempala.
\newblock Fast {Monte-Carlo} algorithms for finding low-rank approximations.
\newblock {\em Journal of the ACM}, 51(6):1025--1041, 2004.

\bibitem{furer2007faster}
Martin F{\"u}rer.
\newblock Faster integer multiplication.
\newblock In {\em Proc. Thirty-Ninth Annual ACM Symposium on Theory of Computing}, pages 57--66, 2007.

\bibitem{galli1996linear}
Giulia Galli.
\newblock Linear scaling methods for electronic structure calculations and quantum molecular dynamics simulations.
\newblock {\em Current Opinion in Solid State and Materials Science}, 1(6):864--874, 1996.

\bibitem{galli1992large}
Giulia Galli and Michele Parrinello.
\newblock Large scale electronic structure calculations.
\newblock {\em Physical Review Letters}, 69(24):3547, 1992.

\bibitem{george1973nested}
Alan George.
\newblock Nested dissection of a regular finite element mesh.
\newblock {\em SIAM Journal on Numerical Analysis}, 10(2):345--363, 1973.

\bibitem{giannozzi2017advanced}
P~Giannozzi, O~Andreussi, T~Brumme, O~Bunau, M~Buongiorno Nardelli, M~Calandra, R~Car, C~Cavazzoni, D~Ceresoli, M~Cococcioni, N~Colonna, I~Carnimeo, A~Dal Corso, S~de~Gironcoli, P~Delugas, R~A DiStasio, A~Ferretti, A~Floris, G~Fratesi, G~Fugallo, R~Gebauer, U~Gerstmann, F~Giustino, T~Gorni, J~Jia, M~Kawamura, H-Y Ko, A~Kokalj, E~Küçükbenli, M~Lazzeri, M~Marsili, N~Marzari, F~Mauri, N~L Nguyen, H-V Nguyen, A~Otero de-la Roza, L~Paulatto, S~Poncé, D~Rocca, R~Sabatini, B~Santra, M~Schlipf, A~P Seitsonen, A~Smogunov, I~Timrov, T~Thonhauser, P~Umari, N~Vast, X~Wu, and S~Baroni.
\newblock Advanced capabilities for materials modelling with {Quantum ESPRESSO}.
\newblock {\em Journal of Physics: Condensed Matter}, 29(46):465901, oct 2017.

\bibitem{giannozzi2009quantum}
Paolo Giannozzi, Stefano Baroni, Nicola Bonini, Matteo Calandra, Roberto Car, Carlo Cavazzoni, Davide Ceresoli, Guido~L Chiarotti, Matteo Cococcioni, Ismaila Dabo, Andrea~Dal Corso, Stefano de~Gironcoli, Stefano Fabris, Guido Fratesi, Ralph Gebauer, Uwe Gerstmann, Christos Gougoussis, Anton Kokalj, Michele Lazzeri, Layla Martin-Samos, Nicola Marzari, Francesco Mauri, Riccardo Mazzarello, Stefano Paolini, Alfredo Pasquarello, Lorenzo Paulatto, Carlo Sbraccia, Sandro Scandolo, Gabriele Sclauzero, Ari~P Seitsonen, Alexander Smogunov, Paolo Umari, and Renata~M Wentzcovitch.
\newblock {QUANTUM ESPRESSO}: a modular and open-source software project for quantum simulations of materials.
\newblock {\em Journal of Physics: Condensed Matter}, 21(39):395502, Sep 2009.

\bibitem{gilbert1986analysis}
John~R Gilbert and Robert~Endre Tarjan.
\newblock The analysis of a nested dissection algorithm.
\newblock {\em Numerische Mathematik}, 50(4):377--404, 1986.

\bibitem{goedecker1999linear}
Stefan Goedecker.
\newblock Linear scaling electronic structure methods.
\newblock {\em Reviews of Modern Physics}, 71(4):1085, 1999.

\bibitem{goedecker2003linear}
Stefan Goedecker and GE~Scuserza.
\newblock Linear scaling electronic structure methods in chemistry and physics.
\newblock {\em Computing in Science \& Engineering}, 5(4):14--21, 2003.

\bibitem{golub2013matrix}
Gene~H Golub and Charles~F Van~Loan.
\newblock {\em Matrix Computations}.
\newblock Johns Hopkins University Press, 2013.

\bibitem{gu1995divide}
Ming Gu and Stanley~C Eisenstat.
\newblock A divide-and-conquer algorithm for the symmetric tridiagonal eigenproblem.
\newblock {\em SIAM Journal on Matrix Analysis and Applications}, 16(1):172--191, 1995.

\bibitem{gu1996efficient}
Ming Gu and Stanley~C Eisenstat.
\newblock Efficient algorithms for computing a strong rank-revealing {QR} factorization.
\newblock {\em SIAM Journal on Scientific Computing}, 17(4):848--869, 1996.

\bibitem{halko2011finding}
Nathan Halko, Per-Gunnar Martinsson, and Joel~A Tropp.
\newblock Finding structure with randomness: Probabilistic algorithms for constructing approximate matrix decompositions.
\newblock {\em SIAM Review}, 53(2):217--288, 2011.

\bibitem{harvey2021integer}
David Harvey and Joris Van Der~Hoeven.
\newblock Integer multiplication in time o(nlog$\backslash$,n).
\newblock {\em Annals of Mathematics}, 193(2):563--617, 2021.

\bibitem{zhao2019}
Qiu He, Bin Yu, Zhaohuai Li, and Yan Zhao.
\newblock {D}ensity {F}unctional {T}heory for {B}attery {M}aterials.
\newblock {\em Energy \& Environmental Materials}, 2(4):264--279, 2019.

\bibitem{higham2002accuracy}
Nicholas~J Higham.
\newblock {\em Accuracy and stability of numerical algorithms}.
\newblock SIAM, 2002.

\bibitem{higham2009cholesky}
Nicholas~J Higham.
\newblock Cholesky factorization.
\newblock {\em Wiley Interdisciplinary Reviews: Computational Statistics}, 1(2):251--254, 2009.

\bibitem{hotelling1933analysis}
Harold Hotelling.
\newblock Analysis of a complex of statistical variables into principal components.
\newblock {\em Journal of educational psychology}, 24(6):417, 1933.

\bibitem{hu2022lora}
Edward~J Hu, yelong shen, Phillip Wallis, Zeyuan Allen-Zhu, Yuanzhi Li, Shean Wang, Lu~Wang, and Weizhu Chen.
\newblock Lo{RA}: Low-rank adaptation of large language models.
\newblock In {\em International Conference on Learning Representations}, 2022.

\bibitem{hutter2014cp2k}
J{\"u}rg Hutter, Marcella Iannuzzi, Florian Schiffmann, and Joost VandeVondele.
\newblock cp2k: atomistic simulations of condensed matter systems.
\newblock {\em Wiley Interdisciplinary Reviews: Computational Molecular Science}, 4(1):15--25, 2014.

\bibitem{ipsen1997computing}
Ilse~CF Ipsen.
\newblock Computing an eigenvector with inverse iteration.
\newblock {\em SIAM Review}, 39(2):254--291, 1997.

\bibitem{ipsen2000absolute}
Ilse~CF Ipsen.
\newblock Absolute and relative perturbation bounds for invariant subspaces of matrices.
\newblock {\em Linear Algebra and its Applications}, 309(1-3):45--56, 2000.

\bibitem{jia2001analysis}
Zhongxiao Jia and GW~Stewart.
\newblock An analysis of the rayleigh--ritz method for approximating eigenspaces.
\newblock {\em Mathematics of computation}, 70(234):637--647, 2001.

\bibitem{johnson1984extensions}
William~B Johnson and Joram Lindenstrauss.
\newblock Extensions of {L}ipschitz mappings into a {H}ilbert space.
\newblock {\em Contemp. Math.}, 26(1):189--206, 1984.

\bibitem{jolliffe2002principal}
Ian~T Jolliffe.
\newblock {\em Principal component analysis for special types of data}.
\newblock Springer, 2002.

\bibitem{kacham2024faster}
Praneeth Kacham and David~P Woodruff.
\newblock {Faster Algorithms for Schatten-p Low Rank Approximation}.
\newblock {\em Approximation, Randomization, and Combinatorial Optimization. Algorithms and Techniques}, 2024.

\bibitem{kahan1975spectra}
William Kahan.
\newblock Spectra of nearly {H}ermitian matrices.
\newblock {\em Proc. American Mathematical Society}, 48(1):11--17, 1975.

\bibitem{kane2014sparser}
Daniel~M Kane and Jelani Nelson.
\newblock Sparser {Johnson-Lindenstrauss} transforms.
\newblock {\em Journal of the ACM}, 61(1):1--23, 2014.

\bibitem{kenney1995matrix}
Charles~S Kenney and Alan~J Laub.
\newblock The matrix sign function.
\newblock {\em IEEE transactions on automatic control}, 40(8):1330--1348, 1995.

\bibitem{kielbasinski1987note}
Andrzej Kie{\l}basi{\'n}ski.
\newblock A note on rounding-error analysis of {C}holesky factorization.
\newblock {\em Linear Algebra and its Applications}, 88:487--494, 1987.

\bibitem{klinkert2020}
Cedric Klinkert, Áron Szabó, Christian Stieger, Davide Campi, Nicola Marzari, and Mathieu Luisier.
\newblock 2-d {M}aterials for {U}ltrascaled {F}ield-{E}ffect {T}ransistors: {O}ne {H}undred {C}andidates under the {A}b {I}nitio {M}icroscope.
\newblock {\em ACS Nano}, 14(7):8605--8615, 2020.

\bibitem{kohn1993density}
Walter Kohn.
\newblock Density functional/{W}annier function theory for systems of very many atoms.
\newblock {\em Chemical Physics Letters}, 208(3-4):167--172, 1993.

\bibitem{kohn1996density}
Walter Kohn.
\newblock Density functional and density matrix method scaling linearly with the number of atoms.
\newblock {\em Physical Review Letters}, 76(17):3168, 1996.

\bibitem{kohn1965self}
Walter Kohn and Lu~Jeu Sham.
\newblock Self-consistent equations including exchange and correlation effects.
\newblock {\em Physical Review}, 140(4A):A1133, 1965.

\bibitem{kresse1996efficient}
Georg Kresse and J{\"u}rgen Furthm{\"u}ller.
\newblock Efficient iterative schemes for ab initio total-energy calculations using a plane-wave basis set.
\newblock {\em Physical Review B}, 54(16):11169--11186, 1996.

\bibitem{kublanovskaya1962some}
Vera~N Kublanovskaya.
\newblock On some algorithms for the solution of the complete eigenvalue problem.
\newblock {\em USSR Computational Mathematics and Mathematical Physics}, 1(3):637--657, 1962.

\bibitem{kyng2016sparsified}
Rasmus Kyng, Yin~Tat Lee, Richard Peng, Sushant Sachdeva, and Daniel~A Spielman.
\newblock Sparsified {C}holesky and {M}ultigrid {S}olvers for {C}onnection {L}aplacians.
\newblock In {\em Proc. 48th Annual ACM Symposium on Theory of Computing}, pages 842--850, 2016.

\bibitem{kyng2016approximate}
Rasmus Kyng and Sushant Sachdeva.
\newblock Approximate {G}aussian {E}limination for {L}aplacians - {F}ast, {S}parse, and {S}imple.
\newblock In {\em Proc. IEEE 57th Annual Symposium on Foundations of Computer Science}, pages 573--582, 2016.

\bibitem{larsen2017optimality}
Kasper~Green Larsen and Jelani Nelson.
\newblock Optimality of the {Johnson-Lindenstrauss} lemma.
\newblock In {\em 2017 IEEE 58th Annual Symposium on Foundations of Computer Science}, pages 633--638. IEEE, 2017.

\bibitem{lipton1979generalized}
Richard~J Lipton, Donald~J Rose, and Robert~Endre Tarjan.
\newblock Generalized nested dissection.
\newblock {\em SIAM Journal on Numerical Analysis}, 16(2):346--358, 1979.

\bibitem{louis2016accelerated}
Anand Louis and Santosh~S Vempala.
\newblock Accelerated {N}ewton iteration for roots of black box polynomials.
\newblock In {\em Proc. IEEE 57th Annual Symposium on Foundations of Computer Science}, pages 732--740. IEEE, 2016.

\bibitem{malhi2004pca}
Arnaz Malhi and Robert~X Gao.
\newblock {PCA-based} feature selection scheme for machine defect classification.
\newblock {\em IEEE transactions on instrumentation and measurement}, 53(6):1517--1525, 2004.

\bibitem{malyshev1989computing}
Alexander~N Malyshev.
\newblock Computing invariant subspaces of a regular linear pencil of matrices.
\newblock {\em Siberian Mathematical Journal}, 30(4):559--567, 1989.

\bibitem{malyshev1993parallel}
Alexander~N Malyshev.
\newblock Parallel algorithm for solving some spectral problems of linear algebra.
\newblock {\em Linear Algebra and its Applications}, 188:489--520, 1993.

\bibitem{mangasarian2005multisurface}
Olvi~L Mangasarian and Edward~W Wild.
\newblock Multisurface proximal support vector machine classification via generalized eigenvalues.
\newblock {\em IEEE transactions on pattern analysis and machine intelligence}, 28(1):69--74, 2005.

\bibitem{martinsson2011randomized}
Per-Gunnar Martinsson, Vladimir Rokhlin, and Mark Tygert.
\newblock A randomized algorithm for the decomposition of matrices.
\newblock {\em Applied and Computational Harmonic Analysis}, 30(1):47--68, 2011.

\bibitem{marzari2012}
Nicola Marzari, Arash~A. Mostofi, Jonathan~R. Yates, Ivo Souza, and David Vanderbilt.
\newblock Maximally localized {W}annier functions: {T}heory and applications.
\newblock {\em Reviews of Modern Physics}, 84(4):1419, 2012.

\bibitem{maurer2019}
Reinhard~J Maurer, Christoph Freysoldt, Anthony~M Reilly, Jan~Gerit Brandenburg, Oliver~T Hofmann, Torbj{\"o}rn Bj{\"o}rkman, S{\'e}bastien Leb{\`e}gue, and Alexandre Tkatchenko.
\newblock Advances in {D}ensity-{F}unctional {C}alculations for {M}aterials {M}odeling.
\newblock {\em Annual Review of Materials Research}, 49(1):1--30, 2019.

\bibitem{mayer2003simple}
Istv{\'a}n Mayer.
\newblock {\em Simple theorems, proofs, and derivations in quantum chemistry}.
\newblock Springer Science \& Business Media, 2003.

\bibitem{meyer2024unreasonable}
Raphael Meyer, Cameron Musco, and Christopher Musco.
\newblock On the unreasonable effectiveness of single vector {K}rylov methods for low-rank approximation.
\newblock In {\em Proc. 2024 Annual ACM-SIAM Symposium on Discrete Algorithms}, pages 811--845. SIAM, 2024.

\bibitem{minami1996local}
Nariyuki Minami.
\newblock Local fluctuation of the spectrum of a multidimensional {A}nderson tight binding model.
\newblock {\em Communications in mathematical physics}, 177:709--725, 1996.

\bibitem{musco2015randomized}
Cameron Musco and Christopher Musco.
\newblock Randomized block {K}rylov methods for stronger and faster approximate singular value decomposition.
\newblock {\em Advances in Neural Information Processing Systems}, 28, 2015.

\bibitem{musco2018stability}
Cameron Musco, Christopher Musco, and Aaron Sidford.
\newblock Stability of the lanczos method for matrix function approximation.
\newblock In {\em Proc. Twenty-Ninth Annual ACM-SIAM Symposium on Discrete Algorithms}, pages 1605--1624. SIAM, 2018.

\bibitem{nakatsukasa2012condition}
Yuji Nakatsukasa.
\newblock On the condition numbers of a multiple eigenvalue of a generalized eigenvalue problem.
\newblock {\em Numerische Mathematik}, 121:531--544, 2012.

\bibitem{nakatsukasa2013stable}
Yuji Nakatsukasa and Nicholas~J Higham.
\newblock Stable and efficient spectral divide and conquer algorithms for the symmetric eigenvalue decomposition and the {SVD}.
\newblock {\em SIAM Journal on Scientific Computing}, 35(3):A1325--A1349, 2013.

\bibitem{nelson2013osnap}
Jelani Nelson and Huy~L Nguy{\^e}n.
\newblock {OSNAP}: Faster numerical linear algebra algorithms via sparser subspace embeddings.
\newblock In {\em Proc. IEEE 54th Annual Symposium on Foundations of Computer Science}, pages 117--126. IEEE, 2013.

\bibitem{ng2001spectral}
Andrew Ng, Michael Jordan, and Yair Weiss.
\newblock On spectral clustering: Analysis and an algorithm.
\newblock {\em Advances in Neural Information Processing Systems}, 14, 2001.

\bibitem{paige1976error}
Christopher~C Paige.
\newblock Error analysis of the lanczos algorithm for tridiagonalizing a symmetric matrix.
\newblock {\em IMA Journal of Applied Mathematics}, 18(3):341--349, 1976.

\bibitem{paige1971computation}
Christopher~Conway Paige.
\newblock {\em The computation of eigenvalues and eigenvectors of very large sparse matrices.}
\newblock PhD thesis, University of London, 1971.

\bibitem{pan1999complexity}
Victor~Y Pan and Zhao~Q Chen.
\newblock The complexity of the matrix eigenproblem.
\newblock In {\em Proc. 31st Annual ACM Symposium on Theory of Computing}, pages 507--516, 1999.

\bibitem{parlett1998symmetric}
Beresford~N Parlett.
\newblock {\em The {S}ymmetric {E}igenvalue {P}roblem}.
\newblock SIAM, 1998.

\bibitem{pearson1901liii}
Karl Pearson.
\newblock Liii. on lines and planes of closest fit to systems of points in space.
\newblock {\em The London, Edinburgh, and Dublin philosophical magazine and journal of science}, 2(11):559--572, 1901.

\bibitem{pokrzywa1981spectra}
Andrzej Pokrzywa.
\newblock Spectra of operators with fixed imaginary parts.
\newblock {\em Proc. American Mathematical Society}, 81(3):359--364, 1981.

\bibitem{rufai2014lossy}
Awwal~Mohammed Rufai, Gholamreza Anbarjafari, and Hasan Demirel.
\newblock Lossy image compression using singular value decomposition and wavelet difference reduction.
\newblock {\em Digital signal processing}, 24:117--123, 2014.

\bibitem{saad2011numerical}
Yousef Saad.
\newblock {\em Numerical methods for large eigenvalue problems: revised edition}.
\newblock SIAM, 2011.

\bibitem{sarlos2006improved}
Tam\'as Sarl\'os.
\newblock Improved approximation algorithms for large matrices via random projections.
\newblock In {\em Proc. 47th Annual Symposium on Foundations of Computer Science}, pages 143--152. IEEE, 2006.

\bibitem{schneider2023fast}
Ryan Schneider.
\newblock When is fast, implicit squaring of $ {A^{-1}B} $ stable?
\newblock {\em arXiv preprint arXiv:2310.00193}, 2023.

\bibitem{scholkopf1998nonlinear}
Bernhard Sch{\"o}lkopf, Alexander Smola, and Klaus-Robert M{\"u}ller.
\newblock Nonlinear component analysis as a kernel eigenvalue problem.
\newblock {\em Neural computation}, 10(5):1299--1319, 1998.

\bibitem{schonhage1971fast}
Arnold Sch{\"o}nhage and Volker Strassen.
\newblock Fast multiplication of large numbers.
\newblock {\em Computing}, 7:281--292, 1971.

\bibitem{shi2000normalized}
Jianbo Shi and Jitendra Malik.
\newblock Normalized cuts and image segmentation.
\newblock {\em IEEE Transactions on pattern analysis and machine intelligence}, 22(8):888--905, 2000.

\bibitem{simchowitz2018tight}
Max Simchowitz, Ahmed El~Alaoui, and Benjamin Recht.
\newblock Tight query complexity lower bounds for {PCA} via finite sample deformed wigner law.
\newblock In {\em Proc. 50th Annual ACM Symposium on Theory of Computing}, pages 1249--1259, 2018.

\bibitem{sobczyk2022approximate}
Aleksandros Sobczyk and Mathieu Luisier.
\newblock {Approximate Euclidean lengths and distances beyond Johnson-Lindenstrauss}.
\newblock {\em Advances in Neural Information Processing Systems}, 35:19357--19369, 2022.

\bibitem{soler2002siesta}
Jos{\'e}~M Soler, Emilio Artacho, Julian~D Gale, Alberto Garc{\'\i}a, Javier Junquera, Pablo Ordej{\'o}n, and Daniel S{\'a}nchez-Portal.
\newblock The {SIESTA} method for ab initio order-{N} materials simulation.
\newblock {\em Journal of Physics: Condensed Matter}, 14(11):2745, 2002.

\bibitem{spielman2004smoothed}
Daniel~A Spielman and Shang-Hua Teng.
\newblock Smoothed analysis of algorithms: Why the simplex algorithm usually takes polynomial time.
\newblock {\em Journal of the ACM}, 51(3):385--463, 2004.

\bibitem{srivastava2023complexity}
Nikhil Srivastava.
\newblock The complexity of diagonalization.
\newblock In {\em Proc. 2023 International Symposium on Symbolic and Algebraic Computation}, pages 1--6, 2023.

\bibitem{stewart1972sensitivity}
Gilbert~W Stewart.
\newblock On the sensitivity of the eigenvalue problem ax=$\lambda$bx.
\newblock {\em SIAM Journal on Numerical Analysis}, 9(4):669--686, 1972.

\bibitem{stewart1973error}
Gilbert~W Stewart.
\newblock Error and perturbation bounds for subspaces associated with certain eigenvalue problems.
\newblock {\em SIAM Review}, 15(4):727--764, 1973.

\bibitem{stewart1977perturbation}
Gilbert~W Stewart.
\newblock On the perturbation of pseudo-inverses, projections and linear least squares problems.
\newblock {\em SIAM Review}, 19(4):634--662, 1977.

\bibitem{stewart1979pertubation}
Gilbert~W Stewart.
\newblock Pertubation bounds for the definite generalized eigenvalue problem.
\newblock {\em Linear Algebra and its Applications}, 23:69--85, 1979.

\bibitem{strassen1969gaussian}
Volker Strassen.
\newblock Gaussian elimination is not optimal.
\newblock {\em Numerische Mathematik}, 13(4):354--356, 1969.

\bibitem{trefethen2005spectra}
Lloyd~N Trefethen and Mark Embree.
\newblock {\em Spectra and {P}seudospectra: {T}he {B}ehavior of {N}onnormal {M}atrices and {O}perators}.
\newblock Princeton University Press, 2005.

\bibitem{vandevondele2012linear}
Joost VandeVondele, Urban Borstnik, and Jurg Hutter.
\newblock Linear scaling self-consistent field calculations with millions of atoms in the condensed phase.
\newblock {\em Journal of Chemical Theory and Computation}, 8(10):3565--3573, 2012.

\bibitem{vogel2016superfast}
James Vogel, Jianlin Xia, Stephen Cauley, and Venkataramanan Balakrishnan.
\newblock Superfast divide-and-conquer method and perturbation analysis for structured eigenvalue solutions.
\newblock {\em SIAM Journal on Scientific Computing}, 38(3):A1358--A1382, 2016.

\bibitem{wegner1981bounds}
Franz Wegner.
\newblock Bounds on the density of states in disordered systems.
\newblock {\em Zeitschrift f{\"u}r Physik B Condensed Matter}, 44(1):9--15, 1981.

\bibitem{weyl1912asymptotische}
Hermann Weyl.
\newblock Das asymptotische {V}erteilungsgesetz der {E}igenwerte linearer partieller {D}ifferentialgleichungen (mit einer {A}nwendung auf die {T}heorie der {H}ohlraumstrahlung).
\newblock {\em Mathematische Annalen}, 71(4):441--479, 1912.

\bibitem{wilkinson1968global}
James~Hardy Wilkinson.
\newblock Global convergene of tridiagonal {QR} algorithm with origin shifts.
\newblock {\em Linear Algebra and its Applications}, 1(3):409--420, 1968.

\bibitem{williams2024new}
Virginia~Vassilevska Williams, Yinzhan Xu, Zixuan Xu, and Renfei Zhou.
\newblock New bounds for matrix multiplication: from alpha to omega.
\newblock In {\em Proc. 2024 Annual ACM-SIAM Symposium on Discrete Algorithms}, pages 3792--3835. SIAM, 2024.

\bibitem{yan2017}
Zewen Xiao and Yanfa Yan.
\newblock Progress in {T}heoretical {S}tudy of {M}etal {H}alide {P}erovskite {S}olar {C}ell {M}aterials.
\newblock {\em Advanced Energy Materials}, 7(22):1701136, 2017.

\bibitem{yang1991direct}
Weitao Yang.
\newblock Direct calculation of electron density in density-functional theory.
\newblock {\em Physical Review Letters}, 66(11):1438, 1991.

\bibitem{zhou2006parallel}
Yunkai Zhou, Yousef Saad, Murilo~L Tiago, and James~R Chelikowsky.
\newblock Parallel self-consistent-field calculations via {C}hebyshev-filtered subspace acceleration.
\newblock {\em Physical Review E}, 74(6):066704, 2006.

\bibitem{zhou2006self}
Yunkai Zhou, Yousef Saad, Murilo~L Tiago, and James~R Chelikowsky.
\newblock Self-consistent-field calculations using {C}hebyshev-filtered subspace iteration.
\newblock {\em Journal of Computational Physics}, 219(1):172--184, 2006.

\end{thebibliography}
\end{document}